\documentclass[11pt,aps,tightenlines,nofootinbib,longbibliography,superscriptaddress, floatfix]{revtex4-1}

\usepackage[LGR,T1]{fontenc}
\usepackage[latin9]{inputenc}
\usepackage{color}
\usepackage{verbatim}
\usepackage{units}
\usepackage{textcomp}
\usepackage{amsmath}
\usepackage{slashed}
\usepackage{amssymb}
\usepackage{graphicx}
\usepackage{setspace}
\usepackage{wasysym}
\usepackage{stmaryrd}
\PassOptionsToPackage{normalem}{ulem}
\usepackage{ulem}
\definecolor{xlinkcolor}{cmyk}{1,1,0,0}
\usepackage[
 colorlinks=true,    
 linkcolor=xlinkcolor,     
 citecolor=xlinkcolor,     
 filecolor=xlinkcolor,  
 urlcolor=xlinkcolor,      
 final=true
]{hyperref}
\usepackage{array, makecell}

\makeatletter

\DeclareRobustCommand{\greektext}{%
  \fontencoding{LGR}\selectfont\def\encodingdefault{LGR}}
\DeclareRobustCommand{\textgreek}[1]{\leavevmode{\greektext #1}}
\ProvideTextCommand{\~}{LGR}[1]{\char126#1}

\newcommand{\lyxmathsym}[1]{\ifmmode\begingroup\def\b@ld{bold}
  \text{\ifx\math@version\b@ld\bfseries\fi#1}\endgroup\else#1\fi}

\providecommand{\tabularnewline}{\\}

\@ifundefined{textcolor}{}
{%
 \definecolor{BLACK}{gray}{0}
 \definecolor{WHITE}{gray}{1}
 \definecolor{RED}{rgb}{1,0,0}
 \definecolor{GREEN}{rgb}{0,1,0}
 \definecolor{BLUE}{rgb}{0,0,1}
 \definecolor{CYAN}{cmyk}{1,0,0,0}
 \definecolor{MAGENTA}{cmyk}{0,1,0,0}
 \definecolor{YELLOW}{cmyk}{0,0,1,0}
}

\newcommand{\Frac}[2]{\frac{\displaystyle #1}{\displaystyle #2}}
\newcommand{\MPQ}{M_{\;{\slash\!\!\!\!\!\text{\tiny PQ}}}}

\newcommand\snowmass{\begin{center}\rule[-0.2in]{\hsize}{0.01in}\\\rule{\hsize}{0.01in}\\
\vskip 0.1in Submitted to the  Proceedings of the US Community Study\\ 
on the Future of Particle Physics (Snowmass 2021)\\ 
\rule{\hsize}{0.01in}\\\rule[+0.2in]{\hsize}{0.01in} \end{center}}

\makeatother

\usepackage{babel}
\begin{document}
\textsf{\scriptsize{}}%
{\scriptsize\par}

\hyphenation{re-con-struct-ed}
\hyphenation{sym-me-tries}
\hyphenation{op-por-tu-ni-ties}
\hyphenation{re-quire-ments}
\hyphenation{ex-per-i-ments}
\hyphenation{elec-tri-cal}
\hyphenation{com-pat-i-ble}
\hyphenation{im-ple-ment-ing}
\hyphenation{en-vi-ron-ment}
\hyphenation{ex-per-i-men-tal}
\hyphenation{con-sid-er-ably}
\hyphenation{com-ple-men-tary}
\hyphenation{sum-ma-rized}
\hyphenation{ap-pro-pri-ate}
\hyphenation{con-ser-va-tion}
\hyphenation{ex-per-i-ment}
\hyphenation{Com-put-ing}
\hyphenation{cor-re-spond-ing}

\snowmass 
\date{\today}

\title{The REDTOP experiment: Rare \texorpdfstring{$\eta/\eta^{\prime}$}{} Decays To Probe New Physics.}
\begin{abstract}

The $\eta$ and $\eta^{\prime}$ mesons are nearly unique in the particle
universe since they are almost Goldstone bosons 
and the dynamics of their
decays are strongly constrained. 
The integrated
$\eta$-meson samples collected in earlier experiments amount to $\sim10^{9}$
events. A new experiment, REDTOP (Rare Eta Decays To Observe Physics
Beyond the Standard Model), is being proposed, with the intent of
collecting a data sample of order 10$^{14}$ $\eta$ (10$^{12}$
$\eta^{\prime}$) for studying very rare decays. Such statistics are
sufficient for investigating several symmetry violations, and for
searching for particles and fields beyond the Standard Model. In this
work we present several studies evaluating REDTOP sensitivity to processes
that couple the Standard Model to New Physics through all four of
the so-called \emph{portals}: the Vector, the Scalar, the Axion and
the Heavy Lepton portal. The sensitivity of the experiment is also
adequate for probing several conservation laws, in particular $CP$,
$T$ and Lepton Universality, and for the determination of the $\eta$
 form factors, which is crucial for the interpretation of the recent
measurement of muon $g-2$.\\[4ex]
{\it Preprint numbers: FERMILAB-FN-1153-AD-PPD-T, LA-UR-22-22208}

\end{abstract}
\author{J.~Elam, A.~Mane}
\affiliation{Argonne National Laboratory, (USA)}
\author{J.~Comfort, P.~Mauskopf, D.~McFarland, L.~Thomas}
\affiliation{Arizona State University, (USA)}
\author{I.~Pedraza, D.~Leon, S.~Escobar, D.~Herrera, D.~Silverio}
\affiliation{Benemerita Universidad Autonoma de Puebla, (Mexico)}
\author{W.~Abdallah}
\affiliation{Department of Mathematics, Faculty of Science, Cairo University, Giza (Egypt)}
\author{D.~Winn}
\affiliation{Fairfield University, (USA)}
\author{M.~Spannowsky}
\affiliation{Durham University, (UK)}
\author{J.~Dey, V.~Di Benedetto, B.~Dobrescu, D.~Fagan, E.~Gianfelice-Wendt,
E.~Hahn, D.~Jensen, C.~Johnstone, J.~Johnstone, J.~Kilmer, 
T.~Kobilarcik, A.~Kronfeld, K.~Krempetz, S.~Los, M.~May, A.~Mazzacane,
N.~Mokhov, W.~Pellico, A.~Pla-Dalmau, V.~Pronskikh, E.~Ramberg, J.
Rauch, L.~Ristori, E.~Schmidt, G.~Sellberg, G.~Tassotto, Y.~D.~Tsai}
\affiliation{Fermi National Accelerator Laboratory, (USA)}
\author{A.~Alqahtani}
\affiliation{Georgetown University, (USA)}
\author{J.~Shi}
\affiliation{Guangdong Provincial Key Laboratory of Nuclear Science, Institute
of Quantum Matter, South China Normal University, Guangzhou 510006,
(China) }
\affiliation{Guangdong-Hong Kong Joint Laboratory of Quantum Matter, Southern Nuclear
Science Computing Center, South China Normal University, Guangzhou
510006, (China)}
\author{R.~Gandhi}
\affiliation{Harish-Chandra Research Institute, HBNI, Jhunsi (India)}
\author{S.~Homiller}
\affiliation{Harvard University, Cambridge, MA (USA)}
\author{X.~Chen, Q.~Hu}
\affiliation{Institute of Modern Physics, Chinese Academy of Sciences,  Lanzhou (China)}
\author{E.~Passemar}
\affiliation{Indiana University (USA)}
\author{P.~S\'anchez-Puertas}
\affiliation{Institut de F\'isica d'Altes Energies - Barcelona (Spain)}
\author{S.~Roy }
\affiliation{Institute of Physics, Sachivalaya Marg, Sainik School Post, Bhubaneswar,
(India)}
\author{C.~Gatto}
\email[Email cgatto@na.infn.it]{[Corresponding author]}

\affiliation{Istituto Nazionale di Fisica Nucleare \textendash{} Sezione di Napoli, (Italy)}
\affiliation{Northern Illinois University, (USA)}
\author{W.~Baldini}
\affiliation{Istituto Nazionale di Fisica Nucleare \textendash{} Sezione di Ferrara,
(Italy)}
\author{R.~Carosi, A.~Kievsky, M.~Viviani}
\affiliation{Istituto Nazionale di Fisica Nucleare \textendash{} Sezione di Pisa,
(Italy)}
\author{W.~Krzemie\'{n}, M.~Silarski, M.~Zielinski}
\affiliation{Institute of Physics, Jagiellonian University, 30-348 Krakow, (Poland)}
\author{D.~Guadagnoli}
\affiliation{Laboratoire d'Annecy-le-Vieux de Physique Th\'eorique, (France)}
\author{D.~S.~M.~Alves, S.~Gonz\'alez-Sol\'is, S.~Pastore}
\affiliation{Los Alamos National Laboratory, (USA)}
\author{M.~Berlowski}
\affiliation{National Centre for Nuclear Research \textendash{} Warsaw, (Poland)}
\author{G.~Blazey, A.~Dychkant, K.~Francis, M.~Syphers, V.~Zutshi, P.~Chintalapati,
T.~Malla, M.~Figora, T.~Fletcher}
\affiliation{Northern Illinois University, (USA)}
\author{A.~Ismail}
\affiliation{Oklahoma State University, (USA)}
\author{D.~Ega{\~n}a-Ugrinovic}
\affiliation{Perimeter Institute for Theoretical Physics, Waterloo, (Canada)}
\author{Y.~Kahn}
\affiliation{Princeton University, Princeton, (USA)}
\author{D.~McKeen}
\affiliation{TRIUMF, (Canada)}
\author{P.~Meade}
\affiliation{Stony Brook University \textendash{} New York, (USA)}
\author{A.~Gutierrez, M.~A.~Hernandez-Ruiz}
\affiliation{Universidad Autonoma de Zacatecas}
\author{R.~Escribano, P.~Masjuan, E.~Royo}
\affiliation{Universitat Aut\'onoma de Barcelona}
\affiliation{Institut de F\'isica d'Altes Energies - Barcelona (Spain)}
\author{B.~Kubis}
\affiliation{Universit\"at Bonn, Helmholtz-Institut f\"ur Strahlen- und Kernphysik (Theorie) and Bethe Center for Theoretical Physics (Germany)}
\author{J.~Jaeckel}
\affiliation{Universit\"at Heidelberg, (Germany)}
\author{L.~E.~Marcucci}
\affiliation{Universita' di Pisa and INFN, (Italy)}
\author{C.~Siligardi, S.~Barbi, C.~Mugoni}
\affiliation{Universita' di Modena e Reggio Emilia, (Italy)}
\author{M.~Guida}
\affiliation{Universita' di Salerno and INFN \textendash{} Sezione di Napoli, (Italy)}
\author{S.~Charlebois, J.~F.~Pratte}
\affiliation{Universit\'e de Sherbrooke, (Canada)}
\author{L.~Harland-Lang}
\affiliation{University of Oxford, (UK)}
\author{J.~M.~Berryman}
\affiliation{University of California Berkeley, (USA)}
\affiliation{Institute for Nuclear Theory, University of Washington, (USA)}
\author{R.~Gardner, P.~Paschos}
\affiliation{University of Chicago, (USA)}
\author{J.~Konisberg}
\affiliation{University of Florida, (USA)}
\author{C.~Mills, Z.~Ye}
\affiliation{University of Illinois Chicago, (USA)}
\author{M.~Murray, C.~Rogan, C.~Royon, N.~Minafra, A.~Novikov, F.~Gautier,
T.~Isidori}
\author{C.~Mills, Z.~Ye}
\affiliation{University of Illinois Chicago, (USA)}
\author{S.~Gardner, X.~Yan}
\affiliation{University of Kentucky, (USA) }
\author{Y.~Onel}
\affiliation{University of Iowa, (USA)}
\author{M.~Pospelov}
\affiliation{University of Minnesota, (USA)}
\author{B.~Batell, A.~Freitas, M.~Rai}
\affiliation{University of Pittsburgh, (USA)}
\author{D.~N.~Gao}
\affiliation{University of Science and Technology of China, (China)}
\author{K.~Maamari}
\affiliation{University of Southern California, (USA)}
\author{A.~Kup\'s\'c}
\affiliation{University of Uppsala, (Sweden)}
\author{B.~Fabela-Enriquez}
\affiliation{Vanderbilt University, (USA)}
\author{A.~Petrov}
\affiliation{Wayne State University, (USA)}
\author{S.~Tulin}
\affiliation{York University, (Canada)}
\collaboration{\textsf{R}EDTOP Collaboration}
\homepage[Homepage:~]{https://redtop.fnal.gov}

\maketitle
\noaffiliation



\tableofcontents{}

\newpage{}

\section*{Executive Summary}
\begin{itemize}
\begin{singlespace}
\item The next decade represents an almost unique opportunity for laboratories
with high intensity proton accelerators to uncover Dark Matter or
New Physics.
\end{singlespace}
\item 
There are strong theoretical reasons to 
search for New Physics  in
the MeV--GeV range.
\item The $\eta$ and $\eta^{\prime}$ mesons are almost unique 
since they carry he same quantum numbers as the
Higgs (except for parity), and have no Standard Model charges. 
Their decays are flavor-conserving and most of them forbidden at leading order (in various symmetry-breaking parameters) within the Standard Model. 
\item Rare decays are therefore enhanced compared to the remaining flavor-neutral mesons. Thus an $\eta/\eta^{\prime}$ factory is an excellent
laboratory for studying rare processes and Beyond Standard Model physics at low energy. 
\item 
A sample  of order $10^{14}$($10^{12}$ ) $\eta(\eta^{\prime})$
mesons can address most of the recent theoretical models. Such an experiment
would have enough sensitivity  
 to explore a very large portion of the
unexplored parameter space for all the four portals connecting the
Dark Sector with the Standard Model.
Lepton Universality and the  $CP$ and $T$ symmetries 
can also be probed with excellent sensitivity.
\item Many other studies can be conducted with such a large
data sample, including, for example, the determination of the $\eta$
form factors, which is crucial to understanding the $(g-2)_\mu$ measurement. 

\item The REDTOP Collaboration is proposing an $\eta/\eta^{\prime}$ factory with
such parameters. No similar experiment exists or is currently planned
by the international community. 
\item 
A full detector simulation and reconstruction
has been implemented to study several processes driven by New Physics,
and many theoretical models have been bench marked. About $5\times10{}^{10}$ background
events have been generated and fully reconstructed to estimate the
sensitivity of REDTOP. This took over three years to complete and
required about $4\times10{}^{7}$ core-hours of computing on the Open
Science Grid. 
\item The Physics case for REDTOP 
is  presented in the first part of this work. 
In Sec.~\ref{subsec:Searches-for-new-particles-and-fileds}
we discuss the four portals and the latest theoretical models  
proposed 
to explain  
outstanding experimental anomalies.
In Sec.~\ref{subsec:Tests of Conservation Laws} we discuss several tests
of conservation laws which could be explored at REDTOP.
\item A description of the experiment, including the sub-detectors, the
trigger systems and the computing model, is presented in the central
part of this document. Radiation damage, detector aging and shielding issues 
have also been considered. They are discussed in some detail in
the appendices.
\end{itemize}
\newpage{}

\section{Introduction}

\label{intro} 

It is  generally accepted that the Standard Model (SM) is not a
complete description of all quantum interactions. The exact nature
of dark energy and dark matter, and the baryon asymmetry of the universe,
the observation of a Universe in accelerating expansion, and neutrino
masses, are among the
very interesting questions that cannot be answered within the framework
of the SM. 

{} There is a strong indication that the physics Beyond
the Standard Model (BSM) could contain new particles and/or force
mediators, which significantly violate some discrete symmetries
of the universe, in particular \emph{CP}.

The High Energy Physics community has engaged in
an unprecedented experimental effort, with the construction of the
LHC and its four detectors, to observe physics
BSM in the High Energy domain. The absence of conclusive evidence
so far may suggest that: a) the New Physics
most immediately accessible to experimenters is at \emph{low}
energies, rather than at high energies, and b) such New Physics is
elusive, in particular it couples to SM matter too faintly to be detected
by experiments at colliders. To detect such interactions,
one may need huge luminosities, which are one of the most attractive
features of fixed-target experiments.%

As a consequence of fact a), a plethora of new theoretical models
have flourished, that extend the SM with light gauge bosons, in the
MeV--GeV mass range (see e.g.,~\citep{PhysRevD.80.095024,Reece_2009,Bjorken_2009}).
Such efforts have, on the one hand, been encouraged by the fact that
recently observed astrophysical anomalies point to such mass range
as a promising area of exploration. In addition, in such mass regime,
otherwise strong astrophysical and cosmological constraints are weakened
or eliminated, while constraints from high energy colliders are, in
most cases, inapplicable.

On the other side, fact b) puts the spotlight on experimental searches
with techniques resting on an integrated luminosity that exceeds by
several orders of magnitudes those currently available at colliders.
As pointed out in Ref.~\cite{BPR2009}, while the characteristic
integrated luminosity for high-energy colliders is of order 10$^{41}$cm$^{-2}$,
the analogue of integrated luminosity for a moderate intensity (namely,
$\sim$10$^{21}$ protons on target (POT)) fixed target experiment
with a 1 mm thick target is of the order of $\sim$10$^{44}$ cm$^{-2}$.
By applying formula (4)  in Ref.~\cite{BPR2009} at a fixed target experiment with E$_{lab}$=1
GeV, one obtains the following comparison between the production rates
for neutral GeV-scale states at LHC and low energy fixed targets:
\begin{equation}
\frac{N_{LHC}}{N_{fixed\:target}}\sim10^{-8+6n}
\end{equation}
where the interaction between the standard and dark matter is assumed
being mediated by marginal or irrelevant operators of dimension $4
+ n$, with $n \ge 0$.

\medskip{}

For the kind of New Physics mentioned above, fixed-target experiments
with hadronic beams thus have well-defined advantages with respect
to high-energy colliders --- let alone the tremendous difference in construction
and operating costs. At the same time, several factors are limiting
the realization of a high intensity fixed target experiment. Two of
the most  limiting factors are: a) the sheer production rate of events
from inelastic interaction of the beam onto the target and, b) the
large background from neutrons (either primary or secondary) which make a signal in the detector. 
Regarding
point a), the technologies implemented in the present generation of
detectors are not fast enough to cope with proton beam intensities
even as modest as few tens of watts. Regarding point b), a neutron
and a photon have very similar signatures in a conventional, single-readout
calorimeter, hindering the ability to disentangle such particles unless novel
detector techniques are implemented. Last, but not least, intense
neutron fluxes could damage quickly a detector if non radiation-hard
materials are used. 

\medskip{}

The design of the REDTOP experiment is based on the above considerations.
REDTOP is a high yield $\eta$/$\eta^{\prime}$-factory, operated in a fixed
target configuration with beam luminosity of order 10$^{34}$ cm$^{-2}$sec$^{-1}$.
The mass range for potential discoveries is approximately {[}14 MeV-950
MeV{]}, limited on the lower side by the resolution of the detector
and on the upper side by the meson mass. 

REDTOP is also a  frontier experiment, aiming at measuring the $\eta/\eta^{\prime}$
decay rates or their asymmetries for very rare processes, and with
a precision several orders of magnitude higher  than the present measurements.
 These decays would provide direct tests  of conservation laws and,
along with other measurements,  will open up new windows to discover
physics beyond the Standard Model.  In addition to searches for  new
physics, the experiment will involve development and first  use of
innovative detectors.  Novel instrumentation will include  a super-light
~\cite{T1059} or an LGAD tracker with unprecedentedly low material
budget~\cite{LGAD2020}, an ADRIANO2 calorimeter, and  a Threshold
Cerenkov Radiator (TCR). An optional Active Muon Polarimeter is being
considered to improve the measurements of the muon polarization. With
the information obtained from the highly granular calorimeter as well
as from the other subdetectors, an extended Particle Flow Analysis
(PFA)~\cite{PFA2006} could be implemented. The 5-D measurement
performed on the showers (energy, space, and time) will facilitate
the disentangling of complex or overlapping events. \smallskip{}

The development of all such detector techniques will require a substantial
effort. On the other hand, future experiments, operating with similar
event rates or requiring similar levels of background rejection, will
certainly benefit from the pioneering R\&D carried by the REDTOP Collaboration.

\section{Motivations for a high luminosity \texorpdfstring{$\eta/\eta^{\prime}$}{} factory\label{sec:Motivations}}

It has been recently noted that ``Light dark matter (LDM) must
be neutral under SM charges, otherwise it would have been discovered
at previous colliders''~\cite{Krnjaic20}. Under such circumstance,
the study of processes originated by particles carrying no SM charges
is, intuitively, more appropriate in LDM searches, as no charged currents
are present, which could potentially interfere with Beyond Standard
Model (BSM) processes. 

The $\eta$ and $\eta^{\prime}$ mesons have been widely studied in the past,
as their special nature has attracted the curiosity of the scientists~\citep{Nefkens_1996}.
The $\eta$ is a Goldstone boson, therefore its QCD dynamics is strongly
constrained by that property. In nature, there are only few Goldstone
bosons. Furthermore, the $\eta$ is, at the same time, an eigenstate
of the \textbf{\textit{C}}\textit{, }\textbf{\textit{P}}\textit{,
}\textbf{\textit{CP, I}}\textit{ }and \textbf{\textit{G}} operators
with all zero eigenvalues (namely: $I^{G}J^{PC}=0^{+}0^{-+}$) which
makes it identical (except for parity) to the vacuum or the Higgs
boson. In that respect, it is a very pure state, carrying no Standard
Model charges and, as noted above, its decays do not involve charge-changing
currents: all decays of $\eta$ and $\eta^{\prime}$ mesons are flavor-conserving.
Therefore, a $\eta/\eta^{\prime}$-factory could be interpreted as a ``low-energy
Higgs-factory'', anticipating much of the exploration achievable
at a high-energy Higgs factory. Any coupling to BSM states, therefore,
does not interfere with Standard Model charge-changing operators (as
it occurs, for example, with mesons carrying flavor). From the experimental
point of view, the $\eta/\eta^{\prime}$ dynamics is particularly favorable
to the exploration of small BSM effects since, as a consequence of the
properties mentioned above, they have an unusually small decay width
($\Gamma_{\eta}$=1.3 KeV vs $\Gamma_{\rho}$=149 MeV, for example).
Electromagnetic and strong decays are suppressed up to order $\mathcal{O}(10^{-6})$
favoring the study of more rare decays, especially those related to
BSM particles and to violation of discrete symmetries. This helps
considerably in reconstructing the kinematic of the event and in reducing
the Standard Model background by requiring that the invariant mass
of the final state particles is consistent with the $\eta$ mass.

Another reason to investigate more precisely the $\eta/\eta^{\prime}$ mesons is that their
structure has never been fully understood. Recent work~\cite{Singh12}
indicates that such mesons are unique among the pseudoscalars as
they have anomalously large masses which are contributed by quarks
only about 80\% of the momentum, leaving considerable room for potential
contribution by New Physics.

\smallskip{}

\begin{figure}[!ht]
\begin{centering}
\includegraphics[width=0.98\textwidth]{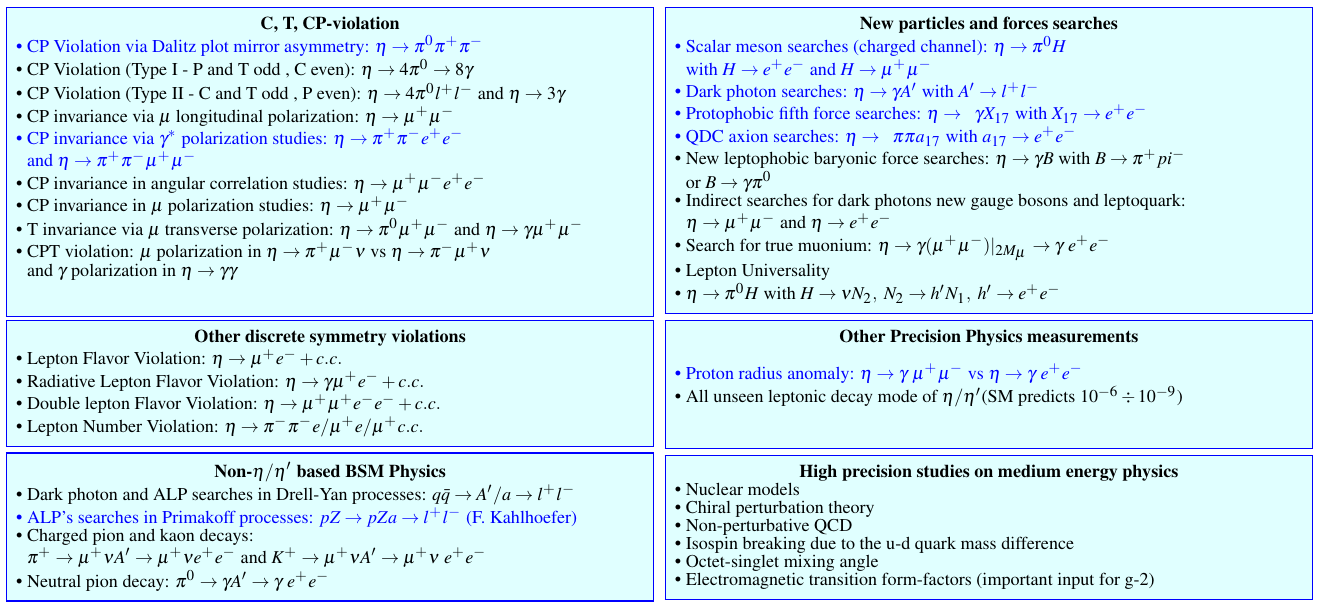} 
\caption{Physics landscape for a $\eta/\eta^{\prime}$ factory.}
\label{fig:physics_landscape}
\par\end{centering}
\end{figure}

A summary of the processes that can be studied at REDTOP for exploring
New Physics is shown in Fig.~\ref{fig:physics_landscape}. The processes
are grouped by their physics topic, and will be discussed in more
details in the next sections.

Considering the present limits on the parameters associated to BSM
physics and the practical limitations of current detector technologies,
the next generation of experiments should be designed with the goal
of producing no less than $10^{13}$ $\eta$ mesons and $10^{11}$
$\eta^{\prime}$ mesons. The physics reach of an experiment with such statistics
is very broad, spanning several aspects of physics BSM. The most relevant
processes to be explored fall into two main fields of research: \emph{Search
for New Particle and Fields}, and \emph{Test of Conservation laws}.
Along with BSM Physics, the availability of such a large sample of
flavor-conserving mesons will also allow probing the isospin violating
sector of low energy QCD to an unprecedented degree of precision.
The large number of processes that could be studied at the proposed
$\eta/\eta^{\prime}$-factory will provide not only a nice scientific laboratory
but also the source of many topics for Ph.D. thesis.\smallskip{}

Few of the BSM processes accessible with an $\eta/\eta^{\prime}$-factory
of REDTOP class have been selected for detailed sensitivity studies.
They will be discussed later in this paper, along with several recent
theoretical models which are explaining some of the anomalies that
have been observed by the experiments. 

\section{BSM physics with an \texorpdfstring{$\eta/\eta^{\prime}$}{} factory\label{sec:Physics-Beyond-the-SM}}

Large samples of $\eta/\eta^{\prime}$ decays 
open new avenues for the study of BSM physics. 
This is 
particularly true for 
weakly coupled hidden sectors, in 
which the new fields are SM singlets, 
as well as studies of  
fundamental symmetries and their breaking.

Models of new hidden sectors are typified by so-called
portals, in which a new field, either a vector, scalar, or a
heavy neutral lepton, appears in an SM gauge-singlet interaction 
of mass dimension four or less. 
As a result, the 
additional interactions do not spoil the UV properties
and hence the renormalizability of the SM. 
Portals of higher mass dimension, such as
axion models, are also very interesting. 
REDTOP  is capable of probing all of  these
portals. 

Turning to symmetry tests, REDTOP offers 
new opportunities for searches for CP violation, as well as for
tests of both lepton flavor violation and universality. 
Within the standard model CP violation is described by one complex phase in the  Cabibbo-Kobayashi-Maskawa 
(CKM) quark-mixing matrix.
All three generations of 
quarks contribute in order to realize a non-zero CP-violating effect. 
It has long been suspected 
new sources of CP violation 
must exist in order to explain 
baryon asymmetry of the universe.
Tantalizing hints of lepton-flavor-universality 
violation have also been seen in $B$-meson decays, and
it is important to search for these effects
in light-quark systems as well. Searches
for lepton-flavor violation in $\eta, \eta^{\prime}$ decays complement searches for
$\mu-e$ conversion in the field of a nucleus, 
for which sensitive searches are being mounted worldwide. 

REDTOP is well suited for all of these studies.
Figure~\ref{fig:physics_landscape} provides  a compact
illustration of the physics possibilities.  
The overview that follows describes these possibilities, 
noting the \emph{golden
channels} that have a higher signal to noise ratio within REDTOP.

\subsection{\label{subsec:Searches-for-new-particles-and-fileds}Searches for
new particles and fields}

One of the most prominent ways to accommodate NP is the so called \emph{``hidden
sector physics}'', comprising new particles with masses below the
electroweak (EW) scale coupled very weakly to the SM world via so-called
\emph{portals}~\cite{BPR2009}. 
Such schema are characterized by 
new particles which are either heavy or interact indirectly with the SM sector.
These \emph{hidden sectors} may be experimentally accessible via particles
in the MeV--GeV mass range, which are coupled to the Standard Model
sectors via renormalizable interactions with dimensionless coupling
constants (the \emph{portals}) or by higher-dimensional operators.
The latter, however, are suppressed by the dimensional couplings
of the order $\Lambda^{-n}$, associated to a new energy scale of
the hidden sector. 

Three such portals are renormalizable within
the Standard Model: the \emph{Vector portal}, the \emph{Scalar portal},
and the \emph{Heavy Neutral Lepton portal}. They differ by the mass
dimension of the SM operator which is coupled to the dark sector. A
fourth portal: the \emph{Pseudoscalar (}or\emph{ Axion) portal}, is
not, in general, renormalizable, and the models falling in this category
are often \emph{Effective Field Theories}. Nonetheless, the discovery
of the Higgs boson indicates that fundamental scalar bosons exist
in nature, justifying the search for more light pseudoscalar particles.
Several classes of models exist, accommodating such additional
states: extensions of the Higgs sector 
\cite{Gunion1990TheHH},
models with extra non-doublet scalars 
\cite{arXiv:1208.2555}, or pseudo-Nambu-Goldstone bosons (PNGB)
of a spontaneously broken U(1) symmetry~\cite{PhysRevD.16.1791,PhysRevLett.40.223}.  
All these models have new light states that couple only weakly  to Standard Model particles.


Several $\eta$/$\eta^{\prime}$-related
processes have been selected to study the sensitivity of REDTOP to
such portals. Some of them could also shed some light on anomalies
observed in recent experiments.  These are discussed below. 
These searches are golden opportunities for REDTOP. 

\subsubsection{Vector portal models}

Several extensions of the standard model are based on the interaction of new light vector
particles, resulting, for example, from extra gauge symmetries. New
vector states can mediate interactions both with the SM fields, and
with extra fields in the dark sector. It is  speculated that
the gauge structure of the SM derives from a larger gauge group, as,
for example, in the Grand Unified Theories (GUTs), where  new
vector states exist. If these particles exist, their mass is expected
to be of the order of $10^{16}$ GeV, an energy well beyond the direct
reach of present accelerators. Other models assume that the SM has
additional gauge structures able to accommodate gauge bosons with
masses below the TeV scale 
\cite{arXiv:0801.1345}.
The current results from LHC experiments has put very strong bounds
on the existence of such new vector states, with the hypothesis that
the coupling of the latter to the SM is large enough. 
To cope with
such observations, more recent theoretical models assume the existence
of relatively light vector states (e.g., in the MeV--GeV mass range)
with small couplings to the SM. This mass range is poorly constrained
by the LHC experiments, and it could be probed easily with dedicated
experiments with high intensity beams such as   $\eta/\eta^{\prime}$-factories.

The \emph{Vector portal} spans several classes of models.
Typical examples are: kinetically mixed dark photons in the GeV mass
range, gauge bosons coupled to baryons, \emph{dark Higgs} bosons generated
through the portal or via \emph{Higgsstrahlung}, \emph{heavy neutral
leptons} (HNL) generated through the portal.
Models currently under study by the REDTOP Collaboration
include 
the \emph{Minimal dark photon model }(kinetically mixed dark photons), the \emph{Leptophobic B boson model}, 
and the \emph{Protophobic Fifth Force model}.

\paragraph{\textbf{\emph{Minimal dark photon model}}}

One of the most popular models in the \emph{Vector portal} is commonly
referred to as: \emph{Minimal dark photon model.} In this case, the
SM is augmented by a single new state $A'$ which couples to visible matter
via a \emph{kinetic mixing parameter $\varepsilon$}~\cite{Holdom86}.
REDTOP could observe new vector particles in the decays: $\eta/\eta^{\prime}\rightarrow\gamma A'\rightarrow\gamma e^{+}e^{-}$
and $\eta/\eta^{\prime}\rightarrow\gamma A'\rightarrow\gamma\mu^{+}\mu^{-}$.

This process has a relatively large branching ratio
$\sim7\times10^{-3}$~\cite{PDGC}. Consequently, REDTOP will be able
to detect a number of such final states with samples larger than $10^{8}$.  %
Two experiments are pursuing dedicated dark photon searches with $e^{+}e^{-}$
colliding beams: the HPS at JLAB~\cite{HPS} and PADME at Laboratori
Nazionali di Frascati~\cite{PADME}. REDTOP, on the other hand, will
perform a similar search with hadron-produced $\eta$ mesons, producing  very different statistical and systematic uncertainties.
Preliminary sensitivity studies on the \emph{Minimal dark photon model}
have been performed as part of CERN's ``Physics Beyond Collider''
program~\cite{PBC19} indicating that REDTOP is sensitive to a large
portion of the unexplored region of the\emph{ $\varepsilon$ }parameter
space. New studies, based on a much better detector that can identify a detached vertices are in progress.

A search for  dark photons can be carried out at an $\eta/\eta^{\prime}$-factory
by looking for final states with a photon and two leptons (Dalitz decay). Considering the process
\begin{equation}
\eta\rightarrow\gamma A'\rightarrow\gamma+l^{+}l^{-}\label{eq:dark_photon}
\end{equation}
in the hypothesis that the mass of this dark photon is smaller than
the mass of the $\eta/\eta^{\prime}$ meson, it will be relatively straightforward
to observe it. 
A detailed study of this process is presented in Sec.~\ref{sec:Sensitivity-Studies-to-physics-BSM}.

\paragraph{\textbf{\emph{Leptophobic B boson model}}}

We consider a model for a leptophobic gauge boson $B$ that couples to baryon number through the following interaction Lagrangian~\cite{Tulin:2014tya}
\begin{equation}
\mathcal{L}_{\rm{int}}=\left(\frac{1}{3}g_{B}+\varepsilon Q_{q}e\right)\bar{q}\gamma^{\mu}qB_{\mu}-\varepsilon e\bar{\ell}\gamma^{\mu}\ell B_{\mu}\,,   
\label{Eq:BbosonLagrangian}
\end{equation}
where $B_{\mu}$ is the new gauge boson field, $g_{B}$ is the new gauge coupling (considered here universal for all quarks $q$) and $\alpha_{B}=g_{B}^{2}/4\pi$ is the fine structure associated to the baryonic force.
This interaction preserves the low-energy symmetries of QCD, i.e., charge conjugation $(C)$, parity $(P)$, and $T$ invariance, as well as isospin and $SU(3)$ flavor symmetry.
In the MeV--GeV mass range, $m_{\pi^{0}}\lesssim m_{B}\lesssim1$ GeV, the $B$ boson decays predominantly to $\pi^{0}\gamma$, or to $\pi^{0}\pi^{+}\pi^{-}$ when kinematically allowed, very much like the $\omega$ meson; in fact, the $\omega$ quantum numbers, $I^{G}(J^{PC})=0^{-}(1^{--})$, can be assigned to the $B$ boson.
In addition, the Lagrangian in Eq.~(\ref{Eq:BbosonLagrangian}) is not completely decoupled from leptons as it contains subleading photon-like couplings to fermions proportional to $\varepsilon=eg_{B}/(4\pi)^{2}$.
This effect allows the purely leptonic decay $B\to e^{+}e^{-}$, which dominates below single pion threshold.

Rare $\eta$ and $\eta^{\prime}$ decays are specially suited to search for $B$ signatures in the MeV--GeV mass range.
Here we concentrate first on the doubly radiative decays $\eta^{(\prime)}\to\pi^{0}\gamma\gamma$ and $\eta^{\prime}\to\eta\gamma\gamma$. 
The current  layout of REDTOP is not sensitive to completly  neutral final states. 
However the discovery potential offered by these processes is very promising and that could strengthen the case for an upgrade of the experiment.
In fact, an improved version of REDTOP is planned, with the $\eta$ being tagged and in which final states with $\gamma$'s and $\pi^{0}$'s could be detected.
In these decays, the new boson would appear as an intermediate state resonance in the decay chain $\eta^{(\prime)}\to B\gamma \to\pi^{0}\gamma\gamma$, thus producing a peak at around $m_{B}$ in the $\pi^{0}\gamma$ invariant mass spectrum.

\begin{figure*}[t]
\centering\includegraphics[scale=0.55]{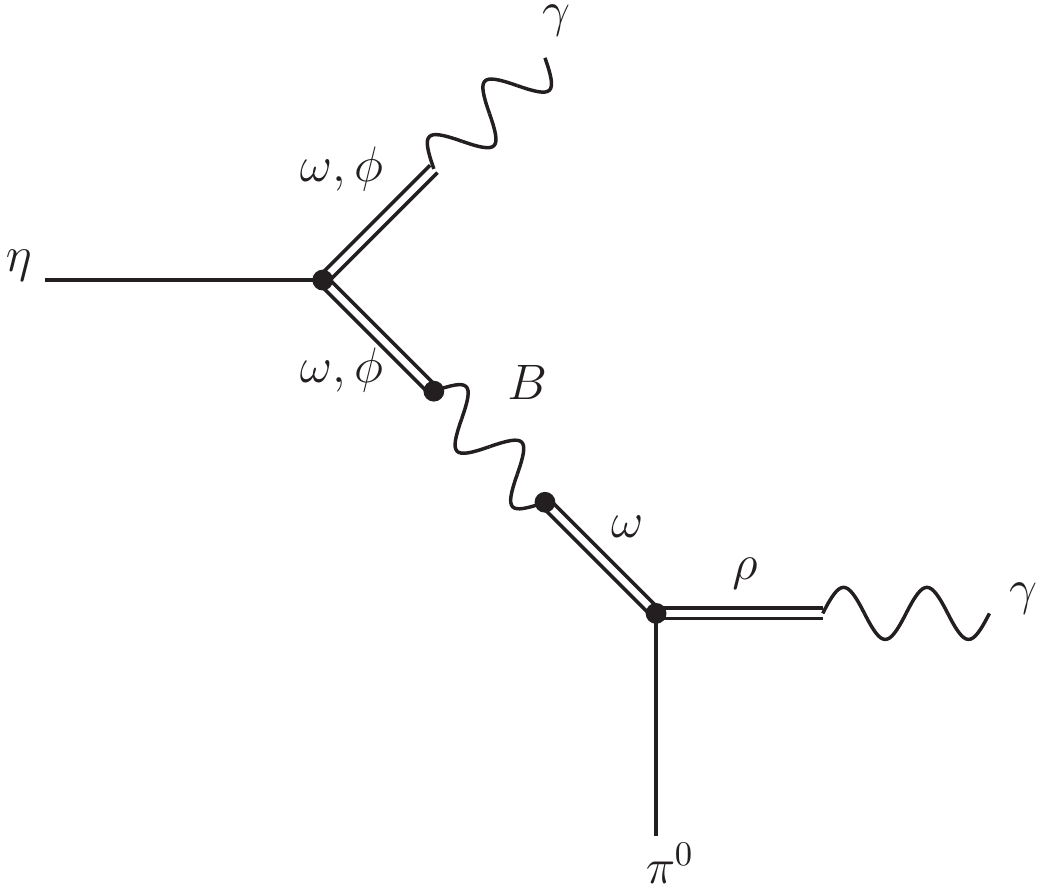}\qquad\includegraphics[scale=0.55]{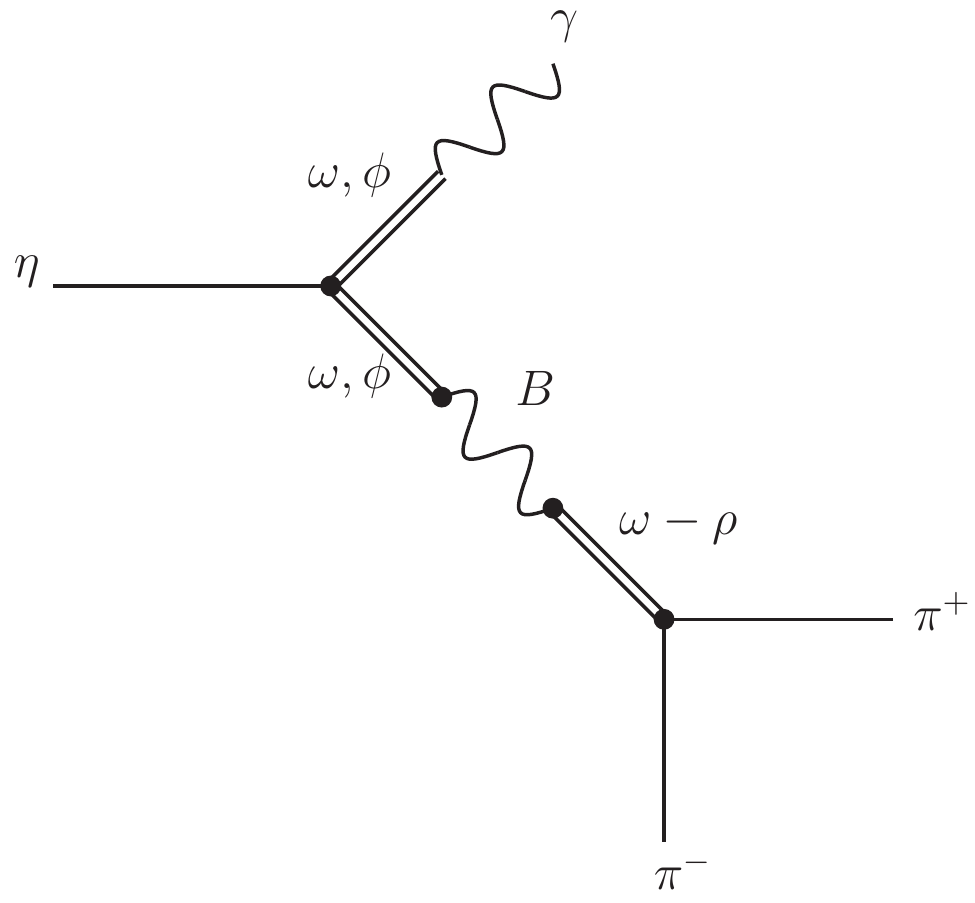}
\caption{Schematic diagram for the $B$ boson exchange mechanism to the decays $\eta\to\pi^{0}\gamma\gamma$ (left) and  $\eta\to\pi^{+}\pi^{-}\gamma$ (right).}
\label{Fig:BbosonExchange} 
\end{figure*}

\begin{sloppypar}
This search requires both experimental precision
as well as a robust SM prediction. 
In~\cite{Escribano:2018cwg}, a VMD framework was used to describe vector-meson exchange contributions to these decays, as well as the L$\sigma$M to describe scalar-meson exchanges.
In analogy to VMD, we now incorporate an intermediate $B$ boson exchange contribution through the Feynman diagram depicted on the left hand side of Fig.\,\ref{Fig:BbosonExchange}.
This contribution can be derived from the standard VMD vector-vector-pseudoscalar and vector-photon Lagrangians, supplemented by an effective vector-$B$ boson vertex.
The standard VMD interaction Lagrangians are given by 
\begin{eqnarray}
\mathcal{L}_{VVP}&=&\frac{G}{\sqrt{2}}\epsilon^{\mu\nu\alpha\beta}{\rm{tr}}\left[\partial_{\mu}V_{\nu}\partial_{\alpha}V_{\beta}P\right]\,, \nonumber\\[1ex]
\mathcal{L}_{V\gamma}&=&-2egf_{\pi}^{2}A^{\mu}{\rm{tr}}\left[QV_{\mu}\right]\,,\label{Eq:Lagrangians1+2}
\end{eqnarray}
where $\epsilon_{\mu\nu\alpha\beta}$ is the totally antisymmetric Levi-Civita tensor, $V^{\mu}$ and $P$ are the matrices for the vector and pseudoscalar meson fields, $A^{\mu}$ is the photon field and, $Q={\rm{diag}}\{2/3,-1/3,-1/3\}$ is 
the quark-charge matrix~\cite{Bramon:1992kr}. 
\end{sloppypar}

The Lagrangian that describes the $VB$ interaction is formally identical to the $V\gamma$ one in Eq.~(\ref{Eq:Lagrangians1+2}), with the substitutions $A^{\mu}\to B^{\mu}$, $e\to g_{B}$ 
and $Q\to{\rm{diag}}\{1/3,1/3,1/3\}$. 
It is given by
\begin{eqnarray}
\label{Eq:Lagrangians3}
\mathcal{L}_{VB}&=&-2\frac{1}{3}g_{B}gf_{\pi}^{2}B^{\mu}{\rm{tr}}\left[V^{\mu}\right]\,.
\end{eqnarray}
From the Lagrangians in Eqs.~(\ref{Eq:Lagrangians1+2}) and (\ref{Eq:Lagrangians3}), it is straightforward to obtain expressions for the $g_{BP\gamma}$ couplings in terms of $g_{B}$, which are found to be:
{
\fontsize{8}{8}{\begin{eqnarray}
g_{B\pi^{0}\gamma}(q^{2})&=&\frac{\sqrt{2}eg_{B}}{4\pi^{2}f_{\pi}}\left(\frac{m_{\omega}^{2}}{m_{\omega}^{2}-q^{2}-im_{\omega}\Gamma_{\omega}}\right)\,,\label{Eq:gBPgammaCouplings1}\\[1ex]
g_{B\eta\gamma}(q^{2})&=&\frac{eg_{B}}{12\pi^{2}f_{\pi}}\frac{1}{\sqrt{6}}\left[(c_{\theta}-\sqrt{2}s_{\theta})\left(\frac{m_{\omega}^{2}}{m_{\omega}^{2}-q^{2}-im_{\omega}\Gamma_{\omega}}\right)+(2c_{\theta}+\sqrt{2}s_{\theta})\left(\frac{m_{\phi}^{2}}{m_{\phi}^{2}-q^{2}-im_{\phi}\Gamma_{\phi}}\right)\right],\label{Eq:gBPgammaCouplings2}\\[1ex]
g_{B\eta^{\prime}\gamma}(q^{2})&=&\frac{eg_{B}}{12\pi^{2}f_{\pi}}\frac{1}{\sqrt{6}}\left[(s_{\theta}+\sqrt{2}c_{\theta})\left(\frac{m_{\omega}^{2}}{m_{\omega}^{2}-q^{2}-im_{\omega}\Gamma_{\omega}}\right)+(2s_{\theta}-\sqrt{2}c_{\theta})\left(\frac{m_{\phi}^{2}}{m_{\phi}^{2}-q^{2}-im_{\phi}\Gamma_{\phi}}\right)\right],
\label{Eq:gBPgammaCouplings3}
\end{eqnarray}
}}
where $\theta$ is the $\eta$-$\eta^{\prime}$ mixing angle, $s_{\theta}\equiv\sin\theta$, and  $c_{\theta}\equiv\cos\theta$. 

Combining the $g_{B\eta\gamma}$ and $g_{B\pi^{0}\gamma}$ couplings from Eqs.~(\ref{Eq:gBPgammaCouplings1}) and (\ref{Eq:gBPgammaCouplings2}) with the propagator of the $B$ boson, one can calculate the $B$ boson exchange contribution to the amplitude of the decay $\eta\to\pi^{0}\gamma\gamma$.
This is given by 
\begin{eqnarray}
\label{Eq:BbosonAmplitude}
{\cal A}^{B\,\mathrm{boson}}_{\eta\to\pi^0\gamma\gamma}=
g_{B\eta\gamma}(t)g_{B\pi^0\gamma}(t)\left[\frac{(P\cdot q_2-m_\eta^2)\{a\}-\{b\}}{m_{B}^{2}-t-im_{B}\Gamma_{B}}+
\bigg\{
\begin{array}{c}
q_2\leftrightarrow q_1\\
t\leftrightarrow u
\end{array}
\bigg\}\right]\ ,
\end{eqnarray}
where
$t,u=(P-q_{2,1})^2=m_\eta^2-2P\cdot q_{2,1}$ are the Mandelstam variables and 
$\{a\}$ and $\{b\}$ are the Lorentz structures. These  are defined as
\begin{equation}
\label{varGamma}
\begin{aligned}
\{a\}&=(\epsilon_1\cdot\epsilon_2)(q_1\cdot q_2)-(\epsilon_1\cdot q_2)(\epsilon_2\cdot q_1) \ , \\
\{b\}&=(\epsilon_1\cdot q_2)(\epsilon_2\cdot P)(P\cdot q_1)+(\epsilon_2\cdot q_1)(\epsilon_1\cdot P)(P\cdot q_2)\\
&-(\epsilon_1\cdot\epsilon_2)(P\cdot q_1)(P\cdot q_2)-(\epsilon_1\cdot P)(\epsilon_2\cdot P)(q_1\cdot q_2)\,,
\end{aligned}
\end{equation}
where $P$ is the four-momentum of the decaying $\eta$ meson, and
$\epsilon_{1,2}$ and $q_{1,2}$ are, respectively, the polarization and four-momentum
vectors of the final photons. 
The amplitudes for the decays $\eta^\prime\to\pi^0\gamma\gamma$ 
and $\eta^\prime\to\eta\gamma\gamma$ have a similar structure to that of Eq.~(\ref{Eq:BbosonAmplitude}), with the replacements $m_\eta^2\to m_{\eta^\prime}^2$, and $g_{B\eta\gamma}g_{B\pi^0\gamma}\to g_{B\eta^\prime\gamma}g_{B\pi^0\gamma}$ for the $\eta^\prime\to\pi^0\gamma\gamma$ decay and
$g_{B\eta\gamma}g_{B\pi^0\gamma}\to g_{B\eta^\prime\gamma}g_{B\eta\gamma}$
for $\eta^\prime\to\eta\gamma\gamma$.

We  also consider $B$ boson signals in the decay $\eta\to\pi^{+}\pi^{-}\gamma$.
In this case, the $B$ boson is produced as in the doubly radiative process discussed above but it decays instead into $\pi^{+}\pi^{-}$ through $\rho$-$\omega$ mixing as depicted in Fig.\,\ref{Fig:BbosonExchange} (right diagram). 
This process  is suppressed since it depends on $\varepsilon$. 
We can write the amplitude for the $\eta(q)\to\pi^{+}(p_{1})\pi^{-}(p_{2})\gamma(k)$ in terms of a scalar function $\mathcal{F}(s,t,u)$ according to
\begin{equation}
{\cal A}(s,t,u)=i\mathcal{F}(s,t,u)\varepsilon_{\mu\nu\alpha\beta}\epsilon^{\mu}(k)p_{1}^{\nu}p_{2}^{\alpha}q^{\beta}\,, 
\end{equation}
with the Mandelstam variables  $s=(q-p_{1})^{2}\,,t=(p_{1}+p_{2})^{2}$ and $u=(q-p_{2})^{2}$. 
In $P$-wave approximation, one has $\mathcal{F}(s,t,u) = F(t)$~\cite{Kubis:2015sga}, and the decay rate is given by
\begin{equation}
\Gamma(\eta\to\pi^{+}\pi^{-}\gamma)=\int_{4m_{\pi}^{2}}^{M_{\eta}^{2}}dt\,\frac{t\sigma^{3}_{\pi}(M_{\eta}^{2}-t)^{3}}{12(8\pi M_{\eta})^{3}}|F(t)|^{2}\,,    
\label{Eq:PartialRate}
\end{equation}
where $\sigma_{\pi}=\sqrt{1-4m_{\pi}^{2}/t}$.  For a discussion of the Standard Model amplitude $F(t)$, see Refs.~\cite{Stollenwerk:2011zz,Kubis:2015sga}.

The $B$ boson exchange contribution to the form factor $F(t)$ in Eq.~(\ref{Eq:PartialRate}) can be written as
\begin{equation}
F(t)=g_{B\eta\gamma}(t)\frac{1}{m_{B}^{2}-t-im_{B}\Gamma_{B}}g_{B\pi\pi}(t)\,,    
\end{equation}
where $g_{B\eta\gamma}(t)$ is given in Eq.~(\ref{Eq:gBPgammaCouplings2}) and $g_{B\pi\pi}(t)$ reads
\begin{equation}
\label{Eq:gBpipicoupling}
g_{B\pi\pi}(t)=\sqrt{4\pi\alpha_{{\rm{em}}}\varepsilon^{2}}|F_{\pi}(t)|\,.    
\end{equation}
The pion form factor in Eq.~(\ref{Eq:gBpipicoupling}) can, in turn, be expressed as
\begin{equation}
F_{\pi}(t)=F_{\rho}(t)\left[1+\frac{1+\delta}{3}\frac{\tilde{\Pi}(t)}{t-m_{\omega}^{2}+im_{\omega}\Gamma_{\omega}}\right]\,,    
\label{Eq:PionFormFactorCorrected}
\end{equation}
where $\delta=2g_{B}/(\varepsilon e)$~\cite{Tulin:2014tya} accounts for the $B$-$\omega$ mixing
and $F_{\rho}(t)=m_{\rho}^{2}/(m_{\rho}^{2}-t-im_{\rho}\Gamma_{\rho})$ is the pion form factor associated to the $\rho$ exchange only. 
For the $\rho$-$\omega$ mixing parameter in Eq.~(\ref{Eq:PionFormFactorCorrected}), we assume $\tilde{\Pi}(t)=\tilde{\Pi}(m_{\omega}^{2})=-3500(300)\,\text{MeV}^2$~\cite{Tulin:2014tya}.

\paragraph{\textbf{\emph{Protophobic Fifth Force model}}}

Another interesting model to challenge has been proposed~\cite{PhysRevLett.117.071803,Feng:2016ysn} 
to explain a $6.8\sigma$ anomaly in the invariant mass
distributions of $e^{+}e^{-}$ pairs produced in $^{8}$Be discrete nuclear
transitions~\cite{Krasznahorkay:2015iga}. The mass of such a gauge boson ($X_{17}$) is determined
to be about 17 MeV, which is below the sensitivity of WASA and KLOE,
but accessible to REDTOP thanks to the slight boost imparted to the
$\eta$ meson in the lab frame. The same fifth force would be able
to reconcile the $3.6\sigma$ discrepancy between the predicted and
measured values of the muon's anomalous 
magnetic moment~\cite{PhysRevLett.117.071803}. In this respect,
REDTOP will be a nice complement to the $(g-2)_\mu$ experiment currently
running at Fermilab~\cite{g-207}. 

The $X_{17}$ has been the subject of much experimental 
and theoretical study, with the NA64 experiment at the CERN SPS, searching for 
$X_{17} \to e^+e^-$ decay, finding only negative 
results~\cite{NA64:2019auh}, with a probe of the
remaining parameter space possible~\cite{NA64:2020xxh}. 
Evidence for $X_{17}$ 
has also been observed in 
$^4$He decay~\cite{Krasznahorkay:2021joi}, and the quantum number selectivity associated with 
its emission from an excited nuclear state supports its 
interpretation as a vector particle~\cite{Krasznahorkay:2021joi,Feng:2020mbt}. However, 
it has been argued $X_{17}$ production is dominated by a non-resonant
process, obviating these conclusions, 
with the nonobservation of $X_{17}$
via bremsstrahlung arguing against a vector interpretation~\cite{Zhang:2020ukq}. 
REDTOP can provide a definitive test of this issue. 
Preliminary studies show  that 
REDTOP has an excellent sensitivity to this model via the production
channel: $\eta/\eta^{\prime}\rightarrow\gamma X_{17}\rightarrow\gamma e^{+}e^{-}$. This sensitivity is mainly due to the high granularity ADRIANO2 calorimeter. 
Particular to the protophobic model is the distinct
couplings that the new gauge boson possesses to 
$u$ and $d$ quarks. In this sense, the model 
represents an explicit example of a generalized 
$B$ boson model. 

We present and contrast the decay width for $\eta \to \gamma X$, where $X$ is a generic gauge boson with couplings to $u$, $d$ and $s$ quarks, in two calculation schemes. 
The first is derived from the traditional triangle anomaly of Adler, Bell and Jackiw (ABJ)~\cite{Adler:1969gk, Bell:1969ts} where one photon leg has been replaced by an $X$. In this scheme, the ratio of branching ratios of the $\eta$ to normal and dark photons is given by 
\begin{equation}
    \label{eq:eta_decay_ABJ}
    \left(\frac{BR_{\eta\to \gamma X}}{BR_{\eta\to \gamma \gamma}}\right)_{\text{ABJ}} = \frac{1}{2\pi \alpha}  \left( 1 - \frac{m_X^2}{m_\eta^2}\right)^3 \left[ \frac{(c_\theta - \sqrt{2} s_\theta)(2\varepsilon_u - \varepsilon_d) + (2c_\theta + \sqrt{2} s_\theta) \varepsilon_s}{c_\theta - 2 \sqrt{2} s_\theta} \right]^2,
\end{equation}
where $m_X$ is the mass of the new gauge boson; $\varepsilon_u$, $\varepsilon_d$ and $\varepsilon_s$ are respectively the up-, down- and strange-quark charges under the new interaction; $\theta \approx -19.5^\circ$ 
is the $\eta$--$\eta^\prime$ mixing angle,  $c_\theta \equiv \cos \theta$, and $s_\theta \equiv \sin \theta$.

Alternatively, the decay width can be calculated in the scheme of vector meson dominance (VMD)~\cite{Fujiwara:1984mp}, in which interactions of the pseudoscalar meson octet are described in terms of a single interaction vertex with the vector meson nonet. The vector mesons then mix kinetically with the SM photon and the new boson $X$. The ratio of branching ratios then becomes 
\begin{eqnarray}
    \label{eq:eta_decay_VMD}
    & \left(\dfrac{BR_{\eta\to \gamma X}}{BR_{\eta\to \gamma \gamma}}\right)_{\text{VMD}} = \dfrac{1}{8\pi \alpha} \left( 1 - \dfrac{m_X^2}{m_\eta^2} \right)^3 \times & \\
    & \left| \dfrac{\left(c_\theta - \sqrt{2} s_\theta\right) \left[3(\varepsilon_u- \varepsilon_d) F_\rho(m_X^2) + (\varepsilon_u + \varepsilon_d) F_\omega(m_X^2) \right] + 2\left(2c_\theta + \sqrt{2} s_\theta\right) \varepsilon_s F_\phi(m_X^2)}{c_\theta - 2\sqrt{2} s_\theta} \right|^2, & \nonumber
\end{eqnarray}
where $F_V(q^2)$ is the vector meson form factor, given by
\begin{equation}
    \label{eq:form_factor}
    F_V(q^2) = \left(1 - \frac{q^2}{m_V^2} - i \frac{\Gamma_V}{m_V}\right)^{-1},
\end{equation}
$m_V$ the vector meson mass and $\Gamma_V$ its corresponding total width. In the limit $m_X\to0$ and $\Gamma_V \to 0$, Eq.~\eqref{eq:eta_decay_VMD} is equivalent to Eq.~\eqref{eq:eta_decay_ABJ}. The case of the leptophobic $B$ boson is recovered in the limit $\varepsilon_u, \, \varepsilon_d, \, \varepsilon_s \to g_B/3$.

\begin{figure}[ht]
    \includegraphics[width=0.9\textwidth]{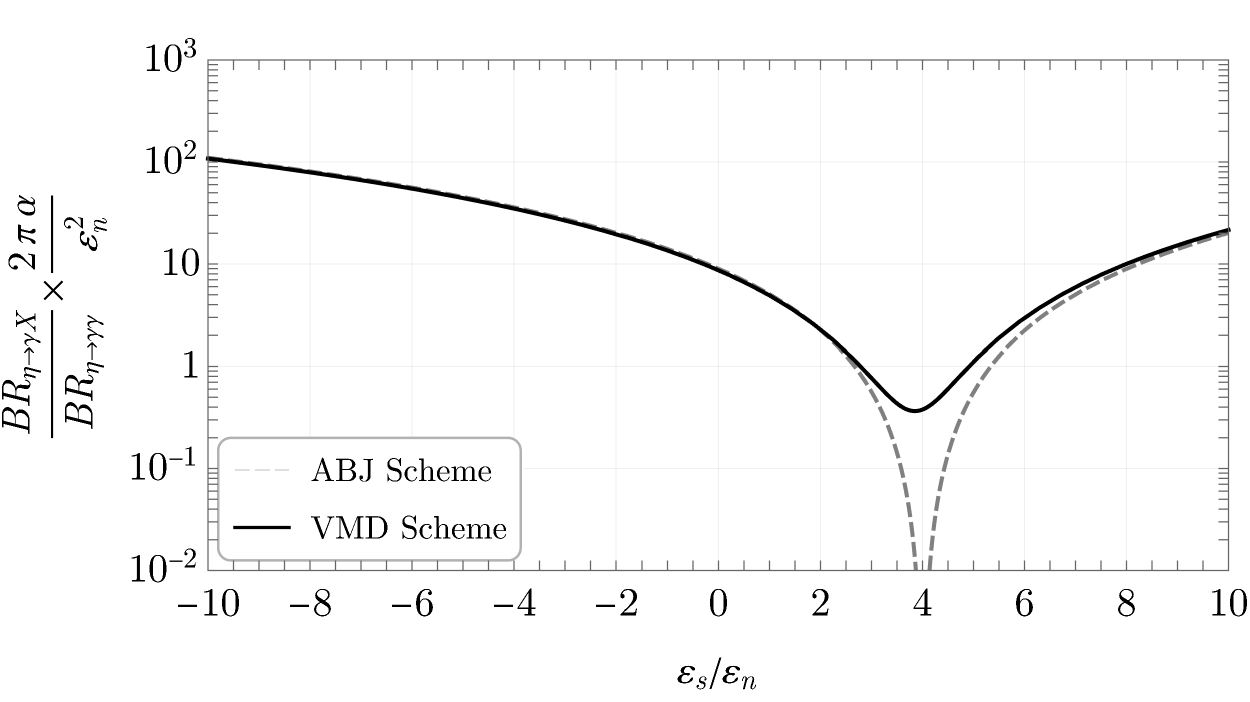}
    \caption{The dependence of $BR_{\eta\to \gamma X}/BR_{\eta\to \gamma \gamma}$ on the ratio of the strange-quark and neutron charges, $\varepsilon_s/\varepsilon_n$, in the protophobic gauge boson model in the limit $m_X \to 0$. The relative strengths of the interactions have been divided out.}
    \label{fig:protophobic_scaling}
\end{figure}

For the case of the protophobic gauge boson,  $2\varepsilon_u = - \varepsilon_d$; and we can define $\varepsilon_n \equiv \varepsilon_u + 2\varepsilon_d = -\varepsilon_u = \varepsilon_d/2$. However, this model is not prescriptive regarding $\varepsilon_s$; therefore, the branching ratio depends on three parameters, which we take to be $\{ m_X, \, \varepsilon_n, \, \varepsilon_s/\varepsilon_n \}$.  Figure~\ref{fig:protophobic_scaling} shows the dependence of the ratio of branching ratios on $\varepsilon_s/\varepsilon_n$ in the limit $m_X \to 0$. We note a cancellation that occurs near $\varepsilon_s/\varepsilon_n \approx 4$; this cancellation is perfect in the ABJ scheme, but the vector boson widths (particularly $\Gamma_\rho$) prevent this from being identically zero in the VMD scheme, even if $m_X$ vanishes.

\subsubsection{Scalar portal models
\label{subsec:Scalar-portal-Models}}

In the so called scalar or Higgs portal, the dark sector couples to
the SM via the Higgs boson or an extension of the latter. Dark scalars
\emph{S} can be explored in REDTOP via $\eta\rightarrow S\rightarrow\pi^{0}e^{+}e^{-}$
and $\eta\rightarrow\pi^{0} S\rightarrow\pi^{0}\mu^{+}\mu^{-}$processes.
Three complementary models are currently under consideration by REDTOP:
the \emph{Minimal dark scalar model, Spontaneous Flavor Violation
model} or \emph{Flavor-Specific Scalar model} (which have similar REDTOP phenomenology), and the\emph{ Two-Higgs doublet model.} In the former, the
dark scalar mimics a light Higgs and, consequently, it couples prevalently
to heavy quarks. The latter has larger coupling, instead, to light
quarks. The predicted branching ratios differ by more than two orders
of magnitude.

\paragraph{\textbf{\emph{Minimal scalar model}}}

The minimal scalar portal model operates with one extra singlet field
$S$ and two types of couplings, $\mu$ and $\lambda$. The mechanism
for the $\eta\to\pi^{0}\mu^{+}\mu^{-}$ (or $e^{+}e^{-}$) decay is
usually described via a 2-photon intermediate state to conserve $C$:
$\eta \to\pi^{0}\gamma\gamma$ along with $\gamma\gamma\to\mu\bar{\mu}$
via a triangle diagram.  Branching ratios are calculated to be of
the order of $10^{-9}$~\cite{NgP92,NgP93,JS02,Escribano:2020rfs}, which should be
well within the sensitivity of REDTOP. Preliminary sensitivity studies
on the \emph{Minimal scalar model }have been performed as part of
CERN's ``Physics Beyond Collider'' program~\cite{PBC19} indicating
that REDTOP has modest sensitivity to $\mu$ and $\lambda$, as it
should be expected from the low quark content of the $\eta$/$\eta^{\prime}$
mesons. New studies, based on a much better performing detector,
also capable of identifying a detached vertex, are in progress.

\paragraph{\textbf{\emph{Spontaneous Flavor Violation model}}}

The limitations in the REDTOP reach to the minimal scalar model are due to the smallness of the scalar's couplings to the $\eta$ and $\eta^{\prime}$ mesons. 
In this minimal model, the scalar couples preferentially to the third generation, while couplings to the light quarks that
compose the $\eta$ and $\eta^{\prime}$ are suppressed. 
A variety of beyond the Standard Model theories share this feature, mostly as a consequence of imposing that the new-physics flavored interactions follow the Standard Model flavor hierarchies in order to avoid stringent bounds from flavor-changing neutral currents (FCNCs)~\cite{DAmbrosio:2002vsn}.
In a recent paper~\cite{Egana-Ugrinovic:2018znw}, however, it was
demonstrated that a novel flavor mechanism called Spontaneous Flavor Violation (SFV) allows to construct natural and well-motivated BSM models, where New Physics may couple preferentially to light quarks while avoiding bounds from FCNCs.
Models with light or heavy scalars based on SFV can be explored both at low-energy experiments~\cite{Egana-Ugrinovic:2019wzj, Batell:2018fqo} and the LHC~\cite{Egana-Ugrinovic:2019dqu, Egana-Ugrinovic:2021uew}.
These two approaches are complimentary, as models containing sub-GeV scalars that couple to light quarks require UV completions that include heavier states, which themselves have sizeable couplings to light quarks.

The implications of finding New Physics with novel flavor hierarchies would have profound consequences for our understanding of the flavor sector. 
Moreover, such New Physics could be important for other fundamental issues, such as the dark matter problem. 
An excellent example that illustrates the broad relevance of looking for extra scalars with novel flavor hierarchies was presented by Batell et al. in~\cite{Batell:2018fqo}.
In this work, it was shown that a light scalar (order 1 GeV and below) coupling preferentially to light quarks may serve as a mediator to the dark sector. 
Batell et al. found that REDTOP would have unique
discovery potential in the mass range $m_K\lesssim m_S \lesssim m_{\eta^{\prime}}$,
with $m_S$ being the mass of the new scalar. The complementarity
with the LHC and the implications for flavor physics were independently explored in Egana-Ugrinovic et al.~\cite{Egana-Ugrinovic:2019dqu}, where the phenomenology of scalars with masses of the order of hundreds of GeV (that arise in UV completions of the model presented in~\cite{Batell:2018fqo})
was explored.

These phenomenological studies have shown the need to further explore New Physics with preferential couplings to light quarks. REDTOP represents an exquisite opportunity to look for such models.


\paragraph{\textbf{\emph{Flavor-Specific and Hadrophilic Scalars}}}


As motivated above, it is of interest to explore new scalars with couplings patterns that are qualitatively distinct to those in the Higgs portal model.
A general effective field theory investigation of scalar mediators with {\it flavor-specific} interactions, i.e., a scalar coupling dominantly to a single SM fermion mass eigenstate, was initiated in Ref.~\cite{Batell:2017kty}. Subsequent work in Ref.~\cite{Batell:2018fqo}  explored the phenomenology of a concrete scenario in which a scalar with hadrophilic couplings mediated interactions with dark matter. Furthermore, Ref.~\cite{Batell:2021xsi} investigated simple renormalizable models of flavor-specific scalars involving a vector-like fermion or scalar doublet, focusing on the complementarity between low and high energy observables. 

\begin{figure}[ht]
\includegraphics[scale=0.6]{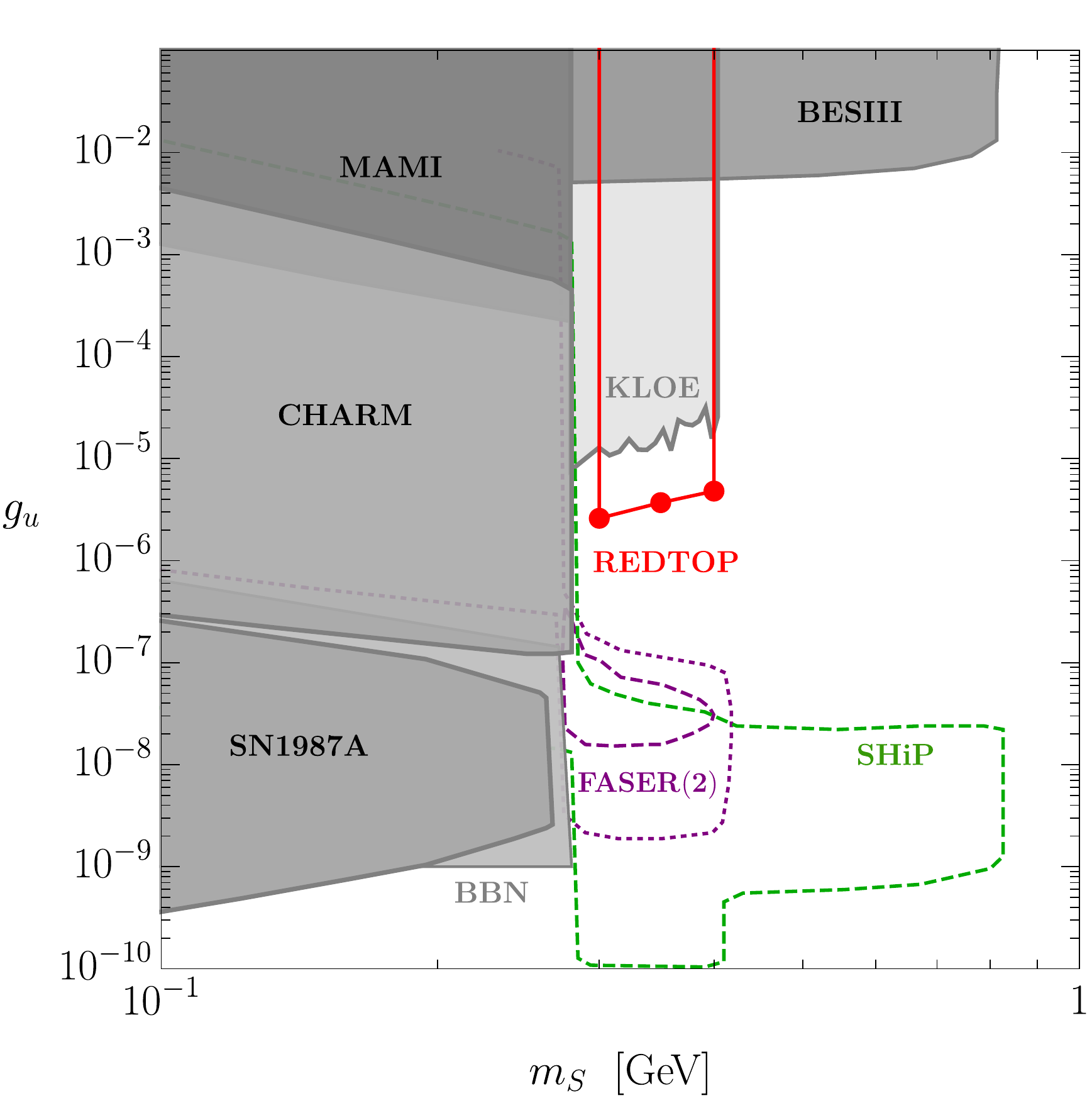}
\caption{REDTOP sensitivity to the the hadrophilic scalar model, Eq.~(\ref{eq:hadrophilic-Lag}), in the $\eta \rightarrow \pi^0 S$, $S\rightarrow \pi^+ \pi^-$ mode. 
We display the projected sensitivity of REDTOP (red) in the $m_S-g_u$ plane to the bump hunt analysis presented in Section~\ref{subsec:hpipi-bumphunt} based on $3.3\times{10}^{18}$ POT for the three mass points $m_S = {300, 350, 400}$ MeV.
Also shown are various existing constraints and projections from other planned or proposed experiments; see Refs.~\citep{Batell:2018fqo,Kling:2021fwx} for further details.
}
\label{fig:dark_scalar}
\end{figure}

As discussed in Ref.~\cite{Batell:2018fqo}, REDTOP has excellent prospects to probe scalars that couple dominantly to first generation quarks. Here we will consider a coupling of a scalar $S$ to up quarks. The low energy Lagrangian is given by
\begin{equation}
\label{eq:hadrophilic-Lag}
{\cal L} \supset \frac{1}{2} (\partial_\mu S)^2  - \frac{1}{2} m_S^2 S^2 - g_u\, S\, \overline u\, u,
\end{equation}
where $m_S$ is the scalar mass and $g_u$ is the effective coupling of the scalar to up quarks.
This model faces strong constraints from cosmology, astrophysics, beam dumps, and past $\eta$ decay searches for $m_S < 2m_\pi$.
However, the constraints are significantly weaker if the scalar mass is above than the two pion threshold. For REDTOP, this singles out $2m_\pi < m_S < m_\eta - m_\pi$ as the mass range of interest in this model. 
The scalar will be produced at REDTOP via $\eta \rightarrow \pi^0 S$, with branching ratio
\begin{eqnarray}
{\rm Br}(\eta \rightarrow \pi^0 S) & = & \frac{c_{S\pi^0\eta}^2 g_u^2 B^2}{16 \pi m_\eta \Gamma_\eta} \lambda^{1/2}\left(1, \frac{m_S^2}{m_\eta^2},\frac{m_{\pi^0}^2}{m_\eta^2}\right),
\label{eq:Breta-Pi-S} 
\end{eqnarray}
where $\lambda(a,b,c)= a^2+b^2+c^2 -2ab-2ac-2 bc$, $B \simeq m_\pi^2/(m_u+m_d) \approx 2.6$ GeV, and the coefficients $c_{S\pi^0\eta} = \frac{1}{\sqrt{3}} \cos \theta - \sqrt{\frac{2}{3}} \sin \theta$ parametrize the effects of $\eta-\eta^{\prime}$ mixing, with $\theta \approx -20^\circ$. Once produced, the scalar will decay promptly via $S\rightarrow \pi^+ \pi^-$, leading to the final state $\pi^0 \pi^+ \pi^-$. 
In Fig.~\ref{fig:dark_scalar} we show the sensitivity of REDTOP to scalars $S$ in this channel, using the results of the $\pi^0 (\pi^+ \pi^-)$ bump hunt analysis presented in Section~\ref{subsec:hpipi-bumphunt} for the three mass points $m_S = {300, 350, 400}$ MeV. 
As can be seen in the plot, REDTOP has the potential to significantly extend the reach in this mass range  beyond the limits from the KLOE experiment. Furthermore, one observes an interesting complementarity with future long-lived particle searches at FASER~\cite{Kling:2021fwx}, FASER2~\cite{Kling:2021fwx} and SHiP~\cite{Batell:2018fqo}, which will probe longer lifetimes and smaller couplings.

\paragraph{\textbf{\emph{Two-Higgs doublet model}}}

This model~\cite{Abdallah:2020vgg}
is one of the simplest possible extensions
of the SM, assuming the existence of a second Higgs doublet $H$ and a dark singlet real scalar~$h'$.
It was initially introduced to explain the anomalies observed by LSND,
MiniBooNE and muon g\textminus 2 experiment, and it is being extended
to the decays of the $\eta\rightarrow\pi^{0}h'/H$ and $\eta^{\prime}\rightarrow\pi^{0}h'/H$. Preliminary calculations indicate that BR($\eta\rightarrow\pi^{0}h'$)$\sim$10$^{-9}$
while BR($\eta^{\prime}\rightarrow\pi^{0}H$)$\sim$10$^{-10}$, in both
cases within REDTOP sensitivity.


A scalar $H$ could be observed in an $\eta/\eta^{\prime}$ final state in association
with a $\pi^{0}$ by detecting the following processes:
\begin{align}
\eta\rightarrow\pi^{0}H\rightarrow\pi^{0}+l^{+}l^{-} , \label{eq:scalar} \\
\eta\rightarrow\pi^{0}H\rightarrow\pi^{0}+\pi^{+}\pi^{-}.\label{eq:scalar-2}
\end{align}
Within the SM, process~\eqref{eq:scalar} can only occur via a two-photon
exchange diagram with a branching ratio of the order of $10^{-9}$.
If such a light particle exists, even with a mass larger than the
$\eta/\eta^{\prime}$ meson, which couples the leptons to the quarks, the
probability for this process could be increased by several orders
of magnitude, changing dramatically the dynamics of the process. Two
groups of theoretical models postulating a BSM light scalar are receiving
great attention lately: the Minimal Extension of the Standard Model
Scalar Sector~\citep{PhysRevD.75.037701,PhysRevD.94.073009} and the
models containing Higgs bosons with large couplings to light quarks~\citep{Batell:2018fqo, Egana-Ugrinovic:2019dqu}.
From the experimental point of view, these models are complementary:
the former predicting large coupling to the $b$-quark and to gluons
but a small one to the light quarks, while the latter predicts a large
coupling to light quarks. An observation at an $\eta/\eta^{\prime}$ factory
of the process~(\ref{eq:scalar}) would be an indication that the
second set of models would be the most likely extension to the SM.
Vice-versa, an observation of a scalar at a $B$-factory but not at REDTOP
would favor the first group of models.

A detailed simulation of the process (\ref{eq:scalar}) and of the
foreseen background, including many instrumental effects, has been
performed by the Collaboration within the ``Physics Beyond Collider''
program~\citep{Alemany:2019vsk} assuming the Minimal Extension of
the Standard Model Scalar model.

An integrated beam flux of $10^{12}$ POT (as available at CERN, see
Sec.~\ref{sec:The-Acceleration-Scheme} below) has been assumed. A
simple ``\emph{bump-hunt}'' analysis was performed, looking at the
invariant mass of di-leptons associated to a prompt photon. The sensitivity
to the $\sin^{2}\theta$ parameter~\citep{PhysRevD.75.037701,PhysRevD.94.073009}
is shown in Fig.~\ref{fig:dark_scalar}. The largest contributing
background was found to be from the 3-body decay $\eta\rightarrow\gamma+l^{+}l^{-}$
where an extra $\gamma$ fakes a $\pi^{0}$ in the final state.

A preliminary sensitivity analysis for several experiments (including
REDTOP) based on the second set of models can be found in Ref.~\citep{Batell:2018fqo}
\vspace{1cm}


\textcolor{black}{LSND and MiniBooNE (MB) observed an excess of electron-like
events over the expected background. These excesses can be understood
in the context of the decay of a heavy neutral lepton ($N_{2}$) produced
via up-scattering of the beam $\nu_{\mu}$ in the detector~\cite{Abdallah:2020vgg}.
Once $N_2$ is produced, it decays instantaneously to another neutral
lepton ($N_1$) and a light scalar~($h'$). The $h'$ subsequently
decays promptly to a collimated $e^+ e^-$ pair producing a  signal
in the detectors. The solution is set in the context of a two-Higgs
doublet model~(2HDM) $(\phi_h,\phi_H)$ and  a dark singlet real scalar
$\phi_{h'}$. After diagonalization of the mass matrix, one
SM-like higgs ($h$) and two other light scalars ($h',H$) have been obtained. In this
model, only the  $\phi_h$ acquires a vacuum expectation value (VEV). Hence, the coupling of
$\phi_H$ with fermions is an independent parameter. The model has
three right-handed neutrinos $N_i$ ($i=1,2,3$) which are responsible
for producing light neutrino masses via a type-I seesaw. Two of the
heavy neutral leptons participate in the production of excess events
in LSND and MB, as mentioned above. The model also resolves the discrepancy
between theory and experiment in the anomalous magnetic moment of
muon via the contributions of the light scalars ($h',H$). Details
of the model are  given in~\cite{Abdallah:2020vgg}. The benchmark
point (BP) of the parameters is shown in Table~\ref{tab-LSND-MB}.}

\textcolor{black}{\begin{table}[t!] \begin{center}  \begin{tabular}{|c|c|c|c|c|c|}   \hline   $m_{N_1}$& $m_{N_2}$ & $m_{N_3}$&$y_{u}^{h'(H)}\!\!\times\!\! 10^{6}$ &$y_{e(\mu)}^{h'}\!\!\times\!\! 10^{4}$&$y_{e(\mu)}^{H}\!\!\times\!\! 10^{4}$ \\   \hline    85\,MeV  & $130$\,MeV & $10$\,GeV&$0.8(8)$ &$0.23(1.6)$&$2.29(15.9)$ \\   \hline     \hline   $m_{h'}$& $m_{H}$ &$\sin\delta$&  $y_{d}^{h'(H)}\!\!\times\!\! 10^{6}$&$y_{\nu_{\!i 2}}^{h'(H)}\!\!\times\!\! 10^{3}$&$\lambda_{N_{\!12}}^{h'(H)}\!\!\times\!\! 10^{3}$ \\[0.05cm]   \hline    17\,MeV  & 750\,MeV &$0.1$& $0.8(8)$&$1.25(12.4)$&$74.6(-7.5)$ \\   \hline  \end{tabular} \caption{Benchmark point for LSND and MB events, and muon $g-2$ calculation.} \label{tab-LSND-MB} \end{center} \end{table}}

\textcolor{black}{The light scalars ($h',H$) couple to $u$ and $d$ quarks.
Hence these scalars could be probed via the decay channels of $\eta\, (\eta^{\prime})$
in REDTOP. The decay amplitudes of $\eta,\eta^\prime \to \pi^0 S$
and $\eta^\prime \to \eta S$, where $S$ refers to the light scalars
$h',H$, are~\cite{Gan:2020aco}:}

\textcolor{black}{\begin{align} \label{eq:Amp_to_scalar} 
\mathcal{A}(\eta^{(\prime)} \to \pi^0 S) &=  - B_0\left[ (\lambda_u - \lambda_d)  \Gamma^{u-d}_{\pi \eta^{(\prime)}}(m_S^2) + (\lambda_u + \lambda_d) \Gamma^{u+d}_{\pi \eta^{(\prime)}}(m_S^2) + \lambda_s  \Gamma^{s}_{\pi \eta^{(\prime)}}(m_S^2) \right], 
\\ \label{eq:Amp_to_scalar_2} \mathcal{A}(\eta^\prime \to \eta S) &=  -B_0 \left\{ (\lambda_u + \lambda_d)  \Gamma^{u+d}_{\eta \eta^{(\prime)}}(m_S^2) + \lambda_s  \Gamma^{s}_{\eta \eta^{(\prime)}}(m_S^2) \right\}.
\end{align}}
The values of $B_0$ and other form factor are given
in~\cite{Gan:2020aco}. In this model $S$ couples exclusively to
both $u$ and $d$ quarks, therefore the  partial widths of $\eta,\eta^\prime \to \pi^0 S$
and $\eta^\prime \to \eta S$ are~\cite{Gan:2020aco} given by:
\begin{align} \label{eq:eta_to_scalar_u1} \Gamma\big(\eta^{(\prime)}\to\pi^0 S\big) &= \frac{|\mathcal{A}(\eta^{(\prime)}\to\pi^0 S)|^2}{16\pi m_{\eta^{(\prime)}}} \lambda^{1/2}\left(1,\frac{m_{\pi^0}^2}{m_{\eta^{(\prime)}}^2}, \frac{m_S^2}{m_{\eta^{(\prime)}}^2}\right)  \,, 
\\ \label{eq:eta_to_scalar_u2} \Gamma\big(\eta^{\prime}\to \eta S\big) &= \frac{|\mathcal{A}(\eta^{\prime}\to \eta S)|^2}{16\pi m_{\eta^{(\prime)}}} \lambda^{1/2}\left(1,\frac{m_{\pi^0}^2}{m_{\eta^{(\prime)}}^2}, \frac{m_S^2}{m_{\eta^{(\prime)}}^2}\right)  \,, 
\end{align}
where $\lambda(a,b,c)=a^2+b^2+c^2-2(ab+ac+bc)$.

\begin{figure}[ht!] 
\includegraphics[width=9cm,height=6cm]{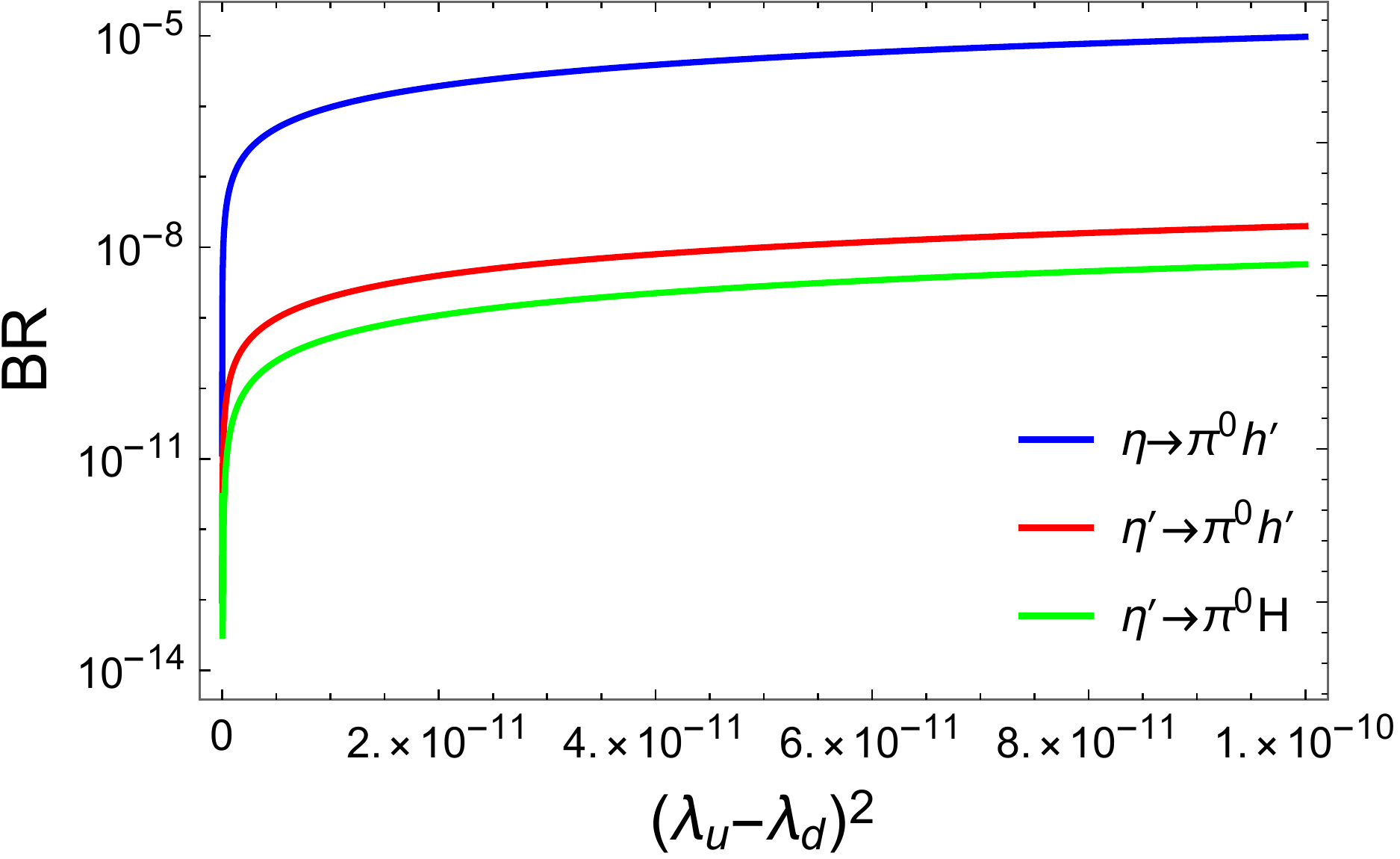} 
\caption{The branching ratios of the decay modes $\eta^{(\prime)} \to \pi^0 S$ as a function of $(\lambda_u-\lambda_d)^2$.} \label{fig:BR} 
\end{figure}

For the benchmark parameter values, as shown in
Table~\ref{tab-LSND-MB}, where $\lambda^{h'(H)}_q=y_{q}^{h'(H)},~q=u,d$,
and $\lambda^{h'(H)}_s=0$, the partial decay widths
and the branching ratios of the decay modes of $\eta$ and $\eta^{\prime}$
are presented in Tables~\ref{tab:REDTOP-1},~\ref{tab:REDTOP-2}.
\begin{table}[t] \begin{center}  \begin{tabular}{|c|c|c|c|c|}   \hline &  $\Gamma(\eta \to \pi^0 S)$ [GeV]&${\rm BR}(\eta \to \pi^0 S)$&  $\Gamma(\eta^{\prime} \to \pi^0 S)$ [GeV]&${\rm BR}(\eta^{\prime} \to \pi^0 S)$   \\   \hline   $S=h'$  & $1.59\times 10^{-17}$ & $1.22\times 10^{-11}$ & $1.91\times 10^{-17}$ & $1.01\times 10^{-13}$ \\   \hline   $S=H$&$0$& $0$&$5.70\times 10^{-16}$& $3.03\times 10^{-12}$ \\[0.05cm]   \hline  \end{tabular} \caption{The partial decay widths and the branching ratios of the decay modes $\eta^{(\prime)} \to \pi^0 S$, where $S=h',H$. } \label{tab:REDTOP-1} \end{center} \vspace{-0.6cm} \end{table}
\begin{table}[t] \begin{center}  \begin{tabular}{|c|c|c|}   \hline &$\Gamma(\eta^\prime \to \eta S)$ [GeV]&${\rm BR}(\eta^\prime \to \eta S)$ \\   \hline   $S=h'$ & $4.54\times 10^{-14}$ &$2.41\times 10^{-10}$ \\   \hline   $S=H$&$0$& $0$\\[0.05cm]   \hline  \end{tabular} \caption{The partial decay width and the branching ratio of the decay mode $\eta^\prime \to \eta S$.} \label{tab:REDTOP-2} \end{center} \end{table}
The $\pi^0 S$ production is proportional to $(\lambda_u-\lambda_d)^2$ which equals to zero for this BP, due to $\lambda^{h'(H)}_u=\lambda^{h'(H)}_d$. This  assumption  of equal couplings to $u$ and $d$ quarks was made for simplicity. However, we now consider a case different from the BP in Table~\ref{tab-LSND-MB} which  assumes unequal couplings, bringing the dominant isovector term into play, but  still keeps the LSND and MB fits intact and also leaves the muon $g-2 $ calculation unaltered. We assume $\lambda^{h'(H)}_u\times 10^{6}=0.86(8.6)$ and $\lambda^{h'(H)}_d\times 10^{6}=0.74(7.4)$, i.e., $(\lambda^{h'(H)}_u-\lambda^{h'(H)}_d)\times 10^{6}=0.12(1.2)$ and $\lambda^{h'(H)}_s=0$. For this case, we have calculated the branching ratios of the decay modes of $\eta^{(\prime)} \to \pi^0 S$ which are shown in Table~\ref{tab:REDTOP-3}. In Fig.~\ref{fig:BR}, we show the branching ratios of the decay modes $\eta^{(\prime)} \to \pi^0 S$ as a function of $(\lambda_u-\lambda_d)^2$.

\begin{table}[t!] \begin{center}  \begin{tabular}{|c|c|c|}   \hline &${\rm BR}(\eta \to \pi^0 S)$& ${\rm BR}(\eta^{\prime} \to \pi^0 S)$  \\   \hline   $S=h'$  & $1.18\times 10^{-9}$&$4.17\times 10^{-12}$  \\   \hline   $S=H$& $0$ & $1.25\times 10^{-10}$\\[0.05cm]   \hline  \end{tabular} \caption{The branching ratios of the decay modes $\eta^{(\prime)} \to \pi^0 S$, where $S=h',H$.} \label{tab:REDTOP-3} \end{center}
\end{table}

\subsubsection{Heavy neutral lepton
portal models}

This portal operates with one or several dark heavy neutral leptons
(HNLs). Among the several models existing under this portal, the\emph{
Two-Higgs doublet model} is the only one considered, at present, by
REDTOP. The process, in this case, would be: $\eta/\eta^{\prime}\rightarrow\pi^{0}H$
with $H\rightarrow\nu N_{2}$ and $N_{2}\rightarrow h'N_{1}$ followed
by $h'\rightarrow e^{+}e^{-}$. The process is identified by the
presence of a $\pi^{0}$ and an $e^{+}e^{-}$ pair in the final state
and a peak in the missing mass of the $\eta/\eta^{\prime}$ spectrum. 

\emph{Two-Higgs doublet model}

The 2HDM discussed in Sec.~\ref{subsec:Scalar-portal-Models}
introduces, besides the $H$ and $h'$ scalars, also two heavy
neutral leptons, $N_{1}$ and $N_{2}$ which represent invisible components
of the HNL portal. This portal could be explored by studying the process
$p+Li\rightarrow\eta+X\;$with$\;\eta\rightarrow\pi^{0}H\;;\;H\rightarrow\nu N_{2}\;;\;N_{2}\rightarrow N_{1}h'\;;\;h'\rightarrow e^{+}e^{-},$ which
is, however, particularly challenging at REDTOP, because the decay
chains contains two undetected particles: $N_{1}$ and $N_{2}$.
The following points may be relevant when considering detectability of the (visible) scalars in REDTOP:  
\begin{itemize} \item In this model~\cite{Abdallah:2020vgg}, the branching ratios of the decay modes of $H$, $h'$, and $N_2$ for the BP in Table~\ref{tab-LSND-MB} are as follows: 
\begin{eqnarray} &&\!\!\!\!\!\!\!\!\!\!\!\!\!\!\!\!\!\!\!\!\!\!{\rm BR}(H \rightarrow \nu_i N_2)= 89.36\%,~{\rm BR}(H \rightarrow N_1 N_2)= 10.17\%,~{\rm BR}(H \rightarrow \mu^+ \mu^-)= 0.46\%, \\ &&\!\!\!\!\!\!\!\!\!\!\!\!\!\!\!\!\!\!\!\!\!\!{\rm BR}(H \rightarrow e^+ e^-)= 0.011\%,~~\! {\rm BR}(H \rightarrow \gamma \gamma)= 3.54 \times 10^{-6}\%,\\ &&\!\!\!\!\!\!\!\!\!\!\!\!\!\!\!\!\!\!\!\!\!\!{\rm BR}(h' \rightarrow e^+ e^-)= 99.9998\%,~~~~~ {\rm BR}(h' \rightarrow  \gamma \gamma)= 2\times 10^{-4}\%,\label{eq:REDTOP-3}\\ &&\!\!\!\!\!\!\!\!\!\!\!\!\!\!\!\!\!\!\!\!\!\!{\rm BR}(N_2 \rightarrow N_1 h')= 99.94\%,~~ {\rm BR}(N_2 \rightarrow \nu_i h')= 0.06\%.~~ 
\label{eq:REDTOP-4}
\end{eqnarray} 
\item It is assumed that $N_1$ could decay to invisible particles and also there is a possibility to tune the parameters such that $N_1$ becomes a long-lived particle.  
\item Thus, an $H$, when produced in the detector, decays with $89.36\%$ BR to $N_2$ and an active SM neutrino. The former, in turn will promptly decay to an $N_1$ and $h'$, leading to a visible $e^+e^-$ pair with missing energy. 
\item For the BP in Table~\ref{tab-LSND-MB}, the total decay width of $H$, $h'$ and $N_2$ are $1.45\times 10^{-5}$~GeV, $3.56\times 10^{-13}$~GeV, and $2.07\times 10^{-5}$~GeV, respectively. Hence all the particles will decay inside the REDTOP.  
\end{itemize}


\subsubsection{The visible QCD axion\label{subsec:Searches-for-QCDaxion}}


The original incarnation of the QCD axion (the so-called `PQWW' axion~\cite{Peccei:1977hh,Peccei1977,Weinberg1978,Wilczek1978}) was a simple Two-Higgs-Doublet Model (2HDM) with a common breaking mechanism for the Electroweak and PQ symmetries. By the late 80s, the parameter space of the PQWW axion was fully excluded by searches for axionic production in hadronic decays and beam dump experiments. Still, many variants of the original QCD axion have been explored (see, e.g.,~\cite{Bardeen:1986yb}), and more recently, Alves and Weiner have shown that a variant of the QCD axion with $m_a\sim\mathcal{O}(1-10)$ MeV and $f_a\sim\mathcal{O}(1-10)$ GeV remains viable~\cite{Alves:2017avw}.

In this variant, the PQ mechanism is implemented by new dynamics at the GeV scale coupling predominantly to the first generation of Standard Model fermions. The resulting QCD axion is short-lived and decays to $e^+e^-$ with a lifetime $\tau_a\lesssim 10^{-13}$ s~\cite{NA64:2021aiq}, avoiding constraints from beam dumps and fixed target experiments, as well as from upper bounds on rare meson/quarkonium decays to a long-lived axion that escapes detection. Furthermore, this axion couples to the first generation of SM quarks with a special relation between the ratios of light quark masses and their PQ charges, namely,
\begin{equation}
 \frac{m_u}{m_d}~\simeq~\frac{Q^{_\text{PQ}}_d}{Q^{_\text{PQ}}_u}~=~\frac{1}{2}\;.\nonumber
 \end{equation}
This relation causes an accidental cancellation of the leading order $\chi$PT contribution to axion-pion mixing,
\begin{equation}\label{thetaAPI}
\theta_{a\pi}~~=~~-~\frac{f_\pi}{f_a}\,\left[\frac{(m_u\,Q^{_\text{PQ}}_u-m_d\,Q^{_\text{PQ}}_d)}{m_u+m_d}~+~\mathcal{O}\left(\frac{m_{ud}}{m_s}\right)\right]~~\simeq~~\frac{~(-0.2\pm30)\times 10^{-4}~}{(f_a/\text{GeV)}}\,,
 \end{equation}
which results in a QCD axion with suppressed isovector couplings. This {\it piophobic} axion is therefore safe from bounds from $\pi^+\rightarrow a\,e^+\nu_e$ mediated via $a-\pi^0$ mixing. In addition, bounds from $K^+\rightarrow\pi^+a$, previously believed to be severe, were shown to suffer from large hadronic uncertainties that preclude the exclusion of significant portions of the piophobic axion parameter space.

There are several motivations for searching for such an axion~\cite{Spier2021}:
\begin{itemize}
\item First and foremost, the QCD axion is tied to the solution of the {\it strong CP problem}, which is one of the most significant puzzles in theoretical physics.
\item In addition, a piophobic QCD axion with mass of $\sim 16-17$ MeV could explain the recent anomalies in {\it isoscalar magnetic} transitions of $^8\text{Be}$ and $^4\text{He}$ nuclei~\cite{Krasznahorkay:2015iga,Krasznahorkay:2019lyl}, while simultaneously explaining the absence of anomalous emissions in {\it isovector} and/or {\it electric} radiative nuclear processes~\cite{Zhang:2020ukq}.
\item Finally, the suppressed axion-pion mixing in (\ref{thetaAPI}) could explain the anomalous rate for $\pi^0\to e^+e^-$ measured by the KTeV collaboration~\cite{Abouzaid:2006kk}.
\end{itemize}

Because the QCD axion couples to quarks and/or gluons, it should invariably lead to new signals in $\eta^{(\prime)}$ decays. The simplest such signal is a contribution to $\eta^{(\prime)}\to e^+e^-$ due to $a-\eta^{(\prime)}$ mixing. Since the SM expectation for $\text{Br}\big(\eta^{(\prime)}\to e^+e^-\big)$ is still two orders of magnitude below current experimental sensitivity, there is significant room for new physics contributions to this final state. REDTOP should have an expected sensitivity to $\text{Br}\big(\eta \to e^+e^-\big)$ with $\mathcal{O}(10^{-10})$ precision, and to $\text{Br}\big(\eta^{\prime}\to e^+e^-\big)$ with $\mathcal{O}(10^{-9})$ precision, which should lead to a sensitivity to $a-\eta^{(\prime)}$ mixing angles of order $\theta_{a\eta}\sim\mathcal{O}(10^{-4})$ and $\theta_{a\eta^\prime}\sim\mathcal{O}(10^{-2})$, respectively (see~\cite{Spier2021}, Sec.\;\ref{sec:Sensitivity-Studies-to-physics-BSM}, and Table\;\ref{table:recoeff_leptonantilepton}).

Firmer evidence for the QCD axion, however, would come as an observation of axio-hadronic $\eta^{(\prime)}\to\pi\pi a$ decays. In fact, a new pseudoscalar particle appearing in $\eta^{(\prime)}$ decays would only be possible if it coupled to quarks and/or gluons. As such, it would affect the QCD topological vacuum by either (i) contributing to or (ii) canceling the strong CP phase $\theta_\text{QCD}$. In case (i), such pseudoscalar would be characterized as a generic `axion-like particle' (ALP), albeit this would be an extremely {\it ad hoc} and fine-tuned scenario since another BSM mechanism would then have to be concocted to cancel the ALP's contribution to $\theta_\text{QCD}$. Therefore, option (ii) would be a more compelling interpretation for such an observation, namely, that a new light pseudoscalar $a$ appearing in $\eta^{(\prime)}\to\pi\pi a$ must be the QCD axion which dynamically relaxes $\theta_\text{QCD}$ to zero and solves the strong CP problem.

\begin{figure}[t]
\centering
\includegraphics[width=0.9\textwidth]{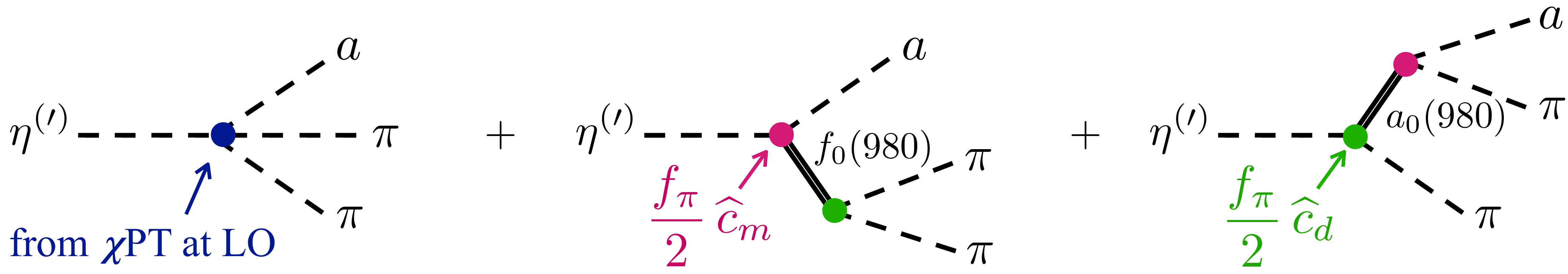}
\caption{Contributions to the amplitude $\mathcal{A}(\eta^{(\prime)}\to\pi\pi a)$ in the framework of Resonance Chiral Theory. Left graph: leading order quartic term. Middle and right graphs: exchange of low-lying scalar resonances. The dimensionless couplings $\widehat{c}_d$ and $\widehat{c}_m$ are expected to deviate from the large-$N_c$ limit of $|\widehat{c}_d|=|\widehat{c}_m|=1$ by $\mathcal{O}(10\%)$.}\label{etaDecayAmplitudes}
\end{figure}

\begin{figure}[t]
\centering
\includegraphics[width=0.6\textwidth]{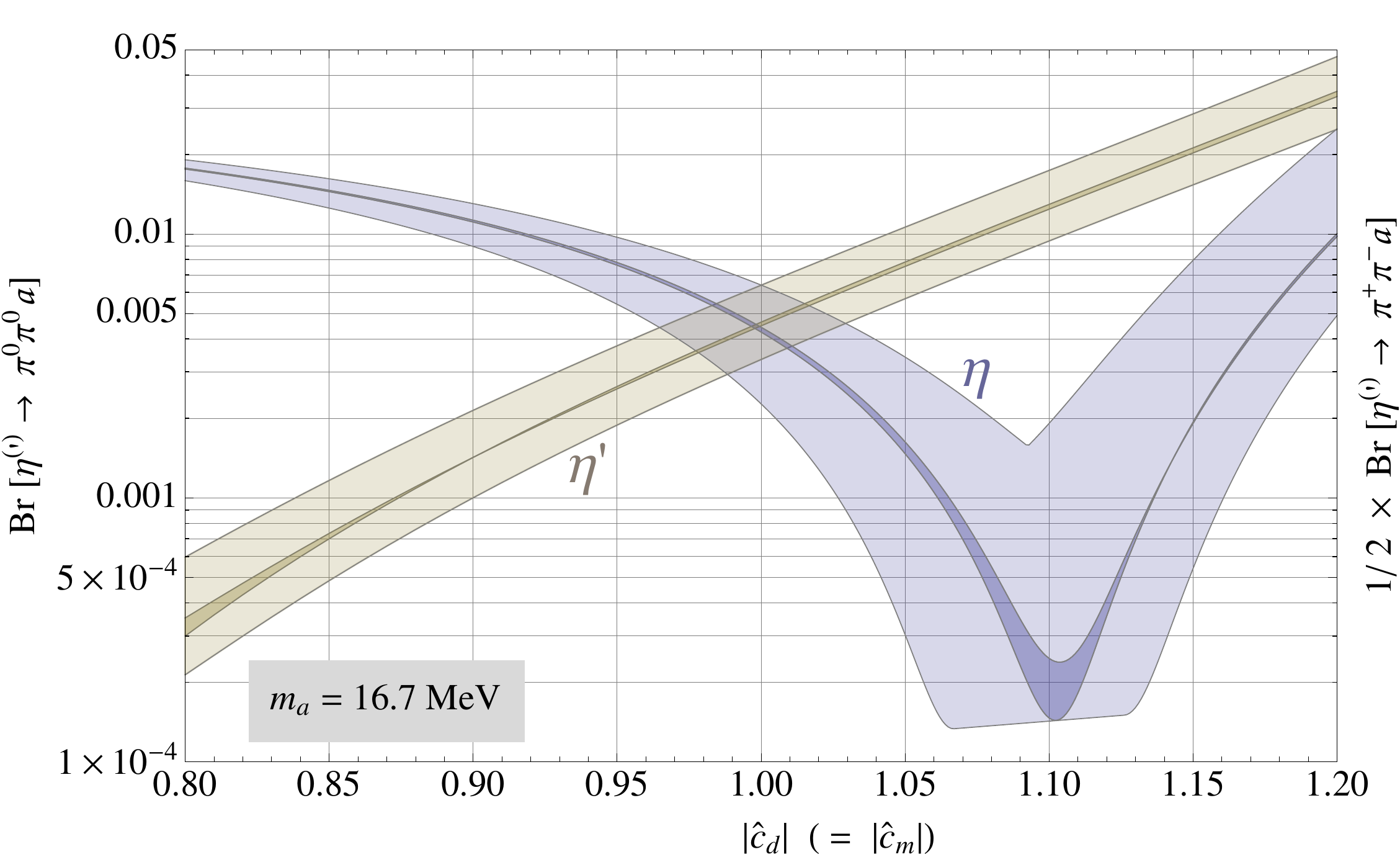}
\caption{Estimated branching ratios for $\eta^{(\prime)}\to\pi\,\pi\,a$ as a function of the scalar octet couplings to the light pseudoscalar mesons (taken from~\cite{Spier2021}). The width of the blue and brown bands result from varying the masses and widths of the scalar resonances, $a_0$ and $f_0$, within their experimental uncertainties. For the narrow bands with darker shading, the scalar masses were fixed to $m_{a_0}=m_{f_0}=980\,$MeV, and their widths were varied within the ranges $\Gamma_{a_0}= (40-100)\,$MeV, $\Gamma_{f_0}= (10-200)\,$MeV. The broader, lighter shaded bands resulted from additionally varying the scalar masses within the ranges $m_{a_0},\,m_{f_0}= (960-1000)\,$MeV.}\label{etaBR}
\end{figure}

Low-energy strong interaction effects, however, introduce large uncertainties to the calculation of axio-hadronic $\eta$ and $\eta^\prime$ decays. In~\cite{Spier2021}, this calculation was performed in the framework of {\it Resonance Chiral Theory} (R$\chi$T), an interpolating effective theory between the low-energy $\chi$PT framework and the microscopic QCD description. R$\chi$T encodes the most prominent features of nonperturbative strong dynamics by incorporating the low-lying QCD resonances and extending the principle of vector meson dominance~\cite{Ecker:1988te}. In this framework, exchanges of $0^{++}$ resonances such as the $a_0(980)$ and $f_0(980)$ contribute to the amplitude for $\eta^{(\prime)}\to \pi\pi a$, see Fig.\,\ref{etaDecayAmplitudes}. These contributions are of the same order of magnitude as the $\chi$PT quartic coupling contribution, and significant destructive interference between these amplitudes can take place within the expected range of couplings between $a_0(980)$, $f_0(980)$ and $\pi$, $a$, $\eta^{(\prime)}$. The resulting branching ratios for axio-hadronic $\eta^{(\prime)}$ decays can vary significantly depending on the degree of destructive interference between amplitudes. For the case of the piophobic QCD axion explaining the $^8\text{Be}$ and $^4\text{He}$ anomalies, the branching ratios for $\eta^{(\prime)}\to \pi\pi a$ are shown in Fig.\,\ref{etaBR} as a function $\widehat{c}_{d}$ and $\widehat{c}_{m}$, which parametrize the couplings between the octet scalars and the chiral mesons. Previous $\eta^{(\prime)}$-decay searches in $\pi\pi e^+e^-$ final states have not been able to probe this scenario. In fact, an excess of 27 events (vs 7.7 expected) in that mass region was observed by CELSIUS/WASA in the $\eta\rightarrow\pi^{+}\pi^{-}e^{+}e^{-}$ process~\cite{WASA07}. A similar excess was also observed by BESIII in the $\eta^{\prime}\rightarrow\pi^{+}\pi^{-}e^{+}e^{-}$ process~\cite{BESIII13}, although it was dismissed as background from $\gamma$-conversion. Despite the two-orders-of-magnitude variation in these branching ratios, they are fully within the expected REDTOP sensitivity to $\eta^{(\prime)}\to \pi\pi (a\to e^+e^-)$ final states. Therefore, REDTOP will definitively probe the remaining parameter space of the visible QCD axion.

\subsubsection{Axion-like particles\label{subsec:Axion-like-particles}}

More generally, axio-hadronic decays of the $\eta$ and $\eta^\prime$ can also probe axion-like particles (ALPs), which have the same types of interactions as the QCD axion but receive an additional, PQ-breaking contribution to their masses. As such, ALP models have a broader parameter space than the QCD axion since the ALP mass and decay constant are independent parameters.

The ALP interactions that contribute most significantly to axio-hadronic $\eta^{(\prime)}$ decays are:
\begin{equation}\label{LalpY}
\mathcal{L}_\text{ALP}^{(Y)}~\supset~-c_{GG}\,\frac{\alpha_s}{4\pi}\,\frac{a}{f_a}G\widetilde{G}~+~\sum_{q=u,d,s}~m_{q}\,~\bar{q}\,e^{i \, c_{q}\frac{a}{f_a}\gamma_5}q~-~\MPQ^2\,\frac{a^2}{2}\,,
\end{equation}
where $a$ denotes an ALP with decay constant $f_a$, $\MPQ$ is a PQ-breaking contribution to the ALP mass, and the quarks $q=u,d,s$ are written in the mass eigenstate basis. Furthermore, (\ref{LalpY}) is defined at the GeV scale, the heavy-flavor quarks $c,b,t$ having been integrated out. If they carry PQ charges $c_c$, $c_b$, and $c_t$, respectively, then they will implicitly contribute to the gluonic ALP coupling through
\begin{equation}\label{cGGcomposition}
c_{GG}~=~c_{GG}^\text{\tiny UV} ~+~ \sum_{\;q=c,b,t} \,\frac{c_{q}}{2}\,.
\end{equation}

By performing ALP-dependent quark chiral rotations, the ALP interactions in (\ref{LalpY}) can be recast in another commonly considered basis in the recent literature:
\begin{subequations}
\begin{alignat}{3}
\mathcal{L}_\text{ALP}^{(\partial)}~\supset~&-\left(c_{GG}+\sum_{q} \,\frac{c_{q}}{2}\right)\frac{\alpha_s}{4\pi}\,\frac{a}{f_a}G\widetilde{G}~+~\frac{\partial_\mu a}{f_a}~ \sum_{q}\, \frac{c_{q}}{2} \;\bar{q}\,\gamma^\mu\gamma_5\, q~-~\MPQ^2\,\frac{a^2}{2}\label{LalpD}\\
&~~+~\sum_{i,j}~\frac{g_2}{\sqrt{2}\,}\;V^{\text{\tiny CKM}}_{ij}~W_\mu^+~~\bar{u}_L^i\, e^{-i \Big(\!\frac{\,c_{u_i}-c_{d_j}\,}{2}\frac{a}{f_a}\gamma_5\!\Big)}\gamma^\mu\,d_L^j ~+~\text{h.c.}\label{LalpW}
\end{alignat}
\end{subequations}
Note that the ``Yukawa basis'' in (\ref{LalpY}) and the ``derivative basis'' in (\ref{LalpD}) are equivalent {\it as long as the weak interactions are neglected}. (The failure to account for the additional terms in (\ref{LalpW}) has led to much confusion in the literature and convoluted treatments of axio-hadronic Kaon decays, {\it e.g.,} \cite{Bauer:2021wjo}.)

However, because axio-hadronic $\eta^{(\prime)}$ decays do not violate flavor, this basis equivalency issue between (\ref{LalpY}) and (\ref{LalpD}) is irrelevant for the decays we will consider in this subsection. For the same reason, more generic derivative ALP couplings to axial and vector flavor-changing neutral currents are also irrelevant. Therefore, for the remainder of this subsection we will adopt the simpler, ``Yukawa basis'' parametrization of the ALP couplings in (\ref{LalpY}).

In order to extract the amplitude for $\eta^{(\prime)}\to\pi\pi a$ from (\ref{LalpY}), three basic steps are needed:
\begin{enumerate}
\item The QCD level ALP-interactions in (\ref{LalpY}) must be mapped into chiral perturbation theory ($\chi$PT), {\it i.e.}, they must be re-expressed in terms of ALP couplings to mesonic degrees of freedom.\label{step1}
\item The mass matrix and kinetic mixing terms must be diagonalized in order to obtain the physical ALP state $a_\text{\tiny phys}$ and the low energy meson states $\pi^0$, $\eta$, and $\eta^\prime$.\label{step2}
\item The $\chi$PT Lagrangian must be re-expressed in terms of the physical states, so that the quartic interactions $\eta^{(\prime)}$-$\pi$-$\pi$-$a_\text{\tiny phys}$ can be obtained.\label{step3}
\end{enumerate} 
 
A proper treatment and execution of steps \ref{step1}$-$\ref{step3} outlined above is still missing in the ALP literature. Firstly, to properly describe the octet-singlet composition of the $\eta$ and the $\eta^{\prime}$, as well as their mixing with the ALP, one needs to go beyond leading order in $\chi$PT. Secondly, (axio)-hadronic $\eta^{(\prime)}$ decays receive significant nonperturbative corrections which require going beyond $\chi$PT,  such as, {\it e.g.,} accounting for strong final state rescattering via dispersive relations, or including the exchange of the low energy QCD resonances via Resonance Chiral Theory. Here, we will not address either of these issues (work in preparation in these directions will appear in \cite{AlvesGonzalezSolis}). We will simply follow recent na{\" i}ve, leading order $\chi$PT treatments to obtain the amplitude for axio-hadronic $\eta^{(\prime)}$ decays. As such, our results should be considered a rough, order-of-magnitude estimate of the rate for $\eta^{(\prime)}\to\pi\pi a$ to qualitatively assess REDTOP's sensitivity reach to the parameter space of hadronic ALPs.

The implementation of step \ref{step1} above at leading order in $\chi$PT amounts to mapping (\ref{LalpY}) into:
\begin{equation}\label{LalpChiPT}
\mathcal{L}_\text{ALP}^{(\chi\text{PT}\text{@LO})}\,=~-\frac{M_0^2}{2}\left(\eta_0-\frac{c_{GG}}{\sqrt{3/2}\,}\frac{f_\pi}{f_a}\,a\right)^2\,+\;\frac{f_\pi^2}{4}\,\text{Tr}\big[2B_0 M_q(a) U\,+\,\text{h.c.}\big]\,-\,\MPQ^2\frac{a^2}{2}\,,
\end{equation}
where $M_0\sim\mathcal{O}$(GeV) parametrizes the large strong anomaly contribution to the mass of the $\eta^\prime$, $B_0=-\langle q\bar{q}\rangle/f_\pi^2$, $M_q(a)$ is the ALP-dependent quark mass matrix,
\begin{equation}
M_q(a)\equiv
\begin{pmatrix} 
\,m_u\,e^{i\,c_u a/f_a}\!\!\! & & \\
  &\!\!\! m_d\,e^{i\,c_d\,a/f_a}\!\!\!  &    \\
  &  &\!\!\! m_s\,\,e^{i\,c_s\,a/f_a}
\end{pmatrix},
\end{equation}
and $U$ denotes the non-linear representation of the chiral meson nonet,
\begin{eqnarray}\label{U80}
U~=~  \text{Exp}\frac{i}{f_\pi}\,
\begin{pmatrix} 
\vspace{0.17cm}
\pi^0+\frac{\eta_8}{\sqrt{3}}+\frac{\eta_0}{\sqrt{3/2}} & \sqrt{2}\,\pi^+ & \sqrt{2}\,K^+ \\
\vspace{0.17cm}
\sqrt{2}\,\pi^-  & -\pi^0+\frac{\eta_8}{\sqrt{3}}+\frac{\eta_0}{\sqrt{3/2}}  &  \sqrt{2}\,K^0  \\
  \sqrt{2}\,K^- & \sqrt{2}\,\overline{K}^0 & -\frac{2\,\eta_8}{\sqrt{3}}+\frac{\eta_0}{\sqrt{3/2}}~
\end{pmatrix}.
\end{eqnarray}
%
%
Even though we are restricting our calculation to the na{\" i}ve leading order $\chi$PT treatment, we will introduce a small improvement by adopting the two-mixing-angle scheme to express the physical $\eta$ and $\eta^\prime$ states in the octet-singlet basis as
\begin{subequations}
\begin{alignat}{2}\label{etaTwoMixingScheme}
\eta_0~=&~\;\frac{f_\pi}{f_0}\,\big(\!\cos\theta_0\,\eta^\prime\,-\,\sin\theta_0\,\eta\big)\,,\\
\eta_8~=&~\;\frac{f_\pi}{f_8}\,\big(\!\sin\theta_8\,\eta^\prime\,+\,\cos\theta_8\,\eta\big)\,.
\end{alignat}
\end{subequations}
For concreteness, we will take the values $\theta_8=-24^\circ$, $\theta_0=-2.5^\circ$, $f_8=1.51\,f_\pi$, and $f_0=1.29\,f_\pi$ from the unconstrained fit in \cite{Escribano:2005qq}.

Step \ref{step2} is the non-trivial part of our calculation, and will involve obtaining the ALP mixing angles $\theta_{a\pi}$, $\theta_{a\eta_8}$, $\theta_{a\eta_0}$ with $\pi^0$, $\eta_8$, $\eta_0$, respectively. Because of our judicious choice of ALP basis with no kinetic mixing terms, our work is greatly simplified: the ALP-meson mixing angles can be simply obtained by diagonalizing the ALP-meson mass matrix in (\ref{LalpChiPT}). While this can be done straightforwardly numerically, a parametric dependence of the mixing angles on $m_a$, $f_a$, $c_q$ and $c_{GG}$ is desirable. This can be obtained by considering the following.

First, in the PQ preserving limit of $\MPQ=0$ and $m_a\ll m_\pi$ ({\it i.e.,} when $a$ is the QCD axion), we have
\begin{equation}
(m_a^{\text{\tiny (PQ)}})^2 ~=~ \frac{(c_u+c_d+c_s+2\,c_{GG})^2}{(1+\epsilon_{\eta\eta^\prime})}\,\frac{m_um_d}{(m_u+m_d)^2}\,\frac{m_\pi^2\, f_\pi^2}{f_a^2}\,,
\end{equation}
and
\begin{subequations}
\begin{alignat}{2}
\theta_{a\pi}^{^{\text{\tiny (PQ)}}} =& -\frac{f_\pi}{f_a}\,\frac{1}{(1+\epsilon_{\eta\eta^\prime})}\,\bigg(\frac{\,(c_u m_u-c_d m_d)\,+\,(c_s+2c_{GG})(m_u-m_d)/2\,}{m_u+m_d}+\epsilon_{\eta\eta^\prime}\,\frac{\,(c_u-c_d)\,}{2}\bigg)\\
\theta_{a\eta_8}^{^\text{\tiny (PQ)}} =&~~ \frac{f_\pi}{f_a}\,\frac{\sqrt{3}}{~2}\;\frac{1}{(1+\epsilon_{\eta\eta^\prime})}\;\bigg(c_s+\frac{2}{3}\,c_{GG}-\epsilon_{\eta\eta^\prime}\frac{\,(c_u+c_d+4c_{GG}/3)+(c_u+c_d-2c_s)\,2m_{\!K}^2/m_{\!\eta^\prime}^2\,}{1+6m_{\!K}^2/m_{\!\eta^\prime}^2}\bigg)\\
\theta_{a\eta_0}^{^\text{\tiny (PQ)}} =&~~\frac{f_\pi}{f_a}\;\bigg(\frac{c_{GG}}{\sqrt{3/2}\,}\,-\,\frac{m_\pi^2}{m_{\eta^\prime}^2}\,\frac{\sqrt{6}\,(c_u+c_d+c_s+2c_{GG})}{1+\epsilon_{\eta\eta^\prime}}\frac{m_um_d}{(m_u+m_d)^2}\bigg)
\end{alignat}
\end{subequations}
where
\begin{equation}
\epsilon_{\eta\eta^\prime}~\equiv~\frac{m_um_d}{m_s(m_u+m_d)}\;\bigg(1+6\,\frac{m_K^2}{m_{\eta^\prime}^2}\bigg)~\simeq~0.04\,.
\end{equation}
This can be generalized to the PQ-breaking case of a generic ALP as
\begin{equation}
m_a^2~=~(m_a^{\text{\tiny (PQ)}})^2 \,+\,\MPQ^2\,,
\end{equation}
and
\begin{subequations}
\begin{alignat}{2}
&\theta_{a\pi} = \theta_{a\pi}^{^\text{\tiny (PQ)}}\bigg(1\,+\,\frac{m_a^2}{m_\pi^2-m_a^2}\bigg)\label{ApiMix}\\
&\theta_{a\eta_8} = \frac{1}{f_8}\,\bigg[\!\cos{\theta_8}\bigg(\frac{f_8\cos{\theta_0}\,\theta_{a\eta_8}^{^\text{\tiny (PQ)}}-f_0\sin{\theta_8}\,\theta_{a\eta_0}^{^\text{\tiny (PQ)}}}{\cos{(\theta_8-\theta_0)}}\bigg)\bigg(1\,+\,\frac{m_a^2}{\,m_\eta^2-m_a^2\,}\bigg)\,+\nonumber\\
&\qquad\qquad\quad\sin{\theta_8}\bigg(\frac{f_0\cos{\theta_8}\,\theta_{a\eta_0}^{^\text{\tiny (PQ)}}+f_8\sin{\theta_0}\,\theta_{a\eta_8}^{^\text{\tiny (PQ)}}}{\cos{(\theta_8-\theta_0)}}\bigg)\bigg(1\,+\,\frac{m_a^2}{\,m_{\eta\prime}^2-m_a^2\,}\bigg)\bigg]\label{Aeta8Mix}\\
&\theta_{a\eta_0} = \frac{1}{f_0}\,\bigg[\!-\sin{\theta_0}\bigg(\frac{f_8\cos{\theta_0}\,\theta_{a\eta_8}^{^\text{\tiny (PQ)}}-f_0\sin{\theta_8}\,\theta_{a\eta_0}^{^\text{\tiny (PQ)}}}{\cos{(\theta_8-\theta_0)}}\bigg)\bigg(1\,+\,\frac{m_a^2}{\,m_\eta^2-m_a^2\,}\bigg)\,+\nonumber\\
&\qquad\qquad\quad~~~\cos{\theta_0}\bigg(\frac{f_0\cos{\theta_8}\,\theta_{a\eta_0}^{^\text{\tiny (PQ)}}+f_8\sin{\theta_0}\,\theta_{a\eta_8}^{^\text{\tiny (PQ)}}}{\cos{(\theta_8-\theta_0)}}\bigg)\bigg(1\,+\,\frac{m_a^2}{\,m_{\eta\prime}^2-m_a^2\,}\bigg)\bigg].\label{Aeta0Mix}
\end{alignat}
\end{subequations}
Expression (\ref{ApiMix}) for the ALP-pion mixing angle $\theta_{a\pi}$ holds as long as $m_a$ is not too close to $m_{\pi^0}$. Similarly, (\ref{Aeta8Mix}) and (\ref{Aeta0Mix}) hold as long as $m_a$ is not too close to either $m_{\eta}$ or $m_{\eta^\prime}$.

We can finally proceed to step \ref{step3} by re-expressing the ALP Lagrangian in terms of the physical ALP:
\begin{equation}
\mathcal{L}_\text{ALP}^{(\chi\text{PT}\text{@LO})}\bigg|_{\pi^0\,\to\,\pi^0+\theta_{a\pi}a~~,~~\eta_8\,\to\,\eta_8+\theta_{a\eta_8}a~~,~~\eta_0\,\to\,\eta_0+\theta_{a\eta_0}a~~,~~a\,\to\, a-\theta_{a\pi}\pi^0-\theta_{a\eta_8}\eta_8-\theta_{a\eta_0}\eta_0}\,,
\end{equation}
from which we obtain the leading order $a$-$\pi$-$\pi$-$\eta^{(\prime)}$ quartic couplings,
\begin{equation}\label{etapipiALPLquartic}
\mathcal{L}_\text{ALP}^{(\chi\text{PT}\text{@LO})}~\supset~\lambda_0\,a\,\left(\pi^+\pi^-+\frac{\pi^0\pi^0}{2}\right)\left(\!\frac{\eta_8}{\sqrt{3}\,}+\frac{\eta_0}{\sqrt{3/2}\,}\!\right)\,,
\end{equation}
where
\begin{equation}\label{etapipiALPquartic}
\lambda_0\,~=~\frac{m_\pi^2}{f_\pi^2}\,\left(\frac{\,(c_um_u+c_dm_d)\,}{(m_u+m_d)}\frac{f_\pi}{f_a}\,+\,\frac{\,(m_u-m_d)\,}{(m_u+m_d)}\,\theta_{a\pi}\,+\,\frac{\theta_{a\eta_{8}}}{\sqrt{3}\,}+\,\frac{\theta_{a\eta_{0}}}{\sqrt{3/2}\,}\right)\,.
\end{equation}

We can now turn our focus specifically to $\eta\to\pi\pi a$.  The amplitude for this decay then follows from (\ref{etapipiALPLquartic}) and (\ref{etapipiALPquartic}) straightforwardly:
\begin{eqnarray}\label{etapipiALPamplitude}
\mathcal{A}_{\eta\to\pi\pi a}~&\equiv&~\mathcal{A}(\eta\to\pi^0\pi^0 a)~=~\mathcal{A}(\eta\to\pi^+\pi^- a)\nonumber\\
&=&~\lambda_0\,\left(\frac{f_\pi}{f_8}\,\frac{\cos\theta_8}{\sqrt{3}}\,-\,\frac{f_\pi}{f_0}\,\frac{\sin\theta_0}{\sqrt{3/2}}\right).
\end{eqnarray}
Since the decay amplitude (\ref{etapipiALPamplitude}) is flat in the Dalitz phase-space, the differential rate for $\eta\to\pi\pi a$ as a function of the ALP 3-momentum $|\vec{p}_a|$ in the $\eta$ rest frame has a closed analytical form:
\begin{eqnarray}
\frac{d\Gamma}{d|\vec{p}_a|}~&=&~\frac{1}{S_{\pi\pi}}\,\frac{|\mathcal{A}_{\eta\to\pi\pi a}|^2}{(4\pi)^3}\;\frac{|\vec{p}_a|^2}{m_\eta\,E_a}\,\sqrt{1-\frac{4\,m_\pi^2}{(p_\eta-p_a)^2}}\\
&=&~\frac{1}{S_{\pi\pi}}\,\frac{|\mathcal{A}_{\eta\to\pi\pi a}|^2}{(4\pi)^3}\;\frac{|\vec{p}_a|^2}{m_\eta}\,\sqrt{\frac{m_\eta^2+m_a^2-2\,m_\eta\sqrt{m_a^2+|\vec{p}_a|^2}-4\,m_\pi^2}{(\,m_\eta^2+m_a^2-2\,m_\eta\sqrt{m_a^2+|\vec{p}_a|^2\,}\,)(\,m_a^2+|\vec{p}_a|^2)}}\nonumber\,,
\end{eqnarray}
where the combinatorial factor $S_{\pi\pi}$ takes the values $S_{\pi^0\pi^0}=2$ and $S_{\pi^+\pi^-\!}=1$.

To assess REDTOP's reach in the ALP parameter space, we consider two common benchmark scenarios that have been broadly adopted in Snowmass studies of ALPs:
\begin{itemize}
\item \emph{Gluon dominance,} which assumes that the ALP couples predominantly to heavy BSM colored fermions at UV scales above $f_a$. Once these fermions are integrated out, the ALP couplings expressed in terms of (\ref{LalpY}) are given by $c_{GG}^\text{\tiny UV}\neq 0$ and $c_q=0$ for all six SM quark flavors.
\item \emph{Quark dominance,} which assumes that the ALP couples predominantly to SM quarks, with no BSM contributions to the gluonic ALP coupling, {\it i.e.}, $c_{GG}^\text{\tiny UV}=0$. We further restrict the parameter space of this benchmark scenario by imposing flavor blindness, {\it i.e.,} we assume that all six quark flavors couple identically to the ALP, as parametrized by a universal coupling $c_q$ in (\ref{LalpY}) and (\ref{cGGcomposition}).
\end{itemize}

\begin{figure}[t]
\centering
\includegraphics[width=0.75\textwidth]{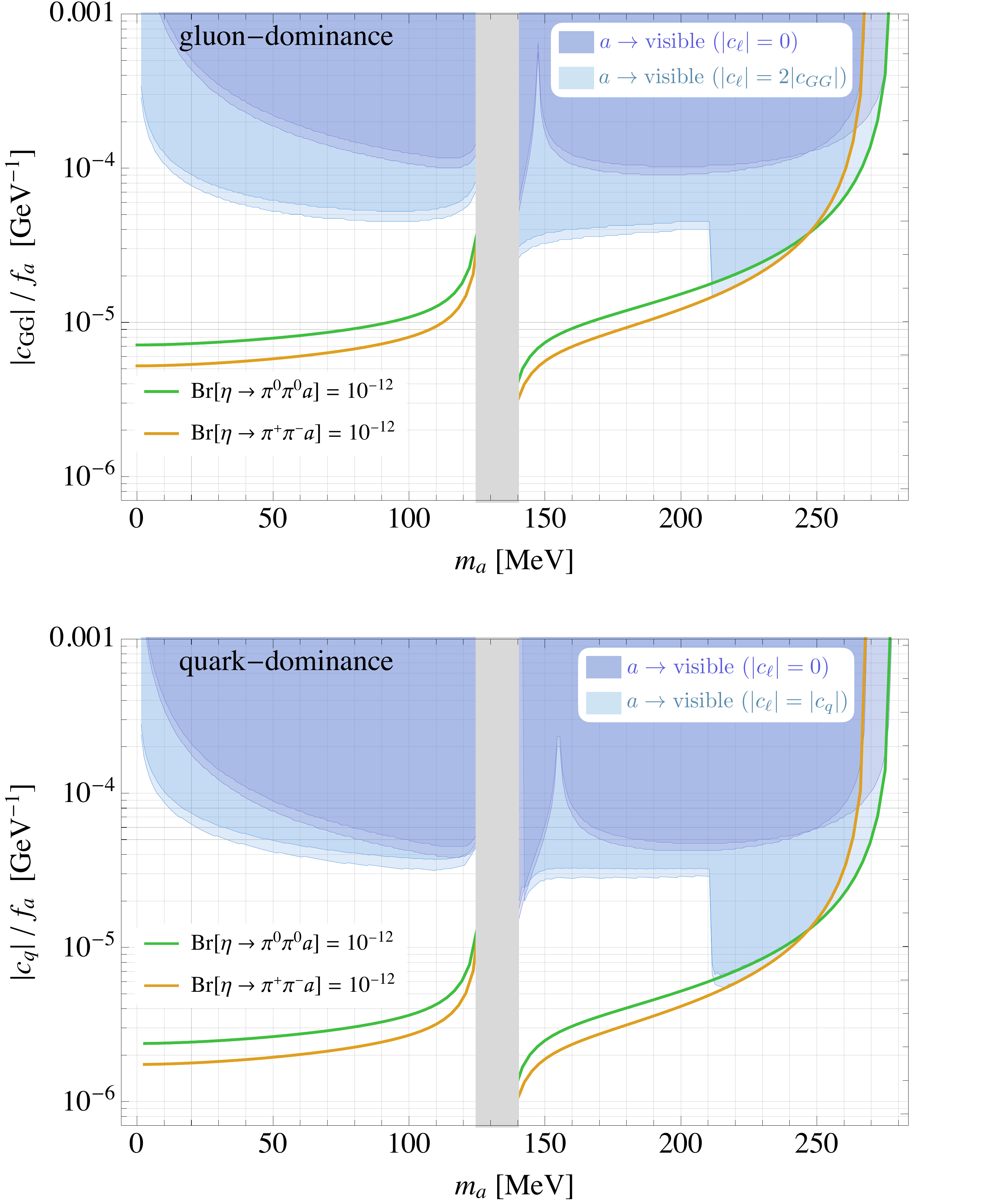}
\caption{Reach to the ALP parameter space by probes of axio-hadronic $\eta$ decays, such as REDTOP, assuming a branching ratio sensitivity of $\text{Br}[\eta\to \pi\pi a] = 10^{-12}$ for all benchmark scenarios. {\bf Top panel:} \emph{gluon dominance} scenario, with the green and orange curves denoting the reach to visible$^*$ and invisible ALP decays; the dark blue shaded region denoting sensitivity to visible$^*$ $a\to\gamma\gamma$ decays of a leptophobic ALP; and the light blue shaded region denoting sensitivity to visible$^*$ $a\to\ell^+\ell^-,\gamma\gamma$ decays of a leptophilic ALP.
~{\bf Bottom panel:} analogous to the top panel, but assuming the \emph{quark dominance} scenario.~~$^*$In these plots, the ALP decay is considered visible if the ALP decay vertex is displaced by less than 100 cm from the $\eta$ decay vertex in the $\eta$ rest frame.}\label{reachPlotALPs}
\end{figure}

Fig.\,\ref{reachPlotALPs} illustrates REDTOP's futuristic reach in the ALP parameter space for the \emph{gluon dominance} and \emph{quark dominance} benchmark scenarios, assuming branching sensitivities of $\text{Br}[\eta\to \pi^0\pi^0 a] = 10^{-12}$ (green curves) and $\text{Br}[\eta\to \pi^+\pi^- a] = 10^{-12}$ (orange curves). It is important to emphasize, however, that these futuristic projections assume that REDTOP would be sensitive to prompt, displaced, and invisible ALP decays alike.

For a more realistic near term sensitivity assuming the current REDTOP detector design, we can also estimate REDTOP's reach to \emph{visibly}-decaying ALPs, {\it i.e.,} ALPs that decay to visible final states within the detector's fiducial volume. We will make the simplified assumption that ALPs decaying within 100 cm of the $\eta$ decay vertex in the $\eta$ rest frame are visibly-decaying ALPs.

To proceed, further considerations are needed regarding the ALP couplings to photons and leptons in addition to its hadronic couplings defined in (\ref{LalpY}):
\begin{equation}\label{LalpYEM}
\mathcal{L}_\text{ALP}^{(Y)}~~\supset~~c_{\gamma\gamma}\,\frac{\alpha}{4\pi}\,\frac{a}{f_a}F\widetilde{F}~+~\sum_{\ell=e,\mu,\tau}~m_{\ell}\,~\bar{\ell}\,e^{i \, c_{\ell}\frac{a}{f_a}\gamma_5}\ell\,.
\end{equation}
In particular, the ALP coupling to photons receives contributions from high UV scale dynamics, as well as from its mixing with the neutral pseudoscalar mesons and its couplings to heavy quarks and charged leptons,
\begin{eqnarray}
c_{\gamma\gamma}~=~c_{\gamma\gamma}^\text{\tiny UV}~&+&~\frac{f_a}{f_\pi}\left(\theta_{a\pi}+\frac{5}{3}\,\theta_{a\eta_{_{ud}}}+\frac{\sqrt{2}}{~3}\,\theta_{a\eta_{_{s}}}\right)\\
&+&~\sum_{q=c,b,t} 3\,c_q\,(Q^\text{\tiny em}_q)^2\;\left|F_{1/2}(\tau_q)\right|~+ \sum_{\ell=e,\mu,\tau} c_\ell\,(Q^\text{\tiny em}_\ell)^2\;\left|F_{1/2}(\tau_\ell)\right|\,,\nonumber
\end{eqnarray}
where $\tau_f\equiv 4m_f^2/m_a^2$, and
\begin{equation}
F_{1/2}(\tau)~\equiv~-\tau f(\tau)\,,\qquad~~ f(\tau)=
\begin{cases}
&\!\!\!\!~~ \left(\sin^{-1}\!\sqrt{1/\tau}\right)^2~~~~~~~~~~~~ \text{if}~~\tau\geq1\,,\\
&\!\!\!\!\! -\frac{1}{4}\left[\ln\left(\frac{1+\sqrt{1-\tau}}{1-\sqrt{1-\tau}}\right)-i\pi\right]^2~~\text{if}~~\tau<1\,.
\end{cases}
\end{equation}

Since we are considering ALPs in the context of axio-hadronic $\eta$ decays, they will never be heavy enough to decay hadronically. Therefore, the ALP decay width will be given by:
\begin{equation}\label{alpWidth}
\Gamma_a~=~\Gamma(a\to\gamma\gamma)~+\;\sum_{\ell}\;\Gamma(a\to\ell^+\ell^-)\,,
\end{equation}
where
\begin{equation}\label{alpPhotonWidth}
\Gamma(a\to\gamma\gamma)~=~\frac{m_a^3}{4\pi}\,\left(\! c_{\gamma\gamma}\;\frac{\alpha}{4\pi f_a}\right)^{\!2}\,,
\end{equation}
and the sum over lepton flavors in (\ref{alpWidth}) of course only runs over leptons for which the decay $a\to\ell^+\ell^-$ is kinematically allowed, {\it i.e.,} for $m_\ell < m_a/2$, so that
\begin{equation}\label{alpLepWidth}
\Gamma(a\to\ell^+\ell^-)~=~\frac{m_a}{\,8\pi\,}\,\left(\frac{c_\ell\,m_\ell}{f_a}\right)^{\!2}\,\sqrt{1-\frac{4\,m_\ell^2}{m_a^2}}\,.
\end{equation}

With (\ref{LalpYEM})-(\ref{alpLepWidth}), we can now complement our definition of the \emph{gluon dominance} and \emph{quark dominance} benchmark scenarios in the following way:
\begin{itemize}
\item \emph{Gluon dominance, leptophobic,}
\begin{equation}
c_{GG}^\text{\tiny UV}\neq 0,~~~~c_{q}= 0~ \forall~q,~~~~c_{\ell}=0~ \forall~\ell,~~~~c_{\gamma\gamma}^\text{\tiny UV}=0.
\end{equation}
\item \emph{Gluon dominance, flavor universal, leptophilic,} 
\begin{equation}
c_{GG}^\text{\tiny UV}\neq 0,~~~~c_q=0~ \forall~q,~~~~c_{\ell}=2c_{GG}^\text{\tiny UV}~ \forall~\ell,~~~~c_{\gamma\gamma}^\text{\tiny UV}=0.
\end{equation}
\item \emph{Quark dominance, flavor universal, leptophobic,}
\begin{equation}
c_{GG}^\text{\tiny UV}= 0,~~~~c_{q}\neq 0~ \forall~q,~~~~c_\ell=0~ \forall~\ell,~~~~c_{\gamma\gamma}^\text{\tiny UV}=0.
\end{equation}
\item \emph{Quark dominance, flavor universal, leptophilic,} 
\begin{equation}
c_{GG}^\text{\tiny UV}= 0,~~~~c_{q}\neq 0~ \forall~q,~~~~c_{\ell}=c_q~ \forall~\ell,~~~~c_{\gamma\gamma}^\text{\tiny UV}=0.
\end{equation}
\end{itemize}

Fig.\,\ref{reachPlotALPs} also illustrates the REDTOP sensitivity reach in these four scenarios (shaded regions) assuming a \emph{visible} branching ratio sensitivity of $\text{Br}[\eta\to \pi\pi (a\to\text{visible})] = 10^{-12}$.

\subsubsection{Probing a BSM origin of the proton radius anomaly}

One of the existing and still unexplained anomalies present in the
Standard Model is related to the measurements of the proton radius
$R_{p}$ with electron and muon probes.

The processes involved are $\eta$-Dalitz decays, and lepton-pair decays:

\begin{equation}
\eta\rightarrow\gamma e^{+}e^{-}\, , \eta\rightarrow\gamma\mu^{+}\mu^{-} \text{ and } \eta\rightarrow e^{+}e^{-}\, , \eta\rightarrow\mu^{+}\mu^{-}  \label{eq:proton_radius}
\end{equation}

\begin{figure}[!htbp]
\centering{}\includegraphics[width=0.6\columnwidth]{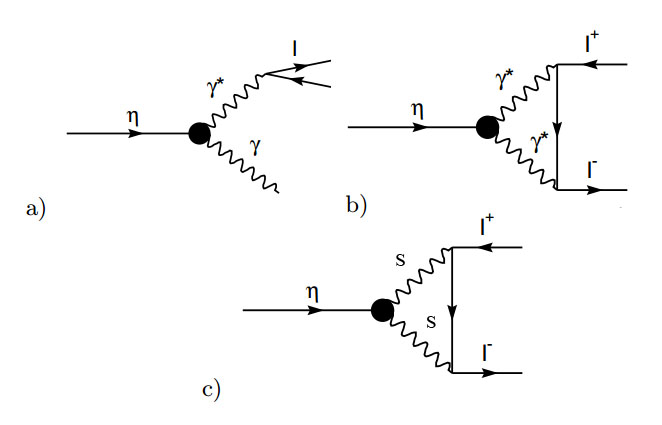}
\caption{\label{fig:rp} Diagrams contributing to the proton radius}
\end{figure}

Other types of $R_{p}$ measurements, using muonic atoms (see, for
example, CODATA-2012), or using elastic scattering of electrons and
muons on hydrogen atoms, have found a discrepancy corresponding to
several sigma in the electron vs muon case. It is worth noting that
such processes occur mainly through the exchange of one virtual photon, but subleading corrections to that---needed to better clarify the situation---may include pseudoscalar-exchange
contribution to the 2S hyperfine splitting in the muonic hydrogen, in a \textit{t-channel} contribution. One would expect the $\pi^0$ to dominate such subleading correction but being a $t$-channel exchange, all pseudoscalars may contribute, 
being the  processes of the type $\eta(\eta{'})\rightarrow\gamma l^{+}l^{-}$ a background to them, as shown in Fig.~\ref{fig:rp} diagram
a), versus the two-photon contribution, as shown in Fig.~\ref{fig:rp} diagram b).

A light scalar particle S with a different coupling with electrons
and muons would mediate this process, as shown in diagram c) of figure
\ref{fig:rp}, would explain this anomaly of the Standard Model. Therefore,
an experiment able to precisely measure the branching ratios of this
particle might help in explaining the $R_{p}$ anomaly.

In REDTOP, the processes (\ref{eq:proton_radius}) are detected simultaneously
and within the same experimental apparatus. Consequently, most of
the systematic errors are common to the two processes and they factor
out in the ratio of the corresponding branching ratios, enhancing
the precision of the overall measurement. $\eta^{'}$ contributions may play a role as well.

\subsection{\label{subsec:Tests of Conservation Laws}Tests of conservation laws}
Conservation laws with their underlying symmetry principles are at
the heart of physics, from the classical space-time conservation laws
of introductory courses through the symmetries and additive quantum
numbers of modern particle physics. The Crystal Ball experiment at
the Brookhaven AGS was able to provide a few times $10^{7}$ $\eta$
(as for the $4\pi^{0}$ decay study). It was subsequently moved to
MAMI, and a goal there is to achieve another order of magnitude in
$\eta$ yield. Other facilities include KLOE (for $\phi(1020)\to\eta\gamma$),
and GlueX at JLAB, all at the few times $10^{7}$ level. Recently,
WASA at COSY reached a milestone, by collecting about $10^{9}$, although
with conventional, background prone, detector technologies, still
insufficient for exploring successfully the realm BSM. To reach the
more exacting levels needed for symmetry violations, the usable $\eta$
flux must be increased by several orders of magnitude.

To achieve this goal, the REDTOP experiment is being designed to provide
a sea change in the number of $\eta$ samples to $1.1\times10^{14}$
and $\eta^{\prime}$ samples to $1\times10^{12}$, along
with a nearly $4\pi$ detector to study a broad range of fore-front
physics. The facility will provide vastly reduced upper limits for
$\eta$ and $\eta^{\prime}$ decays, as well as studies of processes that
can lead to New Physics Beyond the Standard Model.

The light pseudoscalar mesons $\pi^{0}$, $\eta$, and $\eta^{\prime}$ have
very special roles for exploring and testing the conservation laws.
The $\pi^{0}$ has a long history of such tests and has established
tight upper limits of charge ($C$) and lepton flavor ($LF$) violations
\cite{PDGC}. Unlike the isospin $I=1$ for the $\pi^{0}$, all the
additive quantum numbers for the $\eta$ and $\eta^{\prime}$ are zero, and
they differ from the vacuum only in terms of parity. Due to the opposite
$G$ parities of the the $\pi^{0}$ and $\eta$, couplings to strong
interactions are suppressed. Thus, tests of $C$ and $CP$ in electromagnetic
interactions are much more directly accessible in $\eta$ and $\eta^{\prime}$
decays, limited mainly by the flux of such mesons~\cite{Nef94}. In
addition, such decays can provide tests of $P$, $T$, $CT$, $PT$,
and even $CPT$. Among other possibilities are searches for lepton
family violation, leptoquarks, and significant tests of the parameterization
of chiral perturbation theory.

Almost all searches for symmetry violations in $\eta$/$\eta^{\prime}$ decays
are upper limits in the range of $10^{-5}$ or higher~\cite{PDGC}. An
exception is the decay $\eta\to4\pi^{0}$ at $<6.9\times10^{-7}$,
based on $3\times10^{7}$ $\eta$ mesons~\cite{Prak00}. One-sigma errors
have been reported for some asymmetries in the Dalitz distribution
of $\eta\to\pi^{+}\pi^{-}\pi^{0}$ (which are consistent with zero
at the level of $10^{-2}$)~\cite{Amb08}. Most models of symmetry violations
for various decay processes are at or below the level of $10^{-9}$,
typically by several orders of magnitude.\smallskip{}


\emph{CP} violation has been extensively studied in the flavor-changing
decays of the neutral $K$- and $B$-mesons. The origin of the violation
is still not fully understood. The standard model predicts that the
source of \emph{CP} violation is a single phase in the Cabbibo-Kobayashi-Maskawa
(CKM) mixing matrix of quarks couplings. At present, the predictions
based on CKM mechanism are consistent with the observations in $K$ and
$B$ systems, but tensions are arising. We propose to explore other sources of \emph{CP} violation
beyond the CKM mechanism and with flavor-conserving processes, especially through
measurements not bound by EDM limits. Rare $\eta$/$\eta^{\prime}$ decays
provide a good laboratory for that. All decays of the $\eta$/$\eta^{\prime}$
mesons into multiple  photons or into photons plus $\pi^0$s provides
a direct test  of $C$ invariance.  Each photon in the final state,
including the two from $\pi^0$ decay, has $C = -1$.  Because the
$\eta$  has $C = +1$, final states with an odd number of photons
are  forbidden. However, the branching ratio for these processes are
bound by EDM measurements and they would explore aspects of \emph{CP}
violations not accessible even at a $\eta$-factory. We propose, instead,
studying three processes that are not bound by current EDM measurements
and that are probing different operators which would induce a violation of
\emph{CP} from sources Beyond the Standard Model.

Several $\eta$/$\eta^{\prime}$-related processes have been selected to study
the sensitivity of REDTOP to Conservation Laws. These studies are
restricted, for now, only to the CP violation (CP) and Lepton Flavor
violation, where more solid theoretical models exists which are consistent,
at the same time, with bounds from EDM measurements and with outstanding
experimental anomalies. These are discussed below, alongside with
the theoretical models supporting them.

\subsubsection{\label{subsec:CP-Violation-from-Dalitz}\texorpdfstring{\mbox{C and CP violation from Dalitz asymmetries in 
$\eta\rightarrow\pi^{+}\pi^{-}\pi^{0}$}}{}} 


The decay $\eta\to \pi^+\pi^-\pi^0$ can only occur when isospin or/and charge conjugation (C) is/are broken. Thus the interference of a C-conserving but isospin-breaking amplitude with a C-violating one would give rise to a charge asymmetry in the 
Dalitz plot of the 
3$\pi$ final state for this process. 
Since parity P is conserved in this decay, 
the existence of a nonzero charge asymmetry would 
attest to the breaking of C {\it and} CP symmetry.
In contrast, searches for a nonzero permanent electric
dipole moment (EDM) probe the possibility of 
new sources of P {\it and} CP violation. 
Moreover, the SM mechanism of CP violation, 
vis-a-vis the flavor-changing weak interactions 
of quarks, is expected to be 
completely
negligible in this context. 
Although the charge asymmetry observable
was first proposed long ago~\cite{Lee:1965hi,Nauenberg:1965asym,Lee:1965zza}, we revisit
and refine it not only because it is 
a flavor-diagonal, 
C- and CP-violating observable, but it is also, 
moreover, an effect that scales 
linearly, rather than quadratically, 
in the underlying CP-violating
parameter~\cite{gardner2019patterns}. 
Moreover, a study of the new physics sources 
in Standard Model Effective Field Theory (SMEFT), starting from the compilation of Ref.~\cite{Grzadkowski:2010es}, shows
that the sources of CP violation that stem from 
C- or P- violating effects in the 
flavor diagonal sector are distinct~\cite{Shi:2020ffh,gardnershi2022}. 
A charge asymmetry in the 
$B^0 (\bar{B}^0) \to \pi^+ \pi^- \pi^0$ 
Dalitz plot also probes CP violation~\cite{Gard04}, 
and the SMEFT operators in that case are mass dimension
six~\cite{Shi:2020ffh,gardnershi2022}. In our case
the SMEFT operators are mass dimension eight, but could
be of dimension six in numerical size, which would signal
the existence of 
dynamics beyond SMEFT~\cite{Shi:2020ffh,gardnershi2022}. 
(We note Ref.~\cite{Burgess:2021ylu} for 
a discussion of analogous ideas in $b\to c\tau \bar{\nu}$
decay.)
Thus the measurement of the charge
asymmetry 
in $\eta\to \pi^+\pi^-\pi^0$ decay is an 
ideal probe with which to 
explore the possibility of 
physics beyond the SM. 
The REDTOP experimental concept is 
well suited to careful measurements of 
the charge asymmetry 
observable, and we consider it 
a {\em golden channel} for study 
at REDTOP. 

Recently, theoretical work has been  done~\cite{gardner2019patterns,Akdag:2021efj} 
investigating the patterns of C and CP violation in this process from mirror asymmetry breaking of the Dalitz plot. 
As long known, when plotting the Dalitz plot in terms of Mandelstam variables $t\equiv (p_{\pi^-} + p_{\pi^0})^2$ and $u\equiv (p_{\pi^+} + p_{\pi^0})^2$, the charge asymmetry would correspond to a breaking of mirror symmetry, i.e., $t \leftrightarrow u$ exchange, in the Dalitz plot. The charge asymmetry can be probed through the measurement of a left-right asymmetry, $A_{LR}$~\cite{Lay72}
\begin{equation}
    A_{LR} \equiv \frac{N_+ - N_-}{N_+ + N_-},\label{asym_LR}
\end{equation}
where $N_\pm$ is the number of events when $u \stackrel{>}{{}_{<}} t$, i.e., the $\pi^+$ has more (less) energy than the $\pi^-$ in the $\eta$ rest frame. More conveniently, we can describe the Dalitz plot in terms of variables $X$ and $Y$ 
\begin{eqnarray}
 X &\equiv& \sqrt{3} \frac{T_{\pi^+} - T_{\pi^-}}{Q_{\eta}} =\frac{\sqrt{3}}{2 M_{\eta} Q_{\eta}} (u - t), \notag\\
 Y &\equiv& \frac{3 T_{\pi^0}}{Q_{\eta}} - 1 = \frac{3}{2 M_{\eta} Q_{\eta}} [(M_{\eta} - M_{\pi^0})^2 - s] - 1\,, \label{eq:eta3piXY}
\end{eqnarray}
where 
$ Q_{\eta} = T_{\pi^+} + T_{\pi^-} + T_{\pi^0} = M_{\eta} - 2 M_{\pi^+} -
   M_{\pi^0} $, and $T_{\pi^i}$ is the $\pi^i$ kinetic energy in the $\eta$ rest frame. 
   The decay amplitude square can be parametrized in a polynomial expansion around $(X,~Y)=(0,~0)$~\cite{KLOE-2:2016zfv}
   \begin{eqnarray}
 |A (s, t, u)|^2 \!&=&\! N (1 + a Y + b Y^2 + c X + d X^2 + e X Y 
 + f Y^3 + g X^2 Y\nonumber\\
 &+& h X Y^2 + l X^3 + \ldots) \,.\label{amplitude_expansion}
 \end{eqnarray} 
 The C transformation on the decay is equivalent to $t\leftrightarrow u$ in the amplitude. As a result, the appearance of terms that are odd in $X$ would serve as evidence of C and CP violation in this process~\cite{Gard04,gardner2019patterns}.
 
%
Although an experimental determination of non-zero
coefficients for the terms with odd powers of $X$ ($c$,$e$,$h$)
in the study of the $\eta\to \pi^+\pi^- \pi^0$ Dalitz plot
would signal both $C$ and $CP$ violation, it is possible
to gain insight into the isospin structure of the new physics sources through consideration of their 
chiral dynamics. We follow Ref.~\cite{gardner2019patterns} for our discussion. 
The decay amplitude in the SM, working to leading order
in isospin breaking, 
can be expressed as~\cite{Gasser:1984pr,Anisovich:1996tx}
\begin{eqnarray}
 A (s, t, u) &=& - \frac{1}{Q^2} \frac{M_K^2}{M_{\pi}^2} \frac{M_K^2 -
   M_{\pi}^2}{3 \sqrt{3} F_{\pi}^2} M(s,t,u)\,.
\label{defA}
  \end{eqnarray}
Since $C=-(-1)^I$ in $\eta\to 3\pi$ decay~\cite{Lee:1965zza}, the C- and CP-even
transition amplitude with a $\Delta I=1$ isospin-breaking prefactor
must have $I=1$.
The amplitude $M (s, t, u)$ thus
corresponds to the total isospin $I=1$ component  of the $\pi^+\pi^-\pi^0$ state and
can be expressed as~\cite{Anisovich:1996tx,Lanz:2011foc}
  \begin{eqnarray}
 M^{C}_{1} \!(s, t, u) &=& M_0^0 (s) + (s - u) M_1^1 (t) + (s - t) M_1^1 (u) \nonumber\\
 &+& M_2^0 (t) + M_2^0(u) - \frac{2}{3} M_2^0 (s) \,,
\label{anileut}
  \end{eqnarray}
  where $M_I^{\ell}$(z) is 
  an amplitude with $\pi-\pi$ rescattering in the $z$-channel
with isospin $I$ (and orbital angular momentum $\ell$). 
We refer to Sec.~\ref{sec:non-perturbative-QCD}A 
for a nuanced discussion of their
construction, pertinent to a high-precision extraction
of the light quark mass difference. 
The decomposition can be recovered under isospin symmetry
in chiral perturbation theory (ChPT) up to next-to-next-to-leading order (NNLO),
${\cal O}(p^6)$,
because the only absorptive parts that can
appear are in the $\pi-\pi$ $S-$ and $P$-wave amplitudes~\cite{Bijnens:2007pr}.

Since we are considering C and CP violation, additional amplitudes
can appear --- namely, total $I=0$ and $I=2$ amplitudes.
The complete amplitude is thus
\begin{eqnarray}
A(s,t,u)
&=& - \frac{1}{Q^2} \frac{M_K^2}{M_{\pi}^2} \frac{M_K^2 -
   M_{\pi}^2}{3 \sqrt{3} F_{\pi}^2} M^{C}_{1} \!(s,t,u) \nonumber \\
&& + \alpha M^{\not C}_{0} \!(s,t,u)
+ \beta M^{\not C}_{2} \!(s,t,u) \,,
\label{total_amplitude}
\end{eqnarray}
where $\alpha $ and $\beta$ are unknown,
low-energy
constants ---  complex numbers to be determined
by fits to the experimental event populations in the Dalitz plot.
If they are determined to be non-zero, they signal the appearance of
C- and CP-violation. 
Following the expectations of
Watson's theorem~\cite{Watson:1954uc,Gardner:2001rv} we write~\cite{gardner2019patterns}
\begin{equation}
\!\! M^{\not C}_{0} \!(s,t,u) \!= \!(s - t)\! M_{1}^1(u) + (u - s)\! M_{1}^1(t)
 - (u-t)\! M_{1}^1(s)
\label{CPV_isospin0}
\end{equation}
and
\begin{eqnarray}
\!\!\!\!\!\!M^{\not C}_{2} \!(s,t,u) \!&=&\! (s - t) M_1^1 (u) + (u - s) M_1^1 (t)
\nonumber\\
   && \!\!\!+ 2(u-t)M_1^1(s) + \sqrt{5}[M_2^0 (u) - M_2^0 (t)] \,.
\label{CPV_isospin2}
 \end{eqnarray}
In what follows we adopt the NLO analyses of Refs.~\cite{Gasser:1984pr,Bijnens:2007pr} to 
construct particular forms for the 
$M_I^\ell$ and refer to Ref.~\cite{gardner2019patterns} for
all details. The analysis we have outlined here has
been employed in the sensitivity analysis of 
Sec.~\ref{subsec:Tests-of-Conservation-Laws}.
We emphasize that the normalization factor $N$ as in 
Eq.~(\ref{amplitude_expansion}) drops
out in the experimental asymmetry $A_{LR}$, Eq.~(\ref{asym_LR}).
 
\begin{figure}[ht]
\centering
\vspace*{-0.2cm}
\includegraphics[width=0.5\textwidth]{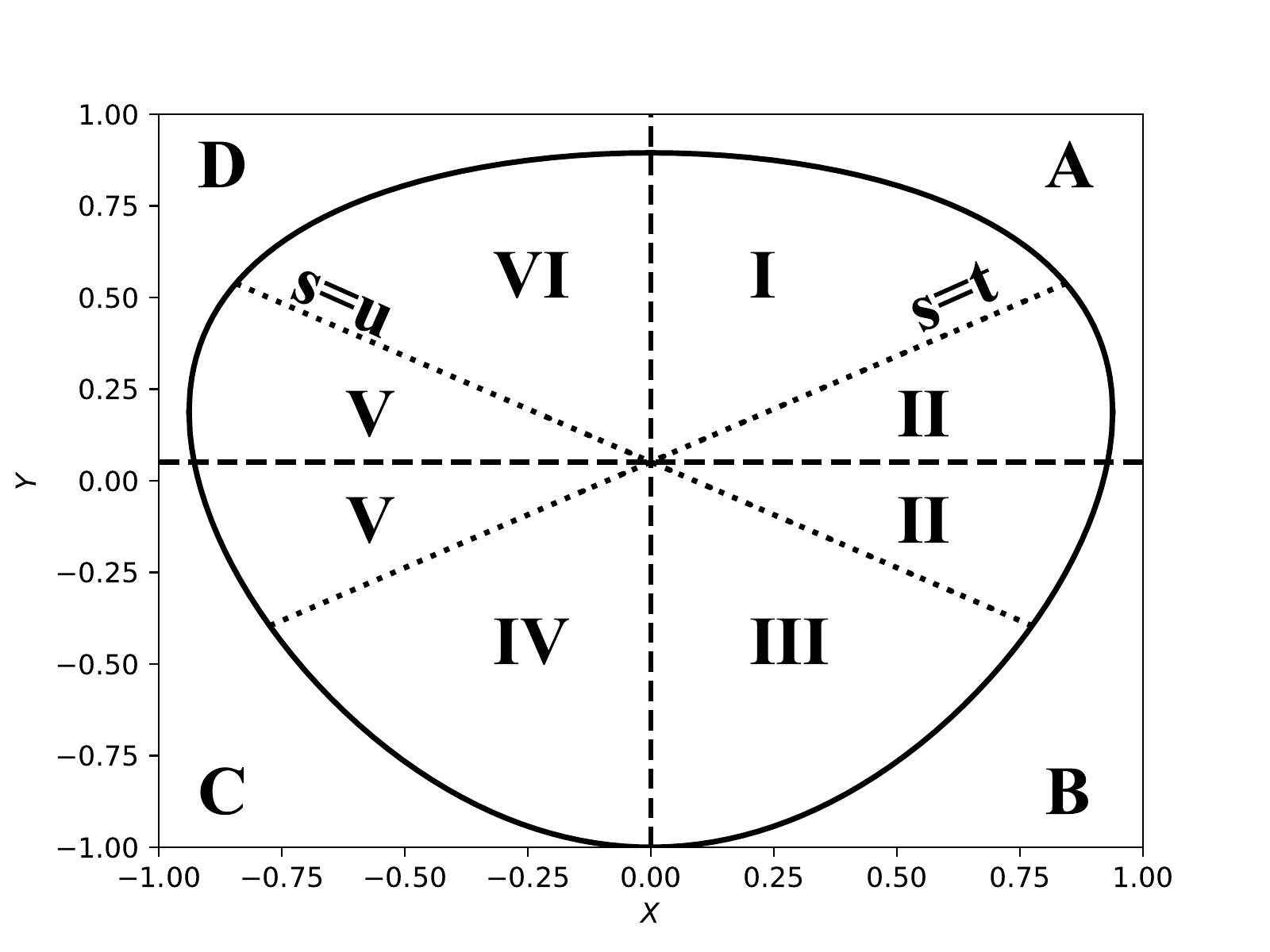}
\vspace*{-0.2cm}
\caption{The Dalitz plot geometry in $\eta\to\pi^+\pi^-\pi^0$ decay, 
 split into regions for probes of its symmetries. 
The solid line is the boundary of the physically accessible region. 
The asymmetry $A_{LR}$, Eq.~(\ref{asym_LR}), compares the population $N_+$ ($X>0$) 
with $N_-$ ($X<0$). The quadrant asymmetry $A_Q$ probes $I=2$ contributions, 
$N_{\rm tot} A_Q \equiv N({\rm A}) + N({\rm C}) - N({\rm B}) - N({\rm D})$~\cite{Lee:1965zza}, 
and the sextant asymmetry $A_S$ probes $I=0$ contributions, 
$N_{\rm tot} A_S \equiv N({\rm I}) + N({\rm III}) + N({\rm V}) - N({\rm II}) 
- N({\rm IV}) - N({\rm VI})$~\cite{Nauenberg:1965,Lee:1965zza}. 
All asymmetries probe C and CP violation. Figure
from Ref.~\cite{gardner2019patterns}.
}\label{fig:Dalitz_As_Aq}
\end{figure}

Besides, it is also possible to measure the asymmetries which can probe the isospin of the CP violating final state: a sextant asymmetry $A_S$ which is sensitive to the $I=0$ state~\cite{Lee:1965zza,Nauenberg:1965asym} and a quadrant asymmetry which is sensitive to the $I=2$ final state~\cite{Lee:1965zza,Lay72}, which are illustrated in Fig.~\ref{fig:Dalitz_As_Aq}. However, an alternate, more sensitive discriminant
of the isospin structure of the new CP-violating sources can be found by studying the pattern of the Dalitz plot in realizing
the left-right asymmetry, Eq.~(\ref{asym_LR})~\cite{gardner2019patterns}.
See also Ref.~\cite{Akdag:2021efj} for a comparison of $C$-violating effects of different isospin in similar $\eta'$ three-body decays.

Analysis of the data is complex. Monte Carlo calculations must include
adjustments for experimental efficiencies and $\pi\pi$ interactions.
The results for the left-right (LR or $X$), quadrant (Q), and sextant
(S) asymmetries for $1.34\times10^{6}$ $\eta\to\pi^{+}\pi^{-}\pi^{0}$
decays~\cite{Amb08} are: \begin{eqnarray} A_{LR} & = & (+0.09\pm0.10\,_{-0.14}^{+0.10})\times10^{-2} \notag\\ A_{Q} & = & (-0.05\pm0.10\,_{-0.05}^{+0.05})\times10^{-2} \notag\\ A_{S} & = & (+0.08\pm0.10\,_{-0.13}^{+0.08})\times10^{-2} \end{eqnarray}
Observation of statistically significant asymmetries would be an evidence of $C$ and $CP$ violation.
REDTOP is particularly suitable for this measurement since the reconstruction
of charged pions in a Optical-TPC does not rely on a magnetic field,
which is usually the largest source of systematic asymmetries. Therefore,
it can vastly improve the accuracy of the measurements and resolve
the discrepancies.

\begin{figure}[!ht]
\includegraphics[scale=0.9]{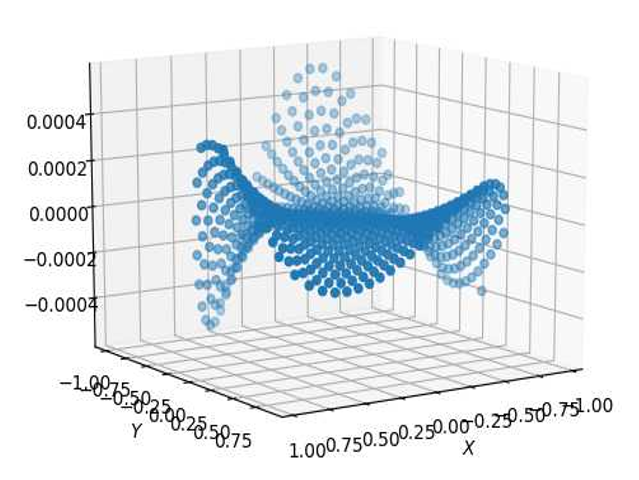} \caption{3-D representation of the Dalitz plot for the $\eta\rightarrow\pi^{+}\pi^{-}\pi^{0}$
process. 
}
\label{fig:dalitz_plot}
\end{figure}
A 3-dimensional representation of the X and Y variables for the $\eta\to \pi^+ \pi^-\pi^0$ process
is shown in Fig.~\ref{fig:dalitz_plot}. The appearance
of terms that are odd in X would indicate both C and CP violation.
The detection of charged pions in REDTOP is based on the measurement
of the Cerenkov angle of the photons radiated in the aerogel (cf.\ Sec.~\ref{subsec:The-LGAD-tracker}). Therefore, the non-uniformity
of the magnetic field, which in general corresponds the largest contribution
to the systematic error in the asymmetry of kinematics variables of
positive and negative charged particles, plays no role at REDTOP.
The expected sensitivity for this process is currently under study
with the REDTOP detector, although it is expected to be higher than
that with more traditional, magnetic spectrometers.

\begin{singlespace}

\subsubsection{\label{subsec:CP-violation-in-eta2pipiee}CP violation in \texorpdfstring{\mbox{$\eta\rightarrow\pi^{+}\pi^{-}e^{+}e^{-}$}}{}}
\end{singlespace}

\emph{}

P and CP violation in the decay of $\eta\to\pi^+\pi^-\gamma$ has
been discussed nearly fifty years before~\cite{HP73}. More recent
study~\cite{GNW02} has analyzed the CP-violating effects in this
decay by considering the photon polarizations, and predicted that
a sizable linear photon polarization could be expected in some new
physics scenarios. In order to avoid measuring the photon polarization,
one can consider, as shown in Ref.~\cite{gao2002cp}, the decay $\eta\rightarrow\pi^+\pi^- e^+ e^-$
resulting from the internal conversion of the photon into an $e^+e^-$
pair, and the CP-violating effects hidden in the polarization of the
photon now can be translated into the CP asymmetry in the angular
correlation of the $e^+ e^-$ plane relative to the $\pi^+\pi^-$
plane. This is actually analogous to the neutral $K$ system, in which
a large CP asymmetry, due to the interference between the parity-conserving
magnetic amplitudes and the parity-violating electric amplitudes of
$K_L\rightarrow \pi^+\pi^-\gamma^*\rightarrow\pi^+\pi^- e^+ e^-$,
has already been predicted theoretically and confirmed experimentally.
Thus the asymmetry in  $\eta\rightarrow\pi^+\pi^- e^+ e^-$ 
transition could be found by analyzing its angular distribution~\cite{gao2002cp},
which is given by \begin{equation}\label{CP1} {A}_{\phi}=\langle sign(\sin\phi \cos \phi)\rangle=
\Frac{\int_0^{2\pi} \frac{d\Gamma( \eta\rightarrow\pi^+\pi^-e^+e^-)}{d\phi}d\phi~ sign(\sin\phi \cos \phi)}{\int_0^{2\pi} 
\frac{d\Gamma(\eta\rightarrow\pi^+\pi^-e^+e^-)}{d\phi}d\phi}, 
\end{equation} where $\phi$ is the angle between the $e^+e^-$ and $\pi^+\pi^-$
planes in the $\eta$ rest frame (see Fig.~\ref{fig:eta2pipiee_phi}).

It is obvious that the asymmetry in the flavor-conserving  $\eta\rightarrow\pi^+\pi^- e^+ e^-$
decay, different from flavoring-changing processes like $K_L\rightarrow \pi^+\pi^- e^+ e^-$,
indicates the presence of non-standard CP-violation. It has been shown
in~\cite{GNW02, gao2002cp} that this asymmetry will arise if a relevant
parity-violating electric transition exists, and such manifestation
of CP violation in some New Physics scenarios might not be bounded by
existing EDM measurements; cf.\ however the discussion in Ref.~\cite{Gan:2020aco}.

Experimentally, the first measurement of such asymmetry has been done
by the KLOE Collaboration in 2009~\cite{KLOE09}. The current best
measurement has been performed by the WASA-at-COSY Collaboration~\cite{WASA16}
and it is consistent with zero. However, the statistical error (based
on the production of $\sim 10^9$ $\eta$-mesons) largely dominates
the measurement. REDTOP's larger statistics will improve on the systematic
error by almost two orders of magnitude, bringing the sensitivity
to a level where CP-violation could be observed. Thus, further experimental
investigation of this asymmetry might be helpful to increase our knowledge
on CP violation, or to impose some interesting constraints on some
theoretical models.

Similar work can be directly generalized to $\eta/\eta^\prime\to \pi^+\pi^-\ell^+\ell^-$
decays including $\ell=e,\mu$. Very recently, the experimental measurement
of such asymmetry in $\eta^\prime\to \pi^+\pi^-e^+ e^-$ has been
performed by the BESIII Collaboration~\cite{BESIII21}.

\emph{}




\subsubsection{\texorpdfstring{Tests of CP invariance via $\gamma^{*}$ polarization studies in $\eta\to \pi^+ \pi^-\gamma^{*}$}{}}

CP-violation can also be investigated with a virtual photon decaying
into a lepton-antilepton pair, as in
\begin{equation}
\eta\rightarrow\pi^{+}\pi^{-}\gamma^{*} \quad \text{with} \quad \gamma^{*}\rightarrow l^{+}+l^{-}, 
\end{equation}
by considering the asymmetry
\begin{equation}
A_{\Phi}=\frac{N(\sin\phi\cos\phi>0)-N(\sin\phi\cos\phi<0)}{N(\sin\phi\cos\phi>0)+N(\sin\phi\cos\phi<0)}
\end{equation}
where $\phi$ is the angle between the decay planes of the lepton-antilepton
pair and the two charged pions. CP invariance requires $A_{\Phi}$
to vanish. At the present, the measurement of such asymmetry performed
by the WASA collaboration~\cite{Adlarson-15} is the best available,
and it is consistent with zero within the measurement errors. Unfortunately,
that measurement is largely dominated by the statistical error, from
the production of only $\sim10^{9}\:\eta$-mesons. The larger statistics
of REDTOP will improve the systematic error by almost two orders of
magnitude.

\subsubsection{CP violation in \texorpdfstring{$\eta\to\mu^+\mu^-$}{}\label{sec:CPVetaTo2L}}

The most general amplitude for $\eta\to\ell^+\ell^-$ can be effectively parametrized as  
\begin{equation}\label{eq:PtoLLamp}     
\mathcal{M} = g_P (\bar{u}i\gamma^5 v) + g_S(\bar{u}v), \end{equation} with $g_{P,S}$ dimensionless parameters. The first contribution is $CP$ even and has been computed in the SM (see~\cite{Sanchez-Puertas:2018tnp,Masjuan:2015cjl} and references therein), while the latter represents a $P$-odd $C$-even contribution, that is negligible in the SM. As such, any indication of a nonzero $g_S$ would be a clear signal of New Physics. A possible way to access this is through $CP$-odd lepton polarization observables, that are of order $\mathcal{O}(g_S g_P)$ and require of muon polarimetry techniques --- otherwise, the first contribution appears at $\mathcal{O}(g_S^2)$.
A general caveat of these searches is that, in general, $P$-odd $C$-even observables are tightly bound by neutron and lepton EDMs, that make such a positive finding a priori unexpected at any foreseeable meson factory. An exception to this was found in Ref.~\cite{Sanchez-Puertas:2018tnp}, where the New Physics were studied using the SMEFT D=6 operators. There it was shown that, while EDM operators (or hadronic $CP$-odd operators inducing $CP$-violating transition form factors) are severely constrained by neutron and lepton EDMs, quark-lepton Fermi operators are less severely constrained for the case of muons (additional constraints exists for electrons~\cite{Yanase:2018qqq}). As such, we foucs in the latter scenario which is the most appealing case (find comments regarding $CP$-violating TFFs in double-Dalitz decays, Sect.~\ref{subsec:Double-Dalitz-decays}). In particular, the single operators of such kind inducing $CP$-violation are $\mathcal{O}_{\ell e q u}^{(1)}$ and $\mathcal{O}_{\ell edq}$~\cite{Grzadkowski:2010es}, whose EDM contribution starts at two loops, providing the necessary suppression to avoid EDM constraints. 
\begin{figure}[!th]     \centering     \includegraphics[width=0.5\textwidth]{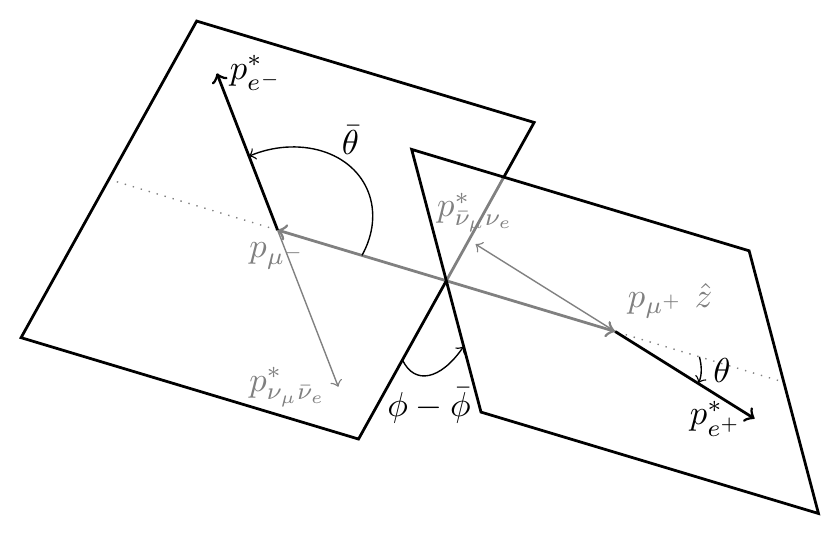}     \caption{Kinematics of the process. The decaying muons' momenta in the $\eta$ rest frame are noted as $p_{\mu^{\pm}}$, while the $e^{\pm}$ momenta, $p^*_{e^{\pm}}$, is shown in the corresponding $\mu^{\pm}$ reference frame along with the momenta of the $\nu\bar{\nu}$ system. The $\hat{z}$ axis is chosen along $p_{\mu^+}$.}     \label{fig:EtaTo2L} \end{figure} Ref.~\cite{Sanchez-Puertas:2018tnp} introduced two different muon's polarization asymmetries,  
\begin{align}     
A_{L} = \frac{N(\cos\theta >0) -N(\cos\theta <0)}{N} = \operatorname{Im}[4.1c_{\ell edq}^{2222} -2.7(c_{\ell equ}^{(1)2211} +c_{\ell edq}^{2211})]\times10^{-2}, \label{eq:AL}\\      
A_{\times} = \frac{N(\sin\Phi >0) -N(\sin\Phi <0)}{N} = \operatorname{Im}[2.5c_{\ell edq}^{2222}-1.6(c_{\ell equ}^{(1)2211} +c_{\ell edq}^{2211})]\times10^{-3}, \label{eq:AT}
\end{align}

with the first one identical if replacing $\theta\to\bar{\theta}$.
Above, $\theta(\bar{\theta})$ represents the polar angle of the $e^+(e^-)$
in the $\mu^+(\mu^-)$ reference frame, with the $\hat{z}$ axis fixed
by the $\mu^+$ direction in the $\eta$ rest frame, see Fig.~\ref{fig:EtaTo2L}.
The angle $\Phi = \phi-\bar{\phi}$ is defined in Fig.~\ref{fig:EtaTo2L}
and reflects the sign of $(\vec{p}_{e^-}\times \vec{p}_{e^+})\cdot \vec{p}_{\mu^+} $.
The right hand side in the equations above provides the contribution
from the aforementioned operators, with $c_{\mathcal{O}}^{(llqq)}$
the corresponding Wilson coefficient for the $l(q)$-th lepton(quark)
generation, respectively. The nEDM bounds derived in Ref.~\cite{Sanchez-Puertas:2018tnp}
(bounds from $D_s$ decays analyzed in Ref.~\cite{Sanchez-Puertas:2019qwm}
are similar, but less severe) updated with the most recent nEDM measurement~\cite{Abel:2020pzs}
imply \begin{equation}     \operatorname{Im}c_{\ell equ}^{(1)2211} < 0.001,      \operatorname{Im}c_{\ell edq}^{2211} < 0.008,     \operatorname{Im}c_{\ell edq}^{2222} < 0.02. \end{equation}
Clearly, the highest sensitivity happens for the coefficient involving
muons and strange quarks, $c_{\ell equ}^{(1)2222}$, and sets the
target sensitivity at REDTOP since one expects $A_L < 10^{-3}$ from
the bounds above.

\subsubsection{CP violation in double Dalitz decays \texorpdfstring{$\eta\to \ell^+\ell^- \ell'^+ \ell'^-$}{}\label{subsec:Double-Dalitz-decays}}

Similar to $\eta\to\mu^+\mu^-$ decays, Ref.~\cite{Sanchez-Puertas:2018tnp}
showed that $CP$ violation effects encoded in either quark/lepton
EDM as well as in $CP$-violating hadronic operators driving $CP$-odd
transition form factors are unexpected due to EDM bounds. In addition,
the sensitivity to quark-lepton $CP$ violating interactions was
studied there, while this involves an $\alpha$ suppression with respect
to the $\eta\to\mu^+\mu^-$ case that arises from the $e^+e^- $ emission:
$\eta\to\mu^+\mu^-\to\mu^+\mu^-\gamma^*\to \mu^+\mu^-e^+e^-$. In
particular, the contribution to the asymmetry was expressed as  
\begin{equation}     
A_{\sin\phi\cos\phi} = \operatorname{Im}[1.9 c_{\ell edq}^{2222} -1.3(c_{\ell equ}^{(1)2211} +c_{\ell edq}^{1122})]\times 10^{-5} -0.2\epsilon_1 +0.0003\epsilon_2 
\end{equation} where $c_{\mathcal{O}}$ are the corresponding Wilson coefficients introduced in Sect.~(\ref{sec:CPVetaTo2L}) and $\epsilon_{1,2}$ characterizes the coupling strength of the $CP$-violating TFFs in Eq.~(\ref{eq:TFF}), see~\cite{Sanchez-Puertas:2018tnp}. Once more, nEDM put constraints on these. In particular, Ref.~\cite{Sanchez-Puertas:2018tnp} showed that, necessarily, $\epsilon_{1} < 3\times 10^{-7}$; $\epsilon_2$ is unconstrained, but there is no theoretical motivation for having $\epsilon_2 \gg \epsilon_1$ (further, possibly more stringent bounds could be derived if the microscopic origin of $CP$ violation is specified). Still, we include them for completeness, as it is the most competitive channel to access $\epsilon_1$. 
\begin{singlespace}

\subsubsection{CP violation in
\texorpdfstring{$\eta\rightarrow\pi^{0}\mu^{+}\mu^{-}$}{} \label{sec:CPVetaToPi02L}}
\end{singlespace}

Yet another opportunity to search for violation of discrete symmetries is found in the $\eta\to\pi^0\mu^+\mu^-$ decay. It is useful to parametrize the corresponding matrix element as~\cite{Escribano:2022wug}
\begin{equation}\label{amplitude}
\mathcal{M} = m_{\mu}(\bar{u}v) F_1(q^2,\bar{q}\cdot k) +(\bar{u}i\gamma^5v) F_2(q^2,\bar{q}\cdot k) +(\bar{u}\slashed{k}v) F_3(q^2,\bar{q}\cdot k)  + i(\bar{u}\slashed{k}\gamma^5v) F_4(q^2,\bar{q}\cdot k) \ ,
\end{equation}
with $q=p_{\mu^+} +p_{\mu^-}$, $\bar{q}=p_{\mu^+} -p_{\mu^-}$, $k = p_{\eta} +p_{\pi^0}$, and the form factors $F_i(q^2,\bar{q}\cdot k)$. 
With these definitions, $F_{1,2}(q^2,\bar{q}\cdot k)$ are $P$ even whereas $F_{3,4}(q^2,\bar{q}\cdot k)$  are $P$ odd. In addition, a $C$ transformation effectively amounts to $F_{1,2,4}(q^2,\bar{q}\cdot k) \to F_{1,2,4}(q^2, -\bar{q}\cdot k)$ and $F_{3}(q^2,\bar{q}\cdot k) \to -F_{3}(q^2, -\bar{q}\cdot k)$. The SM contribution to the above process is vastly dominated by the electromagnetic interactions (thus, $CP$ even) and produces non-vanishing form factors for $i=1,2$, with $F_{1(2)}(q^2, \bar{q}\cdot k)$ an even(odd) function of $(\bar{q}\cdot k)$, see Refs.~\cite{Escribano:2022wug,Escribano:2020rfs} and references therein. 

The $\eta\to\pi^0\mu^+\mu^-$ process is a great testing ground for BSM searches, such as, for example, $P$-odd, $C$-even new-physics effects. In particular, with 3 particles in the final state, this necessarily involves polarization observables, in line with $\eta\to\mu^+\mu^-$ decays, see Sect.~(\ref{sec:CPVetaTo2L}).
The possibility of testing $P$-odd, $C$-even contributions was studied in Ref.~\cite{Escribano:2022wug} using the SMEFT as the general framework to capture physics BSM. Much in the same way as in Sect.~(\ref{sec:CPVetaTo2L}), the less constrained contributions originate from quark-lepton Fermi operators, whose EDM bounds are far less constrained. Specifically, the main contributions arise from the same operators appearing in $\eta\to\mu^+\mu^-$ decays and are associated to the scalar matrix elements 
\begin{equation}
\label{eq:F2smeft}
F_2 = \left[\operatorname{Im} c_{\ell edq}^{2211} \langle 0 | \bar{d}d |\eta\pi^0 \rangle +\operatorname{Im} c_{\ell edq}^{2222} \langle 0 | \bar{s}s |\eta\pi^0\rangle - \operatorname{Im} c_{\ell equ}^{(1)2211} \langle 0 | \bar{u}u |\eta\pi^0\rangle \right]\! /v^{2} \ ,
\end{equation}
that were computed in Ref.~\cite{Escribano:2022wug} within the framework of large-$N_c$~$\chi$PT. Note, in particular, that the second matrix element in Eq.~(\ref{eq:F2smeft}) vanishes in the isospin limit.
Once again, the key polarisation observables are those defined in Eqs.~(\ref{eq:AL},\ref{eq:AT}). Using input from Ref.~\cite{Escribano:2020rfs}, the final results for the asymmetries read
\begin{align}
 \label{eq:ALetapi0}
 A_L^{\eta\to\pi^0\mu^+\mu^-} &{} \!= -0.19(6) \operatorname{Im} c_{\ell equ}^{(1)2211} - 0.19(6) \operatorname{Im} c_{\ell edq}^{2211} - 0.020(9) \operatorname{Im} c_{\ell edq}^{2222} \ , \\[1ex] 
 \label{eq:ATetapi0}
 A_{\times}^{\eta\to\pi^0\mu^+\mu^-} &{} \!= 0.07(2) \operatorname{Im} c_{\ell equ}^{(1)2211} + 0.07(2) \operatorname{Im} c_{\ell edq}^{2211} + 7(3)\times 10^{-3} \operatorname{Im} c_{\ell edq}^{2222} \ .
\end{align}
The comparatively larger suppression for the $s$-quark contribution arises from isospin breaking, that is an $\mathcal{O}(1\%)$ effect, and reduces the sensitivity as compared to $\eta\to\mu^+\mu^-$ decays. Current nEDM bounds imply for the above asymmetries that $A_L < 4\times 10^{-4}$ and $A_{\times} < 1.4\times 10^{-4}$, which should be contrasted with the REDTOP capabilities.

\subsubsection{Lepton flavor violation measurements}


This study is motivated by a recent work on decays of quarkonium states
M with different quantum numbers~\cite{Petr17, Petr19}. The model
could be used to put constraints on the Wilson coefficients of effective
operators describing LFV interactions at low energy scales. Furthermore,
studies of radiative lepton flavor violating meson decays could provide
important complementary access to specific effective operators. REDTOP
will be able to probe the model with the processes: \uline{\mbox{$\eta/\eta^{\prime}\rightarrow e^{+}\mu^{-}+c.c.$}}
and \uline{\mbox{$\eta/\eta^{\prime}\rightarrow\gamma\,e^{+}\mu^{-}+c.c.$}}
This study has been planed for the Snowmass-2022 Summer exercise.

\subsubsection{Tests of lepton flavor universality with \texorpdfstring{$\eta/\eta^{\prime}$}{} mesons} \label{sec:LFU}

For definiteness let us consider the purely leptonic decay $\eta/\eta^{\prime} \to \ell^+ \ell^{(\prime)-}$, starting from the case of leptons of equal flavour. Given the amplitude for such decays, its imaginary and real parts are bounded separately. The former can be unambiguously related to the $\eta \to \gamma \gamma$ decay rate, which is measured. This provides the so-called ``unitarity bound'', which is quoted from~\cite{Drell,Berman:1960zz,Geffen:1965zz} as~\cite{Abegg:1994wx}
\begin{equation}
\label{eq:P_to_ell_ell}
\Gamma(P \to \ell^+ \ell^-) \ge \frac{\alpha^2}{2 \beta} \left[ \left(\frac{m_\ell}{m_P}\right) \ln \left( \frac{1+\beta}{1-\beta} \right) \right]^2 \times \Gamma(P \to \gamma \gamma)~,
\end{equation}
with $P$ denoting either of $\eta, \eta^\prime$ and $\beta = \sqrt{1- 4 m_\ell^2/m_P^2}$.

The real part depends on the $P \to \gamma\gamma$-vertex form factor, which is estimated within quark models or vector meson dominance, and thus involves non-systematic theoretical assumptions. These calculations return values that are typically 30\% {\em larger} than the unitarity lower limit~\cite{Landsberg:1985gaz}. This contribution may be affected by New Physics, for example $t$-channel exchange of a leptoquark, whereby the two quark-lepton currents give rise to the external states in an obvious way. One finds~\cite{Mayer:1990kf} that values for the ratio of mass to coupling constant below several hundred GeV can be excluded for transitions within the 1st generation. For transitions between different generations the excluded region reaches up to 200 TeV based on the upper limit for BR($K_L \to ee$).

The case of leptons with different flavour has the huge advantage of being a null test in the SM. In this case one may consider decays to $\ge 2$ final states, including $\eta^{(\prime)} \to \mu^\pm e^\mp (\gamma \mbox{~or~} \pi^0)$ as well as $\eta^\prime \to \eta \mu^\pm e^\mp$. These decays have been discussed specifically in Ref.~\cite{Hazard:2016fnc}. Although these decays would not {\em directly} probe the LFV couplings hinted at by the ``$B$ anomalies'' (see discussion in~\cite{Glashow:2014iga,Guadagnoli:2015nra}) the probed couplings may actually be related to those from the $B$ anomalies by general effective-theory arguments plus minimal flavour assumptions~\cite{Borsato:2018tcz}. With this in mind, and as shown in table 7 of Ref.~\cite{Hazard:2016fnc}, the constraints placed on the magnitude of LFV $c/\Lambda^2$ couplings range between a few $\times 10^{-3}$ and $10^4$ GeV$^{-2}$. For natural, O(1) WCs this implies that the natural effective scale probed is between about 20 GeV and 6 MeV. Actually, the very same LFV couplings probed by the above decays are also probed by $\mu \to e$ conversion in nuclei, which yields bounds stricter by several orders of magnitude. Ref.~\cite{Gan:2020aco} actually suggests to use the latter to constrain $\eta$ and $\eta^\prime$ decays, by invoking $\eta^{(\prime)}$ exchange mechanisms, and at the price of the ensuing model dependence.

Another set of potential probes of high-energy New Physics are ratios of $\eta$ or $\eta^\prime$ decays to $X \ell^+ \ell^-$, where $\ell$ is different between the numerator and the denominator and $X$ denotes additional final-state particle(s). Such ratios would probe lepton universality. A few considerations are in order however. First, decays induced by the weak interaction are extremely rare, below $10^{-10}$, and thus well below, by many orders of magnitude, existing experimental upper limits. The purely leptonic case is, as mentioned, dominated by intermediate di-photon exchange, and the suppression of this SM mechanism would suggest these decays as good probes of BSM effects. In practice, even assuming tree-level New Physics, modifications to the SM amplitude have to be small, because $Z$-boson exchange per se modifies the rate at the per mil level. Handy formulae can be found in Ref.~\cite{Gan:2020aco}. As a consequence, again, these ratios are way more suited as probes of light, GeV-scale or below, New Physics.

\section{Non-perturbative QCD\label{sec:non-perturbative-QCD}}

Studying the decays of $\eta$ and $\eta^{\prime}$ mesons open a window into the complex QCD dynamics at low energies. For recent reviews on the subject, see e.g.~\cite{Gan:2020aco, JPAC:2021rxu}. 
At these energies, ($m_\eta \sim 548$ MeV and $m_{\eta^{\prime}} \sim 958$ MeV) strong interactions become non-perturbative and the usual series expansion in the strong coupling constant does not provide an appropriate theoretical framework. 
One has to instead rely on non-perturbative techniques such as effective field theory---Chiral Perturbation Theory ($\chi$PT) for light quarks---, dispersion relations, or numerical simulations such as lattice QCD. 

The $\eta$ meson is very peculiar: its main decay channel $\eta \to 3\pi$ violates G-parity, thereby providing an unique access to the light quark masses. It also allows one to test the $\chi PT$ framework and complement it with the dispersive methods. 

The description of $\eta^{\prime}$ decays poses additional challenges. In particular, it requires an extension of the $\chi PT$ framework, many aspects of which still remain to be worked out. 

The REDTOP experiment with the expected $10^{14}$ $\eta$ and $10^{12}$ $\eta^{\prime}$  will provide crucial data to advance this endeavor to the next level of precision. 

\subsection{\texorpdfstring{$\eta \to 3\pi$}{} and light quark mass extraction}
The process $\eta \to 3\pi$ is very interesting because since this decay is forbidden by isospin symmetry---three pions cannot combine to a system with vanishing angular momentum, zero isospin, and even $C$-parity---it offers an unique experimental access to the light quark mass ratio
\begin{equation}
  Q^2 = \frac{m_s^2 - \hat{m}^2}{m_d^2 - m_u^2}\quad\text{where}\quad\hat{m}= \frac{m_u + m_d}{2}. 
\end{equation}
The decay width of $\eta \to 3\pi$ can be seen as a measure for the size of isospin breaking in QCD. 
This is the case because the electromagnetic isospin-breaking effects have been shown to be small~\cite{Bell:1968wta,Sutherland:1966zz,Baur:1995gc,Ditsche:2008cq}.

The extraction of $Q$ is done by comparing the experimental measured decay width with the reduced amplitude $M(s,t,u)$ integrated over the phase space:
\begin{equation}
\Gamma\big(\eta \to \pi^+ \pi^- \pi^0\big) = \frac{1}{Q^4} \frac{M_K^4 (M_K^2-M_\pi^2)^2}{6912 \pi^3 M_\eta^3 M_\pi^4 F_\pi^4} \int_{s_{\rm min}}^{s_{\rm max}} ds 
\int_{u_-(s)}^{u_+(s)} du \left|M(s,t,u) \right|^2\,,
\label{eq:Qexp}
\end{equation}
with $s\equiv (p_{\pi^+} + p_{\pi^-})^2$, $t\equiv (p_{\pi^-} + p_{\pi^0})^2$, and $u\equiv (p_{\pi^+} + p_{\pi^0})^2$, the so-called Mandelstam variables.
The aim is to compute the amplitude $M(s,t,u)$ with the highest possible accuracy. 

This is not an easy task since there are strong rescattering effects among the final-state pions. These were initially calculated perturbatively in $\chi$PT.
The current algebra or leading order (LO) result in $\chi$PT gives $\Gamma(\eta \to \pi^+ \pi^- \pi^0)_{\text{LO}} = 66$~eV~\cite{Osborn:1970nn}. It receives a substantial enhancement 
$\Gamma(\eta \to \pi^+ \pi^- \pi^0)_{\text{NLO}} = 160(50)$~eV~\cite{Gasser:1984pr} due to chiral one-loop corrections. 
The NLO result is still far from the experimental value $\Gamma(\eta \to \pi^+ \pi^- \pi^0)= 300 \pm 12$~eV~\cite{ParticleDataGroup:2020ssz} suggesting a convergence problem. 
Moreover, it has been shown that the two-loop or NNLO  calculation~\cite{Bijnens:2007pr} may lead to a precise numerical prediction 
only after the low-energy constants (LECs) appearing in the amplitude are determined reliably. 
In particular, the role played by the $\mathcal{O}(p^6)$ LECs is non-negligible and they are largely unknown. 

A more accurate approach relies on dispersion relations to evaluate rescattering effects to all 
orders~\cite{Anisovich:1993kn,Anisovich:1996tx,Kambor:1995yc,Walker:1998zz}. 
This is not completely independent of $\chi$PT, because the dispersive representation requires the subtraction constants as input, and they can be calculated in $\chi$PT. 

There has been a renewed interest in $\eta\to3\pi$ dispersive analyses due to new and more precise measurements of this decay. In particular recent measurements of the Dalitz plot of the charged ($\eta\to\pi^+\pi^-\pi^0$) channel by KLOE~\cite{KLOE:2008tdy,KLOE-2:2016zfv} and BESIII~\cite{BESIII:2015fid} and of the neutral channel ($\eta\to\pi^0\pi^0\pi^0$) by A2~\cite{A2:2018pjo} have achieved an impressive level of precision. New measurements are planned by BESSIII and at JLab by GlueX~\cite{JEF-PAC42, Gan:2015nyc} and CLAS~\cite{Amaryan:2013osa}, with completely different systematics and even better accuracy. Given its unprecedented statistics, REDTOP could definitely play a crucial role in measuring this decay. 

The amplitude $M(s,t,u)$ in Eq.~(\ref{eq:Qexp}) is determined using the dispersion-theoretical Khuri--Treiman equations~\cite{Khuri:1960zz, Aitchison:1966lpz}. In this framework, a set of integral equations for the scattering process $\eta \pi \to \pi \pi$ is
established before being analytically continued to describe the decay $\eta \to 3\pi$. The amplitude is decomposed in terms of single-variable functions neglecting $D$-and higher waves:
\begin{align} 
\label{eq:Mdecomp}
	M(s, t, u) = M_0^0(s) + (s-u) M_1^1(t) + (s-t) M_1^1(u) + M_2^0(t) + M_2^0(u) - \frac{2}{3} M_2^0(s) \,.
\end{align}
The functions $M_I^\ell(s)$ have isospin $I$ and angular momentum $\ell$. 
In the context of light mesons, this decomposition is commonly referred to as a \textit{reconstruction theorem}~\cite{Stern:1993rg, Knecht:1995tr, Ananthanarayan:2000cp,
Zdrahal:2008bd}. The latter relies on the observation that up to corrections of order $\mathcal{O}(p^{8})$ (or three loops) in the chiral expansion, partial waves of any meson--meson scattering process with angular momentum $\ell \geqslant 2$ contain no imaginary parts. 
The splitting of the full amplitude into these single-variable functions is not unique: there is some ambiguity in the distribution of the polynomial terms over the various $M_I^\ell$ due to $s+t+u = m_\eta^2 + m_{\pi^+}^2 +  m_{\pi^-}^2 + m_{\pi^0}^2$ being constant. 

The use of analyticity and unitarity allows one to construct dispersion relations for the single-variable functions $M_I^\ell(s)$, and one arrives to
\begin{subequations}\label{eq:MIMIhat}\begin{align} 
\label{eq:MIdispa}
	M_I^\ell(s) & = \Omega_I^\ell(s) \left\{ P_I(s) + \frac{s^{n_I}}{\pi} \int _{4 m_\pi^2}^{\infty} \frac{\text{d} s^\prime}{s^{\prime n_I}}
		\frac{\sin \delta_I^\ell(s^\prime) \hat{M}^\ell_I(s^\prime)}{|\Omega_I^\ell(s^\prime)| (s^\prime - s -i \epsilon)}
		\right\},\\
	\hat M_I^\ell(s) & =\sum_{n,I'} \int_{-1}^{1} \text{d} \cos\theta~\cos^{n} \theta \, c_{n I I'} M_{I'}^\ell\big(t(s,\cos\theta)\big)~,
	\label{eq:MIdispb}
\end{align}\end{subequations}
where $\hat M_I^\ell(s)$ represents the left-handed cuts. 
The explicit forms of the coefficients $c_{nII'}$ can be found e.g. in Refs.~\cite{Anisovich:1996tx,Walker:1998zz}. 
\begin{equation}
\Omega_I^\ell(s) = \exp\Bigg\{\frac{s}{\pi}\int_{4M_\pi^2}^\infty \text{d} s'\frac{\delta_I^\ell(s')}{s'(s'-s-i\epsilon)} \Bigg\} ~, 
\end{equation}
are the so-called Omn\`es functions~\cite{Omnes:1958hv}. 
These functions are enterely given 
in terms of the appropriate pion--pion
phase shift $\delta_I^\ell(s)$. 

\begin{figure}[t]
\centering
\includegraphics[width=0.5\textwidth]{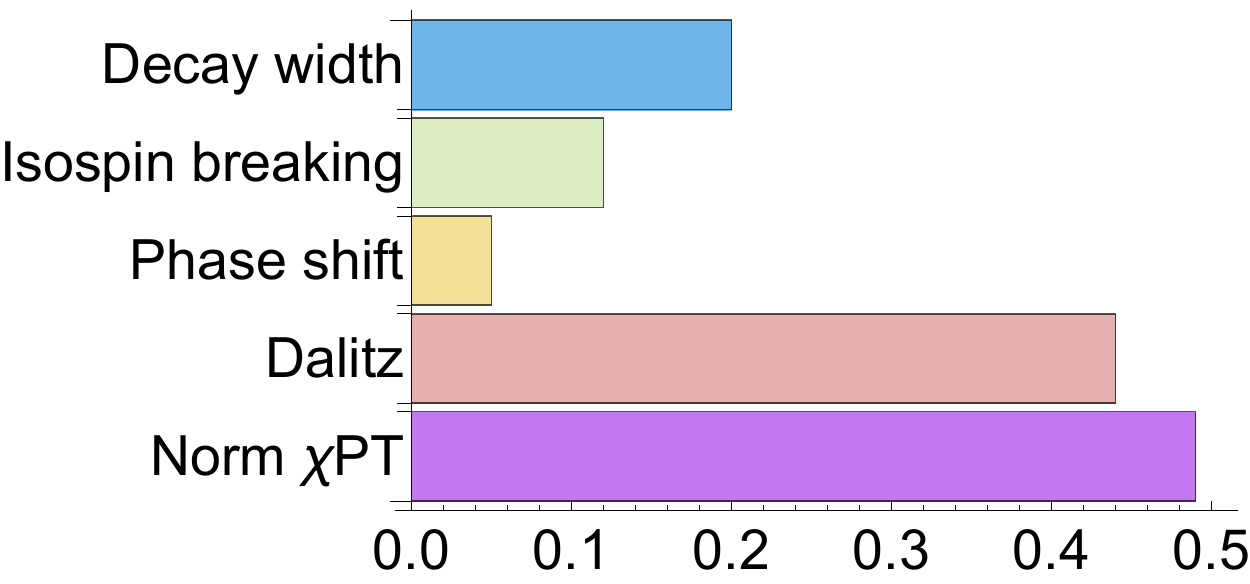}
\caption{\label{fig:results_eta-Qerror} Assessment of the different uncertainties on the quark mass double ratio $Q$, $\delta Q = \pm 0.72_{\rm total} = \pm 0.49_{{\rm Norm~}\chi{\rm PT}} \pm 0.44_{\rm Dalitz} \pm 0.05_{\rm Phase} \pm 0.12_{\rm IB} \pm 0.20_{\Gamma}$ coming from the result $Q = 22.04(72)$ of the analysis of Refs.~\cite{Colangelo:2016jmc,  Colangelo:2018jxw}. 
The final uncertainty due to the partial decay width, $\pm 0.20_\Gamma$, 
is fully dominated by the one in the total width $\Gamma_\eta$, with the branching ratio $\mathcal{B}(\eta\to\pi^+\pi^-\pi^0)=\Gamma(\eta\to\pi^+\pi^-\pi^0)/\Gamma_\eta$
giving a negligible contribution ($\pm 0.067_{\rm BR}$). Figure taken from Ref.~\cite{Gan:2020aco}.}
\end{figure}

In the determination of the amplitude $M(s,t,u)$ there are two sources of uncertainties entering in Eqs.~\eqref{eq:MIdispa} and~\eqref{eq:MIdispb}:
\begin{itemize}
    \item $\delta_I^\ell(s)$: the $\pi \pi$ phase shifts used as inputs. They are inferred in the elastic region from data on $\pi \pi$ scattering amplitudes solving Roy equations, see Refs.~\cite{Ananthanarayan:2000ht,Colangelo:2001df,Garcia-Martin:2011iqs,Caprini:2011ky}. In the inelastic region some assumptions need to be made relying on Froissart bound~\cite{Froissart:1961ux,Martin:1962rt} or Brodsky--Lepage asymptotics~\cite{Lepage:1979zb}. 
    \item The substraction  constants in $P_I(s)$. They are unknown and have to be determined using a combination of experimental information--fits to the Dalitz distributions-- and theory input. In particular since the overall normalization multiplies $1/Q^2$, the quantity that should be extracted from the analysis, it cannot be obtained from data alone and one has to match to $\chi$PT. On the other hand, this matching has to be performed in such a way that the problematic convergence of the chiral expansion is not transferred directly to the dispersive representation. This can be achieved by matching the amplitude around the Adler zeros. As can be seen in Fig.~\ref{fig:results_eta-Qerror} this uncertainty is decomposed as an uncertainty coming from fits to the Dalitz distribution and an uncertainty coming from the matching to $\chi$PT. 
\end{itemize} 
 
Several dispersive analyses~\cite{Schneider:2010hs,Kampf:2011wr,Guo:2015zqa,Guo:2016wsi, Colangelo:2016jmc,Albaladejo:2017hhj,Colangelo:2018jxw} have been performed over the last few years. All these analyses rely on the same theoretical ingredients described above with some subtle differences. For a detailed discussion see Refs.~\cite{Colangelo:2018jxw,Gan:2020aco}. 

\begin{figure}[t]
\includegraphics[width=0.6\textwidth]{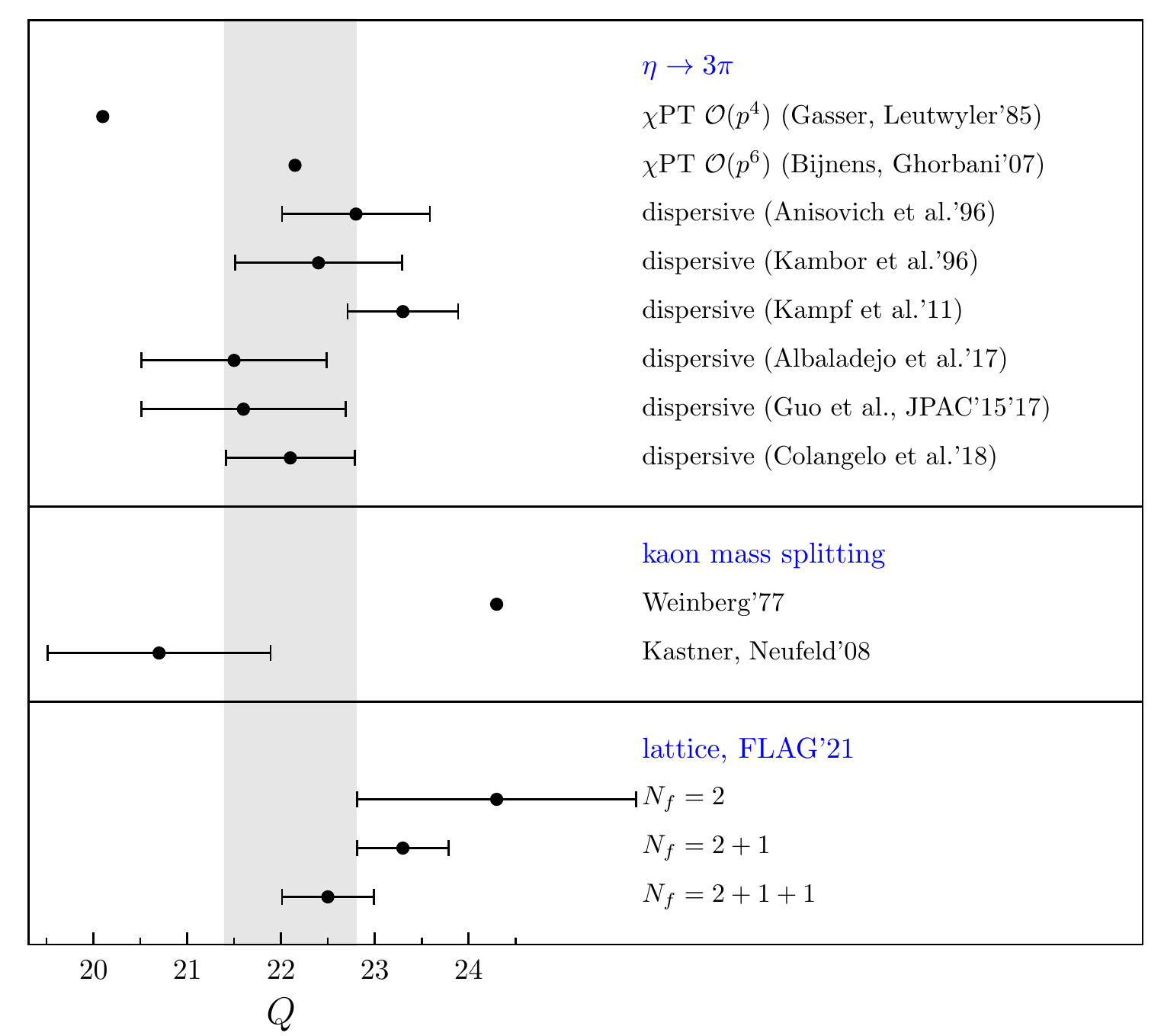}
\caption{\label{fig:results_etaQ}
Different determinations of the light quark mass double ratio, see text. The values for the Kaon mass splitting are taken from Refs.~\cite{Weinberg:1977hb,Kastner:2008ch}. The values from lattice averages are taken from Ref.~\cite{Aoki:2021kgd}. Note that some lattice results entering in the averages are marginally consistent with the results coming from $\eta \to 3\pi$. As discussed in Ref.~\cite{Colangelo:2018jxw} it would be important to understand the origin of this discrepancy. 
Figure adapted from Refs.~\cite{Lanz:2011foc,Colangelo:2018jxw,Gan:2020aco}}
\end{figure}

Figure~\ref{fig:results_etaQ} summarizes the results on the extraction of $Q$ from the different analyses. 
As can be seen there, the extracted values of $Q$ agree very well between the different dispersive analyses allowing $\eta \to 3\pi$ to be the golden plate channel to extract the light quark mass ratios. 

Figure~\ref{fig:results_eta-Qerror} summarizes the different sources of uncertainties entering in the determination of $Q$, see Eq.~\eqref{eq:Qexp}. To the uncertainties coming from the phase shifts and the substraction constants (Dalitz distribution and matching to $\chi$PT) entering in the determination of $M(s,t,u)$ one needs to add an uncertainty from isospin breaking and an uncertainty due to the experimentally measured decay width $\Gamma (\eta \to \pi^+ \pi^- \pi^0)$. 

New more accurate measurements of $\eta \to \pi^+ \pi^- \pi^0$ and $\eta \to 3 \pi^0$ would help to reduce the uncertainties on $Q$, in particular the ones coming from the Dalitz distribution as well as the decay width which are two dominant ones. 
It will also allow to assign a systematic uncertainty in the extraction of the quark mass ratio $Q$ which has not been taken into account so far. 

Note that as emphasized in ~\cite{Gan:2020aco}, for the uncertainty on $\Gamma(\eta \to 3\pi)$ the PDG value for the average of $\Gamma(\eta \to 2 \gamma)$ has been used. It relies on the collider measurements of this partial width only. If the Primakoff measurement were to be taken instead, this would amount to a shift for the central value of $Q$ by $\sim 2 \sigma$. Therefore a new and more precise Primakoff measurement will be important to have. One measurement is ongoing at JLab~\cite{Gan:2014pna}. It would be extremely useful to have a confirmation of this measurement by another experiment like REDTOP.

\subsection{\texorpdfstring{$\eta^{\prime} \to \eta \pi \pi$ and $\eta^{\prime} \to 3\pi$}{}}

Due to the $U(1)_A$ anomaly, the physical $\eta^{\prime}(958)$ meson is not entirely a Goldstone boson in the chiral limit of QCD, and is significantly heavier than the $\eta$ meson also in the real world. The $\eta^{\prime}$, in contrast to the $\eta$, decays strongly even in the isospin limit, via the channels 
$\eta^{\prime} \to \eta \pi \pi$ and $\eta^{\prime} \to 4 \pi $, which renders the width of the $\eta^{\prime}$ two orders of magnitude larger than the one of the $\eta$. However, a combination of small phase space (for the $\eta \pi \pi$ decays) and suppression due to Bose symmetry, angular momentum conservation, and high multiplicity of the final state (for the $4\pi$ decays~\cite{Guo:2011ir}) still leaves the $\eta^{\prime}$ more long-lived than, e.g., the $\omega(782)$ or the $\phi(1020)$.

The decay $\eta^{\prime} \to \eta \pi \pi$ is interesting for several reasons. First, due to the quantum numbers of the pseudoscalar mesons involved, the resonances featuring most prominently therein are the scalars: $G$-parity prevents vectors from contributing. Therefore, this process is particularly suitable for an analysis of the properties of the scalar $f_0(500)$ (or $\sigma$) resonance, even though the $a_0(980)$ is also present and, in fact, widely believed to be dominant~\cite{Escribano:2010wt}. Second, the presence of both $\eta$ and $\eta^{\prime}$ is ideal for studying the mixing properties of these two mesons. Third, and more generally, this decay allows one to test $\chi$PT and its possible extensions such as large-$N_C$ $\chi$PT~\cite{Leutwyler:1997yr,Bickert:2016fgy} and resonance chiral theory (R$\chi$T)~\cite{Escribano:2010wt,Gonzalez-Solis:2018xnw}. Indeed $\chi$PT needs to be generalized to a simultaneous expansion in small momenta, small quark masses, and $1/N_c$ to describe the pseudoscalar nonet including the $\eta^{\prime}$~\cite{Leutwyler:1997yr,Bickert:2016fgy}. $\eta^{\prime} \to \eta \pi \pi$ is a prime channel to test this framework. Finally, this decay is also very important to constrain $\pi \eta$ scattering~\cite{Gonzalez-Solis:2018xnw}. The $\eta^{\prime}$ mass is sufficiently small so that the extraction of $\pi \eta$ scattering is not polluted by intermediate states other than $\pi \pi$ scattering in the crossed channel.  For the investigation of the latter, Khuri--Treiman equations have also been studied for this process~\cite{Isken:2017dkw,Akdag:2021efj}, in analogy to the discussion of $\eta\to3\pi$ in the previous section.

Due to the close similarity of $\eta$ and $\eta^{\prime}$ in all properties except their masses, much of what has been said about the decays $\eta \to 3\pi$ applies equally to $\eta^{\prime} \to 3\pi$: they violate conservation of isospin; electromagnetic effects that could provide isospin breaking are strongly suppressed due to Sutherland's theorem~\cite{Bell:1968wta,Sutherland:1966zz}; as a consequence, they are almost exclusively caused by the light quark mass difference $m_u - m_d$. Indeed, the partial decay widths of all four decays are of the same order of magnitude (of a few hundred eV). However, as the width of the $\eta^{\prime}$ is larger than the one of the $\eta$ by about a factor of 150, the branching ratios for $\eta^{\prime} \to 3\pi$ are much smaller, and hence precise experimental investigations of these decays are much more recent than in the case of the $\eta$ (the decay $\eta^{\prime} \to \pi^+ \pi^- \pi^0$ was only established by the CLEO collaboration in 2008~\cite{CLEO:2008fxt}). Detailed investigations of the decay dynamics are exclusively due to the BESIII collaboration, who have measured the relative branching ratios most precisely~\cite{BESIII:2015you}, studied the $\eta^{\prime} \to 3\pi^0$ Dalitz plot for the first time~\cite{BESIII:2015fid}, and performed an amplitude analysis for both charged and neutral final states~\cite{BESIII:2016tdb}. Interestingly enough, the branching ratio into $3\pi^0$ still shows some tension: while the PDG average is $B(\eta^{\prime} \to 3\pi^0) = 3.57(26) \times 10^{-3}$~\cite{ParticleDataGroup:2020ssz}, dominated by Ref.~\cite{BESIII:2015fid}, the PDG fit of various $\eta^{\prime}$ branching ratios is quoted as $B(\eta^{\prime} \to 3\pi^0) = 2.54(18)\times10^{-3}$. In particular measurements of $B(\eta^{\prime} \to 3\pi^0)/B(\eta^{\prime} \to \eta \pi^0 \pi^0)$ by GAMS-4$\pi$~\cite{Blik:2008zz} seem to point towards smaller values. An independent confirmation of the BESIII result by the REDTOP experiment would therefore be very important. 


\subsection{Form factors and their applications}

The neutral pseudoscalar particles can decay in two photons via their coupling to the electromagnetic current  $J^{\mu}(x) = \sum_q Q_q [\bar{q}(x)\gamma^{\mu}q(x)]$~\cite{Adler:1969gk,Bell:1969ts,Adler:1969er}. Such coupling strength is related to the following---nonperturbative---matrix element
\begin{multline}\label{eq:TFF}
    i \int d^4x e^{iq_1\cdot x} \langle 0 | T\{ J^{\mu}(x), J^{\nu}(0)\} | P(q_1+q_2) \rangle \equiv \epsilon^{\mu\nu\rho\sigma} q_{1\rho} q_{2\sigma} F_{P\gamma^*\gamma^*}(q_1^2,q_2^2) \\
  +[g^{\mu\nu}q_1^2q_2^2
  -q_1^{\mu}q_1^{\nu}q_2^2
  -q_2^{\mu}q_2^{\nu}q_1^2
  +q_1^{\mu}q_2^{\nu}(q_1\cdot q_2) ]F_{P\gamma^*\gamma^*}^{C\!P2}(q_1^2,q_2^2) \\
  +  [g^{\mu\nu}(q_1 \cdot q_2) - q_2^{\mu}q_1^{\nu}]F_{P\gamma^*\gamma^*}^{C\!P1}(q_1^2,q_2^2),
\end{multline}
where $q_{1,2}$ corresponds to the photon's momenta. 
In the equation above, the first tensor structure and associated transition form factor (TFF), $F_{P\gamma^*\gamma^*}(q_1^2,q_2^2)$, corresponds to the standard $CP$ even QCD result, whose properties are discussed in the following. If, however, $CP$ violation is allowed in the QCD sector, two additional form factors $F_{P\gamma^*\gamma^*}^{C\!P1,2}(q_1^2,q_2^2)$ arise, while they are constrained by nEDM bounds (see further details in~\cite{Sanchez-Puertas:2018tnp}).
The function $F_{P\gamma^*\gamma^*}(q_1^2,q_2^2)$ encodes the non-perturbative QCD information that bring such pseudoscalar particles from $q-\bar{q}$ pairs.  TFFs play essentially the same role as the electromagnetic form factors of electrically charged particles but for their netural counterparts. In particular, if one (or both) of the photons is a virtual one, they can probe the hadronic structure of the corresponding pseudoscalar meson, generalizing the notion of electromagnetic form factors, that can be regarded as a Fourier transform of the particle's charge distribution.

\begin{sloppypar}
Their normalization provides the pseudoscalar meson coupling to real photons. 
Remarkably, it can be related to the chiral anomaly~\cite{Adler:1969gk,Bell:1969ts} and can also provide insights into the $\eta-\eta^{\prime}$ mixing~\cite{Leutwyler:1997yr,Feldmann:1999uf,Kaiser:2000gs,Escribano:2005qq,Escribano:2013kba,Escribano:2015nra,Escribano:2015yup}. 
At the other extreme, pQCD predicts at high energies the form factor to behave as $1/q_{1(2)}^{2}$~\cite{Lepage:1979zb,Lepage:1980fj}, whereas it is yet unclear the scale at which pQCD applies~\cite{Escribano:2015nra,Escribano:2015yup}. In between, even though pQCD and ChPT may provide insights on the TFF structure, no full result is known from first principles, thus experimental data is then required.  In this respect, REDTOP could access the singly- and doubly-virtual TFF ($F_{P\gamma^*\gamma^*}(q^2,0)$ and $F_{P\gamma^*\gamma^*}(q_1^2,q_2^2)$) in  Dalitz ($P\to\gamma\ell^+\ell^-$) and double-Dalitz ($P\to\ell^+\ell^-\ell^{'+}\ell^{'-}$) decays, respectively. In this respect, QED radiative corrections have been studied for both cases~\cite{Husek:2017vmo,Kampf:2018wau}.
TFFs play as well an important role in several precision observables at the frontier of the standard model, such as the anomalous magnetic moment of the muon $(g-2)_{\mu}$~\cite{Aoyama:2020ynm}.
\end{sloppypar}

\subsubsection{\texorpdfstring{$P\to\gamma\gamma$}{} decays}

$P\to\gamma\gamma$ decays provide access to the TFF normalization: 
\begin{equation}\label{eq:Pto2g}
    \Gamma_{P\to\gamma\gamma} = \frac{m_{P}^3\alpha^2\pi}{4}|F_{P\gamma^*\gamma^*}(0,0)|^2.
\end{equation}
This is relevant when computing the contribution to the HLbL contribution to the $(g-2)_{\mu}$ (see below). The current best measurements, $\Gamma_{\eta\to2\gamma} = 0.520(20)(13)$~keV and $\Gamma_{\eta^{\prime}\to2\gamma} = 4.17(10)(27)$~keV, come from KLOE-2~\cite{KLOE-2:2012lws} and L3~\cite{L3:1997ocz}, respectively. 
Besides, the decay widths are related to the chiral anomaly and can provide insights to the $\eta-\eta^{\prime}$ mixing~\cite{Leutwyler:1997yr,Feldmann:1999uf,Kaiser:2000gs,Escribano:2005qq,Escribano:2013kba,Escribano:2015nra,Escribano:2015yup}. In particular, they can be expressed as
\begin{equation}
   F_{\eta\gamma^*\gamma^*}(0,0) = \frac{1}{4\pi^2}
    \frac{\hat{c}_8F_{\eta^{\prime}}^0 -\hat{c}_0F_{\eta^{\prime}}^8}{F_{\eta^{\prime}}^0F_{\eta}^8 -F_{\eta^{\prime}}^8F_{\eta}^0}, 
   \qquad F_{\eta^{\prime}\gamma^*\gamma^*}(0,0) = \frac{1}{4\pi^2}
    \frac{\hat{c}_8F_{\eta}^0 -\hat{c}_0F_{\eta}^8}{F_{\eta}^0F_{\eta^{\prime}}^8 -F_{\eta}^8F_{\eta^{\prime}}^0},
\end{equation}
with $F_P^a = \langle 0 | A_{\mu}^a | P(q) \rangle \equiv q_{\mu} F_P^a $ the $\eta,\eta^{\prime}$ decay constants, that can be described in terms of two decay constants and mixing angles,
\begin{equation*}
    F_{\eta}^8 = F_8\cos\theta_8, \quad 
    F_{\eta}^0 = -F_0\sin\theta_0, \quad 
    F_{\eta^{\prime}}^8 = F_8\sin\theta_8, \quad 
    F_{\eta^{\prime}}^0 = F_0\cos\theta_0.
\end{equation*}
Finally, $\hat{c}_8 = \sqrt{1/3}(1+...)$ and $\hat{c}_0 = \sqrt{8/3}(1+...)$, are charge factors, with the ellipses standing for chiral and large-$N_c$ corrections, see~\cite{Escribano:2015yup} for a full detailed description.

\subsubsection{Dalitz decay}

\begin{table}[ht]
    \centering
    \begin{tabular}{c c c c c } \hline
         & $\eta\to\gamma e^+e^-$ & $\eta\to\gamma \mu^+\mu^-$ & $\eta^{\prime}\to\gamma e^+e^-$ & $\eta^{\prime}\to\gamma \mu^+\mu^-$   \\\hline
     BR    & $6.9(4)10^{-3}$ & $3.1(4)10^{-4}$ & $4.91(27)10^{-4}$ & $1.13(28)10^{-4}$ \\ \hline
    \end{tabular}
    \caption{Dalitz decays BR from PDG~\cite{ParticleDataGroup:2020ssz} for the different decay channels.}
    \label{tab:BRdalitz}
\end{table}

The amplitude for the Dalitz decay can be written as
\begin{equation}
    \mathcal{M} = \epsilon^{\mu\nu\rho\sigma}\varepsilon^*_{\mu} k_{\nu}q_{\sigma}[\bar{u}(p_-)\gamma_{\rho}v(p_+)] F_{\eta\gamma^*\gamma^*}(q^2,0)(1/q^{2}),
\end{equation}
where $q=p_+ + p_-$ is the sum of the  lepton momenta, while $\bar{q} = p_+ -p_-$, $k$ is the photon momenta, and $\varepsilon$ the photon polarization vector. As it is clear from above, this process allows to probe the nontrivial structure of pseudoscalar mesons via the $q^2$-dependent form factor. 

Adding over lepton and photon polarizations, we obtain
\begin{equation}
    |\mathcal{M}|^2 = \frac{e^6 |F_{\eta\gamma^*\gamma^*}(q^2,0)|^2}{2q^4} \left[
      (m_{\eta}^2 -q^2)^2(q^2+4m_{\ell}^2) +4s(k\cdot\bar{q})^2
    \right]
\end{equation}
where one should note  that $\bar{q}^2 = 4m_{\ell}^2 -q^2$ and $k^2=0$.
Using $2k\cdot\bar{q}=(m_{\eta}^2 -q^2)\beta_{\ell}\cos\theta$, with $\beta_{\ell}^2= 1 -4m_{\ell}^2/q^2$, and accounting for phase space, the differential decay width can be expressed as 
\begin{align}
    \frac{d\Gamma}{d\cos\theta  dq^2} &{}= \frac{(4\pi\alpha)^3 }{(8\pi)^3 }\bigg(1 -\frac{q^2}{m_{\eta}^2}\bigg)   \times 
    \frac{\beta_{\ell}}{2 q^2 m_{\eta}} (m_{\eta}^2 -q^2)^{2}(2 -\beta_{\ell}^2\sin^2\theta) |F_{\eta\gamma^*\gamma^*}(q^2,0)|^2 \, , \nonumber\\ 
   \frac{d\Gamma}{ dq^2}&{}= \Gamma_{P\to 2\gamma} \frac{2\alpha }{3\pi} \frac{1}{q^2}\beta_{\ell}
    \bigg(1 +\frac{2m_{\ell}^2}{q^2}\bigg) \bigg(1 -\frac{q^2}{m_P^2}\bigg)^{3} 
    |\tilde{F}_{\eta\gamma^*\gamma^*}(q^2,0)|^2,
\end{align}
where the second line carried out the integration over $d\cos\theta$, and the normalized TFF, $\tilde{F}_{\eta\gamma^*\gamma^*}(q^2,0) = F_{\eta\gamma^*\gamma^*}(q^2,0)/F_{\eta\gamma^*\gamma^*}(0,0)$ has been employed, together with the link between the normalization $F_{\eta\gamma^*\gamma}(0)$ and the $2\gamma$ decay amplitude, Eq.~(\ref{eq:Pto2g}). As such, the $q^2$-dependence of the TFF can be accessed in the differential distribution. 

A simple phenomenological parametrization for the kinematic region of interest for $eta$ decays (for $\eta^\prime$ a careful description of the resonance region is required) is given by the monopole function
\begin{equation}\label{eq:etaVMD}
   \tilde{F}_{\eta (\eta^{\prime}) \gamma^*\gamma^*}(s,0) = \frac{\Lambda^{2}}{\Lambda^{2} -s},
\end{equation}
where a typical $\Lambda \simeq m_{\rho}\simeq 770$MeV is expected. While better motivated and more refined approaches have been used based on Pad\'e approximants~\cite{Masjuan:2008fv,Escribano:2013kba,Escribano:2015nra,Escribano:2015yup}, resonance chiral theory~\cite{Czyz:2012nq,Guevara:2018rhj} and dispersion relations~\cite{Hanhart:2013vba,Holz:2015tcg,Holz:2022hwz}, among others,
still the parametrization above can prove useful to compare sensitivities with respect to former experiments. In particular, the most precise measurement of the $\eta-$TFF comes from the A2 Coll.~\cite{A2:2013wad}, which after fitting with Eq.(\ref{eq:etaVMD}) obtained $\Lambda^{-2}=1.95(15)(10)~\textrm{GeV}^{-2}$. Regarding the $\eta^{\prime}$, for $\ell=e$, where the $\rho,\omega$ peaks are kinematically allowed, the monopole ansatz is not really useful, and more sophysticated approaches including resonances are required. The most precise measurements comes from BESIII~\cite{BESIII:2015zpz} and comprises 864 events. 

Importantly, to achieve the necessary precision, radiative corrections are necessary. These have been computed in Ref.~\cite{Husek:2017vmo}.

\subsubsection{Double-Dalitz decays}

\begin{table}[ht]
    \centering
    \begin{tabular}{c c c c c c c} \hline
         & $\eta\to 2e^+2e^-$ & $\eta\to e^+e^-\mu^+\mu^-$ & $\eta\to 2\mu^+2\mu^-$  & $\eta^{\prime}\to 2e^+2e^-$ & $\eta^{\prime}\to e^+e^-\mu^+\mu^-$ & $\eta^{\prime}\to 2\mu^+2\mu^-$  \\\hline
     BR    & $2.71(2)10^{-5}$ & $2.4(1)10^{-6}$ & $4.0(2)10^{-9}$ &  $2.1(5)10^{-6}$ & $6.4(9)10^{-7}$ & $1.7(4)10^{-8}$ \\ \hline
    \end{tabular}
    \caption{Theoretical estimates from~\cite{Escribano:2015vjz}. Only BR$(\eta\to2e^+2e^-) = 2.40(22)10^{-5}$ has been measured~\cite{KLOE-2:2012lws}.}
    \label{tab:BRDDalitz}
\end{table}

Double-Dalitz decays, $P\to\ell^+\ell^-\ell'^+\ell'^-$, where $\ell=e,\mu$ provide access to the doubly-virtual TFF, since both photons are off-shell. In particular, the matrix element can be expressed as
\begin{equation}
  \mathcal{M} = \frac{-e^4 F_{P\gamma^*\gamma^*}(p_{12}^2, p_{34}^2)}{p_{12}^2 p_{34}^2} \epsilon_{\mu\nu\rho\sigma}p_{12}^{\mu}p_{34}^{\rho}
  [\bar{u}(p_1)\gamma^{\nu} v(p_2)]  
  [\bar{u}(p_3)\gamma^{\sigma} v(p_4)]  
\end{equation}
for $\ell\neq\ell'$, while for $\ell = \ell'$ an exchange term exists. For $\ell\neq\ell'$ the decay width can be expressed as 
\begin{equation}
    d\Gamma = \frac{1}{2m_{P}} d\Pi_4 |\mathcal{M}|^2, \quad 
    d\Pi_4 = \frac{\lambda_{12,34}}{2^{14}\pi^6} dp_{12}^2 dp_{34}^2 \lambda_{1,2}d\cos\theta_{12} \lambda_{3,4}d\cos\theta_{34},
\end{equation}
with
\begin{multline}
    |\mathcal{M}|^2 = \frac{e^8 |F_{P\gamma^*\gamma^*}(p_{12}^2, p_{34}^2)|^2}{p_{12}^2 p_{34}^2} m_P^4 \lambda_{12,34}^2 \Big[2  
    -\lambda_{1,2}^2\sin^2\theta_{12}
    -\lambda_{3,4}^2\sin^2\theta_{34} \\
    +\lambda_{1,2}^2\sin^2\theta_{12}\lambda_{3,4}^2\sin^2\theta_{34}\sin^2\phi \Big] ,
\end{multline}
see for instance the definitions in~\cite{Kampf:2018wau},\footnote{In particular,
$\lambda_{i,j}^2 = \sigma((p_i+p_j)^2,p_i^2,p_j^2)/(p_i+p_j)^4$, with  $\sigma(a,b,c) = a^2 +b^2 +c^2 -2ab -2ac -2bc$.}. As such, the differential decay width $d\Gamma/(dp_{12}^2dp_{34}^2)$ provides a natural probe for the doubly-virtual TFF. For $\ell=\ell'$ the distribution is more complicated due to exchange terms.

 Given their $\alpha^4$ suppression, these are extremly rare decays, the reason for which a low-energy measurement of a doubly-virtual TFF has been elusive so far. (To date, only the $\eta^{\prime}$ doubly-virtual TFF has been measured by BaBar~\cite{BaBar:2018zpn} in the deep space-like region using $e^+e^- \to e^+e^-\eta^{\prime} $ events, thus allowing for a test of pQCD.) Indeed, only the $\eta\to 2e^-2e^+$ case has been measured by KLOE~\cite{KLOE:2011qwm} based on 362 events ---insufficient to draw any conclusion on the TFF behavior in terms of doubly virtuallity. Any precise information about the doubly-virtual TFF would be helpful to constrain current approaches used in computing the HLbL contribution to $(g-2)_{\mu}$. In particular, the main challenge would be to distinguish the pQCD behaviour with respect to the naive factorization, this is whether $\tilde{F}_{\eta\gamma^*\gamma^*}(q_1^2,q_2^2) \neq \tilde{F}_{\eta\gamma^*\gamma^*}(q_1^2,0) \times \tilde{F}_{\eta\gamma^*\gamma^*}(0,q_2^2)$, that is expected on the basis of pQCD. (See for instance discussions in~\cite{Escribano:2015vjz} concerning this and estimates for the branching ratio. Radiative corrections have been computed in~\cite{Kampf:2018wau}.)
The double-Dalitz decay is also a basic ingredient for a proper calculation of the $P\to\ell^+\ell^-$ decay~\cite{Masjuan:2015lca,Masjuan:2015cjl}.

\subsubsection{Contribution to the anomalous magnetic moment of the muon, \texorpdfstring{$(g-2)_{\mu}$}{}}

The lightest pseudoscalar mesons play a prominent role in the hadronic light-by-light (HLbL) scattering contribution to the anomalous magnetic moment of the muon, $(g-2)_{\mu}$~\cite{Aoyama:2020ynm}, for which there is at present a $4.2\sigma$ discrepancy between theory and experiment~\cite{Muong-2:2021ojo}. 
In particular, they enter in loops via $\gamma^*\gamma^*\to P \to\gamma^*\gamma^*$ amplitudes, that involve the TFF in each vertex. The calculation requires then a precise estimate for the TFF normalization (that sets the overall scale) as well as a precise description for their $q_{1,2}^2$-dependence at low spacelike ($Q_{1,2}^2 =-q_{1,2}^2$) energies (especially below $Q_{1,2}^2\sim 1~\textrm{GeV}^2$), that renders the loop integral finite. 
 
At $(g-2)_{\mu}$, but also for the $(g-2)_{e}$, TFF enter either at order $\alpha_{em}^2$ in the hadronic vacuum polarization contribution (HVP) or at order $\alpha_{em}^3$ in the HLbL scattering contribution. Currently, the largest uncertaintiy on the SM prediciton for ($g-2_{\mu}$) comes from the HVP. Such HVP is the result of an inclusive measurement nowadays made up from the sum of known exclusive channels from $e^+e^-$ data via a parameterization to these data. Unmeasured channels are guess-estimated via parameterizations that would be constrained using data if would exist. Beyond that, even though the reality deals with $\mu^+ \mu^-$ channels in the $(g-2)_{\mu}$ by definition, all calculations use the aforementioned $e^+e^-$ data, assuming then lepton-flavour universality in all channels.

For HVP, $e^+e^-$ channels with pions in the final state clearly dominate since it is most sensitives to low-energy scales, but $\eta$ and $\eta^{\prime}$ modes are also necessary to reach the precision goal after the recent measurements~\cite{Muong-2:2021ojo}. In that respect, both $\eta$ and $\eta^{\prime}$ single- and double-Dalitz decays including $\mu^+ \mu^-$ channel, $\eta(\eta^{\prime}) \to \pi^+ \pi^- e^+e^-$ or $\mu^+ \mu^-$, $\eta \to 3 \pi (4 \pi) + e^+e^-$, $\eta$ and $\eta^{\prime} \to \pi^0 2 \gamma$, $\eta(\eta^{\prime}) \to \pi \pi \gamma$, shall be included. From the list above, current estimates~\cite{Aoyama:2020ynm} include partial information from $\eta \to e^+e^- \gamma$ and $\eta \to e^+e^- \pi^+ \pi^-$ only.

For HLbL, TFF in the space-like region of the invariant mass of the lepton pair from 0 up to infinity are required. Parameterizations are used and Dalitz-decay data have been shown to be the most constraining information for them as they provide information of the TFF at low energies, the region dominating TFF contributions to HLbL~\cite{Aoyama:2020ynm,Masjuan:2012wy,Masjuan:2012qn,Escribano:2013kba,Escribano:2015nra,Escribano:2015yup,Masjuan:2017tvw}. The normalization of the TFF, $F_{\eta(\eta^{\prime}) \gamma \gamma}(0)$, related to the decay $\eta(\eta^{\prime}) \to \gamma \gamma$ is also essential to get both the central value of the contribution and its error under control.
In contributions to HLbL the doubly-virtual TFF must be used and information for that function is really scarce. Access to $\eta(\eta^{\prime}) \to \mu^+\mu^-$ or $e^+ e^-$ can provide inestimable information for them. If NP contributions are expected in the $(g-2)_{\mu}$ they may also be expected in such pseudoscalar decays into a lepton pair~\cite{Masjuan:2015lca,Masjuan:2015cjl}.

Another aspect for which the current proposal may yield a significant contribution is the comparison between different leponic modes, such as $\eta \to e^+e^- \gamma$ versus $ \eta \to \mu^+\mu^-\gamma$ in order to both investigate LFU and consistency of the whole HVP approach. Let us emphasize that the required $\gamma^*\gamma^*\to P \to\gamma^*\gamma^*$ amplitudes entering in both HVP and HLbL which would have a muon external line, are obtained from $e^+e^-$ data, thus assuming LFU in all modes. 

In summary, precise TFF measurements at low-energies can help improving current estimates for their contributions.




\section{Muon polarimetry at REDTOP\label{sec:muon-polarimetry}}

One striking consequence of the quantum numbers of the $\eta$/$\eta^{\prime}$ mesons is that,
for selected decays, certain net muon polarizations, such as longitudinal or transverse, are highly suppressed in the SM. Such property
open the doors to the exploration of a broad range of effects of New
Physics which would become manifest with a non-null polarization
of the muons obtained from the decay of the $\eta$/$\eta^{\prime}$ mesons.
Muon polarimetry is based on the measurement of the shape of the positron
(electron) spectrum from $\mu\rightarrow e\nu_{\mu}\bar{\nu_{e}}$
decay in flight, which depends on the polarization of the parent muon.
Therefore, it has the following three requirements: a) the detector
has to be made of a material which does not change the initial polarization,
b) the granularity has to be sufficiently small to be able to measure
the angle between the path of the decaying muon and the electron,
c) a magnetic field of known intensity. The implementation of muon
polarimetry is currently under study at REDTOP by exploiting two options.
In the first option, The ADRIANO2 calorimeter is being considered
(see also Sec.~\ref{subsec:The-REDTOP-detector}, possibly equipped
with few layers with smaller tiles to increase the granularity in
the region where muon polarimetry is performed. The material used
for ADRIANO2 tiles (lead glass and scintillating plastic) is amorphous
and it will not change the original polarization of the muon, as it
will occur with crystal calorimeters. The calorimeter is immersed
in a 0.6 T uniform magnetic field, which fulfills requirement c) above.
In the second option, a dedicated polarimeter will be inserted approximately
20 radiation lengths inside the calorimeter. The polarimeter would
be made of thick aluminum plates, where the muon will either stop
or decay in flight, interspersed with gaseous tracking chambers which
will detect the electron. Aluminum is known for its property of not
affecting the original polarization of the muon.

\subsection{CP violation via longitudinal polarization in \texorpdfstring{$\eta\to\mu^+\mu^-$}{}\label{subsec:CP-violation-in-eta-mumu}}

The muon pair in $\eta\to\mu^+\mu^-$ decays can be produced in a $^1S_0$ or a $^3P_0$ state, the former corresponding to a $CP$-conserving transition and the latter to a $CP$-violating one. The most general structure has the form given in Eq.~(\ref{eq:PtoLLamp}). Both contributions are $C$ even, while $g_{P,S}$ terms are $P$ even and $P$ odd, respectively. The polarized decay width can be expressed as~\cite{Sanchez-Puertas:2018tnp}
\begin{multline}
    d\Gamma = \frac{\beta_{\mu}}{16\pi m_{\eta}} \times 
    \frac{m_{\eta}^2}{2}\Big[
    |g_P|^2(1-s^+\!\cdot s^-)
    +|g_S|^2\beta_{\mu}^2[1 - \{2(s^+\!\cdot \hat{\boldsymbol{\beta}}_{\mu^+})(s^-\!\cdot \hat{\boldsymbol{\beta}}_{\mu^+}) - s^+ \! \cdot s^- \}] \\
    +2\operatorname{Re}(g_Pg_S^*)\boldsymbol{\beta}_{\mu^+}\!\cdot(s^-\!\times s^+)
    +2\operatorname{Im}(g_Pg_S^*)\boldsymbol{\beta}_{\mu^+}\!\cdot(s^+ -s^-)
    \Big],
\end{multline}
where $\boldsymbol{\beta}_{\mu^+}$ is the $\mu^+$ velocity, with modulus     $\beta_{\mu} = (1 -4m_{\mu}^2/m_{\eta}^2)^{1/2}$, and $s^{\pm}$ stands for the $\mu^{\pm}$ spin. The expression above allows for an easy connection to the spin density formalism (see Ref.~\cite{Haber:1994pe}) upon $s\to \tau$.

Remarkably, this decay allows for tests of $CP$ violation via longitudinal (and triple-product) spin polarization. In particular, the longitudinal polarization is defined as 
\begin{equation}\label{eq:polL}
    P_L = \frac{N_R -N_L}{N_R +N_L}
\end{equation}
with $N_{R(L)}$ counting the numbers of outgoing $\mu^+$ with positive(negative) helicity. Such observable is of the $C$ even, $P$ odd, $T$ even kind. Paradoxically, this polarization does not prove a test of $CPT$. The $CPT$ theorem merely requires a non-vanishing unitary phase, together with $CP$ violation, to find a nonzero longitudinal polarization. This is indeed the case since the SM amplitude, encapsulated in $g_P$, has a large imaginary part due to the dominant intermediate $2\gamma$ state~\cite{Masjuan:2015cjl}, a feature long recognized~\cite{Sehgal:1969zk}. 
Finally, the triple-product asymmetry can be defined in terms of the $s^+\times s^-$ chirality,
\begin{equation}\label{eq:polchi}
    P_T = \frac{N_{RH} -N_{LH}}{N_{RH} +N_{LH}},
\end{equation}
with $N_{RH(LH)}$ standing for $(s^+\times s^-)\cdot\boldsymbol{\beta}_{\mu^+}$ positive(negative). This is, whether the vectors $\{s^+,s^-\,\boldsymbol{\beta}_{\mu^+} \}$ make up a right-handed or a left-handed system. Such an observable is of the $C$ even, $P$ odd and $T$ odd kind, and thus probes again $CP$ violation.

Any source of $CP$ violation in these decays would clearly point to new physics since, in the SM, $CP$ violation requires the presence of the CKM matrix and thus electroweak physics, necessarily involving weak and CKM suppression factors.

\subsection{CPT violation in transverse polarization}

The transverse muon polarization $P_{T}$ in meson decays is a $T$-odd observable, defined by the projection of the $\mu^{\pm}$ spin transverse to the decay plane. A non-zero value of $P_{T}$ would be a clear evidence for violation of time reversal invariance ($T$)~\cite{Sakharov1967}, since the spurious effects from final state interactions are small~\cite{Zhitnitskii1980}. 

\begin{figure}[!htbp]
\centering{}\includegraphics[width=0.35\columnwidth]{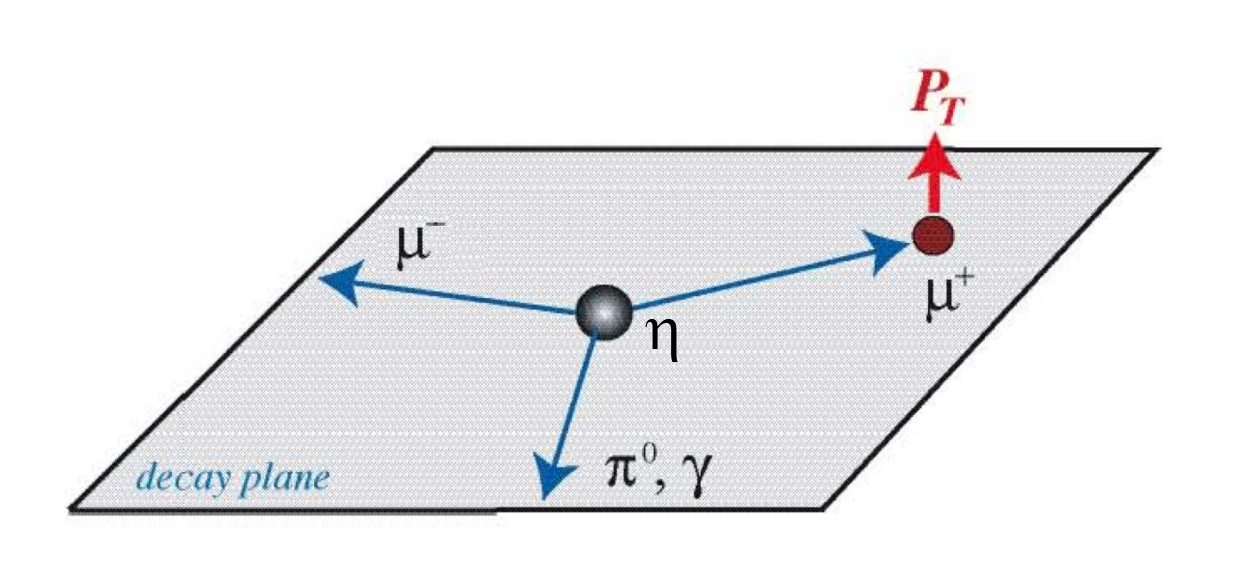}
$\;\;$\includegraphics[width=0.35\columnwidth]{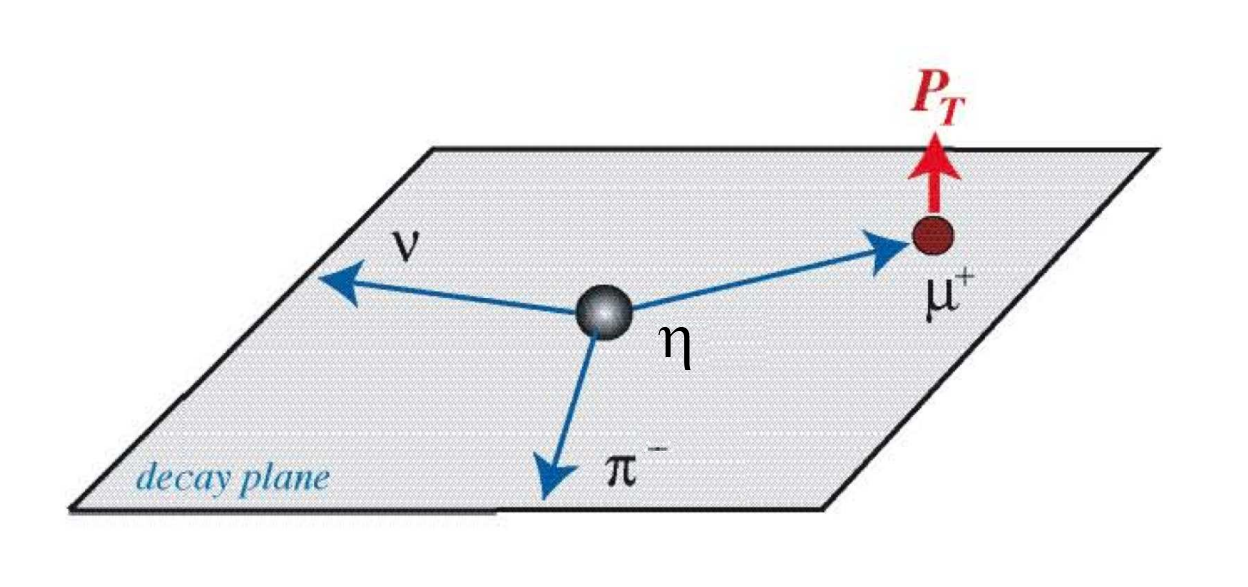}
\caption{\label{fig:eta2muon-pt} Transverse muon polarization P$_{T}$ in
Dalitz (left) and semileptonic (right) $\eta$ decays at rest. }
\end{figure}

Violation of time reversal symmetry ($T$) provides an alternative
means to search for violation of charge conjugation and parity ($CP$)
based on the more general $CPT$ theorem. Sources of $CP$ violation
beyond the Standard Model in the neutral meson sector are one of Sakharov's
criteria~\cite{Sakharov1967} for an explanation of the matter-antimatter
asymmetry observed in the universe. Electroweak theory allows one
to link $T$-odd observables (which change sign under time reversal
transformation) to time reversal symmetry breaking, which can be interpreted
as clear indication of New Physics. The transverse polarization $P_{T}$
of muons in $\eta$/$\eta^{\prime}$ decays is depicted in Fig.~\ref{fig:eta2muon-pt}
for Dalitz (left) and semileptonic (right) decays.


%
%
The Standard Model prediction for $P_{T}$ is extremely small, arising only from higher-order loop contributions
\cite{Bigi1999}. As an example, these have been computed in the context of Kaon decays~\cite{Zhitnitskii1980,Efrosinin2000}, finding them at the $10^{-5}$level --- similar results should apply to $\eta\to\pi^{-}\mu^{+}\nu$ decays. More complex
models involving New Physics such as multi-Higgs doublet models, leptoquark
models or supersymmetric models with R-parity breaking or s-quark
mixing predict much larger values for $P_{T}$ ranging from $10^{-4}$
to $10^{-2}$~\cite{garisto1991non,Wu1997}. As an aside note, the $\eta\rightarrow\pi\mu\nu_{\mu}$
process has not been observed yet.

In Dalitz decays, $\eta\to\gamma\ell^+\ell^-$, the leading order QED and weak contributions do not contribute to such an asymmetry~\cite{Sanchez-Puertas:2018tnp}, such that any positive observation at REDTOP would correspond to physics BSM. Likewise, in $\eta\to\pi^0\mu^+\mu^-$, no contribution to such an asymmetry was found in~\cite{Escribano:2022wug}, such that any positive finding would point to physics BSM.  

In conclusions, any observation of a transverse muon polarization in the 3-body decays of $\eta/\eta^{\prime}$ mesons is an unambiguous indication of $CPT$ violation and physics BSM.

\subsection{Polarization in Dalitz decays}

By contrast to the $\eta\to\mu^+\mu^-$ case, Dalitz decays provide more opportunities regarding muon polarization observables. On the one hand, compared to parity-violating asymmetries in $\eta\to\mu^+\mu^-$ decays (which are necessarily $CP$ odd), parity-violating asymmetries in Dalitz decays are potentially more sensitive to $CP$-even physics. This is related to the fact that $CP$-odd contributions require unitary phases on the basis of the $CPT$ theorem, which are small in Dalitz decays. As an example, $CP$-even weak interactions do contribute to $P_L$ due to $Z$-boson exchange~\cite{Bernabeu:1998hy}. These are however small (note the typical suppression factor $G_F m_{\eta}^2 /\alpha \sim 10^{-5}$ with respect to the QED contribution). In particular, a contribution of $6\times 10^{-7}$ was found in~\cite{Sanchez-Puertas:2018tnp} for $A_L$ (see Eq.~\eqref{eq:AL}), that is beyond REDTOP statistics. Hence any positive finding would point to BSM physics. On the other hand, further asymmetries can be studied, see for instance the transverse plane asymmetry above or those defined in Ref.~\cite{Sanchez-Puertas:2018tnp}.

\subsection{Polarization in \texorpdfstring{$\eta\to\pi^0\mu^+\mu^-$}{} decays}

Similar to Dalitz decays, the richer dynamics of this process allows for further opportunities as compared to the $\eta\to\mu^+\mu^-$ case (note for instance the transverse plane asymmetry among others). However, by contrast to the Dalitz decays, and in the context of parity violating asymmetries, $CP$-violating contributions arise at lower dimension than $CP$-even ones, see Ref.~\cite{Escribano:2022wug}. Further, the SM QED amplitude features a large unitary phase due to the intermediate 2-photon exchange, potentially enhancing the sensitivity to $CP$-odd scenarios over $CP$-even ones. As a final note, the SM $P$-odd, $CP$-even contribution would arise through $\gamma Z$ box diagrams, that are highly suppressed for REDTOP statistics. Hence any positive finding would clearly point to BSM physics.

\newpage{}

\section{\label{sec:The-Experimental-Technique}The REDTOP experiment}

The most efficient way to produce $\eta/\eta^{\prime}$ mesons is 
from nuclear scattering of a proton beam onto a nuclear target. Above
E$_{kin}\approx1.2$ GeV several intra-nuclear baryonic resonances
are created, whose decay produce an $\eta$-meson. The $\eta$ production cross section
increases rapidly above threshold (see left plot in  Figure~\ref{fig:urqmd_eta_production}), and the desired yield can be achieve with a beam energies below 2 GeV.

\subsection{Hadro-production of \texorpdfstring{$\eta$}{} mesons\label{subsec:Hadro-production-of-}}

An important aspect in properly designing the REDTOP detector and the
accelerator complex, is the estimation of the $\eta$-meson production
parameters. 
The most abundant hadro-production mode of $\eta$/$\eta^{\prime}$-mesons
occur via the formation and decay of intranuclear baryonic resonances
($\varDelta$'s, N(1440), N(1535), etc.) in nucleon-nucleon collisions. 
The  cross section for such processes is about seven times larger in $p-n$ collisions than $p-p$ ones. 
Therefore the nuclear target should have a large neutron to proton ratio. 
A target with \emph{ low-Z} minimizes
the formation of nuclear fragmentation products, and helps to reduce the particle multiplicity
of the final state.

\begin{figure}[!ht]
\includegraphics[scale=0.4]{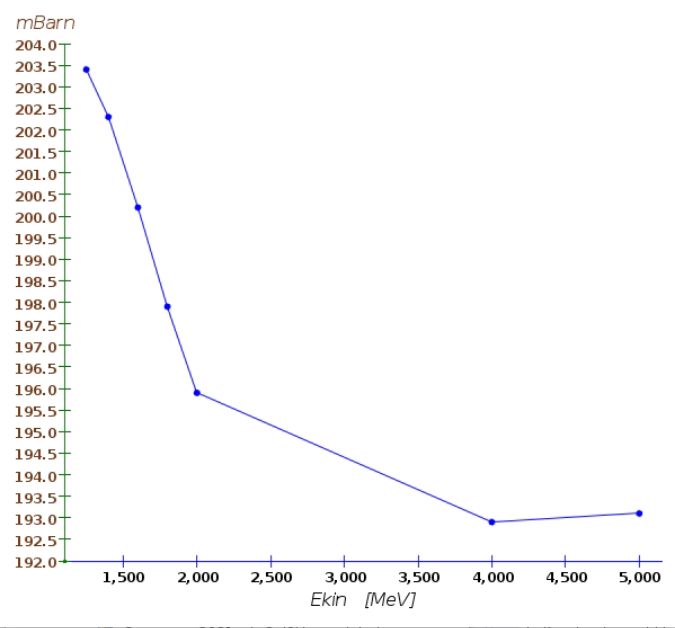}
\includegraphics[scale=0.4]{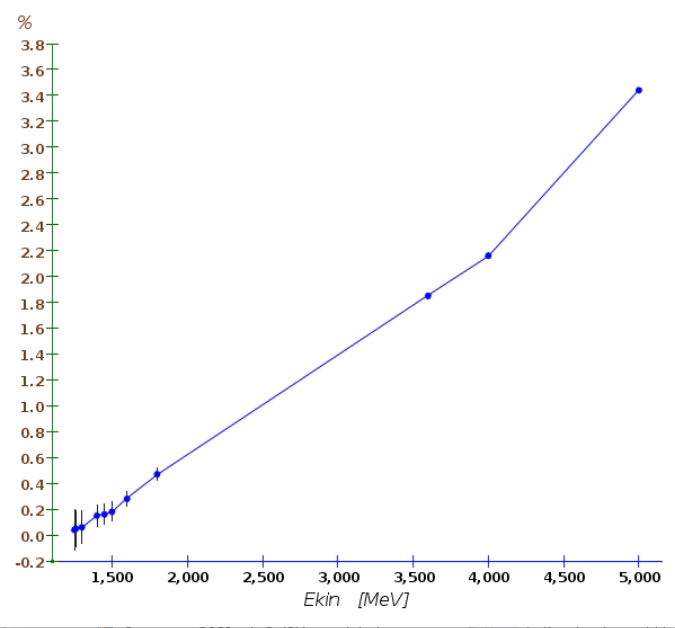}
\caption{Predictions of $p+Li$ cross sections versus beam energy from the \emph{GenieHad}~\cite{GenieHad_2012} model  with
Urqmd +Abla07 scattering models using the Tripathi parametrization
for the inelastic cross section. 
The left panel shows the inelastic cross section while the 
right panel shows the percentage of such collisions that produce an $\eta$ meson.}
\label{fig:urqmd_eta_production}
\end{figure}

The beam energy should be above the threshold  for $\eta$/$\eta^{\prime}$ production of  1.2 GeV  and below the kaon production threshold of 3.3 GeV, since the decay of kaons could mimic the decay of a new long-lived particle. 
Extensive studies have been conducted with the \emph{GenieHad}~\cite{GenieHad_2012}
event-generator framework using multiple beam and target parameters.
Figure~\ref{fig:urqmd_eta_production} shows \emph{GenieHad} predictions for beam energy
dependence
of the total $p+Li$ inelastic cross section (left) 
and the probability to produce an $\eta$ (right). Above threshold,  
the total $\eta$/$\eta^{\prime}$ production cross section increases approximately
linearly with beam energy. 
One consequence of the resonant production mechanism is that the momentum
of the $\eta$/$\eta^{\prime}$ mesons in the lab frame is only sligthly dependent from 
the beam energy, in a interval spanning several hundred MeV above threshold. 
Such studies indicate that a proton beam with E$_{kin}\approx1.8-1.9$
GeV impinging on a thin,
low-Z, high-$\nicefrac{A}{Z}$ material like
Beryllium or Lithium  is an optimal choice for the
experimental program. 


Unfortunately, no experimental data on the $p-Li$
inelastic cross section nor $\eta$/$\eta^{\prime}$ production exist
in this energy range. Furthermore, the abundance of intra-nuclear
resonances complicate the formulation of theoretical models as they
require non-trivial treatment within a non-perturbative approach. 
To minimize the uncertainty in calculating the $\eta$/$\eta^{\prime}$ production
cross section, we used   several models in \emph{GenieHad} to calculate
the cross section and compared few nuclear transport models to describe the
evolution of proton-nucleus collisions. The total $p-target$ inelastic
cross section has been calculated for protons with $E_{kin}$=
1.8 GeV by averaging seven different models and a detailed description of
the target including, besides the lithium core, thin protective epoxy
and copper layers. The results of these calculations are summarized in
Tab.~\ref{table:total-crossections}.

\begin{table}[ht]
\textbf{\footnotesize{}\centering}%
\begin{tabular}{|c|c|c|c|}
\hline 
\textbf{\textit{\footnotesize{}Model}} & \textbf{\textit{\footnotesize{}p-Li cross section}} & \textbf{\textit{\footnotesize{}p-Li}} & \textbf{\textit{\footnotesize{}p-Target}}\tabularnewline
 & \textbf{\textit{\footnotesize{}{[}cm$^{-2}]$}} & \textbf{\textit{\footnotesize{}interaction prob. }} & \textbf{\textit{\footnotesize{}interaction prob. }}\tabularnewline
\hline 
\hline 
\textbf{\footnotesize{}Wellisch \& Axen} & \textbf{\footnotesize{}$2.01\times10^{-25}$} & \textbf{\footnotesize{}$0.710\%
$} & \textbf{\footnotesize{}$0.719\%
$}\tabularnewline
\hline 
\textbf{\footnotesize{}Tripathi Light} & \textbf{\footnotesize{}$1.96\times10^{-25}$} & \textbf{\footnotesize{}$0.693\%
$} & \textbf{\footnotesize{}$0.702\%
$}\tabularnewline
\hline 
\textbf{\footnotesize{}Incl++ } & \textbf{\footnotesize{}$1.60\times10^{-25}$} & \textbf{\footnotesize{}$0.567\%
$} & \textbf{\footnotesize{}$0.574\%
$}\tabularnewline
\hline 
\textbf{\footnotesize{}Sihver et. al } & \textbf{\footnotesize{}$1.51\times10^{-25}$} & \textbf{\footnotesize{}$0.535\%
$} & \textbf{\footnotesize{}$0.543\%
$}\tabularnewline
\hline 
\textbf{\footnotesize{}Barashenkov} & \textbf{\footnotesize{}$1.73\times10^{-25}$} & \textbf{\footnotesize{}$0.612\%
$} & \textbf{\footnotesize{}$0.620\%
$}\tabularnewline
\hline 
\textbf{\footnotesize{}Shen et. al } & \textbf{\footnotesize{}$2.0\times10^{-25}$} & \textbf{\footnotesize{}$0.707\%
$} & \textbf{\footnotesize{}$0.715\%
$}\tabularnewline
\hline 
\textbf{\footnotesize{}Kox et. al } & \textbf{\footnotesize{}$2.98\times10^{-25}$} & \textbf{\footnotesize{}$1.06\%
$} & \textbf{\footnotesize{}$1.07\%
$}\tabularnewline
\hline 
\textbf{\footnotesize{}$Average$} & \textbf{\textit{\footnotesize{}$1.98\pm0.48\times10^{-25}$}} & \textbf{\textit{\footnotesize{}$0.70\pm0.17\%
$}} & \textbf{\textit{\footnotesize{}$0.71\pm0.17\%
$}}\tabularnewline
\hline 
\end{tabular}\caption{Estimate of inelastic cross section and interaction probability for
protons $E_{kin}$= 1.8 GeV on REDTOP targets using different models
in \emph{GenieHad}~\cite{GenieHad_2012}. The error indicates
the standard deviation.}
\label{table:total-crossections}
\end{table}

In order the estimate the $\eta$ yield at REDTOP, the outcome of four $GenieHad$ hadronic
generator models was examined. The $\eta$ production probability
in inelastic $p+Li$ collisions at 1.8 GeV is summarized in in Tab.
\ref{table:eta-production-probability}.
\begin{table}[ht]
\textbf{\footnotesize{}\centering}%
\begin{tabular}{|c|c|}
\hline 
\textbf{\textit{\footnotesize{}Nuclear collision model}} & \textbf{\textit{\footnotesize{}p+Li}}\tabularnewline
 & \textbf{\textit{\footnotesize{}$\eta$ yield}}\tabularnewline
\hline 
\hline 
\textbf{\footnotesize{}Urqmd}~\cite{Bleicher_1999} & \textbf{\footnotesize{}$0.49\%
$}\tabularnewline
\hline 
\textbf{\footnotesize{}Incl++ v6.2~\cite{Incl2020}} & \textbf{\footnotesize{}$1.48\%
$}\tabularnewline
\hline 
\textbf{\footnotesize{}Gibuu v2019}~\cite{BUSS20121} & \textbf{\footnotesize{}$0.74\%
$}\tabularnewline
\hline 
PHSD v 4.0~\cite{Ehehalt:1996uq} & \textbf{\footnotesize{}$0.67\%
$}\tabularnewline
\hline 
Jam v1.9~\cite{Nara2019} & \textbf{\footnotesize{}$0.26\%
$}\tabularnewline
\hline 
\textbf{\footnotesize{}$Average$} & \textbf{\textit{\footnotesize{}$(0.73\pm0.46)\%
$}}\tabularnewline
\hline 
\end{tabular}\caption{Estimate of $\eta$ production probability in inelastic collisions at
$E_{kin}$= 1.8 GeV on REDTOP targets using different models in \emph{GenieHad}~\cite{GenieHad_2012}.
The error indicates the standard deviation.}
\label{table:eta-production-probability}
\end{table}
 The average valued is: $Prob(p+Li\rightarrow\eta\,X)\approx(0.73\pm0.46)\%
$. Assuming $10^{18}$ POT, the yields of $\eta$/ $\eta^{\prime}$ and of
inelastic proton interactions are summarizes in Table~\ref{tab:Expected-yield-at-108POT}.
.

\begin{table}[!ht]
\begin{centering}
\begin{tabular}{|c|c|}
\hline 
 & Expected yield for $E_{kin}$=1.8 GeV\tabularnewline
\hline 
\hline 
$N_{\eta}$ & $5.1\times10^{13}$\tabularnewline
\hline 
$N_{\eta^{\prime}}$ & 0\tabularnewline
\hline 
$N_{ni}$ & $7\times10^{15}$\tabularnewline
\hline 
\end{tabular}$\;\;$%
\begin{tabular}{|c|c|}
\hline 
 & Expected yield $E_{kin}$=3.6 GeV\tabularnewline
\hline 
\hline 
$N_{\eta}$ & $1.8\times10^{14}$\tabularnewline
\hline 
$N_{\eta^{\prime}}$ & $2.4\times10^{11}$\tabularnewline
\hline 
$N_{ni}$ & $9.7\times10^{15}$\tabularnewline
\hline 
\end{tabular}
\par\end{centering}
\caption{\label{tab:Expected-yield-at-108POT}Expected $\eta$ and $\eta^{\prime}$
yield at REDTOP for $10^{18}$ POT (see Sec.~\ref{sec:The-Genie-event}
for details on the estimation method)}
\end{table}
A similar analysis with Urqmd indicates that the inelastic proton
scattering probability on REDTOP target with an $E_{kin}$=3.6 GeV
is 0.68\% (a factor $\sim1.4\times$ than at 1.8 GeV. On the other
side, the fraction of $\eta^{\prime}$ mesons in such events is $2.5\times10^{-5}$.
The corresponding $\eta^{\prime}$ yield is summarized in Table~\ref{tab:Expected-yield-at-108POT}. 

\subsection{Beam and target requirements\label{subsec:Beam-and-target}}

\begin{figure}[!ht]
\includegraphics[scale=0.6]{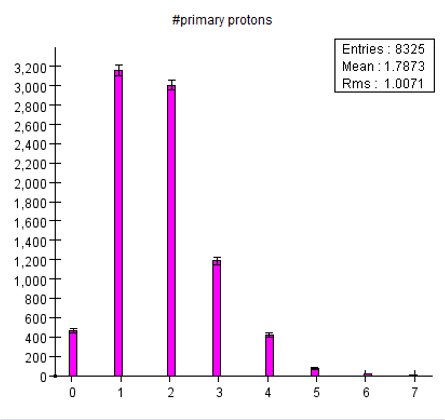} \includegraphics[scale=0.6]{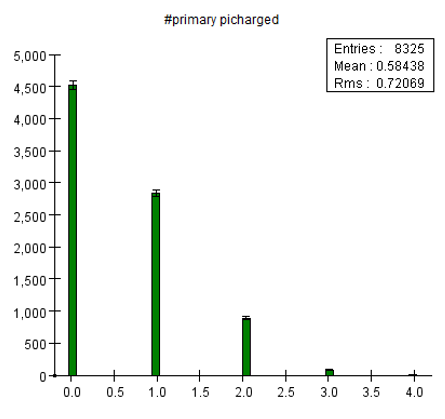}
\caption{Multiplicity of pions, protons and neutrons for p-Li scattering at
1.8 GeV for events without a $\eta$ or $\eta^{\prime}$ meson (\emph{GenieHad}
simulation with Urqmd +Abla07 scattering models).}

\label{fig:urqmd_multiplicity}
\end{figure}

Following the discussion in Sec.~\ref{subsec:Hadro-production-of-}, the
beam energy chosen for the $\eta$ run of REDTOP is 1.8 GeV, where
the neutral kaon production probability for $p+Li$ collisions is
about $1.8\times10^{-4}$. It was also noted that the higher$\nicefrac{A}{Z}$of
the Li, compared to the rest of\emph{ low-Z} nuclei, makes it an ideal
material for REDTOP target systems. For example, the multiplicity
of primary neutrons produced by a Nb target increases more than 20\%
wrt a Be target, while the increase of primary protons is almost 90\%
(from \emph{GenieHad} simulations). Fig.~\ref{fig:urqmd_multiplicity}
summarizes the multiplicity of pions, protons and neutrons for QCD
events non containing and $\eta$ or $\eta^{\prime}$ meson (GenieHad simulation
with Urqmd +Abla07 scattering models). An important requirement for REDTOP is 
minimizing the rescattering of the $\eta$ decay products inside the target
This is achieved by splitting the latter in multiple thinner foils. That would,
on one side, improve the measurement of the z-coordinate of the $\eta$
production vertex, and, on the other side, it minimizes the multiple
scattering affecting the $\eta$ decay products when they escape from
the target. By using 10 foils, spaced 10 \emph{cm} apart, one would
provide the necessary total material budget for the required luminosity,
while the pile-up due to multiple beam interactions within the same
trigger could be easily reduced with vertexing techniques. A similar
study for the $\eta^{\prime}$ case yields an optimal range for E$_{kin}$
of about 3.5-4 GeV.

In conclusion, a CW proton beam delivering $1\times10{}^{11}$POT/sec would generate
a rate of inelastic interaction of about 1 GHz and a $\eta$-meson
yield of $5.1\times10{}^{6}\eta/sec$, corresponding to $5.1\times10{}^{13}\eta/yr.$\footnote{Here we assume that a running time of 1-year corresponds to $10^{7}$
sec} The beam power corresponding to the above parameters is approx. 30
W, of which 1\% (or 300 mW) is absorbed in the target systems. Each
of the 10 foils will have to dissipate about 30 mW, which is easily
achievable by convection of the accumulated heat through the foil holders. These estimate roughly
double for the beam required when running at the $\eta^{\prime}$.

\subsection{Detector requirements\label{subsec:Detector-requirements}}

\begin{figure}[!ht]
\includegraphics[width=8cm,height=6.5cm]{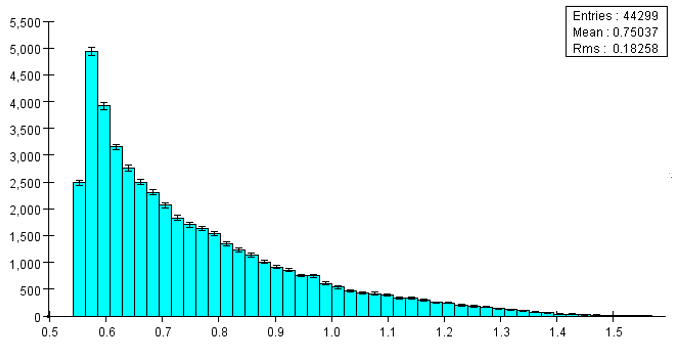} \caption{Total energy of $\eta$ mesons from a $E_{kin}$=1.8 GeV proton beam
scattered on REDTOP targets (\emph{GenieHad} simulation).}

\label{fig:eta_energy}
\end{figure}

The $\eta$ mesons generated with the above beam parameters are almost
at rest in the lab frame, receiving only a small boost in the direction
of the incoming beam (see Fig.~\ref{fig:eta_energy}). This is a consequence
of their production mechanism as they are the decay products of an
intranuclear resonance in the target material. Studies performed with
\emph{GenieHad} have indicated that the $\eta-$spectrum is almost
independent from the beam energies in the {[}1.46-2.1{]} GeV range.
Consequently, an hermetic detector covering the entire solid angle,
is one of the requirements for REDTOP.

A correct identification of the final state particles is, also, of
paramount importance for an experiment exploring rare processes. On
one side, the vast majority of the processes listed in Fig.~\ref{fig:physics_landscape}
have two leptons at at least one photon in the final state. On the
other side, the QCD background, which has a production rate two order
of magnitudes larger, is populated almost invariably by protons, neutrons,
slow pions and nuclear remnants. These are relatively slow particles,
as shown in Fig.~\ref{fig:eta_energy} for the case of protons and
pions. Furthermore, neutrons impinging onto the calorimeter would
easily be mistaken as photons if a proper particle identification
(PID) is not in place. Finally, the identification of$e^{+}e^{-}$
pairs from photons converting in the material upstream the central
tracker will help to reject the large background originating from
QCD production of $\pi^{0}$ particles. The detector is inserted
in a solenoidal magnetic field which bends the charged particles for
proper $P_{t}$ and electric charge measurement.

\begin{figure}[!ht]
\includegraphics[scale=0.7]{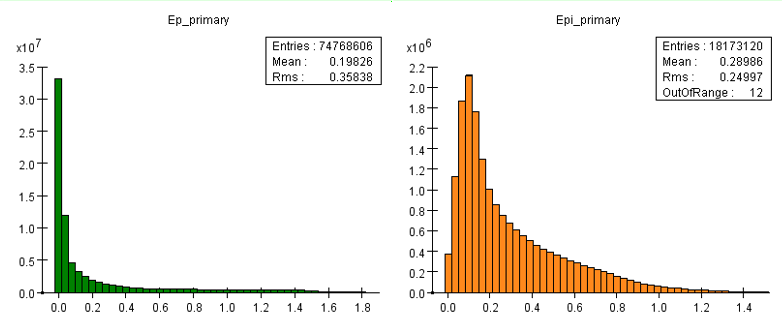} \caption{Momentum of protons (left) and charged pions(right) from inelastic
scattering of a $E_{kin}$=1.8 GeV proton on REDTOP targets (\emph{GenieHad}
simulation).}

\label{fig:bkg_p_pi-1}
\end{figure}
\begin{figure}[!ht]
\includegraphics[width=0.95\textwidth]{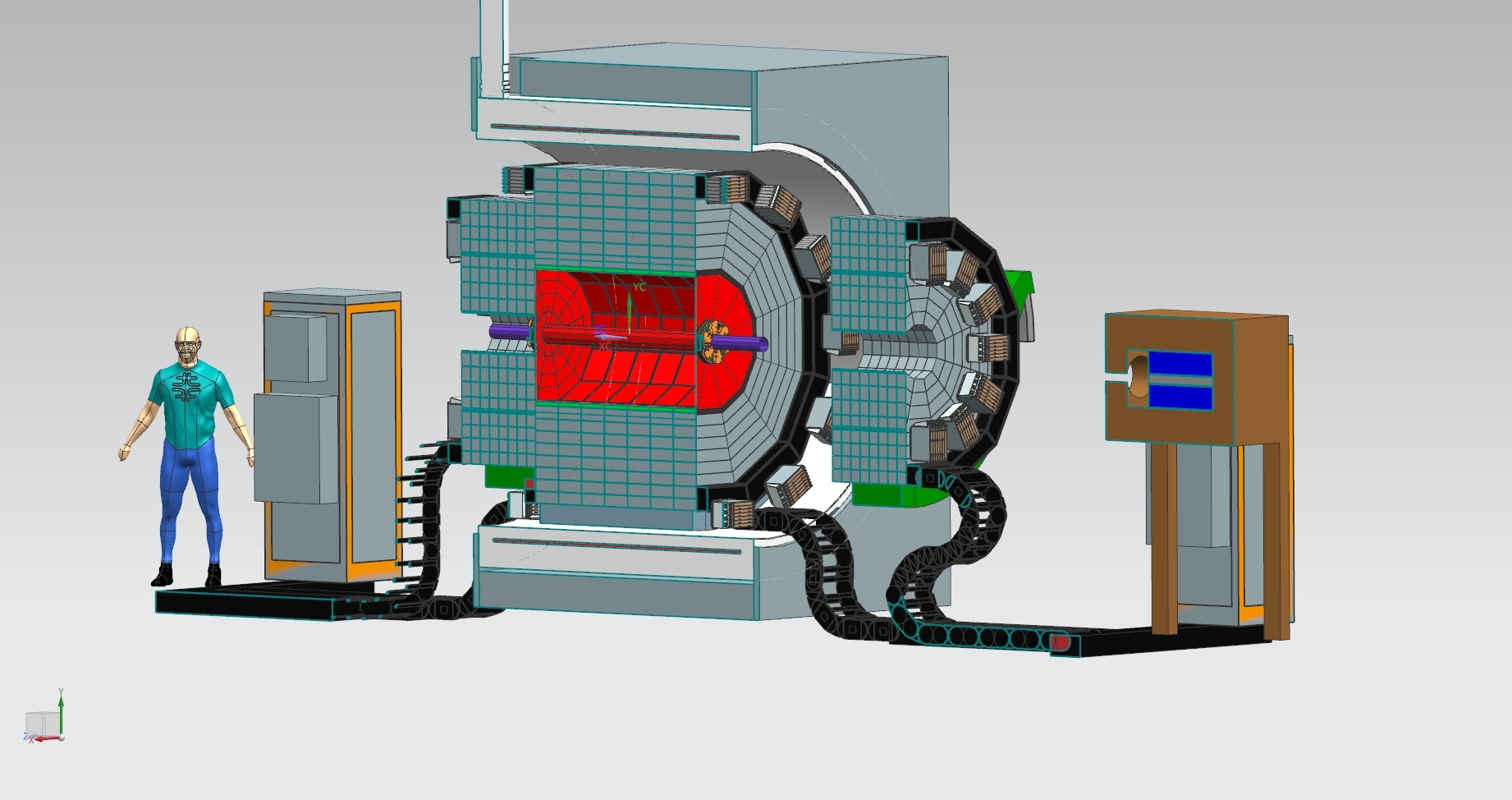} \caption{Cross section of REDTOP detector}

\label{fig:detector-sketch}
\end{figure}

\subsection{The REDTOP detector\label{subsec:The-REDTOP-detector}}

A baseline detector layout has been defined for preliminary studies of sensitivity to a broad range of BSM physics processes. 
Cost optimization  is deferred to at a later stage of the project.
The
individual components of REDTOP detector are briefly summarized below.

\subsubsection{The vertex detector}

The vertex detector has four main tasks: a) identifying events with a
detached secondary vertex. b) contributing to the reconstruction of charged
tracks originating in the target; c) rejecting photons converting
in the target; d) reconstructing track with very low transverse
momentum. 
Two detector technology options are currently being investigated for
the vertex detector: a wafer-scale silicon sensor and a thin scintillating
fiber tracker. 


The most important challenges for REDTOP vertex detector are related
to the material budget and to its proximity to the interaction region.
Keeping the material budget as low as possible is a  necessary requirement to
reduce the multiple scattering,  which would, otherwise,
compromise the momentum resolution and increase the background from
photon conversion. The main requirements for the vertex detector,
derived  from the sensitivity studies to several benchmark processes
(cf.\ Sec.~\ref{sec:Sensitivity-Studies-to-physics-BSM} and \ref{subsec:Tests-of-Conservation-Laws})
are listed below:
\begin{itemize}
\item A spatial resolution near the IP better than 20 \textgreek{m}m;
\item Material budget: $\sim$0.03\% $X_{0}$/layer;
\item Radiation hard up to of $\sim5\times10^{6}$/cm$^{2}$/s ``\textit{1-MeV
equivalent neutron fluence}'' (or $\sim5\times10^{13}$/cm$^{2}$
integrated over 1 year (see, also, Sec.~\ref{sec:Appendix-III:-Radiation-Damage});
\item Coverage: $\sim$96\% of full solid angle. 
\end{itemize}

\paragraph{Vertex detector option I \textendash{} Fiber tracker\label{par:Vertex-detector-option-I}}

A fiber tracker, with an identical technology as implemented for the
LHCb upgrade\citep{Kirn_2017}, is located around the beam pipe, at
a radius r $\approx$5 cm around the z-axis. Three superlayers are stacked
together, interspaced with structural foam   1 cm thick, providing mechanical
stability to the subdetector. Each superlayer consists of  a mat with 5 layers of scintillating
fibers, with a diameter of 250 $\mu m$. The space resolution of each
superlayer is $\sim70\mu m$, as measured by LHCb at a test beam.
The fiber tracker serves two main purposes: a) it provides a vertex
measurement for the reconstructed event, and b) it identifies photons
converting into $e^{+}e^{-}$ pairs inside the OTPC (typically, in
the inner wall of the chamber or in the aerogel). The determination
of the primary vertex of the event identifies the foil where the inelastic
interaction occurred and it provides the position of the $\eta$-mesons. 

\paragraph{Vertex detector option II \textendash{} Wafer-scale silicon sensor
\label{par:Vertex-detector-option-II}}


This option is based on the current R\&D performed by Alice collaboration
for their ITS3 detector. It consists of three-layer of flexible Silicon
Genesis: 20 micron thick wafer, with CMOS sensors. The wafer-scale
sensors are 900 mm long, fabricated using stitching. A total of six
such sensors is needed to cover the innermost region of REDTOP.  
Key benefits of this option are: a) a material
budget corresponding to mere 0.02-0.04\% $X_{0}$; b) an homogeneous
material distribution, resulting in an almost-negligible systematic
error from the material; c) a 3-dimensional hit information. A picture
of the ITS detector proposed by Alice is shown in Fig.~\ref{fig:ITS3}.
\begin{center}
\begin{figure}[!ht]
\centering{}\includegraphics[scale=0.55]{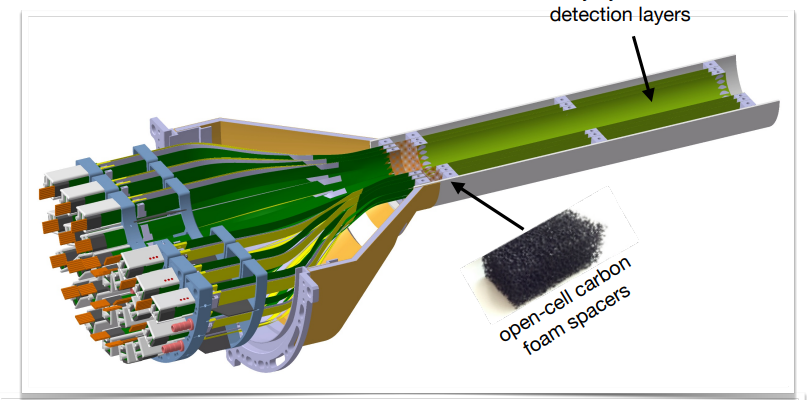}\caption{\label{fig:ITS3}Schematics of Alice proposed ITS3 tracker }
\end{figure}
\par\end{center}

\subsubsection{The central tracker}

\paragraph{Requirements for central tracking}

The two largest challenges for REDTOP tracking are related to the
relatively low momentum of the particles that have to be reconstructed.
These are: a) a very low material budget to reduce the multiple scattering
as much as possible, and b) a time resolution sufficient to contribute 
to the TOF systems for identifying the particles. In other words,
the central tracker needs to be able to reconstruct
charged tracks   in four dimensions. The main requirements
for the central tracker, derived from  the sensitivity studies to several
benchmark processes (cf.\  Sec.~\ref{sec:Sensitivity-Studies-to-physics-BSM}
and \ref{subsec:Tests-of-Conservation-Laws}) are listed below:
\begin{itemize}
\item Momentum resolution: $\sigma_{P_{T}}/P_{T}^{2}\sim2\times10^{-3}$
GeV$^{-1}$at P$_{T}$=1 GeV;
\item Material budget: $\sim$0.1\% $X_{0}$/layer;
\item Time resolutions: $<30psec$;
\item Coverage: $\sim$96\% of full solid angle
\end{itemize}

Two technologies have been considere for REDTOP central tracker: a) low-mass LGAD, and b) an Optical TPC.

\paragraph{\label{subsec:The-LGAD-tracker}Central Tracker option \#1: low-mass
LGAD tracker}

Recent advancements in silicon particle sensors enable precise timing of 
MIP signals, though a sensor design called \emph{Low Gain Avalance Detectors} (LGAD).  
Compared to the p-n junction of a standard semiconductor, the LGAD adds a highly-doped layer.
When depleted, this results in a stronger electric field locally, causing electrons to 
avalanche in that region.  The better signal-to-noise improves the timing resolution 
of the detector.  

Current LGAD sensor technology can meet the timing performance requirements of REDTOP.
The ATLAS and CMS collaborations plan dedicated layers of LGAD detectors for the forward regions
to measure the time of arrival of particles, with a primary motivation being the rejection of 
tracks from pileup events~\cite{ATLAS-HGTD-TDR,Butler:2019rpu}.  Radiation hardness is a primary 
consideration for these detectors, since they must tolerate fluences well over 
$10^{15}$ n$_{\text{eq}}$/cm$^2$.  
Both detectors plan a spatial granularity of $1.3 \times 1.3\ \mu$m$^2$.  

The most recent improvements have been in the spatial segmentation of the detectors.  
Timing resolution of 30 ps and spatial resolution better than 10 $\mu$m has 
been observed in testbeam measurements of AC-coupled LGAD sensors~\cite{Heller:2022aug}. 
LGADs separated by a physical trench (``Trench-isolated'' or ``TI-LGAD'') are also a 
promising direction, with TCT tests showing timing performance on the order of 5 ps
prior to irradiation~\cite{TILGAD-Vienna}.

The majority of the material in the ATLAS and CMS timing layers is cooling and mechanical
support, with the sensor and readout chip are typically only comprising a few percent of the total.
To meet the material budget of REDTOP, only passive cooling can be used.  Carbon-fiber 
support structures similar to those planned for the HL-LHC upgrade of the pixel detector
are strong, lightweight, and have good thermal conductivity.  The system, particularly 
the readout electronics, must be designed to minimize the heat output.  

\paragraph{\label{subsec:The-Optical-TPC}Central Tracker option \#2: The Optical
TPC}

Protons and slow pions account for the largest charged background
at REDTOP. In average 0.6 charged pions and 1.8 protons are produced
every nsec in the target systems by the $E_{kin}$=1.8 GeV proton
beam.

\begin{figure}[!ht]
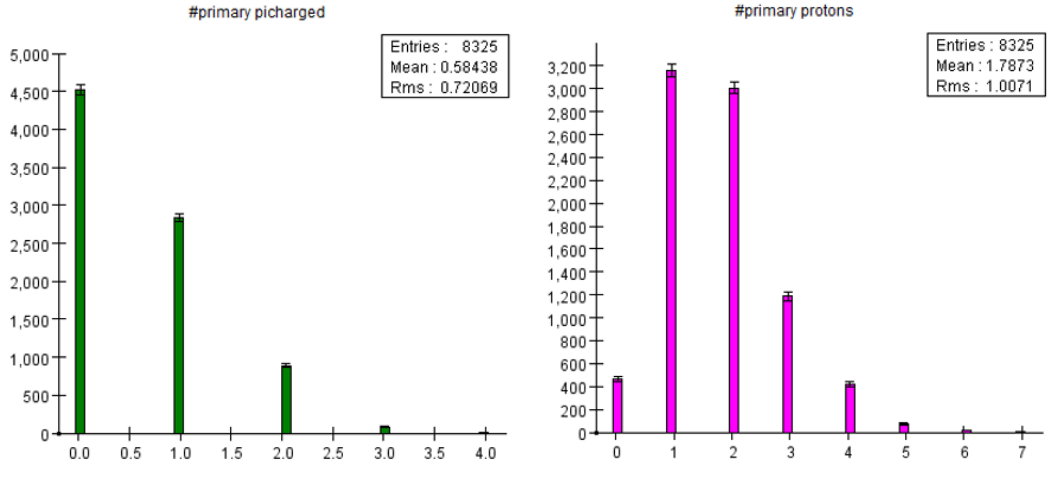

\includegraphics[width=7cm]{pictures/pi_multiplicity} \includegraphics[width=7cm]{pictures/p_multiplicity}
\caption{Charged pions (left) and proton (right) multiplicity from inelastic
scattering of a $E_{kin}$=1.8 GeV proton on REDTOP targets (\emph{GenieHad}
simulation).}

\label{fig:pi_p_multiplicity-1}
\end{figure}
A charged track detector mostly blind to them is required in order
to identify more rare decays. These particles have a consistently
low value of $\beta.$ Consequently, the threshold characteristic
of the Cerenkov effect could be exploited to build a detector insensitive
to this hind of background.

 An Optical TPC\citep{T1059,Oberta}
(OTPC)  (the red dodecagon in FIG.
\ref{fig:detector-sketch}) measures  the transverse momentum and  the $\beta$
of fast particles ( of electrons, muons and fast pions). It is a vessel filled
with gas ($CH_{4}$) with the inner wall covered by an aerogel radiator
with a thickness of $\simeq$3 cm. The external walls of the chamber are instrumented
with optical photo-sensors (LAPPD or sipm's) with a pixel size of
few $mm^{2}$. When an electron or a positron with a $\beta$ above
the Cerenkov threshold of $CH_{4}$ traverses the gas, it bends in
the solenoidal magnetic field while also radiating photons. The shape
of the latter, being collected by the photo-sensors covering the outer
walls of the device, has a specific pattern which provides the $P_{t}$
of the particle. Position and direction of the particle are obtained
by combining the pattern of the light ring, radiated from the aerogel, 
with the information of the impinging point on the photo-sensors.

Slower particles like muons and pions, typically with $\beta$ below
the Cerenkov threshold in the gas, produce a characteristic photon
ring when they cross the aerogel. 
The latter is detected from the OTPC photo-sensors
and reconstructed, providing a measurement of the kinematic parameters
of the particle. Protons and pions from QCD scattering are consistently
below detection sensitivity of the OTPC and do not constitute a sizable
background for the experiment. The magnetic field (0.6 T) also provides
the magnetic bottling preventing these very slow particles particles
from reaching the calorimeter with elevated rates.

\paragraph{\label{subsec:Material budget}Material budget for the baseline tracking
layout}

The  geometry of the baseline detector, consisting of the
fiber tracker, the LGAD central tracker, the Threshold Cerenkov Radiator, and
ADRIANO2 calorimeter, has been implemented in the software framework and
used to perform all physics and detector studies presented in the
second part of this paper. An estimate of the material budget for
the tracking region of the detector (r<50 cm) is shown in Fig.
\ref{fig:Material-budget} as function of the transverse radius. The
plots indicate that the beam pipe contributes with $\simeq$0.2\%~ 
of $X/X_{o}$, the fiber tracker with $\simeq$ 0.3\%~ per layer,
while the central tracker contribution is $\simeq$ 0.2\% per layer.
\begin{center}
\begin{figure}[!th]
\centering{}\includegraphics[width=0.47\textwidth]{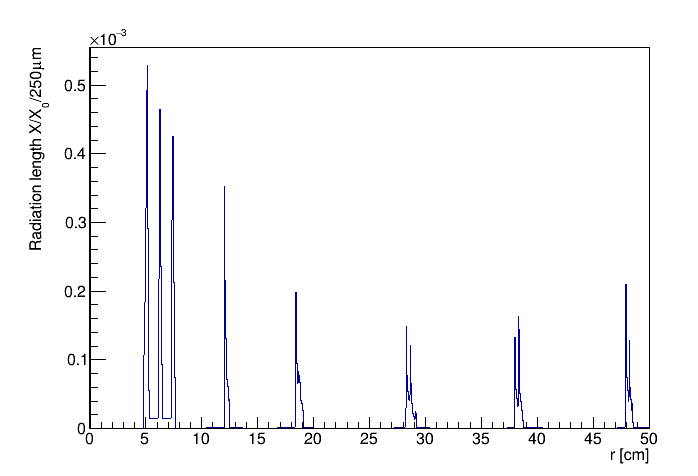}\includegraphics[width=0.47\textwidth]{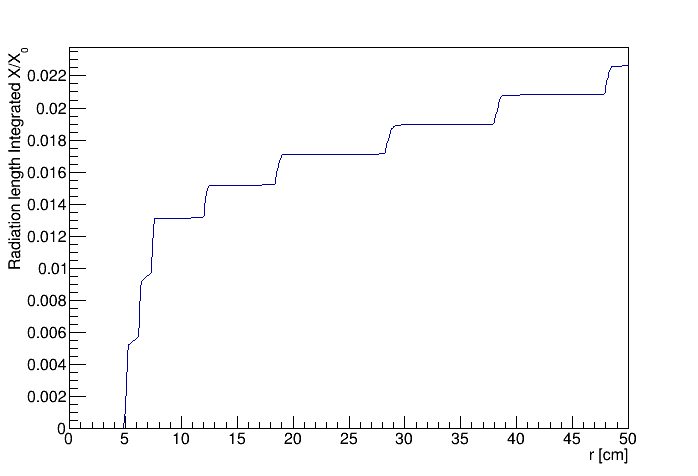}\caption{\label{fig:Material-budget}Plot of $X/X_{o}$ as a function of the
transverse radius for the baseline tracking system of REDTOP. The
calculation is performed in radial steps of 250 $\mu m$ (left). The
integrated material budget is shown in the right plot.}
\end{figure}
\par\end{center}

\subsubsection{Electromagnetic and hadronic calorimeter: ADRIANO2}

\paragraph{Requirements for calorimetry}

REDTOP calorimetry presents several challenging requirements since,
besides measuring the energy and he time of arrival of an impinging
particle, it also participates to all levels of triggers. The latter
implicates that it needs to provide to the trigger system pre-processed
information on a very short time scale.The required performance has
been assessed on the base of the sensitivity to several benchmark
processes. A full discussion of the latter has been carried in Secs.~\ref{sec:Sensitivity-Studies-to-physics-BSM}, \ref{subsec:Tests-of-Conservation-Laws}
and \ref{sec:Sensitivity-to-Non-perturbative}. The main requirements
for REDTOP calorimeter are:
\begin{itemize}
\item Energy resolution: $\sigma_{E}/E=3\%/\sqrt{E}$
\item Particle Identification (PID) with separation efficiency between electromagnetic
and hadronic particles higher than 99\%;
\item Time resolution: $<80psec$ in a single cell with energy deposit >
7 pe;
\item Detector response: within $~100nsec$;
\item Reconstruction efficiency: >50\% for E>20 MeV and >90\% for E>80 MeV.
\end{itemize}
The first requirement is fulfills the need to identify $\pi^{0}$ mesons
produced in the target from the primary interaction, since their decay
products make one of the largest combinatoric background. This could be
accomplished by fully reconstructing the decay of $\pi^{0}$  into $\gamma\gamma$ and
$\gamma e^{+}e^{-}$, which can be achieved with a sufficiently good energy resolution. 
Furthermore, several decays of the $\eta/\eta^{\prime}$
mesons contain $\gamma's$ or $\pi^{0}$'s in their final state. In
those cases, an higher energy resolution will allow a more efficient
reconstruction of events associated to $\eta/\eta^{\prime}$ decays en, consequently,
a better $signal/background$ ratio.

Particle identification is also important to reduce the hadronic background.
As discussed in details in Sec.~\ref{sec:trigger_system}, a fast
PID algorithm has to be implemented already in the Level-0 trigger,
to reduce the number of non-$\eta/\eta^{\prime}$ events to a more manageable
level. As discussed below, a fast particle identification can be implemented
using dual-readout techniques. The requirement on time resolution
is also affects the PID, since baryonic particles (which constitute
the largest fraction of background particles generated in the primary
interaction), have a much slower Time-of-Flight  compared to
the $\eta/\eta^{\prime}$ decay products. A fast detector response is an essential
requirement, considering also that the total expected event rate in the calorimeter
is of  order of 0.7 GHz. Assuming
a pipeline with a depth of fifty , the event needs to be processed
by the Level-0 trigger in less than 100 nsec. 

The technology chosen for filfilling the above requirements is a special
implementation of dual-readout calorimetry called: $ADRIANO2$, 
briefly described below. The calorimeter is divided into two sections.
The innermost section (about 40 cm in depth, or $\sim17\nicefrac{X}{X_{0}}$)
is   a high-granularity, integrally active dual-readout
calorimeter, made of  alternated tiles of lead-glass and scintillating
plastics, optically separated and read-out individually. 

Its main tasks are to reconstruct electromagnetic showers and to measure the energy
of photons and electrons. The outermost section has the same layout
of the innermost section, but it includes plates of passive material sandwiched
with the active layers. The passive material  reduces the nuclear
interaction length so that hadronic showers will be
fully contained in the overall depth of the calorimeter (about 80
cm). Both sections have dual-readout capabilities and participate
in the PID and TOF systems.

\paragraph{The dual-readout technique}

Dual-readout (DR) calorimetry\citep{DREAM_2000,LCWS2015} is a relatively
novel technique aiming at achieving energy compensation in hadronic
showers produced in high energy experiments. One particular implementation:
ADRIANO, where both readout components are integrally active, has
been developed in recent years by T1015 Collaboration\citep{Gatto_2015_1,Gatto_2015_2}
for applications in future lepton colliders or high intensity experiments
requiring low granularity. ADRIANO building block is an integrally
active log made by sandwiching lead glass and scintillating plates,
optically separated and with independent readouts. The Cerenkov and scintillating
lights are captured by wavelength shifting fibers (WLS) coupled
to light sensors.  

One of the advantages of ADRIANO is that, with a proper layout,
the dual-readout method can be used also for electromagnetic calorimeters
as a way to identify a particle by comparing the  response of the   two components.
The particle separation with a dual-readout calorimeter is illustrated
in Fig.~\ref{fig:Plot-of-S} for particles of different species and
same energy (100 MeV). The separation between photons and neutrons,
is better than 4$\sigma$, and it far exceeds the required limit of 99\%
.
\begin{center}
\begin{figure}[!ht]
\centering{}\includegraphics[scale=0.50]{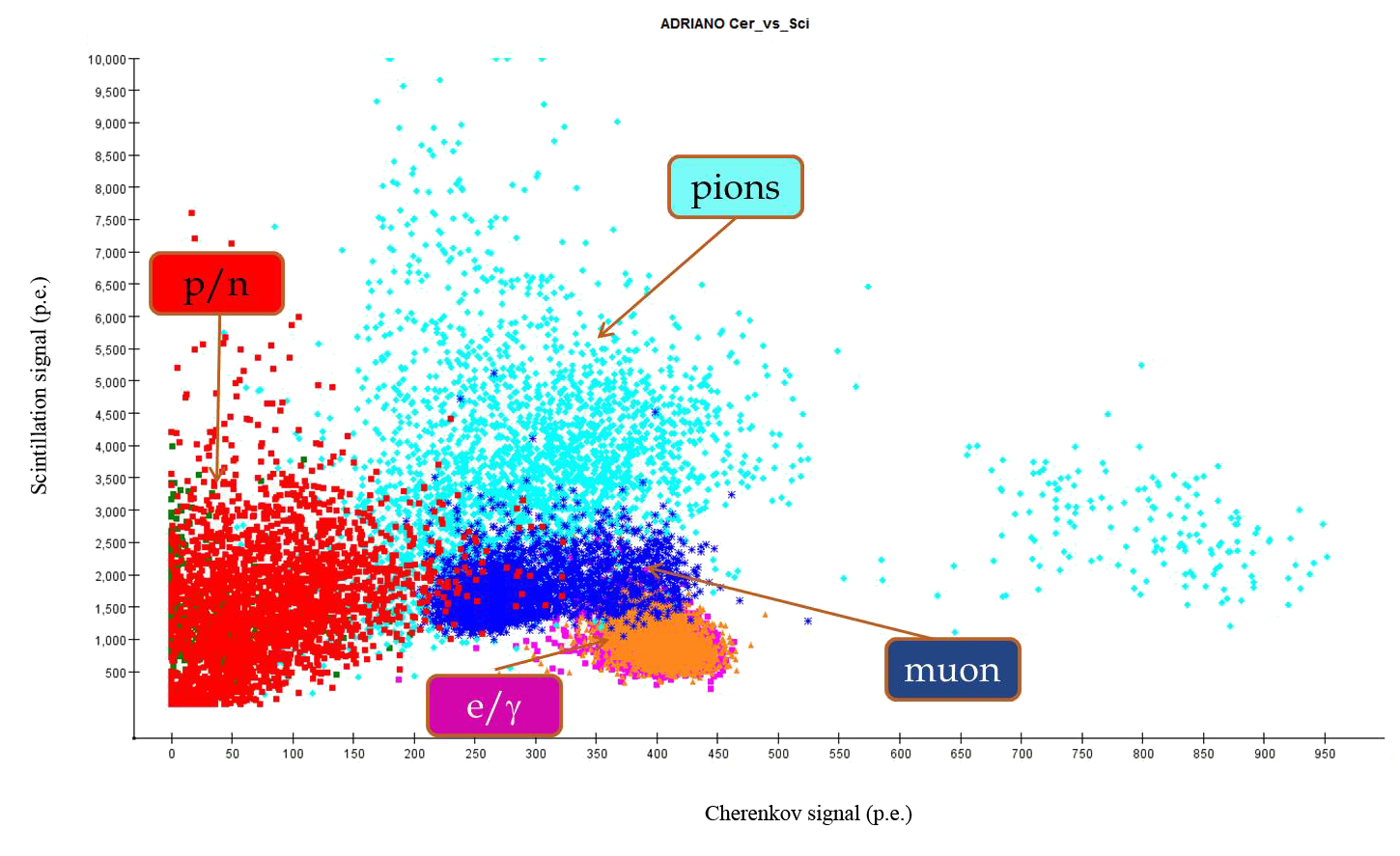}\caption{\label{fig:Plot-of-S}Plot of S vs \v{C} signals for several 100 MeV
particles in ADRIANO (ilcroot simulation)}
\end{figure}
\par\end{center}

REDTOP requires a finer granularity than it could be achieved with
the log based, ADRIANO technique. A higher granularity implementation
is, therefore, being considered: ADRIANO2, where the lead-glass and
the scintillating strips are replaced by small tiles with direct SiPM readout~\citep{TILES2009,TILES2011,MALLA_2019}.
ADRIANO2 is a relatively compact calorimeter, combining Particle Flow
reconstruction (PF) (segmentation in space and potentially in time)
and DR (multiple readout). See reference ~\cite{CALICE2016} and references 
therein for details on the particle Flow technique.
The compound benefits of either algorithms
have a great potential interest for future applications besides
REDTOP\cite{Detector:2784893}. Recent studies by T1604
Collaboration indicate that the small size of the lead glass tiles
and the prompt nature of the Cerenkov light, when captured by a fast
SiPM, meet the timing requirement for the calorimeter discussed above.

\paragraph{The ADRIANO2 Calorimeter}

ADRIANO2 is fabricated with alternate tiles of lead-glass and scintillating
plastic, optically separated and red-out individually. For an optimal
collection  of the Cerenkov light, the lead-glass
tiles (size = 1$\times$3$\times$3 cm$^{3}$) are coated with either
a diffuse or reflecting metal for example Ag, Al, Mo, W, etc. Various
thin film deposition methods such as sputtering, electroplating, evaporation,
chemical vapor deposition (CVD) and atomic layer deposition (ALD)
are currently being explored. Among the various thin film coating
methods, ALD method can offer outstanding control of thickness and
excellent film uniformity. As self-limiting reactions between gaseous
precursor and a solid surface involved in ALD process, therefore,
this method can produce clean thin films in an atomic layer-by-layer
way with precise control over layer thickness and composition. This
feature allows precise coatings to be applied on all exposed surfaces
of complex substrates such as fibers, woven, nanotubes, aerogels,
mesoporous, nanoporous, membranes and solid geometries. For applications
where highly conformal deposition of thin films is required, the acceptable
control can only be achieved with ALD methods. ALD can also be scaled
to coat large batches of 3D parts simultaneously. Considering the
lead-glass cube 3D form factor, here we have explored the ALD material
coating method, which seems to be very promising.

Recent tests with 100 nm thick W and Mo coating by optimized ALD method using
the combination of WF6-Si2H6 or MoF6 \textendash Si2H6 precursors
 and a 10 nm thick composite barrier~\cite{ALD2018}.  The quality of the deposited W or Mo layer were analyzed by x-ray photoelectron spectroscopy (XPS) for
composition and four probe measurements for electrical conductivity.
The photograph of lead-glass cube before and after ALD is shown in
Fig.~\ref{fig:ADRIANO2-cell}.
\begin{center}
\begin{figure}[!ht]
\centering{}\includegraphics[scale=0.6]{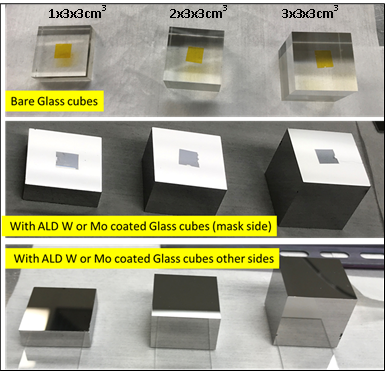}\caption{\label{fig:ADRIANO2-cell}Top row different size bare lead-glass cubes
with kapton tape mask placed on one of the side. Middle row lead-glass
cubes after W or Mo ALD and kapton masked removed. Bottom row non-masked
sides lead-glass cubes coated with W or Mo ALD.}
\end{figure}
\par\end{center}

 Performance of these W and Mo ALD coated cubes were tested at Fermilab
and have shown, so far, excellent results compare to %
{} other coating methods e.g. sputtering, spray paint, evaporation which
were evaluated. %

\paragraph{Readout electronics}


One In order to fulfill the constraints above, stingent requirements
need to be placed on the Front-End Electronics (FEE).


One possible alternative to the analogue sipm readout is based on
the novel technique of $SinglePhotonCountingDetector$ (SPAD) which
reperesent a particular implementation of Digital Photon Counters
(DPC), where the readout electronics is incorporated in the photosensor.
SPAD array in CMOS technologies may offer the following benefits to
fast calorimeters:
\begin{itemize}
\item incorporation of complex functions in the same substrate (e.g. SPAD
masking, counting, TDCs); 
\item optimization of the design of the front-end electronics to preserve
signal integrity (especially useful for timing)
\item easier assembly due to the monolithic structure of the sensor; 
\item significantly reduced temperature sensitivity by at least one-order-of
magnitude
\item reduced afterpulsing \& crosstalk 
\item better linearityand intrinsic timing resolution due to integrated
TDCs ($\simeq$ factor 5) 
\item no analog electronics, no ADCs, no ASICs
\item power consumption, which is especially important in calorimeters like
the one proposed, where the readout-electronic is embedded inside
the volume of the detector and excess heat needs to be removed.
\end{itemize}

\subsubsection{The Threshold Cerenkov Radiator (TCR)}

A thin Cerenkov radiator is inserted between the Central Tracker and the
calorimeter. The detector has two main purposes: a) the detection
of particles above the Cerenkov threshold provided by quartz; b) the measure of
the TOF of such particles. The information from the $TCR$ is also
used in the Level-0 trigger to reduce the contamination of the signal events
from slow baryonic particles. Finally, the $TCR$ complements the
energy measurements of the calorimeter as it operate as a lower density
pre-shower.

The requirements for the $TCR$ are :
\begin{itemize}
\item Time resolutions: $<50psec$;
\item Detector response: within $~100nsec$
\item High-granularity to reduce overlapping showers.
\end{itemize}
The baseline layout of the $TCR$  consists of small tiles made
of borosilicate JGS1 glass, which is transparent  to UV light above $\simeq$200
nm. The latter is abundantly produced by the Cerenkov effect from
all particles above threshold. The momentum threshold for the various
particle species is summarized in Table~\ref{table:TCR-C-threshold}.
\begin{table}[ht]
\textbf{\footnotesize{}\centering}%
\begin{tabular}{|c|c|}
\hline 
\textbf{\textit{\footnotesize{}PID}} & \textbf{\textit{\footnotesize{}Momentum threshold}}\tabularnewline
 & \textbf{\textit{\footnotesize{}{[}MeV/c{]}}}\tabularnewline
\hline 
\hline 
electron & \textbf{\footnotesize{}$0.4$}\tabularnewline
\hline 
muon & \textbf{\footnotesize{}$100$}\tabularnewline
\hline 
pion & \textbf{\footnotesize{}$130$}\tabularnewline
\hline 
proton & \textbf{\footnotesize{}$870$}\tabularnewline
\hline 
\end{tabular}\caption{Cerenkov threshold of $TCR$ detector.}
\label{table:TCR-C-threshold}
\end{table}
The dimension of the tiles is $3\times3\times2\:cm^{3}$, namely, they have the same
footprint of ADRIANO2 tiles. The Cerenkov light is read-out
by a NUV-SiPM optically coupled to each unit. The analog information
from the $TCR$ is also used in the reconstruction of  showers
in the ADRIANO2 calorimeter. 

\subsubsection{The Timing Layer (\texorpdfstring{$TL$, optional}{})}

A further improvement to the PID systems could come from the TOF system
when the requirement on timing resolution of the $TCR$ is lowered to 30 psec or less. Preliminary
studies performed by T1604 collaboration indicate that a such a resolution
could be beyond the limits of the technology adopted for the TCR. An
optional AC-LGAD Timing Layer ($TL$), based on the design planned
for experiments at the Electron Ion Collider (EIC), would be able
to reach the desired goal. The proposed layout for REDTOP would employ
the $TCR$ as supporting structure, contributing to a small fraction of the material budget of the $TL$, with negligible impact on the performance of the ADRIANO2 calorimeter.

A joint effort to develop such a TL for REDTOP and EIC experiments can benefit both experiments, which have similar requirements on the timing resolution and material budget. For example, a cylindrical TL with a radius of about 50 cm is envisioned for EIC experiments that require a single hit timing resolution better than 30 ps, and a material budget per layer less than 1\%. Such a design could be realized by long-strip AC-LGAD sensors with thin sensor active volume, and low-power and low-jitter frontend readout ASICs, which are currently being developed by the eRD112 collaboration for EIC~\cite{eRD112}. 

\subsubsection{The Muon Polarimeter (MuPol, optional)}

The construction of a dedicated muon polarimeter is being consideration for REDTOP in
case the ADRIANO2 performance would not be adequate to for polarization
measurements. The muon polarimeter has the task of stopping the muons
from the $\eta/\eta^{\prime}$ decays and detect the direction of the electron
or positron. The design of such detector is based on the proposal
of the TREK experiment\cite{TREK2010} at J-PARC. The proposed technology
implements the concept of an active polarimeter, which attain higher
detector acceptance and higher sensitivity compared to a conventional
passive polarimeter. The main advantage of an active technique is
that the muon decay vertices will be determined event-by-event, which
make the polarization measurement free from the systematic error associated
with the uncertainties of the muon stopping distribution. 

The polarimeter would consist of two parts: a) a set of passive plates,
where the muon will either stop or decay in flight, interspersed with
b) tracking chambers where the electron or positron would be detected.
A large muon stopping efficiency in relatively small volume requires
higher average material density. On the other side, a clean detection
of positrons without interactions such as bremsstrahlung or annihilation
in flight requires a material with lower average density and the use
of low mass elements. Aluminum and magnesium (or their alloys) are
good candidates as they are also know for their property of not affecting
the original polarization of the muon. An integrated thickness of
$\sim66mm$ of aluminum will have a stopping power of $\sim85\%,$which
is sufficient to perform excellent polarization measurements with
the expected event statistics foreseen at REDTOP.

The elector/positron tracking chambers will be located between the
stopping plates. Two technologies are being considered: a) a gaseous
wire chamber, and b) a fiber tracker. The former has the advantage
of being low cost in terms of construction and readout-electronics.
The latter has the advantage of simplicity and low maintenance. Square
scintillating fibers with sides as large as 2 mm are commercially available
with short lead times and, when glued to the stopping plates, will
contribute to the mechanical robustness of the detector. In both cases,
the coordinate along the wire or fiber is determined by charge or
light division at both ends.

The electron or positron energy will be determined using both the
range in the stopper material and the trajectory curvature in the
field. The tracking is done by taking into account energy loss through
each plate. In a design using aluminum plates, the total material
thickness in the radial direction is $\sim$13 $g/cm^{2}$ to be compared
with the range of electron or positron of 5 $g/cm^{2}$ for 10 MeV
and 15 $g/cm^{2}$ for 25 MeV.

The $MuPol$ will be positioned inside the calorimeter at a depth
of approximately 20 radiation lengths, to avoid any degradation of
the energy measurement in ADRIANO2. The design allow to perform
a measurement of the electron or positron energy as well as its emission angle,
by using the outer section of ADRIANO2, with a significant increase
in the analyzing power of the polarimeter, and, consequently, in a higher sensitivity.
Furthermore the materials in front of the MuPol, essentially lead
glass and scintillating plastics, are amorphous and, unlike the crystals
used in most of the high resolution electromagnetic calorimeters,
do not change or degrade the original polarization of the muon. The
latter will be slow down  before entering the passive
plates of the $MuPol$, with higher probability of being stopped in
the passive metal plates.

Although the stopping material alone is not thick enough to stop an
electron from a muon near the surface, additional range stacks besides
the stopper are useful to discriminate lower energy electrons with
a negative asymmetry. With the magnetic field foreseen for REDTOP
(see next section), the bending radius for a 25 MeV electron is 650
cm. A moderate position resolution of several 100 $\mu$m will be able
to determine the energy of positron with a 10 cm long track with an
accuracy of $\sim$40\%. 

\subsubsection{Superconducting solenoid}

A 0.6 T solenoidal magnetic field is required to measure the $P_{t}$
of the particles with $\beta\approx1$. The field will also magnetically
bottle the very low momentum particles, preventing them from reaching
the calorimeter. The solenoid built for the, now dismantled, Finuda
experiments~\citep{Bert997} matches all operational and dimensional
parameter required for REDTOP. It also fits  through the hatch that
connects the AP50 hall (for the Fermilab running option, cf.\ Appendix
\ref{sec:The-Acceleration-Scheme}) and it can be transported on rails
to its final location. The same magnet could be accomodated even more
easily in the C4 beamline, for a possible run at BNL.

\subsection{The event trigger systems}

\label{sec:trigger_system}

The goal of producing  $5.2\times10^{13}$ $\eta$ mesons per year,
( $5.2\times10^{6}$ $\eta$/s, assuming $10^{7}$ seconds of useful
running time), requires about $7\times10^{8}$ $p$-Li inelastic
collisions per second (cf.\ Table~\ref{table:eta-production-probability}).
This rate can be achieved with a proton beam intensity of $10^{11}$
$p$/s and a Li target of $2\times10^{-2}$ collision lengths ($\approx$
7.7 mm total thickness), possibly subdivided into a number of thinner
layers. Taking into account the total \emph{p}-Li inelastic collision
rate, we estimate that the total rate of events reaching the detector is
 $\sim7\times10^{8}$ Hz . This rate is more than one order of magnitude
larger than the event rate observed in the LHCb experiment~\cite{Lhcb_2021},
indicating that very fast detector technologies need to be implemented
along with a multi-level trigger systems, finely tuned for the expected
event structure. The following estimate are obtained with the detector
layout described in Sec.\ref{sec:Simulation-strategy}. More specifically,
the fiber tracker and the LGAD tracker options have been adopted for this work. 

\subsection{The event structure}

\begin{figure}[!th]
\centering{}\includegraphics[width=0.95\textwidth]{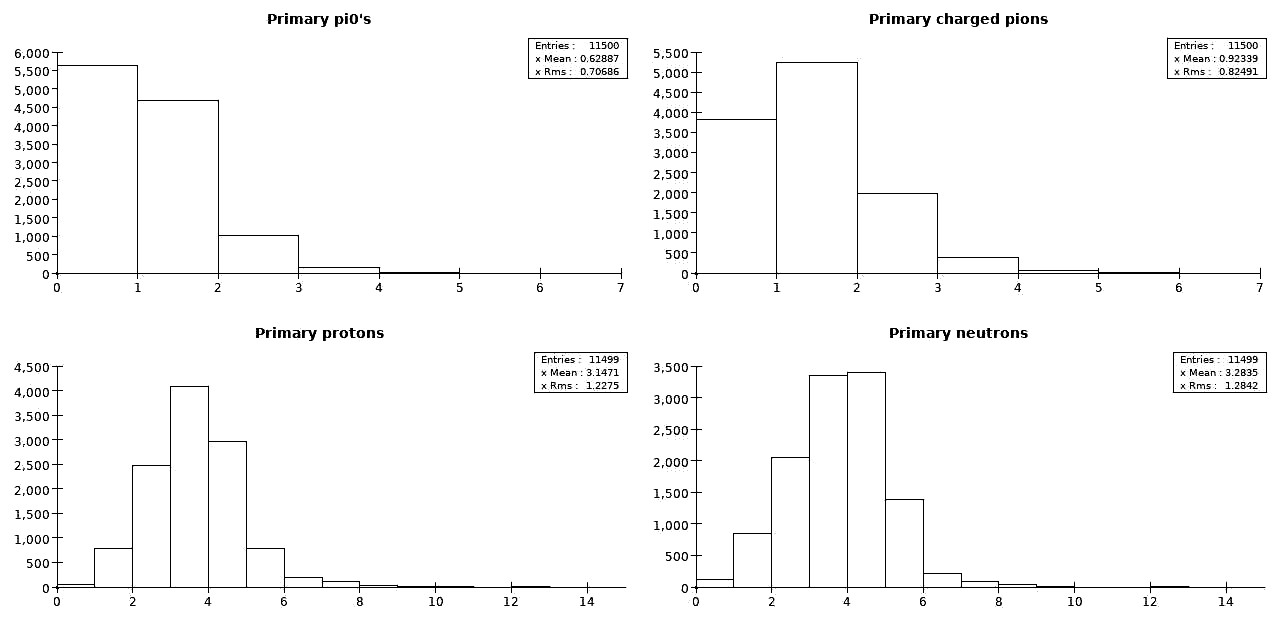}\caption{\label{fig:Mutiplicity}Hadron multiplicity in REDTOP detector. (GenieHad\cite{GenieHad_2012}
simulation with Urqmd 3.4)}
\end{figure}

\begin{figure}[!ht]
\centering{}\includegraphics[width=0.95\textwidth]{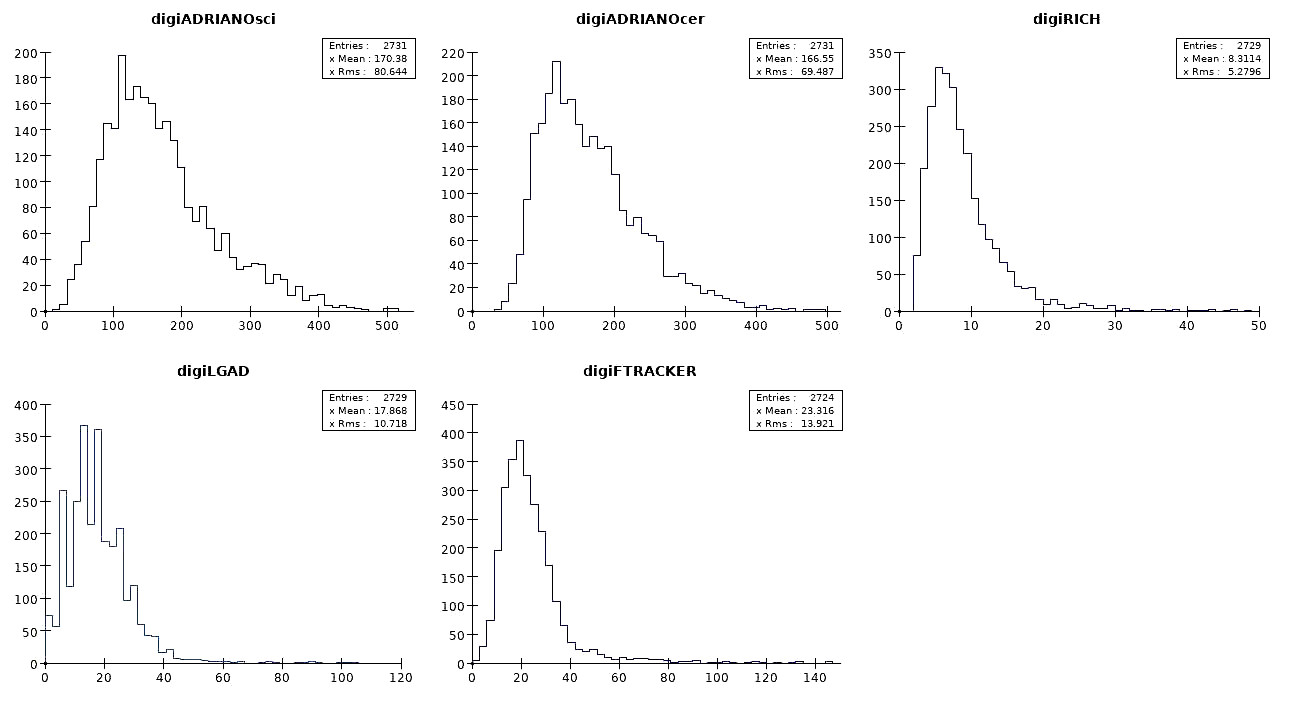}\caption{\label{fig:Occupancy}Background occupancy in REDTOP sub-detectors.
(Geant4 simulation)}
\end{figure}

The vast majority of particles entering REDTOP detector are hadrons
produced by inelastic scattering of the proton beam on nuclear matter.
The multiplicity of those hadrons is shown in Fig.~\ref{fig:Mutiplicity}
($p$+Li$\rightarrow\eta X$ with $\eta\rightarrow3\pi$ and $\eta\rightarrow\gamma\gamma$
are included in the histograms). Such events represent a large fraction of the background,
being generated with a rate $\sim213\times$ that of the $\eta$-meson.
Despite the large background rate, the occupancy of the events is
relatively low, as the kinematics of the final state particles produces
only few hits in the detector. Fig.~\ref{fig:Occupancy} shows the
expected number of digitized hits in each sub-detector of REDTOP ($p$+Li$\rightarrow$$\eta X$
with $\eta\rightarrow3\pi$ and $\eta\rightarrow\gamma\gamma$ are
included in the plots). 
Table ~\ref{table:digits} summarizes the Monte Carlo estimated average size of 
digits recorded in each sub-detector for background events. The innermost
sections account for less than 13\% of the total digitized data.

\begin{table}[!ht]
\centering{\footnotesize{}}%
\begin{tabular}{|c|c|c|c|c|c|c|}
\hline 
\textit{\footnotesize{}Digit} & \textit{\footnotesize{}FTRACKER} & \textit{\footnotesize{}LGAD } & {\footnotesize{}RICH} & \textit{\footnotesize{}ADRIANO2} & \textit{\footnotesize{}ADRIANO2} & \textbf{\footnotesize{}Total}\tabularnewline
\textit{\footnotesize{}occupancy} &  &  &  & {\footnotesize{}Scint} & {\footnotesize{}Cer} & \tabularnewline
\hline 
\hline 
{\footnotesize{}Counts} & {\footnotesize{}$22.3$} & {\footnotesize{}$17.9$} & {\footnotesize{}$8.3$} & {\footnotesize{}$170.4$} & {\footnotesize{}$166.6$} & \textbf{\footnotesize{}386}\tabularnewline
\hline 
\end{tabular}\caption{Average size of digitized data in REDTOP sub-detectors for background events}
\label{table:digits}
\end{table}

On the other hand, the most interesting events for dark matter and BSM physics searches have either two
leptons or faster pions, with a typical detector occupancy which is
50\%-100\% larger than the background events. 

Assuming a 12-bit digitization for charge and time, and an 18-bit address
to identify the struck cell, we estimate that the average size of
digitized information obtained from the detector before any trigger
is applied is about $1.4\times10^{3}$ bytes, corresponding to $9.8\times10^{18}$
bytes/year (or $9.8\times10^{3}$ petabytes/year). A three-level trigger systems
is being considered for REDTOP in order to reduce the data throughput
to a more manageable level. The architecture of REDTOP trigger systems
is described in detail in the following sections.

\subsection{Level-0 trigger}

The selection performed at the Level-0 trigger is based on simple global features
of the events produced by $p$-Li inelastic collisions. All the events
interesting for New Physics have at least either two leptons or two
pions in their final state, in association with at least one energetic
photon in the calorimeter. The purpose of the Level-0 trigger is 
rejecting non-interesting events with a time lag of few tens of
nanoseconds. The leptons and pions originating from the decay of the
$\eta$ meson are more energetic than those generated by the proton-nucleus
interaction, and usually their velocity is above the TCR threshold. The corresponding TOF is, also, smaller. Therefore, the fast Cerenkov signals produced by the LGAD,
the TCR and by the ADRIANO2 sub-detectors are the primary input to the Level-0 trigger
system. Several requirements need to be met to accept the events:
\begin{itemize}
\item a minimum Cerenkov energy from ADRIANO2 integrated over the whole
detector;
\item at least two clusters in the TCR with energy above threshold and
TOF below a defined value;
\item the integrated scintillation/Cerenkov signals ratio from ADRIANO2 must be compatible with an electromagnetic shower.
\end{itemize}
The above requirements are intended to reject  the background induced by slower baryons (usually
below the Cerenkov threshold of TCR and ADRIANO2) and low-energy gamma which
converted in the innermost region of the detector, while accepting
the faster leptons and pions from the decay of the $\eta$ meson.

No tracking information is used at the Level-0 trigger, since that would imply implementing
some sort of pattern recognition,  and, consequently, a longer processing time.
The above information, instead, is readily available from the FEE of each
detector, and it can be processed in real time by a custom FPGA module.
A fast extraction of the scintillation/Cerenkov  ratio from
ADRIANO2 could be achieved with Field-Programmable Analog Arrays (FPAA) 
System-on-Chip platforms~\cite{Asler20}. The analog processing with FPAA's will
greatly reduce processing latency as well as power consumption. Such
a strategy would provide programmable access to analog pulse samples
covering the high C/S regions at the start of the pulse and lo C/S
regions at the end of the pulse.

The acceptance of the Level-0 trigger for $p$-Li inelastic collisions
(background), and several signal events is shown in Table~\ref{table:triggerLevel0}.

\begin{table}[!ht]
\centering{\footnotesize{}}%
\begin{tabular}{|c|c|c|c|c|}
\hline 
{\footnotesize{}Inelastic collisions} & \textit{\footnotesize{}Input event rate} & \textit{\footnotesize{}Average event size} & \textit{\footnotesize{}Input data rate} & \textit{\footnotesize{}Event acceptance}\tabularnewline
 & \textit{\footnotesize{}Hz} & \textit{\footnotesize{}bytes} & \textit{\footnotesize{}bytes/s} & \tabularnewline
\hline 
\hline 
{\footnotesize{}Background} & {\footnotesize{}$7\times10^{8}$} & {\footnotesize{}$1.4\times10^{3}$} & {\footnotesize{}$9.8\times10^{11}$} & {\footnotesize{}21.7\%}\tabularnewline
\hline 
{\footnotesize{}$\eta\to\pi^{+}\pi^{-}\pi^{0}$} & {\footnotesize{}$9.4\times10^{5}$} & {\footnotesize{}$1.5\times10^{3}$} & {\footnotesize{}$1.4\times10^{9}$} & {\footnotesize{}48.6\%}\tabularnewline
\hline 
{\footnotesize{}$\eta\to\gamma e^{+}e^{-}$ } & {\footnotesize{}$2.8\times10^{4}$} & {\footnotesize{}$1.5\times10^{3}$} & {\footnotesize{}$4.2\times10^{7}$} & {\footnotesize{}80.6\%}\tabularnewline
\hline 
\end{tabular}\caption{Level-0 trigger acceptance, event rate, and throughput for inelastic collisions and for typical
$\eta$ final states with and without leptons.}
\label{table:triggerLevel0}
\end{table}
Table~\ref{table:digits-1} summarizes average
digit sizes,  estimated with Monte Carlo techniques, recorded in each sub-detector for QCD background, $p+Li\rightarrow\eta X$
with $\eta\rightarrow3\pi$ and $\eta\rightarrow\gamma\gamma$ events
surviving the Level-0 trigger. The innermost sections account for
less than 17\% of the total digitized data.

\begin{table}[!ht]
\centering{\footnotesize{}}%
\begin{tabular}{|c|c|c|c|c|c|c|}
\hline 
\textit{\footnotesize{}Digit} & \textit{\footnotesize{}FTRACKER} & \textit{\footnotesize{}LGAD} & {\footnotesize{}RICH} & \textit{\footnotesize{}ADRIANO2} & \textit{\footnotesize{}ADRIANO2} & \textbf{\footnotesize{}Total}\tabularnewline
\textit{\footnotesize{}occupancy} &  &  &  & {\footnotesize{}Scint} & {\footnotesize{}Cer} & \tabularnewline
\hline 
\hline 
{\footnotesize{}Counts} & {\footnotesize{}$31.9$} & {\footnotesize{}$27.0$} & {\footnotesize{}$10.3$} & {\footnotesize{}$166.6$} & {\footnotesize{}$180.4$} & \textbf{\footnotesize{}416}\tabularnewline
\hline 
\end{tabular}\caption{Average digits in REDTOP sub-detectors for background events surviving
the L0-trigger}
\label{table:digits-1}
\end{table}
Our studies suggest that, with the Level-0 trigger configured as described
above, we can achieve the needed rejection factor of about 20\% while
preserving a satisfactory efficiency on a large set of interesting
physics processes. 

The data rate into Level-0 is obtained from Monte Carlo simulations
with an estimated reduction factor of $\sim4.6$ of the event rate and an average
event size of $\sim$1.5 Kb, after zero suppression and compression
by the DAC. Assuming a 12-bit digitization for charge and time and
an 18-bit address to identify the struck cell, a collision rate
of 700~MHz, we conclude that the average rate of the
data sent to the Level-1 trigger is about $\sim$230 Gb/s. Such a
data rate can be comfortably transmitted by a network of a few hundred
optical fiber links. Because all data from the detector are continuously
digitized in the front-end and immediately transmitted to the following
stages, trigger latency is not expected to be a problem, at least to first order.
Therefore the Level-0 logic can be heavily pipelined. Although a new
event will arrive, on average, every 1.4 ns, the time taken to make
a decision on a specific event can be much longer, possibly of the order of hundreds of nanoseconds.

\subsection{Level-1 trigger}

In contrast with the global aspects of Level-0 input, the Level-1 rejection
is based on local information obtained directly from  the sub-detectors.
That requires the implementation of at least  a low-level
pattern recognition and clusterization of the hits. The algorithms 
implemented for the present studies are intended to reduce the
background from baryons as well as from photons converting into an
$e^{+}e^{-}$ pair in the innermost region of the detector. The set
of requirements for the Level-1 trigger are listed below, grouped by the specific type of background targeted:

\paragraph*{Baryon rejection}
\begin{itemize}
\item at least three clusters in the LGAD tracker, separated by a distance of 4.5
mm or larger;
\item at least two groups of hits with three LGAD clusters and one TCR cluster,
consistent with oppositely charged particles, and with TOF below a
defined threshold; 
\item the integrated energy in the TCR for positive and negative tracks
has to exceed a defined threshold;
\item at least three ADRIANO2 clusters with scintillation/Cerenkov ratio
consistent with a non-baryonic particle. 
\end{itemize}

\paragraph*{$\gamma\to e^{+} e^{-}$ rejection}
\begin{itemize}
\item No clusters in the FTRACKER, associated to two oppositely charged tracks,
separated by less than 3 mm in the x-y plane and 12 mm in the z-direction;
\item No clusters in the LGAD associated to two oppositely charged tracks separated
by a distance shorter than of 4.5;
\end{itemize}
The particle identification in ADRIANO2, at the Level-1 trigger, could
be implemented in a two-layer Artificial Neural Network (ANN) with
a dedicated Field-Programmable Analog Arrays, receiving the input 
from the C/S system implemented
in the Level-0 trigger. The choice of the sample to send to the ANN
can be programmed in order to fine tune the trigger system and to satisfy the requirements of a particular physics run. The technique
would be capable to optimize the C/S separation between hadronic and
electromagnetic showers at run time. The output of the ANN would be
digitized and recorded with the channel, with an estimated reduction
of information of about one-order of magnitude. 

The acceptance of the Level-1 trigger for $p$-Li inelastic collisions
(background), and for several topologies of signal events, is shown in Table~\ref{table:triggerLevel1}.
Assuming a 12-bit digitization for charge and time and an 18-bit address
to identify the struck cell, the  Level-1 input data rate is 
$\sim$230 Gb/s. Factorizing the event reduction of $\sim60$ (cf.\ 
Table~\ref{table:triggerLevel1}), we conclude that the expected average rate
of data sent to the Level-2 trigger is about $\sim$3.8 Gb/s. Such
a data rate can be comfortably transmitted by a network of a few hundred
optical fiber links. A significant challenge of the REDTOP trigger
will be to design specialized processors to achieve the needed rejection
factor of under 2\% by reconstructing proto-tracks with high efficiency,
at an event rate of 100 KHz. We believe that this selectivity can
only be achieved with specialized hardware, possibly based on massive
use of FPGAs.

To gain more time to process each event, we can adopt a time multiplexing
strategy. We anticipate that events will be distributed to a bank
of identical processors in a round-robin fashion. The larger the number
of processor, the more time each one of them will have to process
one event. A time multiplexing of factor of 10 will allow each processor
5 microseconds, on average, to process each event, which seems adequate
for the task.

As a comparison, a demonstrator of the Level-1 tracking trigger for
CMS Phase II has recently shown that ten identical, FPGA-based, processors,
housed in a single ATCA crate and operated in a time-multiplexing
mode, can process events coming at a rate of 40~MHz from a unit corresponding
to 1/48 of the CMS tracker. Each unit contained the order of 500 hits
and was carried by about 400 fibers. All the tracks with a $P_{t}$
above 3~GeV/c could be reconstructed with high efficiency and a latency
of a few microseconds. The rates processed and the complexity of the
problem are on the same level as expected for REDTOP Level-1 trigger.

\begin{table}[!ht]
\centering{\footnotesize{}}%
\begin{tabular}{|c|c|c|c|c|}
\hline 
\textit{\footnotesize{}Process} & \textit{\footnotesize{}Input event rate} & \textit{\footnotesize{}Average event size} & \textit{\footnotesize{}Input data rate} & \textit{\footnotesize{}Event acceptance}\tabularnewline
 & \textit{\footnotesize{}Hz} & \textit{\footnotesize{}bytes} & \textit{\footnotesize{}bytes/s} & \tabularnewline
\hline 
\hline 
{\footnotesize{}Inelastic collisions} & {\footnotesize{}$1.5\times10^{8}$} & {\footnotesize{}$1.5\times10^{3}$} & {\footnotesize{}$2.3\times10^{11}$} & {\footnotesize{}1.7\%}\tabularnewline
\hline 
{\footnotesize{}$\eta\to\pi^{+}\pi^{-}\pi^{0}$} & {\footnotesize{}$4.6\times10^{5}$} & {\footnotesize{}$1.6\times10^{3}$} & {\footnotesize{}$6.9\times10^{8}$} & {\footnotesize{}10.9\%}\tabularnewline
\hline 
{\footnotesize{}$\eta\to\gamma e^{+}e^{-}$} & {\footnotesize{}$2.3\times10^{4}$} & {\footnotesize{}$1.6\times10^{3}$} & {\footnotesize{}$3.6\times10^{7}$} & {\footnotesize{}64.6\%}\tabularnewline
\hline 
\end{tabular}\caption{Level-1 trigger acceptance for inelastic collisions and for a typical
$\eta$ final states with and without leptons.}
\label{table:triggerLevel1}
\end{table}
Table~\ref{table:digits-1-1} summarizes the Monte Carlo-estimated
average size of digits which would be recorded in each sub-detector by QCD background, $p$+Li$\rightarrow$$\eta X$
with $\eta\rightarrow3\pi$ and $\eta\rightarrow\gamma\gamma$ events
surviving the Level-1 trigger. We note that the innermost regions of the detector account for
less than 17\% of the total digitized data.

\begin{table}[!ht]
\centering{\footnotesize{}}%
\begin{tabular}{|c|c|c|c|c|c|c|}
\hline 
\textit{\footnotesize{}Digit} & \textit{\footnotesize{}FTRACKER} & \textit{\footnotesize{}LGAD} & {\footnotesize{}RICH} & \textit{\footnotesize{}ADRIANO2} & \textit{\footnotesize{}ADRIANO2} & \textbf{\footnotesize{}Total}\tabularnewline
\textit{\footnotesize{}occupancy} &  &  &  & {\footnotesize{}Scint} & {\footnotesize{}Cer} & \tabularnewline
\hline 
\hline 
{\footnotesize{}Counts} & {\footnotesize{}$34.1$} & {\footnotesize{}$28.9$} & {\footnotesize{}$10.7$} & {\footnotesize{}$178.0$} & {\footnotesize{}$193.2$} & \textbf{\footnotesize{}446}\tabularnewline
\hline 
\end{tabular}\caption{Average digits in REDTOP sub-detectors for background events surviving
the L1-trigger}
\label{table:digits-1-1}
\end{table}

\subsection{Level-2 trigger}

The Level-2 trigger aims at positively identify the underlying physics
process and, therefore, it relies heavily on the topology of the final state. 
The algorithm implemented requires a fully reconstructed event, including:
a) the identification of the target foil where the primary interaction occurred,
b) the approximate reconstruction of tracks and calorimetric showers, c) identification
of potential secondary vertexes. The requirement on the  final state topology
for discriminating the event at the Level-2 trigger are:
\begin{itemize}
\item at least two fully identified leptons;
\item two oppositely charged pions and two calorimetric showers;
\item four pions;
\item any two oppositely charged tracks with a secondary vertex detached
by more than 5 cm from the primary interaction.
\end{itemize}
The acceptance of the Level-2 trigger for $p$-Li inelastic collisions
(background), and several topologies of signal events is shown in Table~\ref{table:triggerLevel2}.
The Level-2 processor farm will receive 2.5~MHz of events, equivalent
to a data rate of $\sim$3.8~GB/s, from Level~1. These events need
to be reconstructed, filtered and formatted for permanent storage.
We assume that this task can be completed by using less than 100 ms
of CPU time and that, consequently, a farm of 2000 CPUs should be
adequate for the job.

We note that this same processor farm can be used for ``data production'' when
the experiment is not taking data. 

Factorizing an event reduction of $\sim4.5$ as being estimated with Monte Carlo simulations (cf.\ also  Table~\ref{table:triggerLevel2}),
we conclude that the average output data rate sent to permanent storage
is $\sim$0.9 Gb/s or about$\sim9$ PB/year, which we consider manageable.

\begin{table}[!ht]
\centering{\footnotesize{}}%
\begin{tabular}{|c|c|c|c|c|}
\hline 
\textit{\footnotesize{}Process} & \textit{\footnotesize{}Input event rate} & \textit{\footnotesize{}Average event size} & \textit{\footnotesize{}Input data rate} & \textit{\footnotesize{}Event acceptance}\tabularnewline
 & \textit{\footnotesize{}Hz} & \textit{\footnotesize{}bytes} & \textit{\footnotesize{}bytes/s} & \tabularnewline
\hline 
\hline 
{\footnotesize{}Inelastic collisions} & {\footnotesize{}$2.5\times10^{6}$} & {\footnotesize{}$1.5\times10^{3}$} & {\footnotesize{}$3.8\times10^{9}$} & {\footnotesize{}22.2\%}\tabularnewline
\hline 
{\footnotesize{}$\eta\to\pi^{+}\pi^{-}\pi^{0}$} & {\footnotesize{}$5.0\times10^{4}$} & {\footnotesize{}$1.7\times10^{3}$} & {\footnotesize{}$8.5\times10^{7}$} & {\footnotesize{}82.7\%}\tabularnewline
\hline 
{\footnotesize{}$\eta\to\gamma e^{+}e^{-}$ } & {\footnotesize{}$1.5\times10^{4}$} & {\footnotesize{}$1.7\times10^{3}$} & {\footnotesize{}$2.5\times10^{7}$} & {\footnotesize{}94.3\%}\tabularnewline
\hline 
\end{tabular}\caption{Level-2 trigger acceptance for inelastic collisions and for typical
$\eta$ final states with and without leptons.}
\label{table:triggerLevel2}
\end{table}
Table~\ref{table:digits-1-1-1} summarizes the Monte Carlo estimated
average digits recorded in each sub-detector for QCD background, $p$+Li$\rightarrow$$\eta X$
with $\eta\rightarrow3\pi$ and $\eta\rightarrow\gamma\gamma$ events
surviving the Level-2 trigger. 

\begin{table}[!ht]
\centering{\footnotesize{}}%
\begin{tabular}{|c|c|c|c|c|c|c|}
\hline 
\textit{\footnotesize{}Digit} & \textit{\footnotesize{}FTRACKER} & \textit{\footnotesize{}LGAD} & {\footnotesize{}RICH} & \textit{\footnotesize{}ADRIANO2} & \textit{\footnotesize{}ADRIANO2} & \textbf{\footnotesize{}Total}\tabularnewline
\textit{\footnotesize{}occupancy} &  &  &  & {\footnotesize{}Scint} & {\footnotesize{}Cer} & \tabularnewline
\hline 
\hline 
{\footnotesize{}Counts} & {\footnotesize{}$36.1$} & {\footnotesize{}$30.6$} & {\footnotesize{}$11.6$} & {\footnotesize{}$190.8$} & {\footnotesize{}$206.0$} & \textbf{\footnotesize{}475}\tabularnewline
\hline 
\end{tabular}\caption{Average digits in REDTOP sub-detectors for background events surviving
the L2-trigger}
\label{table:digits-1-1-1}
\end{table}

\subsection{Digitize and Compress: summary of trigger performance}

The task of the REDTOP trigger systems is to reduce the event rate from the
$p$-Li total inelastic collision rate of$\sim7\times10^{8}$ Hz
down to about $3$ MHz of events to be permanently recorded. Assuming
a 12-bit digitization for charge and time and an 18-bit address to
identify the struck cell, we conclude that the average size of the
final event surviving all levels of trigger is about $1.6\times10^{4}$
bytes. A summary of the expected data throughput is presented in Table
\ref{table:triggerLevels}. Such yield will produce an output data
rate of $\sim$0.9 Gb/s or about$\sim9$ PB/year, which we consider
manageable.

The needed $\sim2.3\times10^{3}$ reduction in event rate is achieved
by three trigger stages. These three stages are preceded by a digitization
and compression (DAC) stage. The DAC stage is directly connected to
the front-end of the detector. Level-0 and Level-1 are located off the detector.
A fiber optics network delivers data from the DAC to Level-0 and
from Level-0 to Level-1. The events filtered by Level-1 are delivered
to Level-2, a processor farm that performs event building, reconstruction,
formatting, and classification. A further rate reduction can possibly
be achieved at Level-2, before permanent recording, if needed. Table~\ref{table:triggerLevels}
summarizes data and event rates into and out of the different stages.

\begin{table}[!ht]
\centering{\footnotesize{}}%
\begin{tabular}{|c|c|c|c|c|}
\hline 
\textit{\footnotesize{}Trigger} & \textit{\footnotesize{}Input event rate} & \textit{\footnotesize{}Event size} & \textit{\footnotesize{}Input data rate} & \textit{\footnotesize{}Event rejection}\tabularnewline
\textit{\footnotesize{}stage} & \textit{\footnotesize{}Hz} & \textit{\footnotesize{}bytes} & \textit{\footnotesize{}bytes/s} & \tabularnewline
\hline 
\hline 
{\footnotesize{}Level 0} & {\footnotesize{}$7.\times10^{8}$} & {\footnotesize{}$1.4\times10^{3}$} & {\footnotesize{}$9.8\times10^{11}$} & {\footnotesize{}$\sim$4.6}\tabularnewline
\hline 
{\footnotesize{}Level 1} & {\footnotesize{}$1.5\times10^{8}$} & {\footnotesize{}$1.5\times10^{3}$} & {\footnotesize{}$2.3\times10^{11}$} & {\footnotesize{}$\sim$60}\tabularnewline
\hline 
{\footnotesize{}Level 2} & {\footnotesize{}$2.5\times10^{6}$} & {\footnotesize{}$1.5\times10^{3}$} & {\footnotesize{}$3.8\times10^{9}$} & {\footnotesize{}$\sim$4.5}\tabularnewline
\hline 
{\footnotesize{}Storage} & {\footnotesize{}$0.56\times10^{6}$} & {\footnotesize{}$1.6\times10^{3}$} & {\footnotesize{}$0.9\times10^{9}$} & \tabularnewline
\hline 
\end{tabular}\caption{Data and event rates for different stages}
\label{table:triggerLevels}
\end{table}
Future improvement in the detector design are expected to reduce
further the background. In particular, we foresee the adoption of a
ultra-thin vertex detector (cf.\ Sec.~\ref{par:Vertex-detector-option-II}) which will improve considerably the
rejection factor of the Level-1 trigger. 

\subsection{The Computing Model}

The computing model for REDTOP is adapted from what is currently implemented for several High Energy Physics experiments out of a DOE or NSF major facility. It aims to paste together the pieces of an end-to-end infrastructure and software that enables the delivery of the final data-products to the researchers at the different institutional research partners. 

The REDTOP experiment operates in a typical beam environment where the inelastic interaction rate of the proton beam with the target is ~1 GHz. Level-0 and Level-1, implemented in hardware, will reduce such a large rate by a factor of $\sim104$ before sending  the data to a local compute-farm for the Level-2 trigger and preliminary reconstruction. We expect approximately 2.5 PB of production-quality experiment data and approximately 2 PB of processed data to be generated each year via the acquisition and reconstruction of the experimental output together with the required simulations. In order to accommodate these storage requirements, the Fermilab (FNAL) facility is needed to provide long-term tape storage during each year of operation and a disk allocation on FNAL's dCache for staging and distributing data elsewhere. In addition to FNAL, REDTOP will a leverage an allocation on the Open Science Connect service which can provide 100 TB-scale ephemeral storage (Stash) for jobs to the Open Science pool.

In terms of compute requirements, REDTOP employs a single-core computational workflow which has proven to be well suited for the distributed High Throughput Computing (dHTC) environment of the Open Science Farbic of services. The computing model here aims to facilitate reconstruction of the expected flow of data from the full experimental apparatus along with the suite of required Monte Carlo simulations. Extrapolating from REDTOP jobs currently running on Open Science pool, it is estimated that the we would need approximately 90 million core hours annually; 55 million core-hours for Monte Carlo jobs and 35 million core-hours for data reconstruction jobs.

With the above considerations in mind, we assume that the output data stream from the Level-2 farm will be staged at Fermilab's dCache storage and, eventually, pre-processed on site. The process will require an allocation on the lab's General Purpose Grid (GPGrid) with local file access on dCache enabled via a POSIX-like interface over an NFS mount. dCache will primarily serve as a high speed front-end ephemeral storage to provide access to data stored on tape. Since direct access to tape is limited to on-site infrastructure, staging on dCache first enables downstream access by off-site resources. In order to increase the flexibility of the model and to offer more optimal implementations to the participating institutions, the following options are considered and discussed below.

\begin{figure}[!th]
\centering{}\includegraphics[scale=0.3]{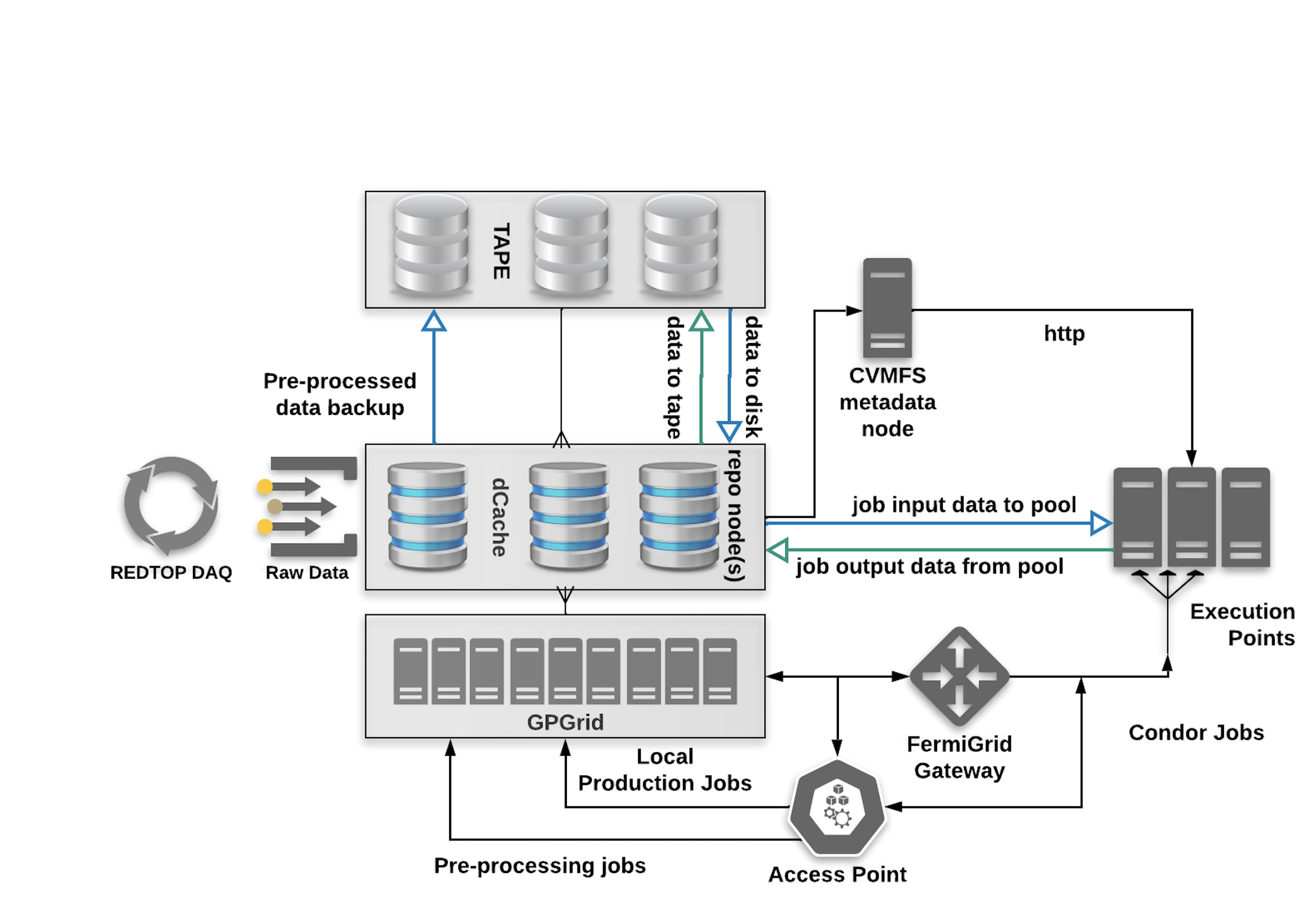}\caption{\label{fig:compmodel1} The REDTOP compute workflow out of Fermilab. Experimental data out of the REDTOP DAQ are stored on FNAL's dCache from where they can be pre-processed for level-2 triggers and preliminary reconstruction on the local cluster. Data are then stored on tape. An access point at FNAL, an HTCondor submit node, can submit jobs to the Open Science pool. Data in this scheme are delivered directly from dCache after being staged first on tape. 
}
\end{figure}

The collaboration will process the bulk of the experimental results via HTCondor jobs submitted to FermiGrid or Open Science pool from a lab access point (submit node). An example of a job submission workflow for REDTOP at FNAL is shown in Fig.~\ref{fig:compmodel1}. A submit host sends jobs to the pool while data are delivered to the remote compute sites (execution points) by first staging them on dCache and then transferring them to the remote execution points via a protocol like XRootD or http. It is proposed that Fermilab also deploys a data origin repository for REDTOP in the distributed CernVM File System (CVMFS) which will deliver input data to remote compute sites using OSG's federation of caches~\cite{Weitzel2019StashCacheAD}. The OSG data federation (OSDF) service replicates data needed by a project on a cache in regional proximity to a compute site. When a user job requests a dataset or a software bundle via a collaboration specific CVMFS path in the execution script, it fetches the data from the nearest cache instead of the origin. This minimizes the flight path to the execution point (worker node) and improves the runtime efficiency of the job. The software stack, which includes the GenieHad and Geant4 Monte Carlo software as well as the reconstruction framework, can also be distributed out of the same repository, however there is an alternative option to host the software origin in the OASIS repository - OSG's Application Installation Service. Any output data designated for long term storage can then be archived back to tape at Fermilab by first transferring the files back onto dCache using the same protocols as above.

\begin{figure}[!th]
\centering{}\includegraphics[scale=0.6]{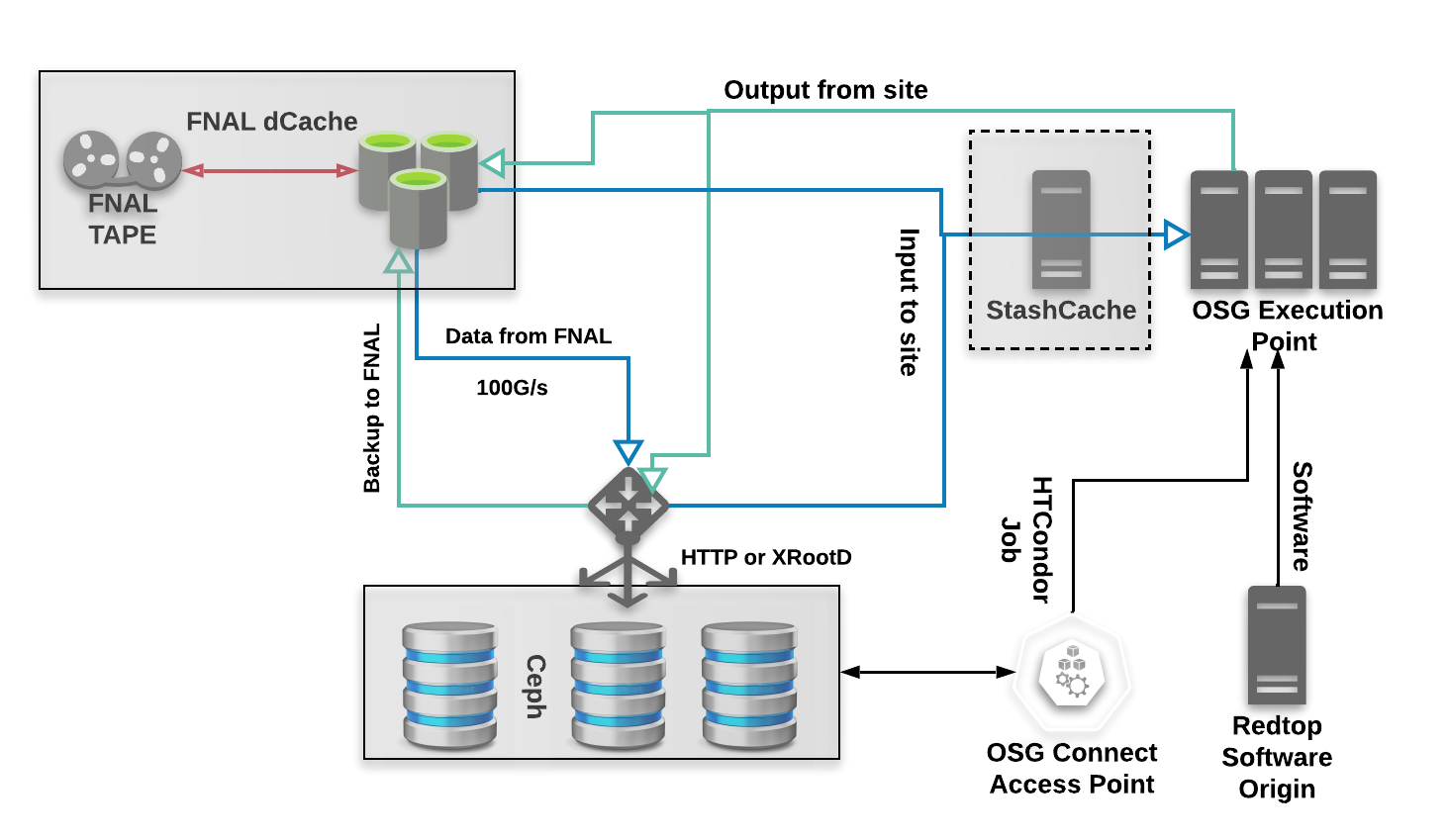}\caption{\label{fig:compmodel2} The REDTOP compute workflow using an Open Science Connect access point, a HTCondor job submit node. This entry point provides a broader access to the research community for job submissions to the Open Science pool. Data from FNAL can be pre-staged on the OSG Connect storage - Stash - although that is not strictly necessary if the user has credentials to access Fermilab resources as in this case the job on the execution point can request data directly from the FNAL's dCache disk.   
}
\end{figure}

The collaboration will continue to leverage the OSG Connect entry points at the University of Chicago to launch reconstruction or simulation jobs to the Open Science pool. OSG Connect provides a collaborative environment for a multi-institutional partnership such as REDTOP, where researchers from the individual institutions can request accounts to the access point via an online portal. Users can then join the REDTOP project, gain access to the project's shared directory on the OSG distributed storage (Stash) and submit jobs to the Open Science pool. Stash, a Ceph filesystem storage system, provides temporary storage capacity for OSG projects and can distribute data to remote execution points in the Open Science pool over CVMFS or via a direct copy using XRootD  or http. 

REDTOP has the option to make use of Stash to pre-stage input files and to store the processed output files from the jobs before distributing them to various endpoints of the collaboration including archiving them back to tape at FNAL. Figure~\ref{fig:compmodel2} shows such an example of a data processing workflow using OSG Connect. Users submit HTCondor jobs to the Open Science pool from the submit host (the access point). Input data can be delivered to the execution point either from stash or via a gfal-copy from dCache at Fermilab. Similarly, the output data can be copied back to stash or to Fermilab's dCache and eventually to tape. Note that user access to the files at Fermilab will require them to authenticate with their credentials to the storage endpoint.

While not in the critical path to the processing campaign of the experimental data, the REDTOP collaboration can augment the availability of computing cycles by fostering the creation of a dedicated resource pool via allocations on facilities at member institutions. This is similar to the general purpose Open Science pool; user submitted HTCondor jobs run as payloads in pilots launched to these dedicated execution points by the Open Science gWMS factory via entry points (gateways) that can be hosted by OSG or deployed locally at the institutional edge infrastructure. 

The approach described above constitutes a minimalist compute-storage model that will meet the requirements of REDTOP if the appropriate scale of the annual storage requirements is met at Fermilab. A distributed storage model can also be considered which requires participating institutions to contribute storage to the collaboration. Such model requires the introduction of a data management appliance/service that will coordinate the data placement into the different storage endpoints (RSE: Rucio Storage Endpoint) and manage the mobility across sites and execution points at the Open Science Pool. 

\begin{figure}[!th]
\centering{}\includegraphics[width=0.95\textwidth]{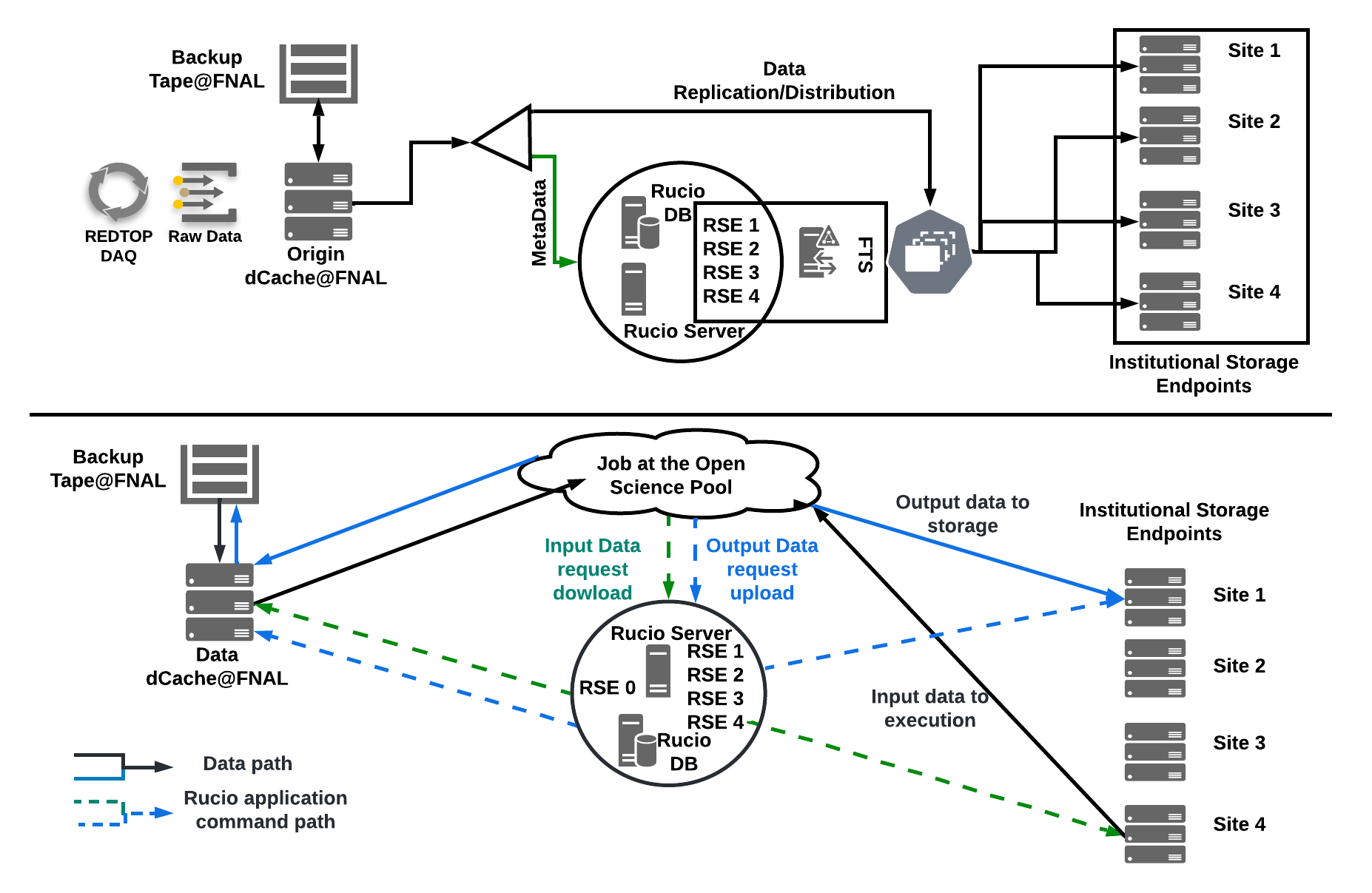}\caption{\label{fig:compmodel3}A compute-storage model for REDTOP where a data management appliance, Rucio, is used to distribute data from an Origin (FNAL) to dedicated institutional storage end-points. When jobs at the Open Science Pool request an input file, they query the Rucio server for a download. The server then coordinates the proper download location of the file to the execution point. The process is reversed when uploading files to the storage-end points (RSEs) which also updates the catalog maintained by Rucio.
}
\end{figure}

This model is briefly described in the upper panel of Fig.~\ref{fig:compmodel3}. Data collected by the experiment are ingested into the Rucio database and distributed into various storage end-points using the File Transfer Service (FTS). The metadata stored in the Rucio database include information such as the size, institutional location, scope of the file and directory path of a file. This process facilitates the distribution of the entire experimental output, reducing the storage footprint on the primary site (FNAL) which can still maintain post level-2 data on tape. Portions of the experimental output will then be readily available to individual institutions for on-site reconstruction or analysis besides the capability to deliver the data to jobs at the Open Science pool on demand. 

This scenario is described in the lower panel of Fig.~\ref{fig:compmodel3}, where an HTCondor job running at a remote execution point can request an input file from the Rucio server which mediates the delivery of the file to the job from the appropriate RSE that has it. The process can be reversed by uploading an output file back to a storage end-point which also updates the dataset catalog in the Rucio database. This process can be considered supplementary to the capabilities offered by the OSDF service of caches. The latter delivers datasets in proximity to a nearby execution points but only in the context of a HTCondor job submitted by a user from an access point such an FNAL or OSG Connect. A data management service like Rucio will afford the opportunity to the collaboration's research partners to process data in local facilities or dedicated Open Science execution points even in the absence of a nearby OSDF cache.

\section{\label{sec:The-Simulation-Framework}The Simulation Framework}

\subsection{\label{sec:The-Genie-event}The event generator: \texorpdfstring{$GenieHad$}{}}

An experiment aimed at detecting faint signals of
New Physics in a harsh hadronic environment presents several challenges
concerning the knowledge of the detector response and of the topology
of the physics events. Considering the significant number of inelastic nuclear collisions
produced in REDTOP targets, a reliable simulation of the hadronic
interactions between the proton beam and the nucleons in the target
is of paramount importance to estimate the background and, consequently,
the sensitivity to New Physics. Among the simulation packages presently
available to the experimenters in High Energy and Nuclear physics,
Geant4 is by far the most widely used and its particle transport mechanism is, probably, unsurpassed. However,
processes occurring at energies below $\sim10$ GeV need specific hadronic generators and framework for the following reasons.
Hadronic interactions governing the beam-target scattering in
that energy range occurs predominantly via the formation and the decay
of intra-nuclear baryonic resonances. The treatment of such processes
is very complex, and it requires a non-perturbative approach.
Furthermore,
excited nuclei could de-excite into nucleons and nuclear remnants,
which could, subsequently, interact with the detector.

The approach followed
in the present work is of separating the simulation of the primary
interaction from the transport of the secondary particles throughout
the detector. More specifically, the primary interaction between the
beam and the nuclear matter contained in the target is simulated with a specialized
event generator which would reproduce the scattering and the de-excitation
of the nucleus according to one or more nuclear transport models, evaporation/fragmentation/fission and final de-excitation models. In the following
step, those particles and ions are propagated throughout the material volumes
of the detector using the Geant4 package.

Several standalone intranuclear scattering and transport models have
been implemented in the last few years to simulate the interaction
between a hadron or an ion and a nucleus. Given the complexity of the
underlying models, those programs are usually developed and maintained
by nuclear theorists.
From the experimental
point of view the packages currently available have two important
deficiencies:
\begin{enumerate}
\item The nuclear target is an isolated nucleus surrounded by vacuum (namely,
re-scattering in the target is not considered);
\item Except for one case\cite{Incl2020}, the nucleons generated in the
final state is considered as isolated (i.e, un-aggregated) particles,
rather than as clusters of, eventually excited, ions.
\end{enumerate}
To overcome such limitations, the $GenieHad$\cite{GenieHad_2012}
event generator package has been developed. $GenieHad$ is an extension
of the $Genie$ framework~\cite{Andreopoulos:2009rq} that is widely
used in the neutrino community, as it relies on its geometry package and
its architecture. The heart of \emph{$GenieHad$} is a collection
of twenty modules interfacing external hadronic simulation packages
\cite{BUSS20121,Incl2020,Bleicher_1999,Nara2019,ESPP_2018,Ehehalt:1996uq,Sjstrand:208575} to $Genie$.
The primary nuclear interaction is deferred by the corresponding $GenieHad$
interface to the external hadronic generator which performs the
scattering according to the parameters specified by the user. Once
the scattering engine has completed its process, the final products
are returned to $GenieHad$ which prepares them for the following steps.
These are, usually, a) the clusterization of the individual nucleons
into nuclear remnants or fissioned ions, alpha particles, etc, b)
the de-excitation and/or evaporation of eventual remnants with positive
binding energy. Such final steps are performed by a set of dedicated
simulation packages. A list of the nuclear,
electromagnetic, clusterization and de-excitation models
currently implemented in $GenieHad$ is presented, in Tables
\ref{table:geniehad-interfaces-had},
\ref{table:geniehad-interfaces-evap},
\ref{table:geniehad-interfaces-em}, and
\ref{table:geniehad-interfaces-clus}.

\begin{table}[!ht]
\textbf{\footnotesize{}\centering}{\scriptsize{}}%
\begin{tabular}{|c|c|c|}
\hline 
{\scriptsize{}Package} & {\scriptsize{}Model} & {\scriptsize{}Type}\tabularnewline
 &  & \tabularnewline
\hline 
\hline 
{\scriptsize{}Urqmd~\cite{Bleicher_1999}} & {\scriptsize{}QMD} & {\scriptsize{}Microscopic many body approach}\tabularnewline
\hline 
{\scriptsize{}Incl++ v6.2~\cite{Incl2020}} & {\scriptsize{}INCL} & {\scriptsize{}Intranuclear cascade}\tabularnewline
\hline 
{\scriptsize{}Gibuu v2019~\cite{BUSS20121}} & {\scriptsize{}BUU} & {\scriptsize{}time evolution of Kadanoff\textendash Baym-equations}\tabularnewline
\hline 
{\scriptsize{}PHSD v 4.0~\cite{Ehehalt:1996uq}} & {\scriptsize{}HSD} & {\scriptsize{}covariant transport with NJL-type Lagrangian}\tabularnewline
\hline 
{\scriptsize{}Jam v1.9~\cite{Nara2019}} & {\scriptsize{}Cascade/RQMD.RMF/BUU} & {\scriptsize{}Multi-model - hybrid approach}\tabularnewline
\hline 
{\scriptsize{}Dpmjet-III~\cite{dpmjet-iii2005}} & {\scriptsize{}Dual Parton/ perturbative QCD} & {\scriptsize{}Multi-model approach}\tabularnewline
\hline 
{\scriptsize{}Pythia 7, 8\cite{Sjstrand:208575}} & {\scriptsize{}LUND} & {\scriptsize{}string hadronization model}\tabularnewline
\hline 
{\scriptsize{}IAEA tables\cite{iaea}} & {\scriptsize{}LUT of measured cross sections} & {\scriptsize{}Look-up tables based on ENDF (by IAEA)}\tabularnewline
\hline 
{\scriptsize{}Intranuke\cite{GENIE:2021wox}} & {\scriptsize{}Parametric} & \tabularnewline
\hline 
{\scriptsize{}ALPACA\cite{Alpaca2019}} & {\scriptsize{}Alpaca} & {\scriptsize{}Bremsstrahlung of Axion-Like-Particles (ALPs) }\tabularnewline
\hline 
\end{tabular}\caption{Nuclear scattering models available in \emph{GenieHad}~\cite{GenieHad_2012}. }
\label{table:geniehad-interfaces-had}
\end{table}
\begin{table}[!ht]
\textbf{\footnotesize{}\centering}{\scriptsize{}}%
\begin{tabular}{|c|c|c|c|c|}
\hline 
{\scriptsize{}Package} & {\scriptsize{}$\gamma$ } & {\scriptsize{}N, p, He} & {\scriptsize{}IMF} & {\scriptsize{}Fission}\tabularnewline
 & {\scriptsize{}emission} & {\scriptsize{}model} & {\scriptsize{}Model} & {\scriptsize{}model}\tabularnewline
\hline 
\hline 
{\scriptsize{}Abla07\cite{kelic2009abla07}} & {\scriptsize{}yes} & {\scriptsize{}Weisskopf-Erwig} & {\scriptsize{}no} & {\scriptsize{}Fokker-Plank}\tabularnewline
\hline 
{\scriptsize{}Abla++~\cite{abla++2015}} & {\scriptsize{}yes} & {\scriptsize{}Weisskopf-Erwig} & {\scriptsize{}no} & {\scriptsize{}Fokker-Plank}\tabularnewline
\hline 
{\scriptsize{}GEM~\cite{FURIHATA2000251}} & {\scriptsize{}no} & {\scriptsize{}Weisskopf-Erwig} & {\scriptsize{}Weisskopf-Erwig} & {\scriptsize{}parametric}\tabularnewline
\hline 
{\scriptsize{}Gemini++~\cite{Gemini2001}} & {\scriptsize{}no} & {\scriptsize{}Hauser-Feshbach} & {\scriptsize{}Transition state} & {\scriptsize{}Transition state model}\tabularnewline
\hline 
{\scriptsize{}Evapor~\cite{KAW2012}} & {\scriptsize{}no} & {\scriptsize{}Weisskopf-Erwig} & {\scriptsize{}no} & {\scriptsize{}parametric}\tabularnewline
\hline 
{\scriptsize{}SMM~\cite{SMM2020}} & {\scriptsize{}no} & {\scriptsize{}Weisskopf-Erwig} & {\scriptsize{}Weisskopf-Erwig} & {\scriptsize{}Statistical Multifragmentation}\tabularnewline
\hline 
\end{tabular}\caption{De-excitation, evaporation, fission models available in \emph{GenieHad}~\cite{GenieHad_2012}. }
\label{table:geniehad-interfaces-evap}
\end{table}

\begin{table}[!ht]
\textbf{\footnotesize{}\centering}%
\begin{tabular}{|cc|}
\hline 
\textbf{\textit{\footnotesize{}Package}} & \textbf{\textit{\footnotesize{}Type}}\tabularnewline
 & \tabularnewline
\hline 
\hline 
{\small{}Geant4 EM physics~\cite{Agostinelli:2002hh}} & {\small{}Parametric}\tabularnewline
\hline 
{\small{}LELAPS~\cite{langeveld2005fast}} & {\small{}Parametrized shower simulation}\tabularnewline
\hline 
\end{tabular}\caption{Electromagnetic interaction models available in \emph{GenieHad}~\cite{GenieHad_2012}. }
\label{table:geniehad-interfaces-em}
\end{table}
\begin{table}[!ht]
\textbf{\footnotesize{}\centering}%
\begin{tabular}{|c|c|}
\hline 
\textbf{\textit{\footnotesize{}Package}} & Model\tabularnewline
 & \tabularnewline
\hline 
\hline 
{\small{}Coalescence\cite{Bleicher_1999}} & {\small{}GEM based Phase-Space Coalescence}\tabularnewline
\hline 
{\small{}KAW\cite{KAW2012}} & {\small{}Urqmd based Phase-Space Coalescence}\tabularnewline
\hline 
\end{tabular}\caption{Clusterizer models available in \emph{GenieHad}~\cite{GenieHad_2012}. }
\label{table:geniehad-interfaces-clus}
\end{table}
In the  $GenieHad$'s architecture, because of the approach described above, the environment of the
event (geometry of the apparatus, materials, beam profile and composition,
targets, etc.)  is taken  care of by the underlying
framework (namely, $Genie$), while the scattering at the microscopic level is performed by the external engine, which is called at run-time according
to the selected transport model. One of the main advantage of this approach
is the possibility to compare the outcome of different nuclear
transport models under identical conditions, and in particular,
within the same geometry and the same beam parameters.
By repeating the simulation and comparing the results from several models, one can evaluate the robustness of the procedure
and estimate the uncertainties of  observables.\medskip{}

In the contest of the studies performed for this work, $GenieHad$
was employed to generate signal and background events according
to physics of the SM. In order to study physics BSM, the $\eta$ and $\eta^{\prime}$
mesons produced (along with other mesons and baryons generated
by the fragmentation of the target), are subsequently decayed
by a special $Geniehad$ module reproducing the process under study.
Nineteen such modules have been developed, each incorporating the decay
amplitude for that particular theoretical model. More details on the
event sets being produced can be found in the respective subsections of Sec. \ref{sec:Sensitivity-Studies-to-physics-BSM}
through \ref{sec:Sensitivity-to-Non-perturbative}

\subsection{The particle transport software: \texorpdfstring{$slic$}{}}

The second step of the simulation consists of transporting the particles and 
nuclear remnants generated in $GenieHad$ throughout the target and the
the detector.
This task is performed by Geant4, via the $slic$ package, which manages 
the geometry and the event input-output. $slic$\cite{slic} was
initially developed by the SID Collaboration within the ILC framework.
It is currently used for the simulation of the events by the HPS experiment\cite{HPS}
at JLAB. A very detailed geometry is implemented in a gdml format
and an extended gdml schema is used for the segmentation
of  sensitive detectors. For the sake of this work, most of the passive
material, relevant for the determination of the multiple  scattering, is also taken into account.
Furthermore, REDTOP uses
a proprietary version of $slic$ including, among other things,  the physics of optical photons,
which is necessary for the correct simulations of the signal of ADRIANO2 calorimeter.
The results of this step is a collection of ``{\em hits}'' which carry the information of the energy deposition and its location 
in the sensitive volumes of the detector.

\subsection{The reconstruction and analysis software: \texorpdfstring{$lcsim$}{}}

The software framework which implements the simulation of the response of the experimental apparatus, after 
the particle transport and the energy deposition in the apparatus is completed by Geant4, is $lcsim$\cite{lcsim}.
This software framework is a java based package, initially developed
by SID Collaboration within the ILC framework and also used
by the HPS experiment\cite{HPS} at JLAB.
The reconstruction in \emph{lcsim} proceeds according to
the following steps:
\begin{itemize}
\item Digitization of the hits produced by Geant4, based on the simulated response of the front-end electronics;
\item Simulation of the algorithms implemented in the Level-0 trigger for
fast rejection of the background;
\item Pattern recognition;
\item Simulation of the algorithms implemented in the Level-1 trigger;
\item Track reconstruction from the digits collected in the tracking system;
\item Shower reconstruction in the $TCR$ and $ADRIANO2$;
\item Simulation of the algorithms implemented in the Level-2 trigger for
rejection of the background based on the event topology;
\item Final analysis of the event.
\end{itemize}
Further details are given in the next section.

\subsection{\label{sec:Simulation-strategy}The simulation strategy}

This section describes in detail the strategy implemented in the simulations, along with the parameters used.

\subsubsection{Beam parameters}

A collimated proton beam with a Gaussian energy spread and  transverse profile
was implemented in $Genie$ + $GenieHad$. In order to preserve the
validity of the results of the  studies with a broader range of accelerator
configurations, we have used the following, very conservative, beam
parameters:
\begin{itemize}
\item $E_{kin}$ = 1800 MeV
\item $\sigma_{E_{kin}}$= 0.9\%
\item $\sigma_{x},\,\sigma_{y}$ = 1 mm, 5 mm
\item Beam halo: 4.5 mm $\times$ 4.5 mm
\end{itemize}

\subsubsection{Digitization}

The  first step of the reconstruction performs the digitization of the hits found in  the four sub-detectors. The following experimental
effects are taken into account for each of them:

\paragraph{Fiber tracker}
\begin{itemize}
\item Efficiency of  individual fibers;
\item Photo-detector efficiency ($pde$) and  energy response of the SiPM;
\item Amplification and discrimination of the signal.
\end{itemize}

\paragraph{Central tracker (LGAD tracker)}
\begin{itemize}
\item Single pixel efficiency;
\item Diffusion and cross-talk among nearby pixels;
\item Amplification and discrimination of the signal;
\item Smearing of the time of formation of the hit.
\end{itemize}

\paragraph{Threshold Cerenkov Radiator}
\begin{itemize}
\item Parametrization of the detector response based on the results of T1604
test beam.
\end{itemize}

\paragraph{ADRIANO2}
\begin{itemize}
\item Parametrization of the detector response based on the  results of T1604
experiment.
\end{itemize}

\subsubsection{Simulation of the trigger systems}

The algorithms for the three trigger levels have been fully implemented
in $lcsim$. A detailed description of REDTOP trigger system can be
found in Sec.~\ref{sec:trigger_system}. 

\subsubsection{Pattern recognition and PID}

Given the novelty of the detector technologies proposed, REDTOP does
not yet have, as of this writing, a full pattern recognition in place. Consequently,
a cheating strategy has been implemented, where the digits are assigned to
a particle using the Monte Carlo truth. The same situation holds for the PID,
which is assigned to each reconstructed particle using a LUT, storing
realistic values of efficiency and mis-identification probabilities. The latter have been evaluated with Monte Carlo techniques and known experimental results.

\subsubsection{Track reconstruction in the Fiber Tracker and in the Central Tracker
(helix fit and Kalman filter)}

A full helix fit of the tracks in a solenoidal magnetic field is performed
in the tracking systems. Multiple scattering and energy loss in materials
traversed by the tracks are taken into account in the covariance matrix.
Only those tracks successfully fitted and with a satisfactory $\chi^{2}$
are passed to the next step of the reconstruction. In addition to the helix fit, a full Kalman filter has been implemented, although it has not been used for the studies presented here.

\subsubsection{Shower reconstruction in the TCR and ADRIANO2 calorimeter
(parametric)}

The TCR, which is located immediately in front of ADRIANO2,  has also the functions of a pre-shower for the latter.
As mentioned in Sec.~\ref{sec:Simulation-strategy}, several test
beams have been conducted with ADRIANO2 prototypes by T1604 Collaboration.
To speed up the simulations for the studies presented in this work,
a parametric response of the detector has been implemented, based
on the results of those tests. 


\newpage{}

\section{\label{sec:Sensitivity-Studies-to-physics-BSM}Sensitivity Studies to Physics Beyond the Standard Model}

In this section we present the results of the physics studies performed to estimate the sensitivity of REDTOP to several benchmark
BSM processes and several theoretical models. The studies are based
on the Snowmass-2021 detector layout,  described in some details in Sec.~\ref{sec:The-Simulation-Framework}.
Such studies have two goals: a) presenting the physics case for a $\eta/\eta^{\prime}$-factory; b) finding possible improvements
of the experimental apparatus and in which way they would impact most strongly the discovery
potential of REDTOP. This section is organized as follows. In Sec.~\ref{subsec:Estimation-of-the-sensitivity-curves}
we derive the formulae used to determine the branching ratio
sensitivity. Sec.~\ref{subsec:Generation-of-signal} and \ref{subsec:Standard-model-background}
describe the methodology used to simulate the signal and background
samples. In Sec.~\ref{subsec:Sensitivity-to-the-vector} through \ref{subsec:Sensitivity-to-the-heavylepton}
we focus on the sensitivity studies for the four portals connecting
the Dark Sector with the Standard Model. We proceed first by evaluating
the sensitivity to the branching ratio of the relevant processes, and
then, we apply those results to benchmark several theoretical
models. In Sec.~\ref{subsec:Tests-of-Conservation-Laws} we present
our results on sensitivity studies relevant for testing the discrete symmetries
of the Standard Model. The symmetries considered in this work are: $C$ and $CP$, Lepton
Flavor Violation ($LFV$), and Lepton Universality. Finally, in Sec.
\ref{subsec:Discussion-of-the-results} we discuss the results, and
try to identify the possible improvements to the experiment.

\subsection{Estimation of the sensitivity curves\label{subsec:Estimation-of-the-sensitivity-curves}}

In this section we derive the formulae needed
to estimate the branching ratio sensitivity as well as to a generic theoretical models.
More specifically, assuming that the model is based on a (set of)
parameter(s) $\bar{g}$, usually the coupling constant(s) of the amplitude(s)
involved, by sensitivity we define the minimum value(s) of $\bar{g}$
measurable at REDTOP with a signal which is three times larger
than the background fluctuations. For an experiment at a meson factory, where the $\eta/\eta^{\prime}$ are hadro-produced, we expect a large background from QCD contaminating
most of the final states. The sensitivity is define as the value of  $\bar{g}$   such that:
\begin{equation}
S_{min}(\bar{g})=3\times\sqrt{B}\label{eq:Master1}
\end{equation}
where: $S_{min}$ is the minimum number of signal events reconstructed
and surviving the final selection, while $B$ is the number of background
events.

If$N_{\eta}$ ($N_{\eta^{\prime}}$) is the number of $\eta$ ($\eta^{\prime})$ mesons
produced and $N_{ni}$ is the number of inelastic nuclear
interactions between the beam and the target, then:

\begin{eqnarray}
S_{min}(\bar{g})=N_{\eta}\times Br(\bar{g})\times\epsilon_{reco}\label{eq:MasterS}\,,\\[1ex]
B=N_{ni}\times\epsilon_{bkg}\label{eq:MasterB}
\end{eqnarray}

where $Br(\bar{g})$ and $\epsilon_{reco}$ are, respectively, the
branching ratio and the reconstruction efficiency for that
channel, while $\epsilon_{bkg}$ is the probability that a Standard
Model event will survive all analysis cuts. From. Eqs. \ref{eq:Master1}-\ref{eq:MasterB},
we define the branching ratio sensitivity as:

\begin{equation}
sensitivity(Br(\bar{g}))=\frac{3\times\sqrt{N_{ni}\times\epsilon_{bkg}}}{N_{\eta}\times\epsilon_{reco}}\label{eq:MasterS-final}
\end{equation}

Assuming an incoming beam with $10^{11}$ POT/sec and an integrated
proton yield of $3.3\times10^{18}$ POT (that, according to the analysis performed in Sec.~\ref{subsec:Hadro-production-of-},
is achievable with an effective running time of 12 months), the expected
total yield of $\eta$/$\eta^{\prime}$ and of inelastic proton collisions, are summarizes
in Table~\ref{tab:Expected-yield-at}. In those calculations, we conservatively assumed a value of 0.49\% for the production
probability, as obtained from the Urqmd model in $GenieHad$, instead of the  value of 0.70\% obtained by
averaging the results from several generators (cf.\ Tab.\ref{table:eta-production-probability}).
This is also consistent with our final choice of $Urqmd$ for the generation
of the events for the sensitivity studies presented below (see, also,
Sec.~\ref{subsec:Standard-model-background}).

\begin{table}
\begin{centering}
\begin{tabular}{|c|c|}
\hline 
 & Total yield for $E_{kin}$=1.8 GeV\tabularnewline
\hline 
\hline 
$N_{\eta}$ & $1.1\times10^{14}$\tabularnewline
\hline 
$N_{\eta^{\prime}}$ & $0$\tabularnewline
\hline 
$N_{ni}$ & $2.5\times10^{16}$\tabularnewline
\hline 
\end{tabular}$\;\;$%
\begin{tabular}{|c|c|}
\hline 
 & Total yield for $E_{kin}$=3.6 GeV\tabularnewline
\hline 
\hline 
$N_{\eta}$ & $5.9\times10^{14}$\tabularnewline
\hline 
$N_{\eta^{\prime}}$ & $7.9\times10^{11}$\tabularnewline
\hline 
$N_{ni}$ & $3.2\times10^{16}$\tabularnewline
\hline 
\end{tabular}
\par\end{centering}
\caption{\label{tab:Expected-yield-at}Expected event yield at REDTOP for $3.3\times10^{18}$
POT, estimated with the Urqmd module of $GenieHad$
(see Sec.~\ref{sec:The-Genie-event} for details on the estimation
method).}
\end{table}
The values of $\epsilon_{reco}$ and $\epsilon_{bkg}$ have been determined for each
of the analyses presented in the remaining of this article. Once a
master formula for $Br(\bar{g})$ is obtained for a particular theoretical
model, then Eq.~\eqref{eq:MasterS-final} can be used to extract the
experimental sensitivity to the parameter(s) $\bar{g}$.

\subsection{Standard Model background\label{subsec:Standard-model-background}}

As discussed in Sec.~\ref{sec:The-Genie-event}, several nuclear
scattering generators are available in $GenieHad$. For the studies
presented in this work, we have extensively examined two of them: $Urqmd$
and $Incl++$. We generated about $2\times10^{9}$ $p+Li\rightarrow X$
inclusive inelastic events with each generator, and analyzed their
outcome. We concluded that our sensitivity studies are not appreciable
affected by which of the two is chosen. We decided to use $Urqmd$ to generate the remaining event sample, as, in spite of being computationally  more intensive than $Incl++$, 
it is widely regarded by the intermediate energy community the most
effective in simulating the nuclear scattering at the energies of
interest for REDTOP. An integrated sample of about $5\times10^{10}$
events was finally produced, corresponding to $2\times10^{12}$ POT,
or 1/500,000 of the expected beam integrated current. The
final nucleons from $Urqmd$ were, subseqently, clusterized into nuclear remnants by the KAW 
clusterizer\cite{KAW2012} and, finally, de-excite and evaporated
with  $ABLA\,v7$~\cite{kelic2009abla07}. 

At present, only few decays of the $\eta/\eta^{\prime}$ mesons currently
implemented in Geant4. Therefore, the missing exclusive $\eta/\eta^{\prime}$
processes needed to estimate the signal and the background were individually
generated using $GenieHad$. %

The events generated with $GenieHad$ were, subsequently, processed
with Geant4, to simulate the response of the detector. The digitization
and reconstruction strategy are described in detail in Sec.~\ref{sec:Simulation-strategy}.
Finally, the background was merged with the signal during
the analysis step, and re-weighted to take into account eventual differences
in the number of POT's.

\subsection{Generation of signal events\label{subsec:Generation-of-signal}}

The studies discussed in the present work refer only to the production
and decay of the $\eta$-meson. Similar studies for the $\eta^{\prime}$-meson
production and decay are planned, and will be published in a future
work.

The $Urqmd$ sample described above sample consists of about $2.35\times10^{8}$
$p+Li\rightarrow\eta+X$ events which were extracted and used as input
to $GenieHad$ to generate exclusive decays corresponding to the process under study. 
As already stressed above, the vast majority of $\eta$-mesons is produced by the decay
of intra-nuclear resonances with masses below 2 GeV. Therefore, those mesons have only
a modest boost in the laboratory frame. As shown in Fig.~\ref{fig:eta_energy}, about 40\% of the  $\eta^{\prime}$s have
a kinetic energy less than 100 MeV. In order to optimize the computing
time, sub-samples of the full $\eta$-set were prepared for the individual
analyses, while the entire background sample was used. The $\eta$-meson was further decayed in $GenieHad$, with
a decay amplitude derived from the theoretical model being benchmarked.
The full simulation process is described in detail in Sec.~\ref{sec:Simulation-strategy}. 

\section{Sensitivity to the Vector portal\label{subsec:Sensitivity-to-the-vector}}

The \emph{Vector portal} can be probed via radiative decays of the
$\eta$-meson and its subsequent decay into a lepton-antilepton $\ell\bar{\ell}$
pair or into two pions. The following processes have been considered in this work:
\begin{itemize}
\item $p+Li\rightarrow\eta+X\;with\;\eta\rightarrow\gamma A'\;and\;A'\rightarrow e^{+}e^{-}$
\item $p+Li\rightarrow\eta+X\;with\;\eta\rightarrow\gamma A'\;and\;A'\rightarrow\mu^{+}\mu^{-}$
\item $p+Li\rightarrow\eta+X\;with\;\eta\rightarrow\gamma A'\;and\;A'\rightarrow\pi^{+}\pi^{-}$
\end{itemize}
Two different analysis were performed, aiming at testing the performance
of different components of the detector: a $bump-hunt$ and a $detached-vertex$
analysis.

\subsection{\texorpdfstring{$A'\rightarrow e^{+}e^{-}$}{}: Bump-hunt analysis}

This study aims at evaluating the sensitivity of the detector in the
assumption that the vector particle has a decay length not resolvable by
the tracking system. In this case, the process is identified only
from its kinematics, in particular by observing a bump in the invariant
mass of the $\ell\bar{\ell}$ or the $\pi^{+}\pi^{-}$ system. 

The vector boson $A'$ was generated in a mass range between 17 MeV
and 500 MeV, and decayed promptly in $GenieHad$, by setting
the vertex of the di-lepton in the point where the proton struck the
target. This final state has a large background contribution on the
low side of the kinematically allowed mass range. This arises mostly
from the process $\eta\rightarrow\gamma\gamma$, when one of the
photons converts into a $e^{+}e^{-}$ pair in the detector material,
possibly associated to a $\pi^{0}$ decaying into two photons or into $\gamma e^{+}e^{-}$
with a kinematics compatible with the $\eta$ mass. On the higher mass
range, the largest background is from the process $\eta\rightarrow\gamma e^{+}e^{-}$
which proceeds via a three-body phase space kinematics. 
This process generates
an irreducible non-resonant component under the resonant bump. Another
source of combinatoric background is due to the production of two
or more $\pi^{0}$'s, where one $\pi^{0}$ decays into $\gamma e^{+}e^{-}$
and a second $\pi^{0}$ decays into $\gamma\gamma$. As shown in Fig.
\ref{fig:Mutiplicity}, the probability that a 1.8 GeV proton impinging
on a lithium target produces two or more $\pi^{0}$'s is about
10\%. Therefore, without a proper reconstruction and analysis, the background potentially faking the signal could be of
order $\mathcal{O}(10{}^{13})$. 

The full chain of generation-simulation-reconstruction-analysis was
repeated for each event set. 
Information from the  TOF and PID systems contributes in reduces
the background  at the trigger level, by a factor $\mathcal{\sim}10{}^{4})$.
Very generic requirements on the quality of reconstructed particles
were applied in the data analysis. In particular, no
cuts were applied to the reconstructed $e^{+}e^{-}$ vertex. Neutral
pions, decaying into $\gamma e^{+}e^{-}$ and into $\gamma\gamma$, where
reconstructed by considering all combinations of photons, electrons
and positrons with an invariant mass within 5 MeV from the $\pi^{0}$
mass. This requirement is effective in rejecting most of the combinatoric
background. Similarly, photons
converting into a $e^{+}e^{-}$ pair were reconstructed by requiring
that the invariant mass of the $e^{+}e^{-}$ system was less than
4 MeV. Finally, the events were required to have a topology consistent
with a $\gamma e^{+}e^{-}$ final state, and an invariant mass compatible
with the $\eta$ mass.

\begin{figure}[!ht]
\includegraphics[width=15cm]{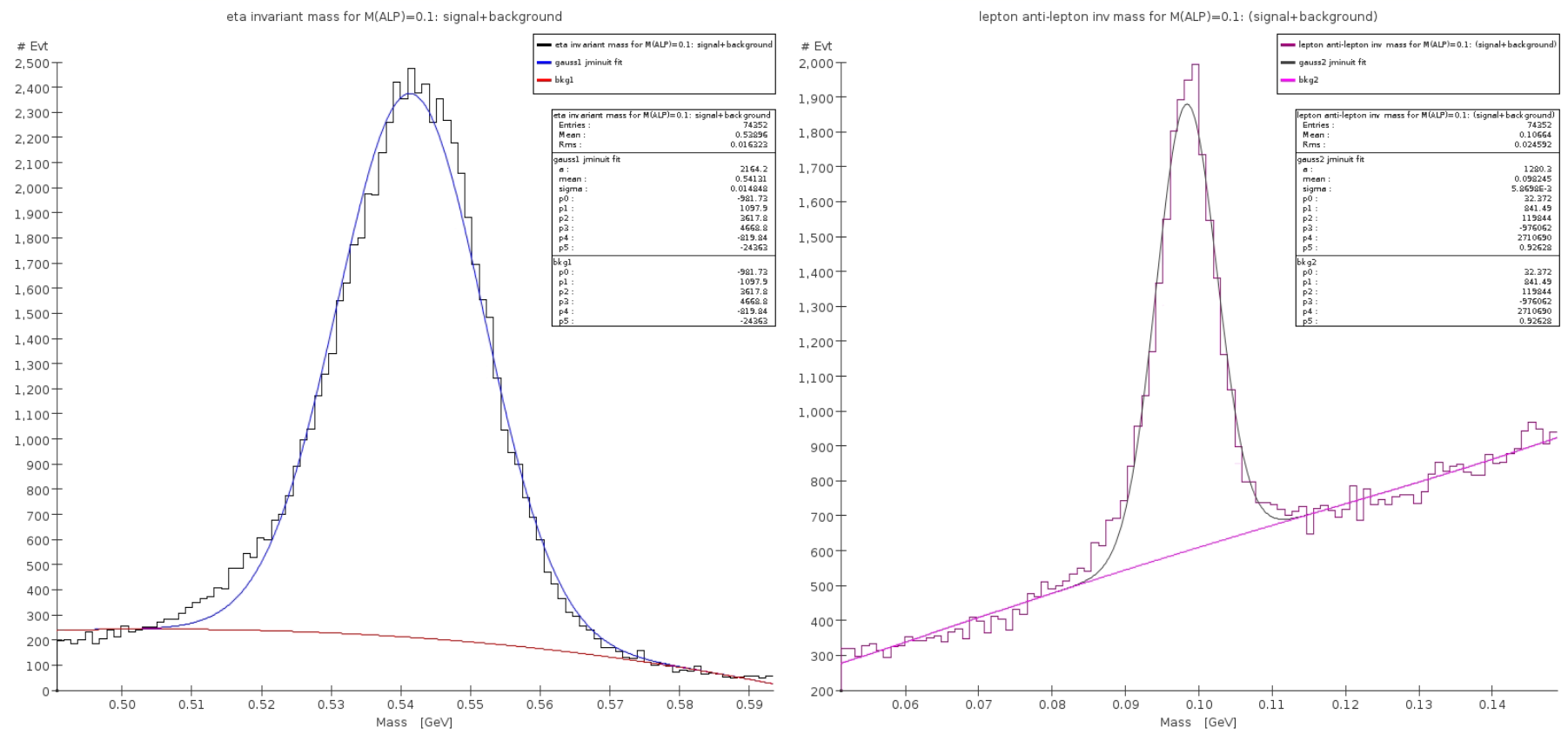} 

\caption{Invariant mass of $\gamma e^{+}e^{-}$ (left) and of the $e^{+}e^{-}$
system (right) for a vector boson with mass 100 MeV merged with the Urqmd generated background. See text for an explanation
of the fitting procedure.}

\label{fig:eta2gammaAp_mass}
\end{figure}

The total reconstruction efficiency for this process was found to
be between $\sim$9\% and $\sim$22\% for the signal samples and of
order $\mathcal{O}(10^{-9})$ for the Urqmd background, with 
a strong dependence on the  $e^{+}e^{-}$ invariant mass. For illustrative
purposes, Fig.~\ref{fig:eta2gammaAp_mass} shows the invariant mass
distribution of the reconstructed $\gamma e^{+}e^{-}$ and of the di-lepton
system for the signal merged with the Urqmd background, assuming a BR($\eta\rightarrow\gamma A'$)=
$2.5\times10^{-4}$ and an $\eta$ sample of $2\times10^{8}$ events (corresponding
to $1.8\times10^{-6}$ of the full integrated luminosity). The distribution
was fitted using the sum of a Gaussian and a 5th-order polynomial.
The branching ratio sensitivity for this process was calculated according
to Eqs. \ref{eq:Master1}-\ref{eq:MasterB}. 

\begin{figure}[!ht]
\includegraphics[width=7cm]{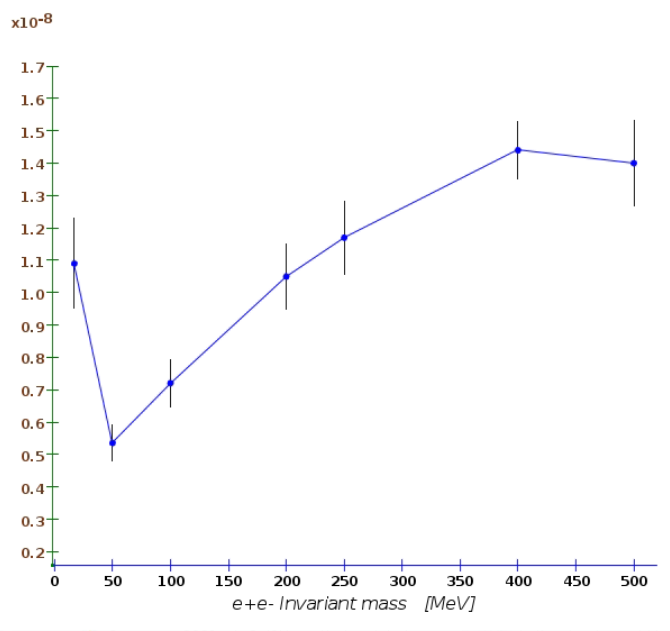} \includegraphics[width=7cm]{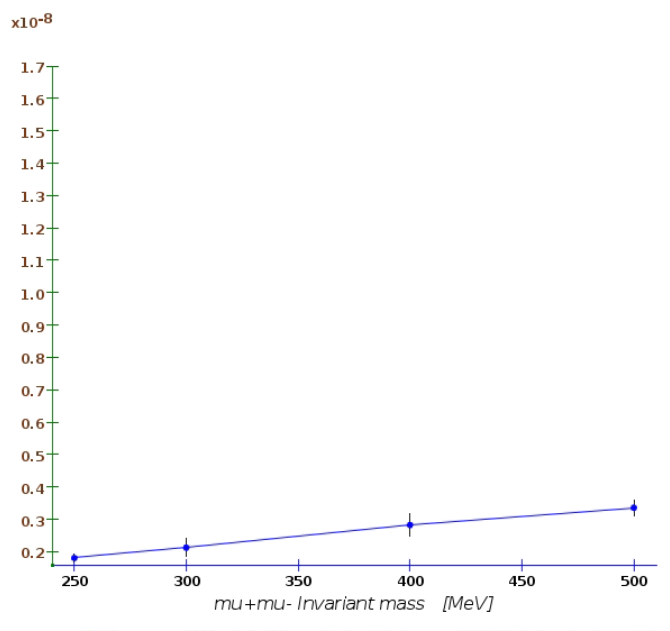}
\caption{Branching ratio sensitivity for the process $\eta\rightarrow\gamma A'\;;\;A'\rightarrow e^{+}e^{-}$(left)
and $\eta\rightarrow\gamma A'\;;\;A'\rightarrow\mu^{+}\mu^{-}$ (right)
as a function of the mass of the vector boson $A'$.}

\label{fig:eta2gammaAp_ee_br}
\end{figure}

The resulting values are shown in Fig.~\ref{fig:eta2gammaAp_ee_br}.
The error bars are statistical only. The effect of the cuts applied to reduce
the background from photon conversion is clearly visible, as they degrade the  sensitivity for lower values of the $e^{+}e^{-}$ invariant mass.

\subsection{\texorpdfstring{$A'\rightarrow e^{+}e^{-}$}{}: Detached-vertex analysis}

This study aimed at evaluating the sensitivity of the detector to 
events with a long-lived particle decaying into a lepton pair and  a
secondary vertex detached from the $\eta$ production vertex. Since
no Standard Model process is known to be responsible for a similar
event topology, the rejection of the background with the $detached-vertex$ analysis
is expected to improve considerably compared to the $bump-hunt$ analysis.
The vector boson $A'$ was generated within a mass range between 17 MeV
and 500 MeV. For each value of mass, $c\tau$ of the vector boson
was varied from 20 mm to 150 mm. A total of 28 event sets were generated
and fully reconstructed. The analysis of the kinematics
follows the same guidelines as for the $bump-hunt$ analysis. Additional
cuts on the $\chi^{2}$ from the fit of two charged tracks to a common vertex and on the distance between the
primary and secondary vertexes were applied. The goal of those cuts was to remove events
with particles originating from the $\eta$ production point. Since
the $\beta\gamma$ boost of the $A'$ depends on its mass, the cuts above
were optimized for each individual event set.

\begin{figure}[!ht]
\includegraphics[width=7cm]{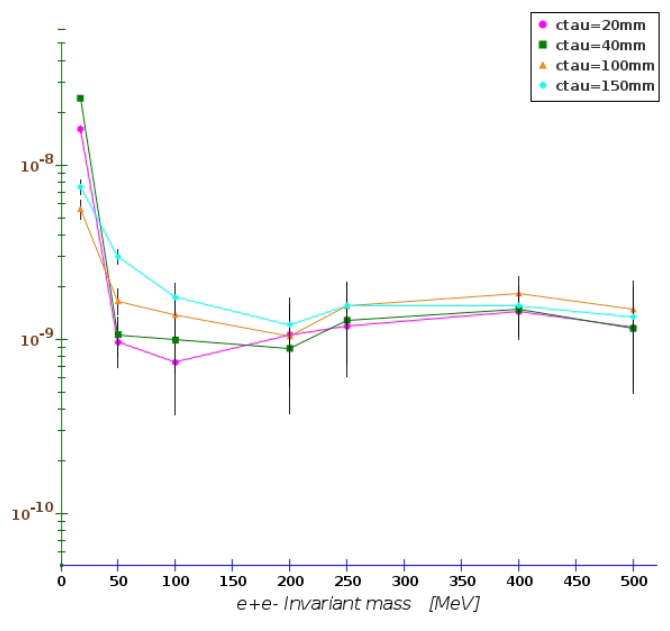} \includegraphics[width=7cm]{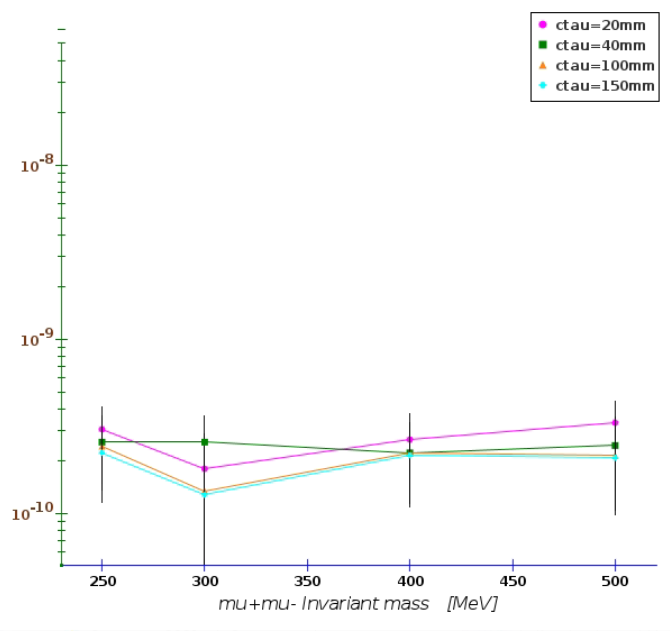}
\caption{Branching ratio sensitivity for the process $\eta\rightarrow\gamma A'\;;\;A'\rightarrow e^{+}e^{-}$
(left) and $\eta\rightarrow\gamma A'\;;\;A'\rightarrow\mu^{+}\mu^{-}$
(right) as a function of the mass and $c\tau$ of the a long-lived
vector boson $A'$.}

\label{fig:eta2gammaAp_ee_br-vtx}
\end{figure}

The final reconstruction efficiency for this process, including the additional
vertex cuts, was found to be between $\sim$2\% and $\sim$10\% for
the signal samples and of order $\mathcal{O}(10^{-10})$ for the Urqmd
background. The resulting branching ratio sensitivity is shown
in Fig.~\ref{fig:eta2gammaAp_ee_br-vtx}, as a function of the invariant
mass of the vector boson. The error bars are statistical only. Even for
the detached-vertex analysis, the lower $e^{+}e^{-}$ invariant mass
region shows a lower sensitivity. This is a consequence of the
larger $\beta\gamma$ factor of bosons with lighter masses, 
which boosts the secondary vertex toward
the outermost region of the detector. When that happens, 
the detector records fewer hits, which is detrimental to the reconstruction efficiency. 
The kinematic cut aiming at removing converting photons
also  
worsens the sensitivity. The poorer reconstruction
efficiency is reflected in a lower branching ratio sensitivity in
the low-mass region, which ranges between $5.5\times10^{-9}$ to $2.4\times10^{-8}$.
For the remaining mass regions the branching ratio sensitivity is
of order $1\times10^{-10}$.

\subsection{\texorpdfstring{$A'\rightarrow\mu^{+}\mu^{-}$}{}: Bump-hunt analysis}

The rationale for studying this decay mode is because it probes
the lepton flavor dependence of the $A'$ vector. If  coupling of the $A'$ was different for electrons and muons that 
could explain some anomalies recently observed, in particular that
related to the measurement of muon $g-2$. 
Furthermore, we expect  for
this final state a reduced  contribution to the background
from converting photons: $\gamma\rightarrow e^{+}e^{-}$ and from $\pi^{0}$
decays, and, consequently, a higher branching ratio sensitivity. 
Four event sets were generated in a mass ranging between
250 and 500 MeV. The largest background  to this channel was found to originate from mis-identified pions, mistakenly
reconstructed as muons. The $\pi$/$\mu$ mis-identification probability
for the detector configuration considered here, has a conservative value of
$\simeq$ 3.5\% (or, $\sim$0.12\% for mis-identifying both
leptons). Since the probability of generating two charged pions in
the primary interaction interaction is almost 11\% (see, also, Fig.~\ref{fig:urqmd_multiplicity}),
and the probability of having at least one $\pi^{0}$ is 58\%, we
expect that about $\sim1.9\times10^{11}$ events could potentially fake
a $\eta\rightarrow\gamma\mu^{+}\mu^{-}$ final state. 

\begin{figure}[!ht]
\includegraphics[width=12cm]{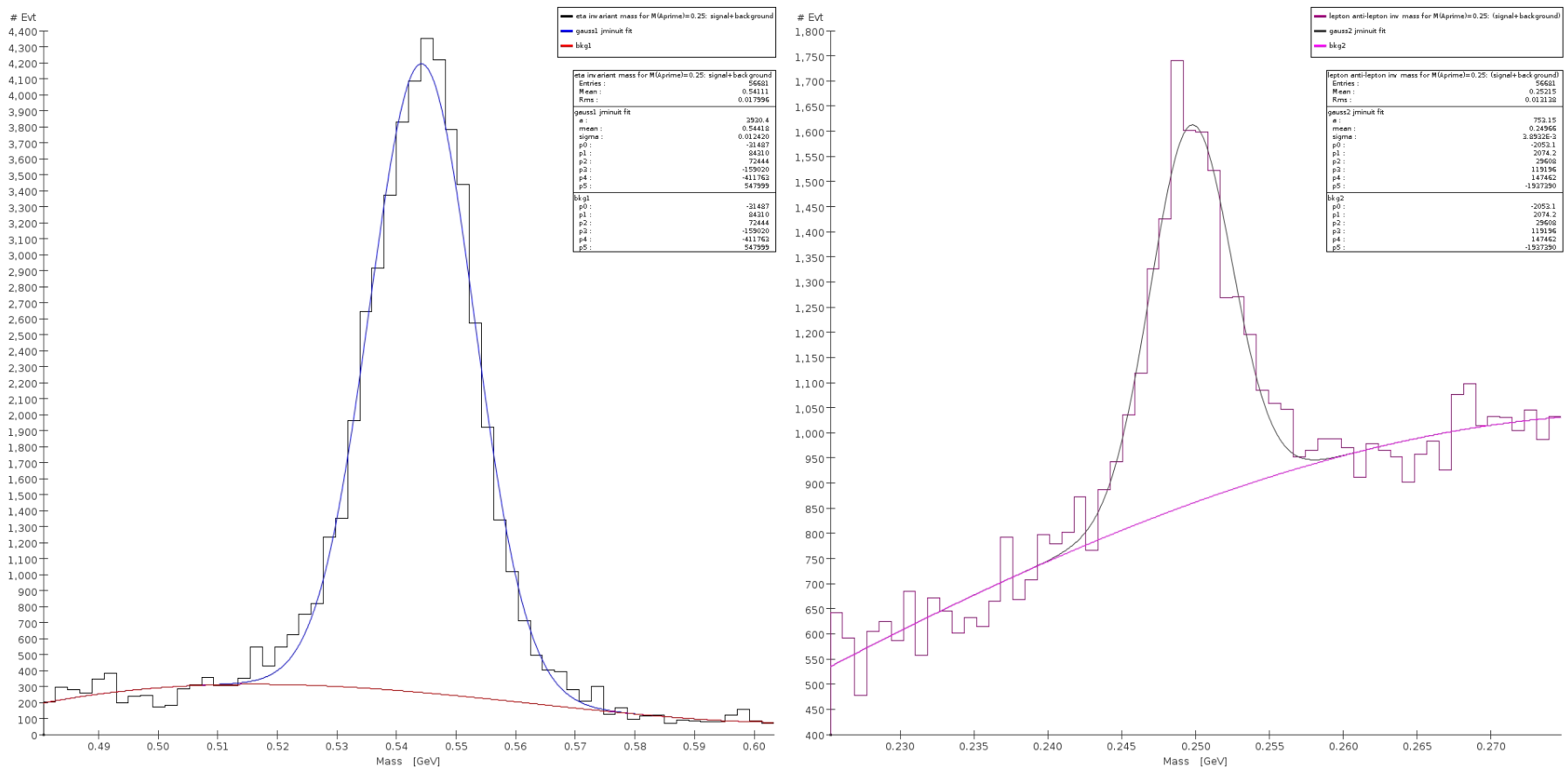} 

\caption{Invariant mass of $\gamma\mu^{+}\mu^{-}$ (left) and of the $\mu^{+}\mu^{-}$
system (right) for a vector with mass 250 MeV. The plot includes
the Urqmd generated background. See text for an explanation of the
fitting procedure.}

\label{fig:eta2gammaAp_mumu_mass}
\end{figure}

The full chain of generation-simulation-reconstruction-analysis was
repeated for each event set. Very generic requirements on the quality
of reconstructed particles were applied to signal and background. Neutral pions, decaying into $\gamma e^{+}e^{-}$ and $\gamma\gamma$,
where reconstructed by considering all combinations of photons, electrons
and positrons with an invariant mass within 5 MeV from $\pi^{0}$
mass. Finally, the events were required to have a topology consistent
with a $\gamma\mu^{+}\mu^{-}$ final state, and an invariant mass
compatible with the $\eta$ mass. The largest background
was found for values of the $\mu^{+}\mu^{-}$ invariant mass corresponding
to the middle of the kinematically allowed range. The final reconstruction
efficiency for this process was found to be between $\sim$16\% and
$\sim$42\% for the signal samples and of order $\mathcal{O}(10^{-8})$
for the Urqmd background. For illustrative purposes of the detector
performance, Fig.~\ref{fig:eta2gammaAp_mumu_mass} shows the invariant
mass distribution of the reconstructed $\gamma\mu^{+}\mu^{-}$ and
di-lepton system for the signal and the Urqmd background, assuming
a BR($\eta\rightarrow\gamma A'$)= $9\times10^{-6}$ and an $\eta$
sample of $5.4\times10^{9}$ events (corresponding to $\sim4.9\times10^{-5}$
of the full integrated luminosity). The distribution was fitted using
the sum of a Gaussian distribution and a 5th-order polynomial. The branching ratio
sensitivity for this process was calculated according to Eqs. \ref{eq:Master1}-\ref{eq:MasterB}. 

The resulting branching ratio sensitivity is summarized in the right
plot of Fig.~\ref{fig:eta2gammaAp_ee_br}, as a function of the invariant
mass of the short-lived vector boson. The best sensitivity
is observed near the lower kinematic limit of the $\mu^{+}\mu^{-}$
invariant mass, where the feed trough from background is approximately
half of that observed in the central region. 

\subsection{\texorpdfstring{$A'\rightarrow\mu^{+}\mu^{-}$}{}: Detached-vertex analysis}

The study for this channel was carried for  masses of the vector boson in the
range between 250 MeV and 500 MeV. For each value of the mass, the $c\tau$
of the resonance was varied from 20 mm to 150 mm. A total of 16 event
sets were generated and fully reconstructed. This analysis follows the same guidelines of the $bump-hunt$
analysis. Additional
cuts on the $\chi^{2}$ from the fit of two charged tracks to a common vertex and on the distance between the
primary and secondary vertexes were applied. The goal of those cuts was to remove events
with particles originating from the $\eta$ production point. Since
the $\beta\gamma$ boost of the $A'$ depends on its mass, the cuts above
were optimized for each individual event set.

The final reconstruction efficiency for this process was found to be between $\sim$6\% and $\sim$24\% for
the signal  and of order $\mathcal{O}(10^{-11})$ for the Urqmd
background. The resulting branching ratio sensitivity is shown
in the right plot of Fig.~\ref{fig:eta2gammaAp_ee_br-vtx}, as a function
of the invariant mass and of $c\tau$ of the a long-lived
of the vector boson. The $\beta\gamma$ boost for $\mu^{+}\mu^{-}$ invariant
masses close to the lower limit of the kinematically allowed range
is larger and the secondary vertex occurs in the outermost region
of the detector. For such events  fewer hits are recorded by the tracking detectors. Nonetheless,
the branching ratio sensitivity is mostly uniform in the $\mu^{+}\mu^{-}$
invariant mass range considered in this study. 

\subsection{\texorpdfstring{$B\rightarrow\pi^{+}\pi^{-}$}{}: Bump-hunt analysis}

In some theoretical models the coupling of the vector boson to
leptons is suppressed, as it occurs, for example with the ``\emph{Leptophobic
B boson} model''. In those cases, the preferred decay mode of the
vector boson is into pions or photon pairs. Therefore, in addition
to the leptonic decay modes, we have considered also the decay mode
of a vector boson (in this case, often referred to as \emph{``B}
boson''), into a pair of charged pions. From the experimental point
of view, analogously to the $A'\rightarrow\mu^{+}\mu^{-}$ decay,
this channel has a reduced background contribution from converting
photons $\gamma\rightarrow e^{+}e^{-}$ and from $\pi^{0}$ decay.
On the other side, the Standard Model background is expected to be
particularly challenging, due to the large yield of two-pion events
from nuclear scattering of the beam onto the target, accompanied by
one or more photons from $\pi^{0}$ and $\eta$ decays.

The reconstruction of this process is further complicated by the fact that the detector
configuration adopted for these studies, along with the algorithms implemented
in the trigger systems, are optimized for the detection of leptons and for the rejection of hadrons. Consequently, we expect a
somehow lower sensitivity compared to the other channels. Besides
the combinatorics background, the relatively large branching ratios
of the $\eta\rightarrow\pi^{+}\pi^{-}\pi^{0}$ and $\eta\rightarrow\pi^{+}\pi^{-}\gamma$,
represents an irreducible, non-resonant background to searches of New
Physics with \emph{bump-hunt} techniques. 

The study for this final state was carried for masses of the B boson
ranging between 300 MeV and 500 MeV. Each sample consists of $3.5\times10^{5}$
$\eta\rightarrow\gamma B\;;\;B\rightarrow\pi^{+}\pi^{-}$ events.
The full chain of generation-simulation-reconstruction-analysis was
repeated for each event set. Very generic requirements on the quality
of reconstructed particles were applied to the signal and background
samples. We found, as expected, that the largest background is due
to purely combinatorics hadronic events (namely, events without $\eta$ mesons)
containing two pions (cf.\ 
Fig.\ref{fig:pi_p_multiplicity-1}) and one photon. The background
was  larger for values of $M(B)$ where the invariant mass of
the $\pi^{+}\pi^{-}$ pair is in the lower and central region of the
kinematically allowed range. In this region, in fact, the combinatoric background mimics
more closely the kinematics of the $\eta\rightarrow\gamma B\;;\;B\rightarrow\pi^{+}\pi^{-}$
process. Neutral pions, decaying into $\gamma e^{+}e^{-}$ and $\gamma\gamma$,
were reconstructed by considering all combinations of photons, electrons
and positrons with an invariant mass within 5 MeV from $\pi^{0}$
mass, and removed from the event. Finally, the reconstructed particles
were required to have a topology consistent with a $\gamma\pi^{+}\pi^{-}$
final state, and an invariant mass compatible with the $\eta$ mass.

\begin{figure}[!ht]
\includegraphics[width=12cm]{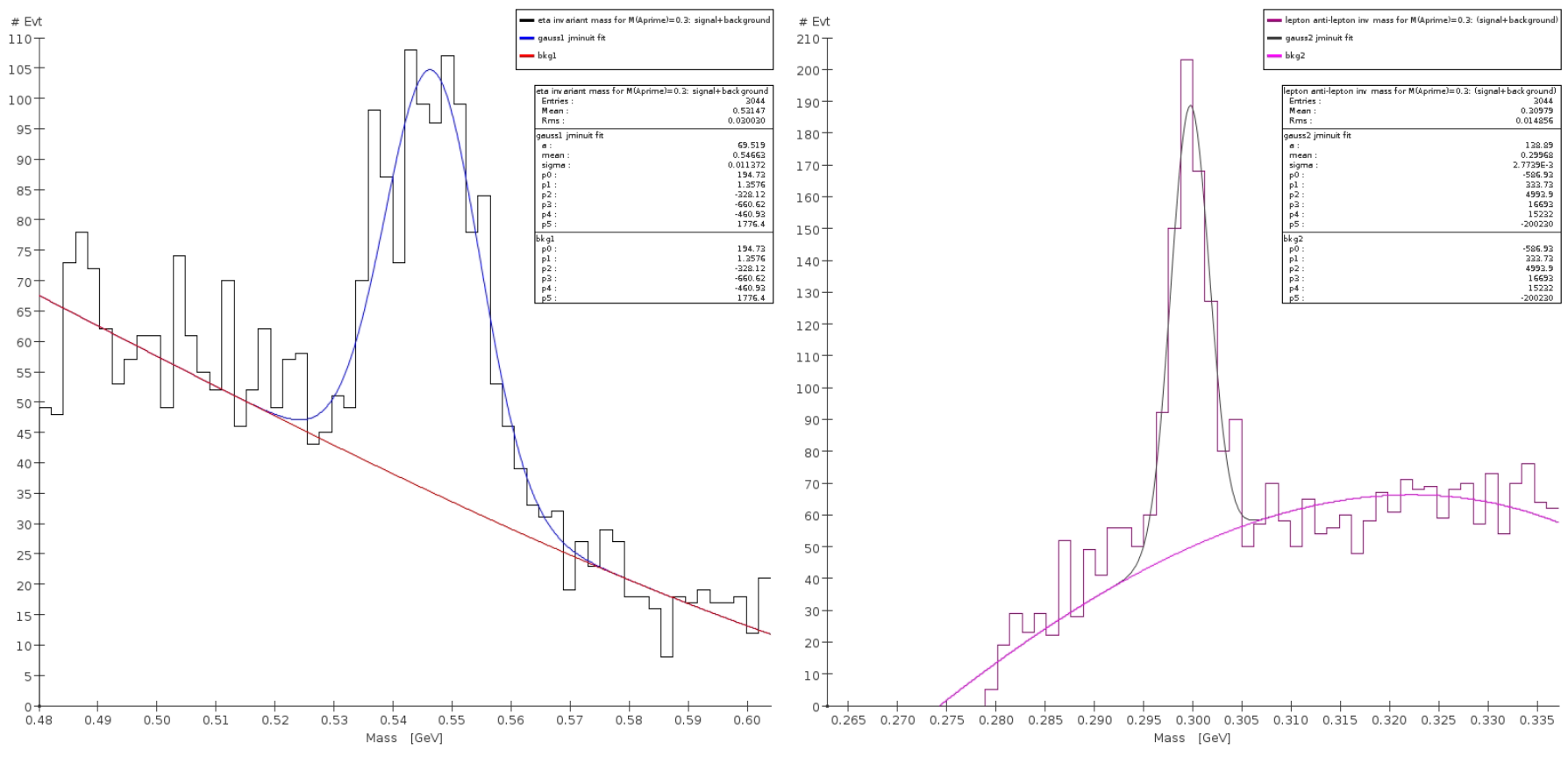} 

\caption{Invariant mass of $\gamma\pi^{+}\pi^{-}$ (left) and of the $\pi^{+}\pi^{-}$
system (right) for a vector boson with mass 300 MeV. The plot includes
the Urqmd generated background. See text for an explanation of the
fitting procedure.}
\label{fig:eta2gammaAp_pipi_mass}
\end{figure}

The total reconstruction efficiency for this process was found equal
to $\sim$2\% for the signal samples and of order $\mathcal{O}(10^{-7})$
for the Urqmd background. For illustrative purposes, Fig.~\ref{fig:eta2gammaAp_pipi_mass} shows the invariant
mass distribution of the reconstructed $\gamma\pi^{+}\pi^{-}$ and
$\pi^{+}\pi^{-}$ system for the signal and the Urqmd background,
assuming a BR($\eta\rightarrow\gamma B$)= $5\times10^{-2}$ and an
$\eta$ sample of $5.4\times10^{9}$ events (corresponding to $\sim4.9\times10^{-5}$
of the full integrated luminosity). The distribution was fitted using
the sum of a Gaussian and a 5th-order polynomial. The branching ratio
sensitivity for this process was calculated according to Eqs. \ref{eq:Master1}-\ref{eq:MasterB}. 

\begin{figure}[!ht]
\includegraphics[width=7cm]{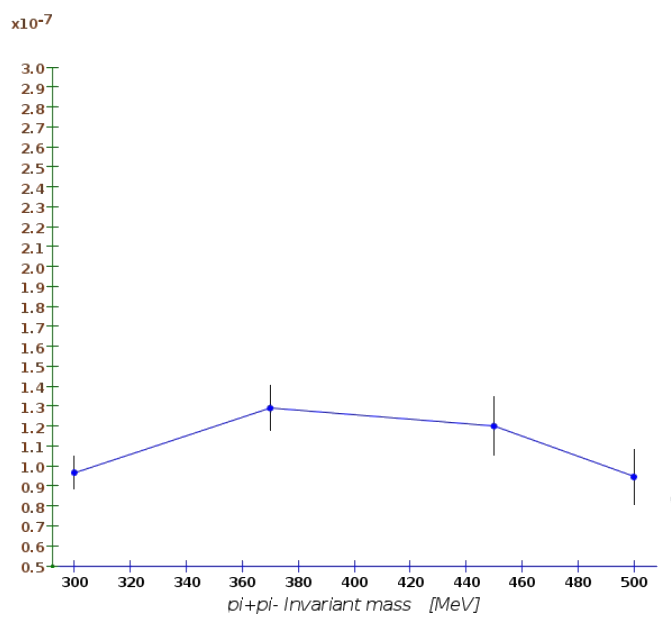} \caption{Branching ratio sensitivity for the process $\eta\rightarrow\gamma B\;;\;B\rightarrow\pi^{+}\pi^{-}$
with the \emph{bump-hunt} analysis 
as a function of the mass 
of the vector
boson $B$.}

\label{fig:eta2gammaAp_pipi_br}
\end{figure}
The resulting branching ratio sensitivity is summarized in 
Fig.~\ref{fig:eta2gammaAp_pipi_br}. The error bars are statistical
only. 
 The best sensitivity is observed near the upper and lower kinematic
limits of the $\pi^{+}\pi^{-}$ invariant mass, where the feed trough
from background is approximately 30\%-40\% lower than that observed
in the central region. 

\subsection{\texorpdfstring{$B\rightarrow\pi^{+}\pi^{-}$}{}: Detached-vertex analysis}
The study for this final state was carried for vector masses in the
range between 300 MeV and 500 MeV. For each value of mass, the $c\tau$
of the resonance was varied from 20 mm to 150 mm. A total of 16 event
sets were generated and fully reconstructed. The analysis for the
kinematic variables follows the same guidelines of the $bump-hunt$
analysis. A further cut on the $\chi^{2}$ of vertex fit and on the
distance between the primary and secondary vertexes were applied in
order to remove events with particles originating from the $\eta$
production point. Since the $\beta\gamma$ boost of the $B$ boson
depends on its mass, those cuts were optimized for each individual
event set.

The final reconstruction efficiency for this process with the additional
vertex cuts was found to be between $\sim$1\% and $\sim$1.5\% for
the signal samples and of order $\mathcal{O}(10^{-9})$ for the Urqmd
background.
The $\beta\gamma$ boost for $\mu^{+}\mu^{-}$ invariant
masses closer to the lower limit of the kinematically allowed range
is larger and the secondary vertex occurs in the outermost region
of the detector, with fewer hits in both tracking detectors. Nonetheless,
the branching ratio sensitivity is mostly uniform in the $\pi^{+}\pi^{-}$
invariant mass range considered in this study.

\subsection{Sensitivity to selected theoretical models}

In this section we consider four distinct
theoretical models, discussed in details in Sec.~\ref{sec:Physics-Beyond-the-SM}, for which New Physics appears through the vector portal. More specifically,  the branching ratio sensitivity obtained in the previous sections is used to
determine the sensitivity to the corresponding coupling constants. 

\subsubsection{Minimal Dark Photon Model }

The relevant parameter of this model is the kinetic mixing $\epsilon^{2}$
defined in Sec.~\ref{fig:Mutiplicity}. The master formula for $Br(\epsilon^{2})$
(cf.\ Eq.~\eqref{eq:MasterS-final}) was obtained from Ref.~\citeauthor{BPR2009}:

\begin{equation}
BR_{\eta\rightarrow\gamma A'}=BR_{\eta\rightarrow\gamma\gamma}\times2\epsilon^{2}\left(1-\frac{m_{A'}^{2}}{m_{\eta}^{2}}\right)\label{eq:Br2epsilon}
\end{equation}

The partial widths of the \emph{A'} into leptons are~\citet{Ilten2016}:

\begin{equation}
\Gamma_{A'\rightarrow\ell^{+}\ell^{-}}=\epsilon^{2}\alpha_{EM}\frac{m_{A'}}{3}\left(1+2\nicefrac{m_{l}^{2}}{m_{A'}^{2}}\right)\sqrt{1-4\nicefrac{m_{l}^{2}}{m_{A'}^{2}}}\label{eq:darkphoton-partial-widths}
\end{equation}

where $\ell=e,\mu,\tau.$ and $m_{A'}>2m_{\ell}$. The total width
can be expressed as:

\begin{equation}
\Gamma_{A'}=\sum_{\ell}\Gamma_{A'\rightarrow\ell^{+}\ell^{-}}+\Gamma_{A'\rightarrow hadrons}+\Gamma_{A'\rightarrow invisible}.\label{eq:darkphoton-total-width}
\end{equation}
Following Ref.~\citet{Ilten2016}, for this work, we assume that:

\[
\Gamma_{A'\rightarrow hadrons}=\Gamma_{A'\rightarrow\mu^{+}\mu^{-}}\Re_{\mu}(A'),\qquad\Gamma_{A'\rightarrow invisible}=0
\]
where $\Re_{\mu}(A')=\nicefrac{\sigma_{e^{+}e^{-}\rightarrow hadrons}}{\sigma_{e^{+}e^{-}\rightarrow\mu^{+}\mu^{-}}}$ ~\cite{ParticleDataGroup:2020ssz}.
A consequence of Eqs. (\ref{eq:darkphoton-partial-widths}) and \ref{eq:darkphoton-total-width}
is that the value of $\epsilon^{2}$ and the lifetime of the $A'$
are directly interconnected. Therefore, in a \emph{detached-vertex}
analysis the region of $\epsilon^{2}$ which could be explored is
bound from above by the size of the fiducial volume where the secondary
vertex of the dark photon is being reconstructed. For this work, we
have conservatively used a fiducial volume consisting in a sphere
with radius equal to 1 cm.

\begin{figure}[!ht]
\includegraphics[width=7cm]{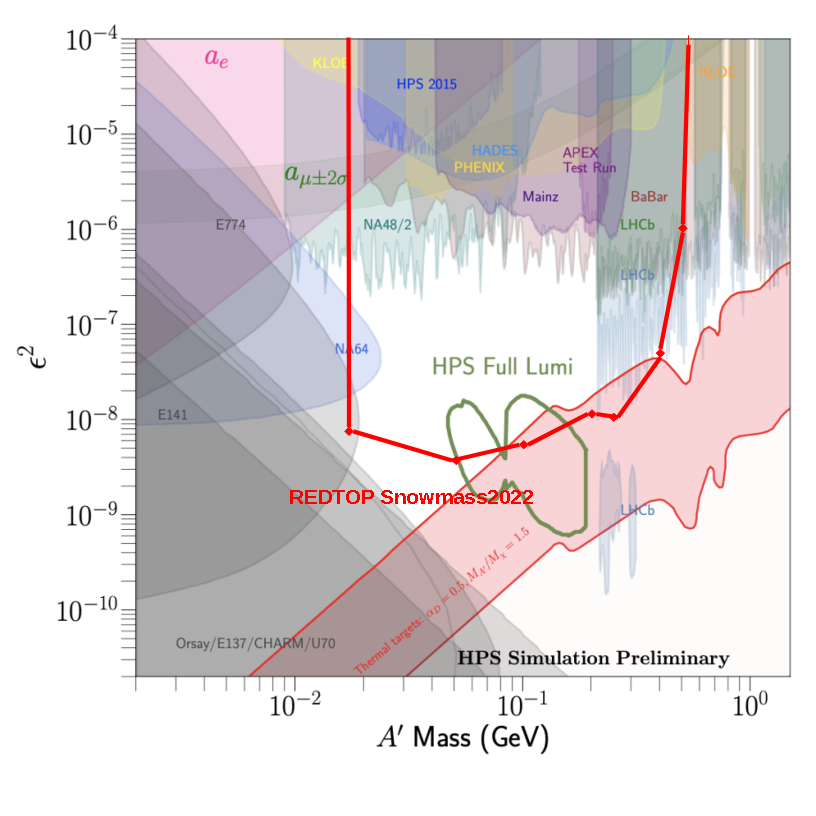} \includegraphics[width=7cm]{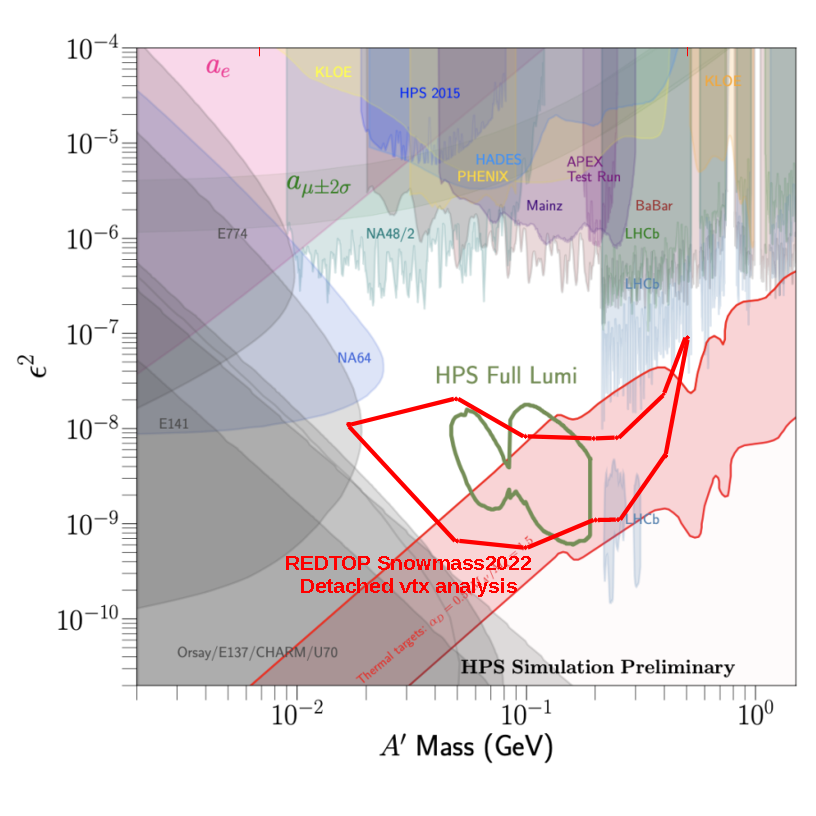}
\caption{Sensitivity to to $\varepsilon{}^{2}$ for the processes $\eta\rightarrow\gamma A'$
for integrated beam flux of $3.3\times10^{18}$ POT. Left plot: \emph{bump-hunt}
analysis. Right plot: \emph{detached-vertex} analysis). }

\label{fig:eta2gammaAp_eeepsilon}
\end{figure}

Inserting in Eqs. \ref{eq:Br2epsilon}-\ref{eq:darkphoton-total-width}, the values of the branching ratio sensitivity obtained for each
value of the $A'$ mass and $c\tau$ considered in this study, we
show the plot for the corresponding sensitivity of $\varepsilon^{2}$
in in Fig.~\ref{fig:eta2gammaAp_eeepsilon} for the $bump-hunt$ and
the detached-vertex analyses. 

\subsubsection{Leptophobic B boson Model }

In this section we place limits on the $B$ boson parameters.
Let us first consider the $\eta^{(\prime)}\to\pi^{0}\gamma\gamma$ and $\eta^{\prime}\to\eta\gamma\gamma$ decay channels.
On the experimental side, 
the CrystalBall collaboration at AGS in 2008~\cite{Prakhov:2008zz} and the A2 collaboration at the Mainzer Microtron (MAMI) in 2014~\cite{A2atMAMI:2014zdf} published the first measurements of the diphoton energy spectrum $d\Gamma(\eta\to\pi^{0}\gamma\gamma)/dm_{\gamma\gamma}^{2}$, along with the associated branching ratio (BR).
They reported the measurements $\mbox{BR}=(2.21\pm 0.24\pm 0.47)\times 10^{-4}$ and $\mbox{BR}=(2.54\pm 0.27)\times 10^{-4}$, respectively, based on the analysis of $\sim1.2\times10^{3}$ $\eta\to\pi^{0}\gamma\gamma$ decay events.
Surprisingly low in comparison with all previous measurements is
the 2006 result reported by the KLOE collaboration~\cite{KLOE:2005hln},
$\mbox{BR}=(0.84\pm 0.27\pm 0.14)\times 10^{-4}$, based on a sample of only $68\pm 23$ events.
Very recently, however, the KLOE collaboration has presented a new analysis, finding a preliminary value of $\mbox{BR}=(1.23\pm 0.14)\times 10^{-4}$ based on the analysis of $\sim1.4\times10^{3}$ $\eta\to\pi^{0}\gamma\gamma$ events~\cite{Cao:2022rxo}.
For the $\eta^\prime\to\pi^0\gamma\gamma$ decay, the BESIII collaboration recently reported for the first time
the $m_{\gamma\gamma}^{2}$ invariant mass distribution
\cite{BESIII:2016oet} and the measured branching fraction, 
$\mbox{BR}=(3.20\pm 0.07\pm 0.23)\times10^{-3}$.
Finally, for the $\eta^\prime\to\eta\gamma\gamma$ decay, 
a measurement of $\mbox{BR}<1.33\times 10^{-4}$ at 90\% CL 
has been provided, again for the first time, by the BESIII collaboration~\cite{BESIII:2019ofm}.
This experimental information is important to place limits on the $B$ boson parameters.

\begin{figure*}[!ht]
\includegraphics[width=0.95\textwidth]{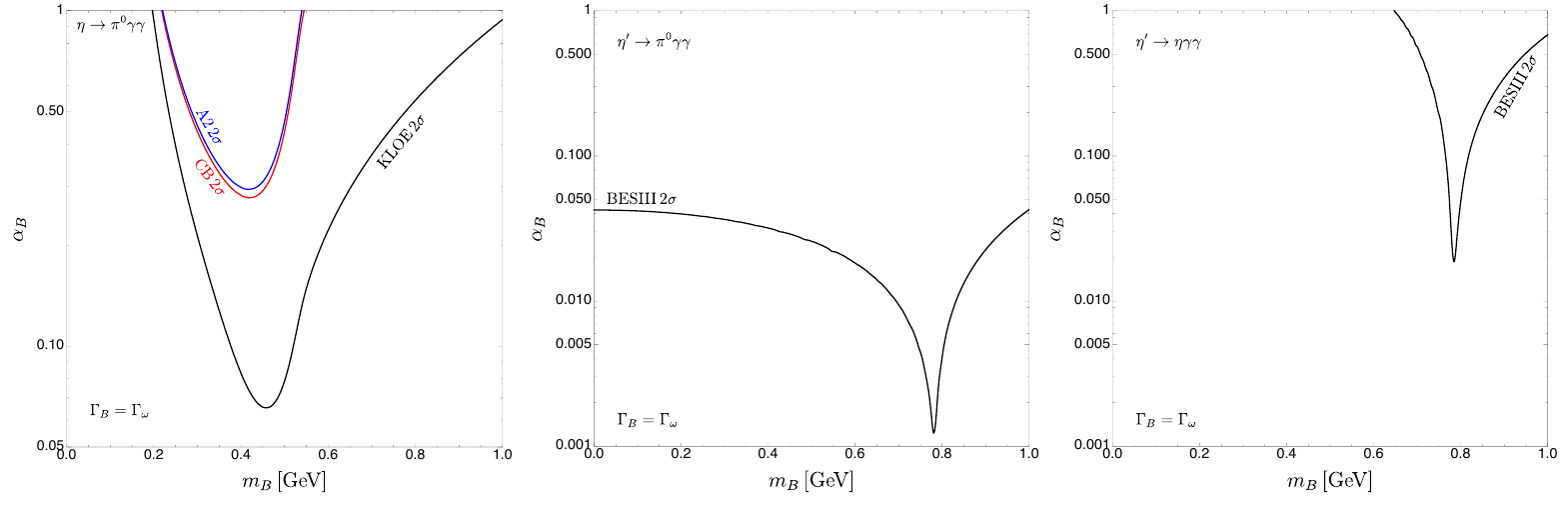}
\caption{Limits on the leptophobic $B$ boson mass  $m_{B}$ and coupling $\alpha_{B}$ from $\eta\to\pi^{0}\gamma\gamma$ (left), $\eta^{\prime}\to\pi^{0}\gamma\gamma$ (middle) and $\eta^{\prime}\to\eta\gamma\gamma$ (right).}
\label{Fig:ExclusionPlot} 
\end{figure*}

On the theoretical side, the $B$ boson exchange contribution to the amplitude of these decays, defined in Eq.~(\ref{Eq:BbosonAmplitude}), depends on three parameters: the baryonic fine structure constant, $\alpha_{B}$, and the $B$ boson resonance parameters, mass and width, $m_{B}$ and $\Gamma_{B}$.
Adding this to the Standard Model contribution~\cite{Escribano:2018cwg}, along with the available experimental data, allows us to place limits on the $B$ boson parameters~\cite{inpreparationEscribanoSGSRoyo}.
These limits are shown in Fig.\,\ref{Fig:ExclusionPlot} in terms of $\alpha_{B}$ and $m_{B}$.
In this figure, we require the amplitude not to exceed the corresponding branching ratios at $2\sigma$, and fixed the $B$ boson width to the decay width of the $\omega$ meson for illustration purposes.
These results differ somewhat from~\cite{Tulin:2014tya}, which assumed no contribution from the Standard Model and the narrow width approximation, i.e., $\mbox{BR}(\eta\to\pi^{0}\gamma\gamma)=\mbox{BR}(\eta\to B\gamma)\times \mbox{BR}(B\to\pi^{0}\gamma)$, with $\mbox{BR}(B\to\pi^{0}\gamma)=1$.

We next consider $B$ boson contributions to the decay $\eta\to\pi^{+}\pi^{-}\gamma$. 
The relevant parameter of this model is the couplings $\alpha_{B}$ defined above. 
The master formula for $Br(\alpha_{B})$  was obtained by numerically integrating Eq.~(\ref{Eq:PartialRate}) over the Mandelstam variable $t$. 
For simplicity, we have fixed the $B$ boson width to the width of the $\omega$ meson.

\begin{figure}[!ht]
\includegraphics[width=7cm]{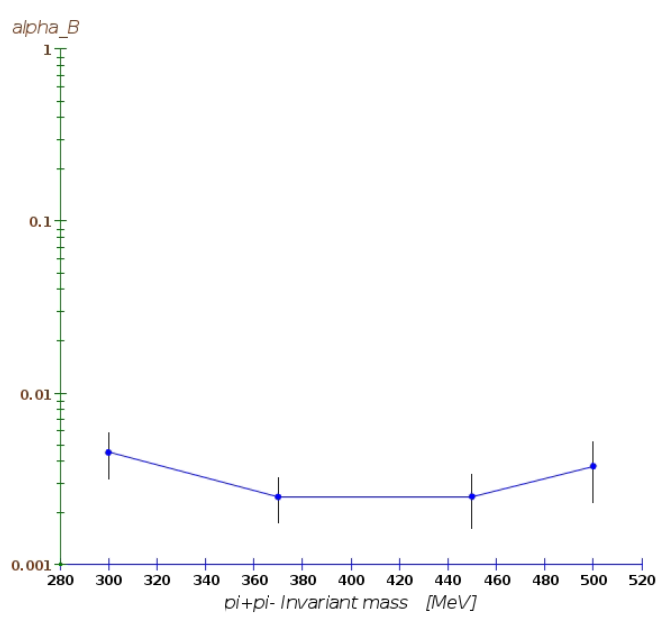}\caption{Sensitivity to $\alpha_{B}$ for the processes $\eta\rightarrow\gamma B\;;\;B\rightarrow\pi^{+}\pi^{-}$
for the \emph{bump-hunt} analysis
as a function of the mass of the short-lived
$B$ boson.}

\label{fig:eta2gammaAp_pipi_gB}
\end{figure}

Inserting in the master formula the values of the branching ratio sensitivity obtained for each value of the $B$ mass  
considered, we obtain the corresponding sensitivity
to $\alpha_{B}$. This is summarized in Fig.~\ref{fig:eta2gammaAp_pipi_gB}
for the $bump-hunt$ analysis only, since the leptophobic $B$ boson is predicted to have a short lifetime.

\subsubsection{Protophobic Fifth Force Model }

The relevant parameter of this model is the $\varepsilon_{n}^{2}$ defined
in Eqs.~\eqref{eq:eta_decay_ABJ} and \eqref{eq:eta_decay_VMD}; .
For simplicity, we will set $\varepsilon_s \equiv 0$, but we remind
the reader that this can have a sizeable impact on the branching ratio
(cf.~Fig.~\eqref{fig:protophobic_scaling}). The branching ratios
become: 
\begin{equation}
\left(\dfrac{BR_{\eta\to\gamma X}}{BR_{\eta\to\gamma\gamma}}\right)_{\text{ABJ}}=\dfrac{8\varepsilon_{n}^{2}}{\pi\alpha}\left(1-\dfrac{m_{X}^{2}}{m_{\eta}^{2}}\right)^{3}\left[\dfrac{c_{\theta}-\sqrt{2}s_{\theta}}{c_{\theta}-2\sqrt{2}s_{\theta}}\right]^{2},\label{eq:protophobic-abj}
\end{equation}
\begin{equation}
\left(\dfrac{BR_{\eta\to\gamma X}}{BR_{\eta\to\gamma\gamma}}\right)_{\text{VMD}}=\dfrac{\varepsilon_{n}^{2}}{8\pi\alpha}\left(1-\dfrac{m_{X}^{2}}{m_{\eta}^{2}}\right)^{3}\left[\dfrac{c_{\theta}-\sqrt{2}s_{\theta}}{c_{\theta}-2\sqrt{2}s_{\theta}}\right]^{2}\left|9F_{\rho}(m_{X}^{2})-F_{\omega}(m_{X}^{2})\right|^{2}\label{eq:protophobic-vmd}
\end{equation}

These expressions are nearly identical (at the $\sim3-4\%$ level)
for $m_X \ll m_\rho, \, m_\omega$, but the branching ratio in the
ABJ scheme is smaller by a factor of $\sim2-3$ near threshold, where
the form factors become important. Inserting in Eq.~\eqref{eq:protophobic-abj}
and \ref{eq:protophobic-vmd} the values of the branching ratio sensitivity
obtained for the point at 17 MeV in the analyses above, we derive
the sensitivity for the parameter $\varepsilon_{n}^{2}$. The values
obtained for the two model considered in this discussion are summarized
in Table~\ref{table:sensitivity-epsilon_n}.

\begin{table}[!ht]
\centering\textcolor{black}{\small{}}%
\begin{tabular}{|c||c|}
\hline 
\textit{\textcolor{black}{\footnotesize{}Scheme}} & $\varepsilon_{n}^{2}$\tabularnewline
 & \textbf{\textcolor{black}{\footnotesize{}Sensitivity}}\tabularnewline
\hline 
\hline 
\textcolor{black}{\footnotesize{}$ABJ$} & \textbf{\footnotesize{}$1.41\pm0.17\times10^{-10}$}\tabularnewline
\hline 
\textcolor{black}{\footnotesize{}VMD } & \textbf{\footnotesize{}$1.46\pm0.17\times10^{-10}$}\tabularnewline
\hline 
\end{tabular}\caption{Sensitivity to $\varepsilon_n$ for a 17-MeV protophobic gauge boson
model in the ABJ and VMD schemes. We assume that $X$ decays to neutrinos
contribute negligibly to its total width, i.e, that $BR(X\to e^+ e^-) \approx 1$;
the sensitivity scales as the inverse of the decay branching ratio
for this channel.}
\label{table:sensitivity-epsilon_n}
\end{table}


\section{Sensitivity to the Scalar portal\label{subsec:Sensitivity-to-the-scalar}}

The \emph{Scalar portal} can be probed via  decays of the
$\eta$-meson with a $\pi^{0}$ meson in the final state.
Three processes have been considered for these studies:
\begin{itemize}
\item $p+Li\rightarrow\eta+X\;with\;\eta\rightarrow\pi^{0}H\;and\;h\rightarrow e^{+}e^{-}$
\item $p+Li\rightarrow\eta+X\;with\;\eta\rightarrow\pi^{0}H\;and\;h\rightarrow\mu^{+}\mu^{-}$
\item $p+Li\rightarrow\eta+X\;with\;\eta\rightarrow\pi^{0}H\;and\;h\rightarrow\pi^{+}\pi^{-}$
\end{itemize}
As for the case of the \emph{Vector portal}, a \emph{bump-hunt} and
a \emph{detached-vertex} analysis were performed to test the performance
of different components of the detector.

\subsection{\texorpdfstring{$h\rightarrow e^{+}e^{-}$: $Bump-hunt$}{} analysis}

This study aims at evaluating the sensitivity of the detector in the
assumption that scalar boson has a decay length not resolvable by
the tracking system. In this case, the process is identified exclusively
through its kinematics, in particular by observing a bump in the invariant
mass of the $\ell\bar{\ell}$ or the $\pi^{+}\pi^{-}$ system. 

The scalar boson \emph{h} was generated with a mass ranging between 17
MeV and 400 MeV. The \emph{h}
was decayed promptly within \emph{ GenieHad}, by setting the vertex of
the di-lepton in the point where the proton struck the target. This
final state has a large  contribution from background on the lower side of
the kinematically allowed mass range. The process responsible for that is the decay
$\eta\rightarrow\gamma\gamma$, when one of the photons converts into
a $e^{+}e^{-}$ pair and an extra accidental or
misidentified $\pi^{0}$ is found in the event, with kinematics compatible
with the $\eta$ mass. Another large contribution to the background
originates from the decay of the $\eta$ meson: $\eta\rightarrow\gamma e^{+}e^{-}$, 
when the radiative photon is paired to another photon to fake a $\pi^{0}$ and the kinematics 
of the event is compatible
with a di-lepton mass near the $h$ mass. 

The full chain of generation-simulation-reconstruction-analysis was
repeated for each event set.  The information from TOF and PID systems contributes in reducing
the background by a factor $\mathcal{\sim}10{}^{4})$.
Very generic requirements on the quality of reconstructed particles
were applied to the signal and background samples. No cuts were applied
to the reconstructed $e^{+}e^{-}$ vertex. Neutral pions, decaying
into $\gamma e^{+}e^{-}$ and $\gamma\gamma$, where reconstructed
by considering all combinations of photons, electrons and positrons
with an invariant mass within 5 MeV from $\pi^{0}$ mass. This requirement
is able to reject most of the combinatoric background with no $\eta$
mesons in the final state. Similarly, photons converting into a $e^{+}e^{-}$
pair where reconstructed by requiring that the invariant mass of the
$e^{+}e^{-}$ system was lower than 5 MeV. 

\begin{figure}[!ht]
\includegraphics[scale=0.3]{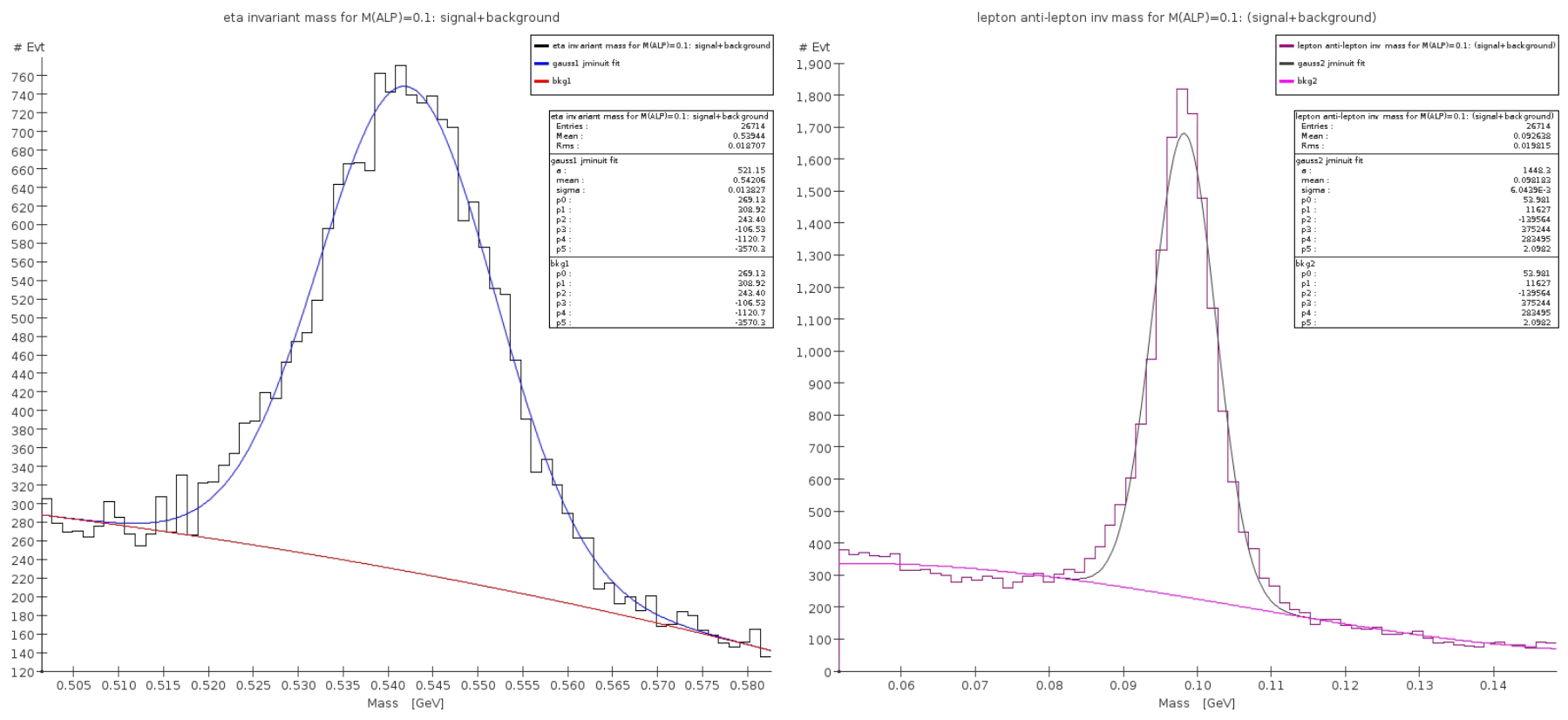} 

\caption{Invariant mass of $\pi^{0}e^{+}e^{-}$ (left) and of the $e^{+}e^{-}$
system (right) for a scalar with mass 100 MeV. The plot includes
the Urqmd generated background. See text for an explanation of the
fitting procedure.}

\label{fig:eta2pi0H_mass}
\end{figure}

\begin{figure}[!ht]
\includegraphics[width=7cm]{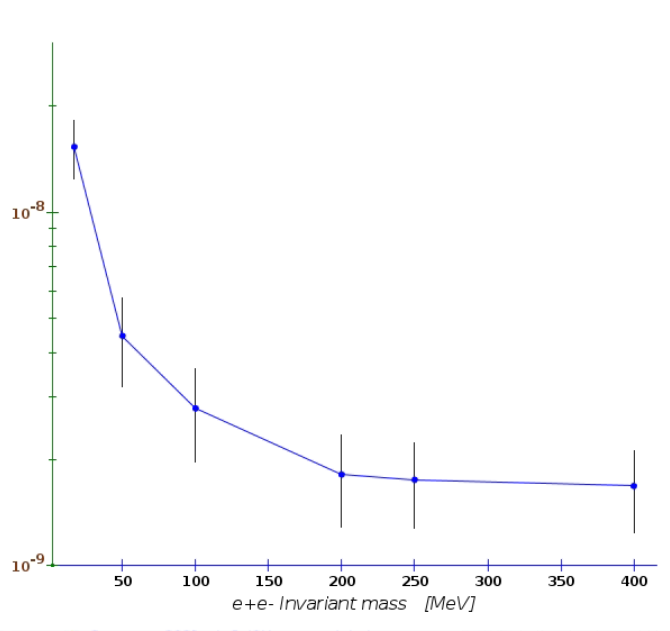} \includegraphics[width=7cm]{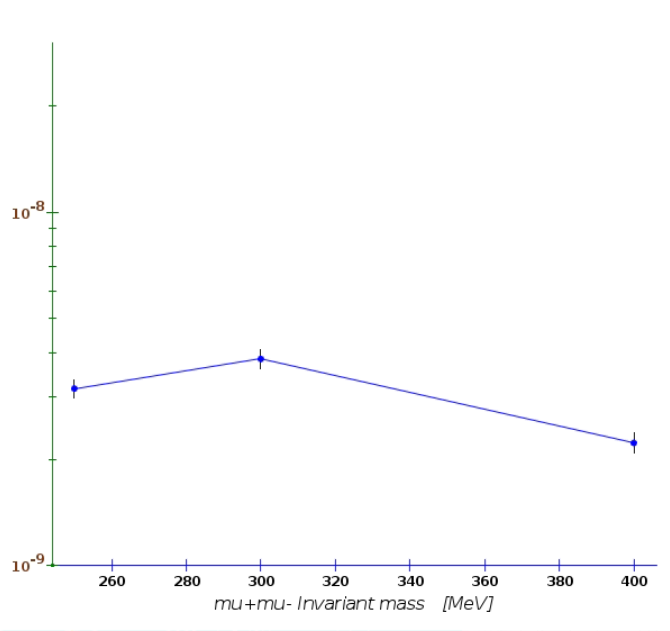}
\caption{Branching ratio sensitivity for the process $\eta\rightarrow\pi^{0}h\;;\;h\rightarrow e^{+}e^{-}$
(left) and $\eta\rightarrow\pi^{0}h\;;\;h\rightarrow\mu^{+}\mu^{-}$
(right) as a function of the mass of the scalar boson $h$.}

\label{fig:eta2pi0H_ee_br}
\end{figure}

The total reconstruction efficiencies for this process was found to be between
6\% and 16\% for the signal and
in the range $10^{-9}-10^{-8}$ for the Urqmd background. For illustrative
purposes, Fig.~\ref{fig:eta2pi0H_mass}
shows the invariant mass distribution of the reconstructed $\pi^{0}e^{+}e^{-}$
and di-lepton system for the signal and the Urqmd background, assuming
BR($\eta\rightarrow\pi^{0}H$)= $2.5\times10^{-4}$ and an $\eta$
sample of $2\times10^{8}$ (corresponding to $1.8\times10^{-6}$ of
the full integrated luminosity). The distribution was fitted using
the sum of a Gaussian and a 5th-order polynomial. The branching ratio
sensitivity for this process was calculated according to Eqs. \ref{eq:Master1}-\ref{eq:MasterB}. It is summarized in Fig.~\ref{fig:eta2pi0H_ee_br},
as a function of the invariant mass of the scalar boson.

\subsection{\texorpdfstring{$h\rightarrow e^{+}e^{-}$}{}: Detached-vertex analysis}

This study aims at evaluating the sensitivity of the detector to
$\eta$ mesons decaying into long-lived particles, further decaying into a $\mu^{+}\mu^{-}$
pair with a secondary vertex detached from the $\eta$ production point.
As noted previously, no Standard Model process is known to be responsible
for a similar event topology. Consequently, we expected that the rejection
of the background would improve considerably compared to a
\emph{bump-hunt} analysis. The scalar \emph{h} was generated within
a mass range between 17 MeV and 400 MeV. For each value of the mass, 
$c\tau$ of the resonance was varied from 20 mm to 150 mm. A total of
24 event sets were generated and fully reconstructed.The analysis of the kinematics
follows the same guidelines as for the $bump-hunt$ analysis. Additional
cuts on the $\chi^{2}$ from the fit of two charged tracks to a common vertex and on the distance between the
primary and secondary vertexes were applied. 
The goal of those cuts was to remove events
with particles originating from the $\eta$ production point. Since
the $\beta\gamma$ boost of the $A'$ depends on its mass, the cuts above
were optimized for each individual event set.

\begin{figure}[!ht]
\includegraphics[width=7cm]{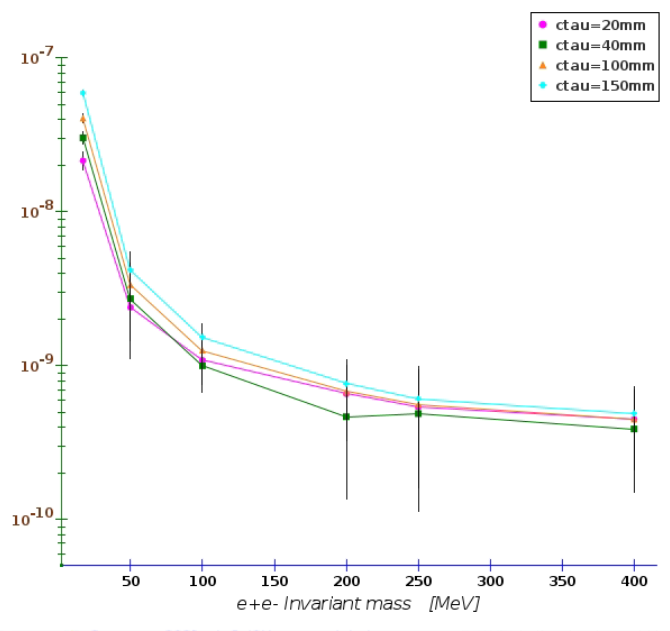} \includegraphics[width=7cm]{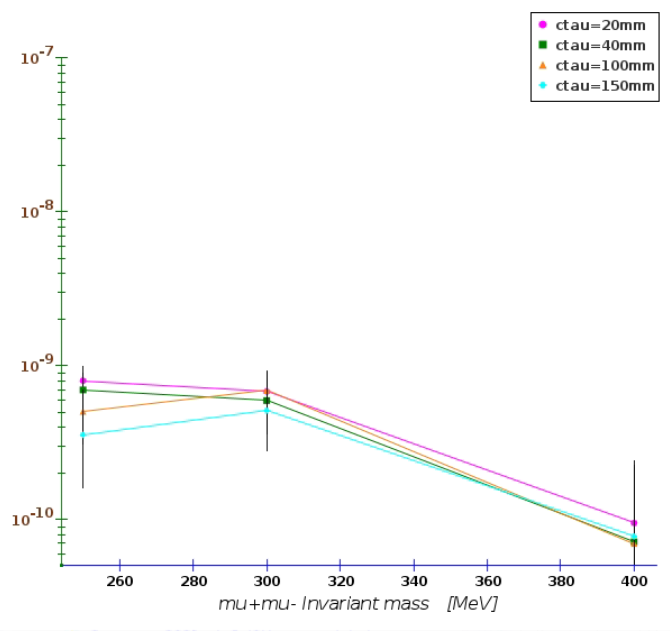}
\caption{Branching ratio sensitivity for the process $\eta\rightarrow\pi^{0}h\;and\;h\rightarrow e^{+}e^{-}$
(left) and $\eta\rightarrow\pi^{0}h\;;\;h\rightarrow\mu^{+}\mu^{-}$
(right) as a function of the mass and $c\tau$ of a long-lived
scalar boson $h$.}

\label{fig:eta2pi0H_ee_br-vtx}
\end{figure}

The total reconstruction efficiency for this process, including the additional
vertex cuts, was found to be between $\sim$1\textbackslash\% and
$\sim$9\textbackslash\% for the signal samples and of order $\mathcal{O}(10^{-10})$
for the Urqmd background. The lower reconstruction sensitivity is
due to the extra cuts on detached vertex. The resulting branching
ratio sensitivity curve is shown in Fig.~\ref{fig:eta2pi0H_ee_br-vtx},
as a function of the invariant mass and $c\tau$ of the long-lived
 scalar boson. The lower
$e^{+}e^{-}$ invariant mass region shows a degraded sensitivity.
This is a consequence of the larger\emph{ $\beta\gamma$} factor,
boosting the secondary vertex toward the outermost region of the detector,
were the reconstruction efficiency is lower. 
The kinematic cut that removes converting photons also degrades the sensitivity.
The lower reconstruction efficiency is reflected in a lower branching
ratio sensitivity in the low-mass region, the former being of order $\mathcal{O}(10^{-8})$.
For the remaining mass region, the branching ratio sensitivity is
close to $1\times10^{-9}$

\subsection{\texorpdfstring{$h\rightarrow\mu^{+}\mu^{-}$}{}: Bump-hunt analysis\label{subsec::pi0hmumu-Bump-hunt-analysis}}

As already observed for the Vector portal case, the study of
this decay mode gives the opportunity to
probe lepton universality and to explore a complementary, and possibly,
reduced, background source. We expect  for this final state a reduced
background contribution from converting photons $\gamma\rightarrow e^{+}e^{-}$
and from $\pi^{0}$ decays, and, possibly, a higher branching
ratio sensitivity. 
Four event sets were generated
with the $h$ mass ranging between 250 MeV and 500 MeV. The largest background
contribution to this final state was found to originate from mis-identified
pions, mistakenly reconstructed as muons. The $\pi$/$\mu$ mis-identification
probability for the detector layout considered in this work has a
conservative value of $\simeq$~3.5\% (or, $\simeq$ 0.12\%
for mis-identifying both leptons). Since the probability of generating
two charged pions in the primary interaction is almost 11\% (see, also, Fig.~\ref{fig:urqmd_multiplicity}),
and the probability of having at least one $\pi^{0}$ is 58\%, we
expect that about $\sim1.9\times10^{11}$ events could possibly fake
a $\eta\rightarrow\pi^{0}\mu^{+}\mu^{-}$ process. The second largest
background was due to the process: $\eta\rightarrow\gamma\mu^{+}\mu^{-}$,
when either an accidental/misidentified  photon is paired
with the radiative photon, and they both  fake a $\pi^{0}$. 

\begin{figure}[!ht]
\includegraphics[scale=0.3]{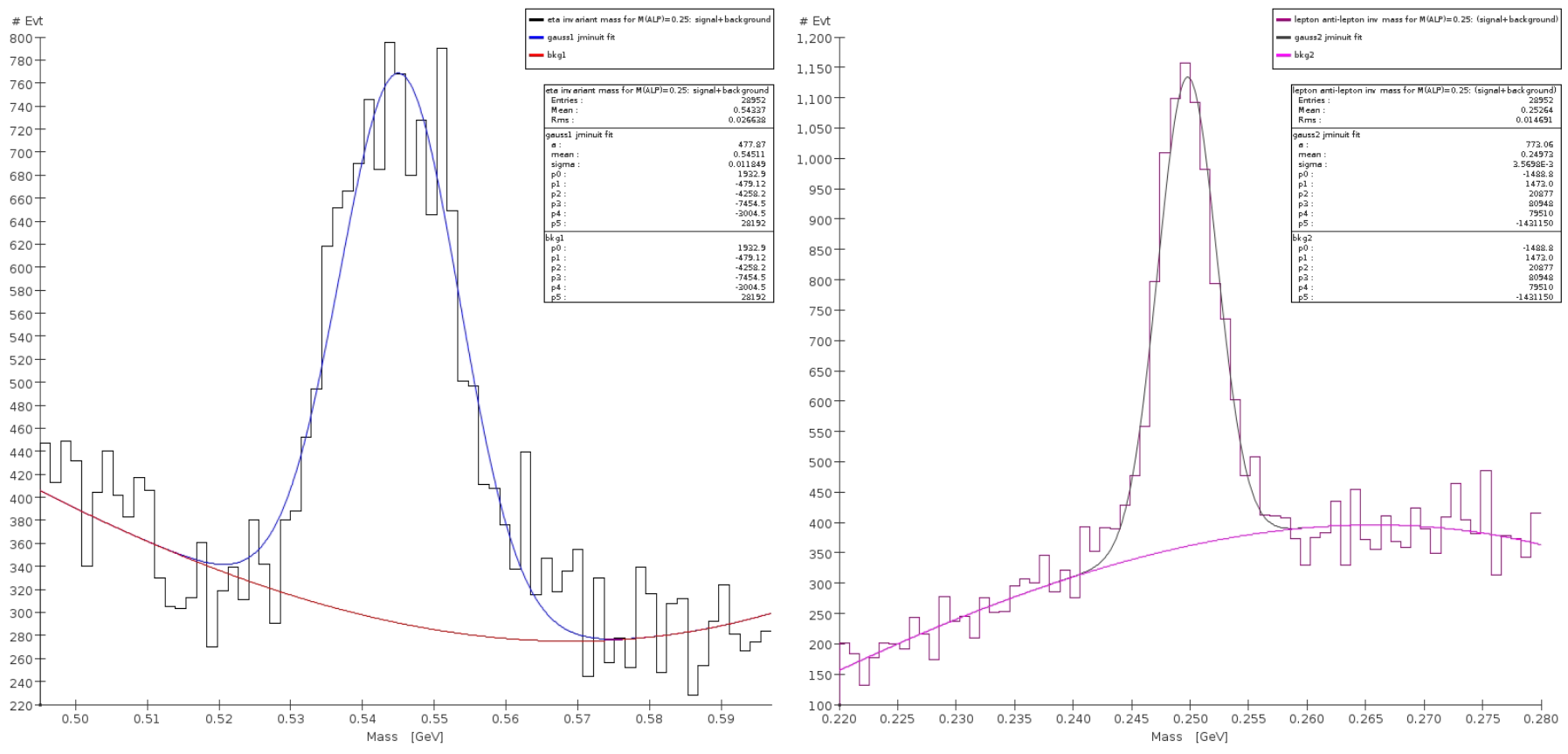} 

\caption{Invariant mass of $\pi^{0}\mu^{+}\mu^{-}$ (left) and of the $\mu^{+}\mu^{-}$
system (right) for a scalar boson $h$ with M($h$)=250 MeV. The plot includes
the Urqmd generated background. See text for an explanation of the
fitting procedure.}

\label{fig:eta2pi0H_mumu_mass}
\end{figure}

The full chain of generation-simulation-reconstruction-analysis was
repeated for each event set. Neutral pions, decaying into $\gamma e^{+}e^{-}$
and $\gamma\gamma$, where reconstructed by considering all combinations
of photons, electrons and positrons with an invariant mass within
5 MeV from $\pi^{0}$ mass. Finally, the events were required to have
a topology consistent with a $\pi^{0}\mu^{+}\mu^{-}$ final state,
and an invariant mass compatible with the $\eta$ mass. The largest
 background was found for values of the $\mu^{+}\mu^{-}$
invariant mass values near the middle of the kinematically allowed
range. The final reconstruction efficiency for this process was found
to be between $\sim$14\textbackslash\% and $\sim$18\textbackslash\%
for the signal and of order $\mathcal{O}(10^{-8})$ for the
Urqmd background. For illustrative purposes,
Fig.~\ref{fig:eta2pi0H_mumu_mass} shows the invariant mass distribution
of the reconstructed $\pi^{0}\mu^{+}\mu^{-}$ and di-lepton system
for the signal and the Urqmd background, assuming a BR($\eta\rightarrow\pi^{0}h$)=
$5\times10^{-5}$ and an $\eta$ sample of $2\times10^{8}$ (corresponding
to $\sim1.8\times10^{-6}$ of the full integrated luminosity). The
distribution was fitted using the sum of a Gaussian and a 5th-order
polynomial. 

The branching ratio sensitivity estimated for this channel is shown in the right
plot of Fig.~\ref{fig:eta2pi0H_ee_br}, as a function of the invariant
mass of the scalar boson. The highest sensitivity is observed near the upper
kinematic limit of the $\mu^{+}\mu^{-}$ invariant mass, where the
feed trough from combinatorics background is approximately half 
that observed in the central region. 

\subsection{\texorpdfstring{$h\rightarrow\mu^{+}\mu^{-}$}{}: Detached-vertex analysis}

The study for this final state was carried for masses of the scalar boson in the
range between 250 MeV and 400 MeV. For each value ofthe  mass, the $c\tau$
of the resonance was varied from 20 mm to 150 mm. A total of 12 event
sets were generated and fully reconstructed. The analysis for the
kinematic variables follows the same guidelines of the $bump-hunt$
analysis. Further cuts on the $\chi^{2}$ of a fit to the secondary vertex and on the
distance between the primary and secondary vertexes were applied.
Since the $\beta\gamma$ boost of the particle depends on the mass
of the boson $h$, the vertex cuts were individually optimized for each  event
set.

The final reconstruction efficiency for this analysis was found to be between $\sim$5\textbackslash\% and
$\sim$10\textbackslash\% for the signal samples and of order $\mathcal{O}(10^{-10})$
for the Urqmd background. The resulting branching ratio sensitivity
is shown in the right plot of Fig.~\ref{fig:eta2pi0H_ee_br-vtx},
as a function of the invariant mass and of $c\tau$ of the long-lived
scalar $h$. The $\beta\gamma$ boost for lower values of the $\mu^{+}\mu^{-}$ invariant
mass  
is larger and the secondary vertex occurs in the outermost region
of the detector, where the tracking systems has a lower reconstruction efficiency. That is reflected in a lower branching ratio
sensitivity in the low-mass region of the kinematically allowed range.

\subsection{\label{subsec:hpipi-bumphunt}\texorpdfstring{$h\rightarrow\pi^{+}\pi^{-}$: $Bump-hunt$}{}
analysis}

The decay of the scalar boson \emph{h} in non-leptonic mode is relevant
for probing so called \emph{``hadrophilic}'' theoretical models,
where the dark scalar or mediator couples more strongly to quarks rather
than to leptons. 
For the present work, the detector layout and the algorithms
implemented in the trigger, are optimized for the detection of final state containing leptons. Consequently, we expect a  lower sensitivity
compared to the other channels considered in this work. Furthermore, the relatively large
branching ratio of the $\eta\rightarrow\pi^{+}\pi^{-}\pi^{0}$ process,
represents an irreducible non-resonant background to searches of New
Physics with \emph{bump-hunt} techniques. 

The study of this final state was carried for masses of the scalar $h$
in the range between 300 MeV and 400 MeV. Each sample consists of $2\times10^{5}$
$\eta\rightarrow\pi^{0}h\;;\;h\rightarrow\pi^{+}\pi^{-}$ events. 
The largest background was found to originate from events
containing three pions (cf.\ Fig.\ref{fig:pi_p_multiplicity-1}).
This background accounts for approximately 90\% of the total. The
remaining contribution was, in large part, originating from the three-body
decay: $\eta\rightarrow\pi^{+}\pi^{-}\pi^{0}$, when one of the pions
was mis-reconstructed or when an extra pion from the primary beam-target
interaction was mistakenly associated to the decay of the $\eta$ meson. The background
was larger for values of $M(h)$ where the invariant mass of
the $\pi^{+}\pi^{-}$ system is in the central region of the kinematically
allowed range, where the combinatoric background mentioned above mimics more closely the
kinematics of the $\eta\rightarrow\pi^{0}h\;;\;h\rightarrow\pi^{+}\pi^{-}$
process. 

The full chain of generation-simulation-reconstruction-analysis was
repeated for each event set. Very generic requirements on the quality
of reconstructed particles were applied to signal and background
samples. Neutral pions, decaying into $\gamma e^{+}e^{-}$ and $\gamma\gamma$,
where reconstructed by considering all combinations of photons, electrons
and positrons with an invariant mass within 5 MeV from $\pi^{0}$
mass. Finally, the events were required to have a topology consistent
with a $\gamma\mu^{+}\mu^{-}$ final state, and an invariant mass
compatible with the $\eta$ mass.

\begin{figure}[!ht]
\includegraphics[scale=0.3]{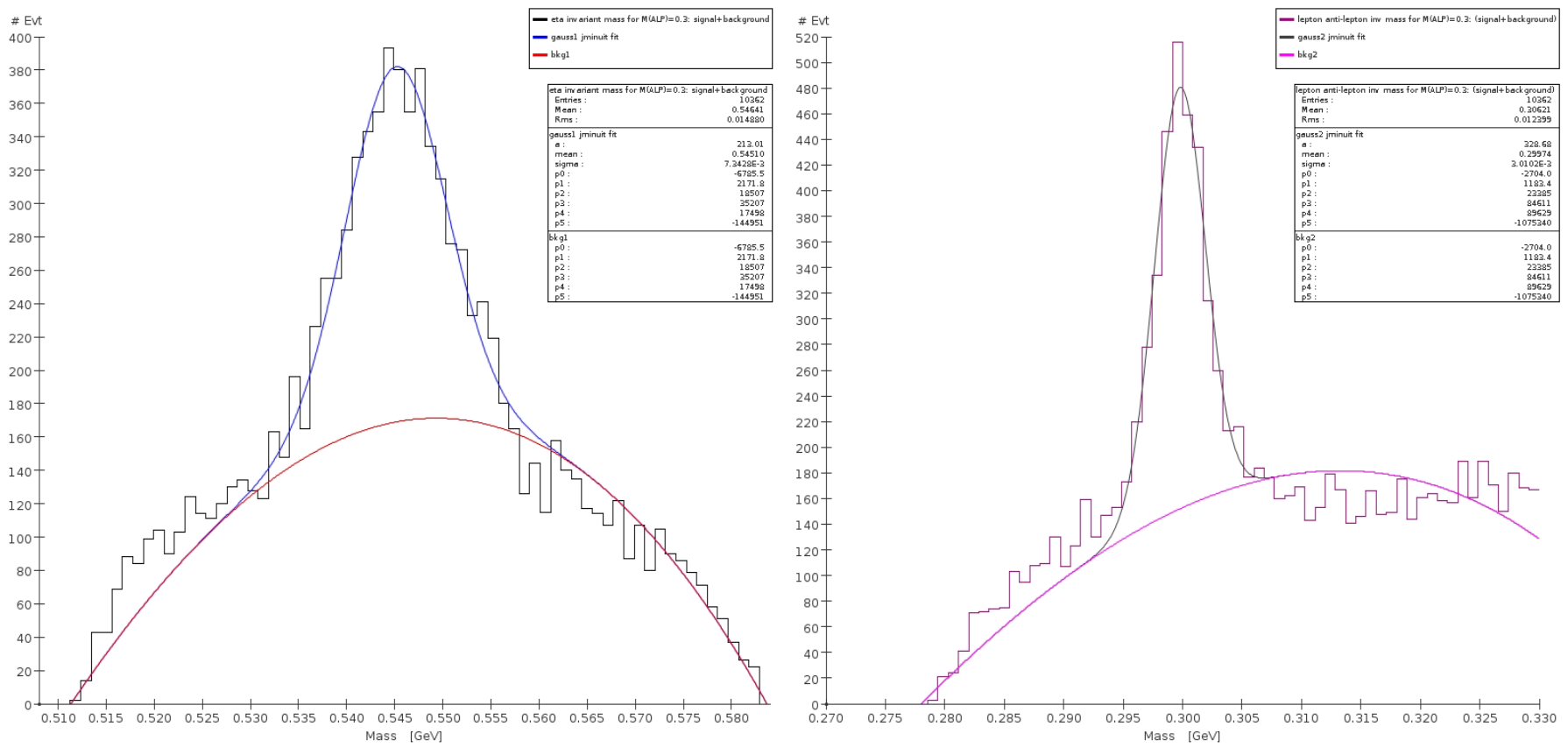} 

\caption{Invariant mass of $\pi^{0}\pi^{+}\pi^{-}$ (left) and of the $\pi^{+}\pi^{-}$
system (right) for a scalar boson $h$  with mass M($h$)=300 MeV. The plot includes
the Urqmd generated background. See text for an explanation of the
fitting procedure.}

\label{fig:eta2pi0H_pipi_mass}
\end{figure}

The reconstruction efficiencies for this process was found to be between
1.5\% and 2\% for the $h$ signal and a relatively large $\sim10^{-6}$
for the Urqmd background. For illustrative purposes, Fig.~\ref{fig:eta2pi0H_pipi_mass} shows the invariant
mass distribution of the reconstructed $\pi^{0}\pi^{+}\pi^{-}$ and
$\pi^{+}\pi^{-}$ systems for the signal and the Urqmd background,
assuming a BR($\eta\rightarrow\pi^{0}h$) = $5\times10^{-2}$ and an
$\eta$ sample with $1\times10^{7}$ events (corresponding to $0.9\times10^{-7}$
of the full integrated luminosity). The distribution was fitted using
the sum of a Gaussian and a 5th-order polynomial. 
As expected, the combinatoric background accumulated under the $\eta$ peak, is much worse than for any other channel examined.

\begin{figure}[!ht]
\includegraphics[width=7cm]{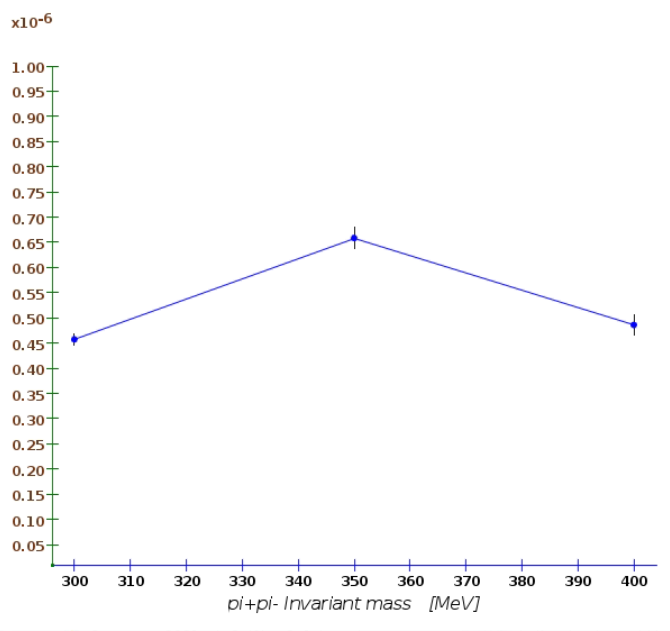} \includegraphics[width=8cm]{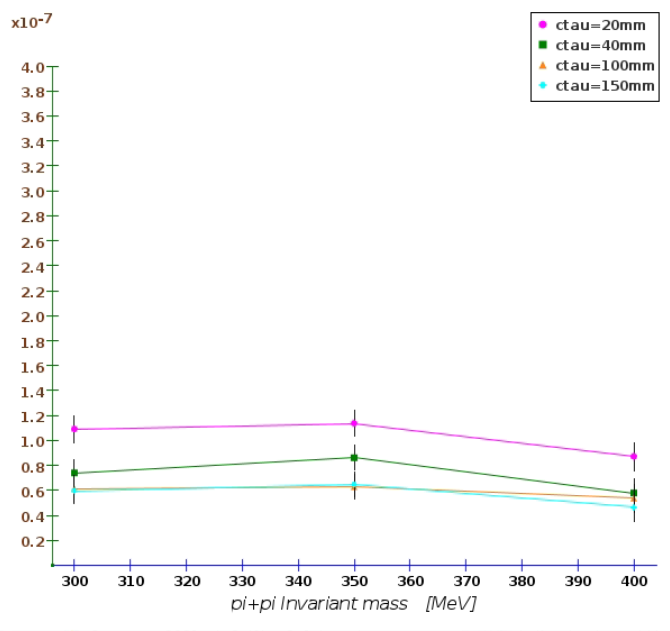}
\caption{Branching ratio sensitivity for the process $\eta\rightarrow\pi^{0}h\;and\;h\rightarrow\pi^{+}\pi^{-}$
as a function of the mass of the scalar short-lived scalar $h$ (left)
and of the mass and $c\tau$ of a long-lived scalar $h$.}

\label{fig:eta2pi0H_pipi_br}
\end{figure}

The branching ratio sensitivity for this process was, then, calculated
according to Eqs. \ref{eq:Master1}-\ref{eq:MasterB}. The resulting
sensitivity curve is shown in the left plot of Fig.~\ref{fig:eta2pi0H_pipi_br},
as a function of the invariant mass of the scalar $h$. Sensitivity is
worse in the central region, as expected from the discussion above.

\subsection{\label{subsec:hpipi-detachedvertex}\texorpdfstring{$h\rightarrow\pi^{+}\pi^{-}$}{}:
\emph{Detached-}vertex analysis}

The experimental sensitivity to the $\eta\rightarrow\pi^{0}h\; with h\rightarrow\pi^{+}\pi^{-}$
channel improves considerably when the scalar boson travels inside the detector
and it decays far from the interaction point. As already noted,
the  background from the Standard Model is dramatically reduced when constraints on
a detached secondary vertex are imposed. The scalar $h$ was generated
within a mass range between 300 MeV and 400 MeV. For each value of
mass, the $c\tau$ of the resonance was varied from 20 mm to 15 0mm.
A total of 12 event sets were generated and fully reconstructed. The
analysis of the kinematics of the event follows the same guidelines of
the $bump-hunt$ analysis. Further cuts on the $\chi^{2}$ of fit on the secondary vertex
and on the distance between the primary and secondary vertexes
were applied. Since the $\beta\gamma$ boost of the particle depends
on the mass of the $h$, those cuts were optimized for each individual
event set. 
A rejection factor of about $100\times$ was obtained for the background
with respect to a purely kinematic analysis.

The final reconstruction efficiency for this process, including the additional
vertex cuts, was found to range between $\sim$8\textbackslash\% and
$\sim$15\textbackslash\% for the signal and of order $\mathcal{O}(10^{-8})$
for the Urqmd background. The resulting branching ratio sensitivity
is shown in the right plot of Fig.~\ref{fig:eta2pi0H_pipi_br}, as
a function of the invariant mass and of $c\tau$ of the long-lived
scalar $h$. The effect of the vertex cuts, tuned to remove the combinatoric background,
is clearly visible, with an improvement of the experimental sensitivity
of almost one order of magnitude compared to the case of a short-lived scalar boson. 
We observe, once more, that the sensitivity is lower in the central region of the kinematically
allowed mass range.

\subsection{Sensitivity to selected theoretical models}

In this section, we consider three distinct
theoretical models, discussed in details in Sec.~\ref{sec:Physics-Beyond-the-SM}, for which New Physics appears through the scalar portal. More specifically,  the branching ratio sensitivity obtained in the previous sections is used to
determine the sensitivity to the corresponding coupling constants. 

\subsubsection{Hadrophilic Scalar Mediator (including Spontaneous Flavor Violation models)}

\begin{figure}[!ht]
\includegraphics[width=7cm]{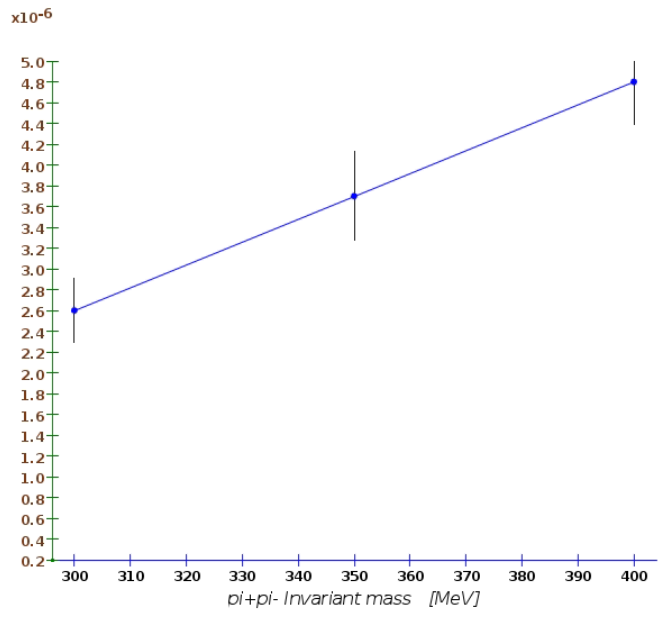} 
\caption{Sensitivity to $g_{u}$ for the process $\eta\rightarrow\pi^{0}h\;and\;h\rightarrow\pi^{+}\pi^{-}$
for the bump-hunt as a
function of the mass of the short-lived scalar $S$.}

\label{fig:eta2pi0H_pipi_gu}
\end{figure}

The $bump-hunt$ search in the  $\pi^0 (\pi^+ \pi^-)$ mode discussed in this section is directly applicable to the hadrophilic scalar mediator model described above in Section~\ref{subsec:Scalar-portal-Models}; see Eq.~(\ref{eq:hadrophilic-Lag}). In this model, the scalar $S$ is produced  via $\eta \rightarrow \pi^0 S$ and subsequently decays via $S\rightarrow \pi^+ \pi^-$. The scalar decays promptly in the parameter region of interest to REDTOP.  
Using the branching ratio formula for ${\rm Br}(\eta \rightarrow \pi^0 S)$ given in Eq.~(\ref{eq:Breta-Pi-S}), we can translate the limits presented in the left panel of Fig.~\ref{fig:eta2pi0H_pipi_br} to limits in the mass-coupling $(m_S-g_u)$ plane. 
The result is shown in Figure.~\ref{fig:eta2pi0H_pipi_gu} and in Figure.~\ref{fig:dark_scalar} (in the latter, superimposed on the parameter space explored by previous experiments). The sensitivity curve indicates that 
REDTOP will probe couplings of order $g_u \sim {\rm few} \times 10^{-6}$, extending the reach by a factor of a few beyond previous recasted bounds from the KLOE experiment~\cite{KLOE-2:2016zfv,Batell:2018fqo}. Furthermore, it is worth highlighting that REDTOP nicely complements the sensitivity from proposed long-lived particle searches at FASER~\cite{Kling:2021fwx}, FASER2~\cite{Kling:2021fwx} and SHiP~\cite{Batell:2018fqo}, which will be able to explore longer lifetimes and smaller couplings.

\subsubsection{Two-Higgs doublet model }

The parameters $\lambda_u$ and $\lambda_d$ for this model, along
with the branching ratios predicted for the decay channel $\eta^{(\prime)} \to \pi^0 S$
are discussed in Sec.~\ref{subsec:Scalar-portal-Models}. For the Two-Higgs
doublet model considered in this work~\cite{Abdallah:2020vgg}, the
branching ratio of $\eta\to\pi^0 h'$, in the assumption that $\lambda_u = \lambda_d$,
is predicted to be $1.22\times10^{-11}$, which is below REDTOP sensitivity
in the present run. The situation is different for the case when $\lambda_{u}\neq\lambda_{d}$.
It is known that the strange and non-strange isoscalar transition
form factors are suppressed by $\epsilon \sim 0.012$. Therefore,
the $\pi^0 S$ production is dominated by the isovector form factor
and the $\pi^0 S$ production is proportional to $(\lambda_u-\lambda_d)^2$.
Thus, the branching ratio of $\eta\to \pi^0 S$ approximately equals
to:
\begin{equation}
{\rm BR}(\eta\to\pi^{0}S)\simeq\frac{2}{3}B_{0}^{2}\frac{(\lambda_{u}-\lambda_{d})^{2}}{\Gamma_{\eta}^{{\rm tot}}}\frac{\sqrt{\tilde{E}_{\eta}^{2}-m_{\pi^{0}}^{2}}}{8\pi m_{\eta}^{2}},\label{eq:BR__eta_decay_to_scalar-Approx}
\end{equation}

where 
\begin{equation}
\tilde{E}_{\eta}=\frac{m_{\eta}^{2}+m_{\pi^{0}}^{2}-m_{S}^{2}}{2m_{\eta}}.\label{eq:E_eta}
\end{equation}
In this model~\cite{Abdallah:2020vgg}, for non-zero difference of $(\lambda_u-\lambda_d)^2$,
BR$(\eta \to \pi^0 h')$ range is of order ${\cal O}(10^{-11}-10^{-5})$
with $m_{h'} \simeq17$~MeV. From the data presented in Fig.~\ref{fig:eta2pi0H_ee_br-vtx},
we can conclude that REDTOP is sensitive to a slight mismatch in $\lambda_u$
and $\lambda_d$  values that keeps the LSND and MB fit intact. 

Inserting the values obtained for each value of the $S$ mass into
the formula for ${\rm BR}(\eta \to \pi^0 S)$ in Eq.~(\ref{eq:BR__eta_decay_to_scalar-Approx}),
the corresponding sensitivity for the parameter $(\lambda_u-\lambda_d)^2$
is shown in Table~\ref{table:delta-lambda-sensitivity} for the two
analysis techniques we have used.

\begin{table}[!ht]
\centering\textcolor{black}{\scriptsize{}}%
\begin{tabular}{|c|c|c||c|}
\hline 
\textit{\textcolor{black}{\scriptsize{}Process}} & \textcolor{black}{\scriptsize{}$m_{S}$} & \textit{\textcolor{black}{\scriptsize{}Analysis}} & \textcolor{black}{\scriptsize{}$(\lambda_{u}-\lambda_{d})^{2}$}\tabularnewline
 &  &  & \textcolor{black}{\scriptsize{}sensitivity}\tabularnewline
\hline 
\hline 
\textcolor{black}{\scriptsize{}$\eta$$\rightarrow\pi^{0}S\;;\;S\rightarrow e^{+}e^{-}$} & \textcolor{black}{\scriptsize{}17 MeV} & \textcolor{black}{\scriptsize{}bump hunt} & \textcolor{black}{\scriptsize{}$2.0\times10^{-13}$}\tabularnewline
\hline 
\textcolor{black}{\scriptsize{}$\eta$$\rightarrow\pi^{0}S\;;\;S\rightarrow\mu^{+}\mu^{-}$} & \textcolor{black}{\scriptsize{}17 MeV} & \textcolor{black}{\scriptsize{}detached vertex} & \textcolor{black}{\scriptsize{}$3.2\times10^{-13}$}\tabularnewline
\hline 
\end{tabular}\caption{Sensitivity to $(\lambda_u-\lambda_d)^2$ for the process $\eta \to \pi^0 S$
and $S \to e^+e^-$ and $S \to \mu^+\mu^-$.}
\label{table:delta-lambda-sensitivity}
\end{table}

\subsubsection{Minimal scalar model }

The theoretical aspects of \emph{Minimal Scalar model} have been discussed in detail in Sec.
\ref{subsec:Scalar-portal-Models}. The relevant parameter in this
case is the mixing angle $\theta^{2}$~\cite{Oconnel2007} between
the Standard Model Higgs and its lighter counterpart. In this study
we consider two parametrizations for the branching ratio, as they
bring to numerically appreciable differences in the results.

\paragraph{Tulin parametrization}

\begin{figure}[!ht]
\includegraphics[width=7cm]{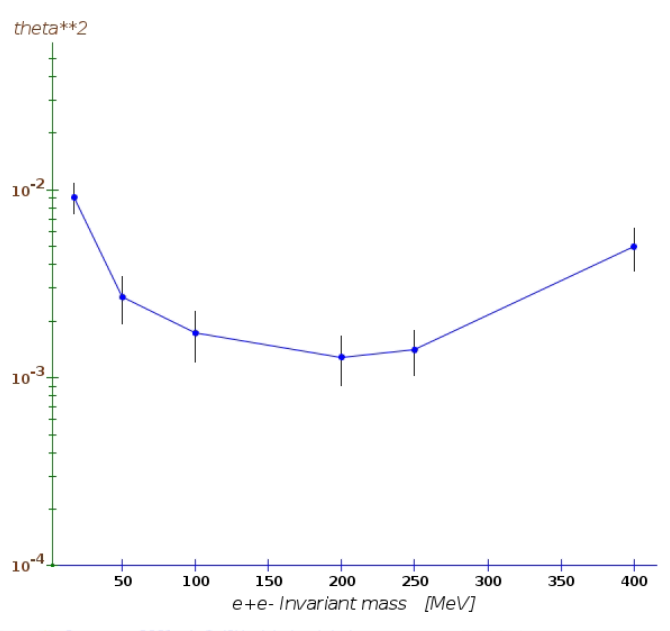} \includegraphics[width=7cm]{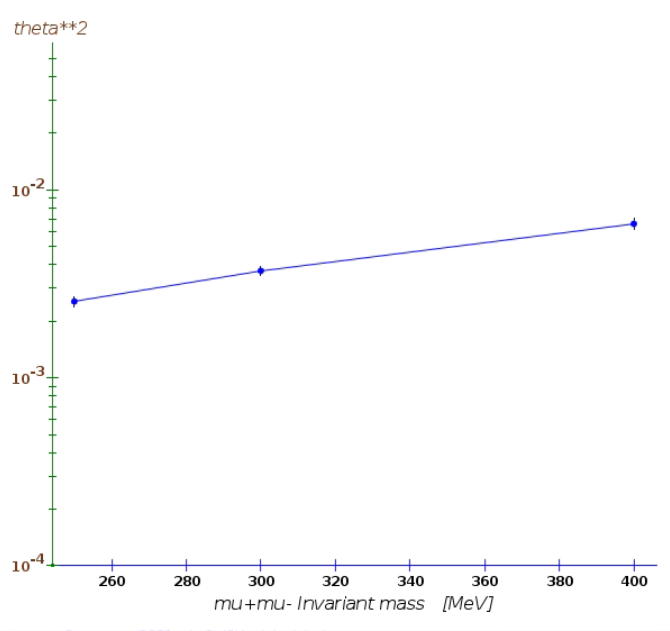}
\caption{Sensitivity to $\theta{}^{2}$ in the Tulin approach\cite{Tulin:2014tya}
for the bump-hunt analysis of the processes $\eta\rightarrow\pi^{0}h\;and\;h\rightarrow e^{+}e^{-}$
(left) and $\eta\rightarrow\pi^{0}h\;and\;h\rightarrow\mu^{+}\mu^{-}$
(right) as a function of the mass of a short-lived scalar $h$.}

\label{fig:eta2pi0H_ee_theta-1}
\end{figure}

\begin{figure}[!ht]
\includegraphics[width=7cm]{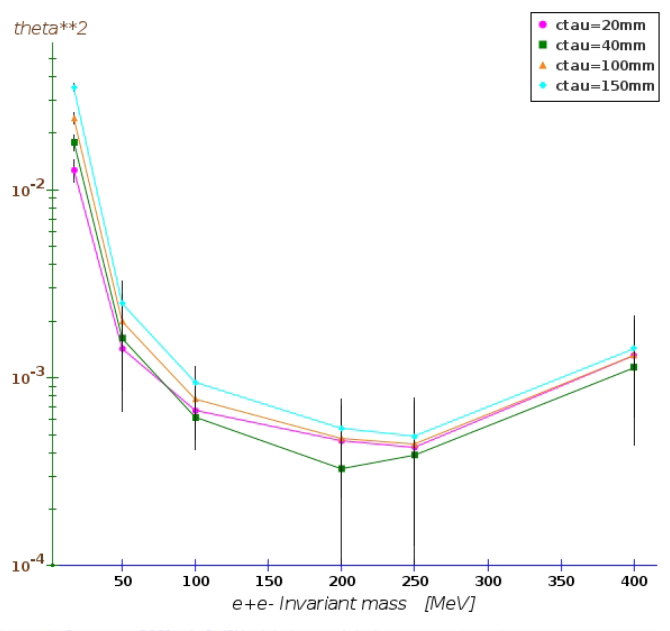}
\includegraphics[width=7cm]{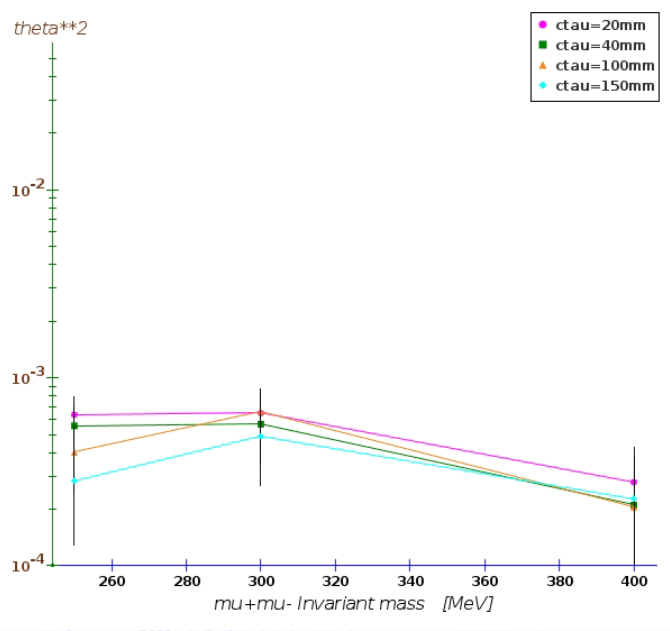}
\caption{Sensitivity to $\theta{}^{2}$ for the bump-hunt analysis of the processes
$\eta\rightarrow\pi^{0}h\;and\;h\rightarrow e^{+}e^{-}$ (left) and
$\eta\rightarrow\pi^{0}h\;and\;h\rightarrow\mu^{+}\mu^{-}$ (right)
as a function of the mass and $c\tau$ of the a long-lived scalar
$h$.}

\label{fig:eta2pi0H_ee_theta-vtx-1}
\end{figure}

The master formula used for this parametrization of $Br(\vartheta^{2})$
is \cite{Tulin:2014tya}:

\begin{equation}
BR_{\eta\rightarrow\pi^{0}h}\eqsim1.8\times10^{-6}\times\lambda^{\nicefrac{1}{2}}\left(1,\frac{M_{\pi_{0}}^{2}}{M_{\eta}^{2}},\frac{M_{H}^{2}}{M_{\eta}^{2}}\right)\label{eq:BR2theta-2}
\end{equation}

where the function $\lambda$ has been defined in Sec.~\ref{subsec:Scalar-portal-Models}.
Inserting in Eq.~\eqref{eq:BR2theta-2} the values of the branching
ratio sensitivity derived for each value of the $h$ mass, we obtain
the corresponding plots of sensitivity for the parameter $\theta^{2}$.
They are shown in Fig.~\ref{fig:eta2pi0H_ee_theta-1} for the bump-hunt
analyses of the two leptonic final states 
and in Fig.~\ref{fig:eta2pi0H_ee_theta-vtx-1} for the two detached-vertex
analyses.

\paragraph{Pospelov parametrization}

\begin{figure}[!ht]
\includegraphics[width=7cm]{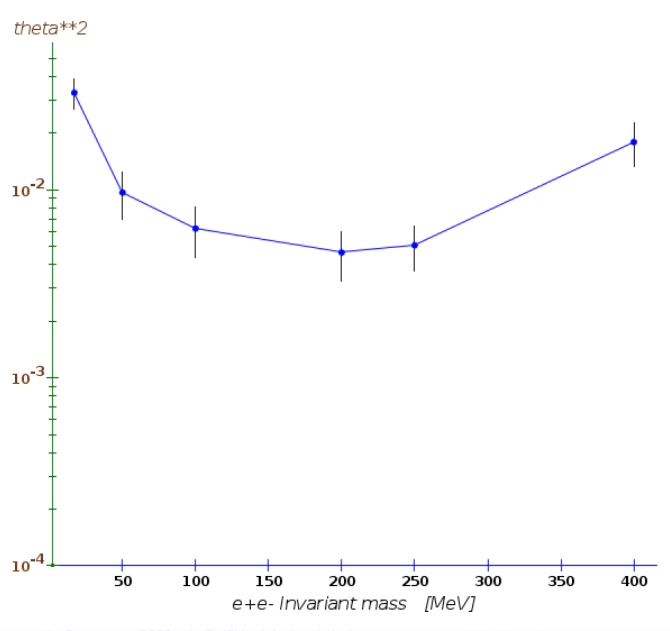} \includegraphics[width=7cm]{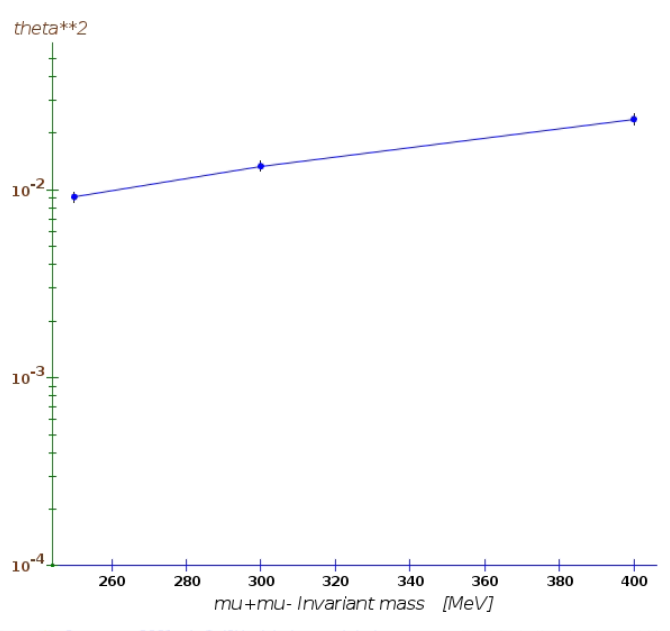}
\caption{Sensitivity to $\theta{}^{2}$ in the Pospelov approach \cite{Pospelov2018}
for the bump-hunt analysis of the processes $\eta\rightarrow\pi^{0}h\; \;h\rightarrow e^{+}e^{-}$
(left) and $\eta\rightarrow\pi^{0}h\; \;h\rightarrow\mu^{+}\mu^{-}$
(right) as a function of the mass of a short-lived scalar $h$.}

\label{fig:eta2pi0H_ee_theta}
\end{figure}

\begin{figure}[!ht]
\includegraphics[width=7cm]{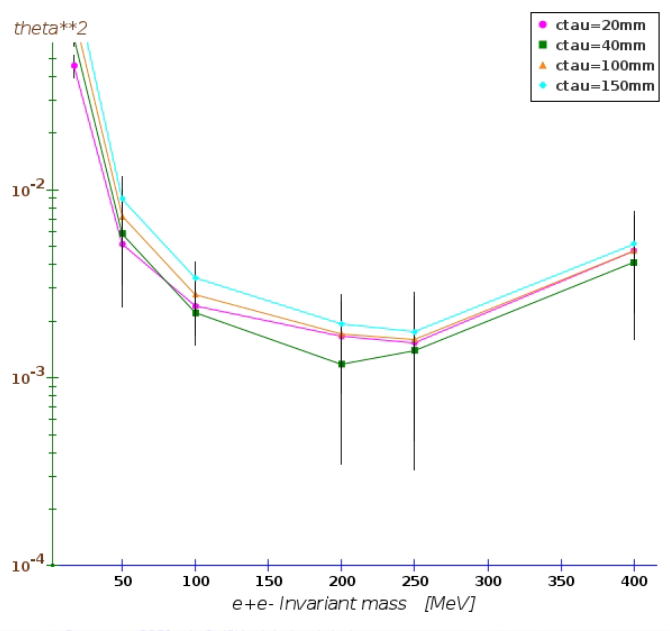} \includegraphics[width=7cm]{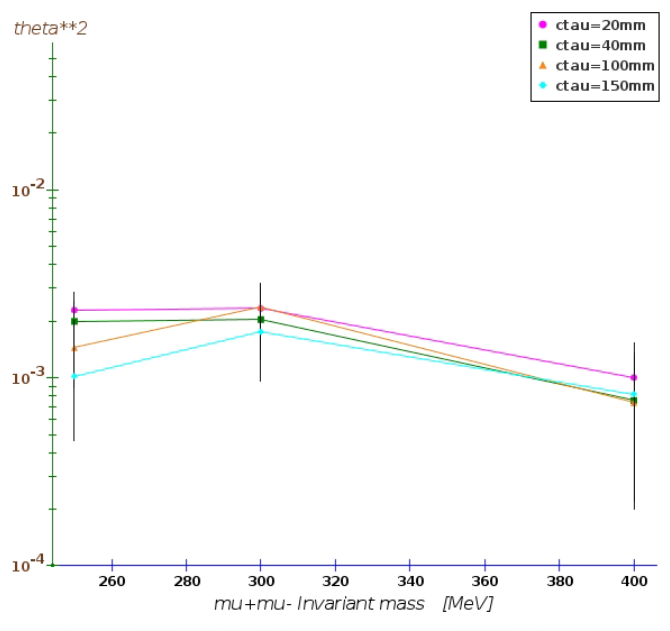}
\caption{Sensitivity to $\theta{}^{2}$ for the bump-hunt analysis of the processes
$\eta\rightarrow\pi^{0}h\; \;h\rightarrow e^{+}e^{-}$ (left) and
$\eta\rightarrow\pi^{0}h\; \;h\rightarrow\mu^{+}\mu^{-}$ (right)
as a function of the mass and $c\tau$ of the a long-lived scalar
$h$.}

\label{fig:eta2pi0H_ee_theta-vtx}
\end{figure}

In this case, see Ref \cite{Pospelov2018}, the master formula for $Br(\vartheta^{2})$ (cf.\ Eq.
\ref{eq:MasterS-final}) is given by:  

\begin{equation}
BR_{\eta\rightarrow\pi^{0}h}=\frac{\theta^{2}p_{h}\,g_{\eta\pi^{0}h}^{2}}{8\pi m_{\eta}^{2}\Gamma_{\eta}}\label{eq:BR2theta}
\end{equation}

where $\Gamma_{\eta\ensuremath{}}$ is the total width, and $p_{h}$ is
the  momentum of \emph{h} particle in the $\eta$ rest
frame. The squared amplitude of the process is represented by the
parameter $g_{\eta\pi^{0}H}^{2}$:

\begin{equation}
g_{\eta\pi^{0}h}^{2}=\sqrt{\nicefrac{1}{3}}\times\frac{m_{\pi}^{2}}{\upsilon}\times\frac{m_{d}-m_{u}}{m_{d}+m_{u}}\label{eq:scalar-1}
\end{equation}

where: $\upsilon=246\,GeV$. Inserting in Eq.~\eqref{eq:BR2theta} the
values of the branching ratio sensitivity derived for each value of
the $h$ mass, we obtain the plots for the corresponding sensitivity
to the parameter $\theta^{2}$. They are shown in Fig.~\ref{fig:eta2pi0H_ee_theta}
for the two \emph{bump-hunt} analyses 
and in Fig.~\ref{fig:eta2pi0H_ee_theta-vtx} for the two \emph{detached-vertex}
analyses.

\section{Sensitivity to the Pseudoscalar portal\label{subsec:Sensitivity-to-the-pseudoscalar}}

The \emph{Pseudoscalar portal} is a very rich sector of BSM physics and several novel theoretical models are predicting the existence of
a new pseudoscalar state. Among others, the long-sought
\cite{Peccei1977}, but never confirmed, Peccei-Quinn model introduces an axion
particle which could explain several anomalies recently observed by experiments.
Another important class of models falling into this sector is are those which have  \emph{Axion-Like Particles} (\emph{ALP}'s). These models have 
extensions of the axion particle with relaxed constraints. The \emph{Pseudoscalar portal} can be probed
at REDTOP via decays of the $\eta$-meson associated to two charged
or neutral pions. In this work, we have considered  leptonic, radiative, and
hadronic decays of the  pseudoscalar boson \emph{a}. 

\subsection{Leptonic decays: \texorpdfstring{$a\rightarrow e^{+}e^{-}$ and $a\rightarrow\mu^{+}\mu^{-}$}{}
final states}

The processes under study with a leptonic decay mode of  the pseudoscalar boson \emph{a} are:
\begin{itemize}
\item $p+Li\rightarrow\eta+X\;with\;\eta\rightarrow\pi^{+}\pi^{-}a\;and\;a\rightarrow e^{+}e^{-}$
\item $p+Li\rightarrow\eta+X\;with\;\eta\rightarrow\pi^{0}\pi^{0}a\;and\;a\rightarrow e^{+}e^{-}$
\end{itemize}
Two different analysis were performed, aiming at testing the performance
of different components of the detector: a \emph{bump-hunt} and a
\emph{detached-vertex} analysis.

\subsubsection{\texorpdfstring{$\eta\rightarrow\pi^{+}\pi^{-}a$:  {\it bump-hunt}}{} analysis}

The same considerations already done for the \emph{Vector} and \emph{Scalar
portals}  hold also for the \emph{Pseudoscalar portal}. The
pseudoscalar boson \emph{a} was generated in a mass range between
17 MeV and 250 MeV. The boson was decayed promptly in \emph{GenieHad},
by setting the vertex of the di-lepton in the $\eta$ meson production point. 
We observed, for this final state, a large background contribution
for low mass values for the di-lepton pair originating from photon conversion:
$\gamma\rightarrow e^{+}e^{-}$, in the detector material. 
The photon is usually generated  from
 unreconstructed $\pi^{0}(\eta)\rightarrow\gamma\gamma$ decays of the  $\pi^{0}$
 and  $\eta$ mesons. 
On the other side, the higher mass range is relatively background free. The second
largest background originates from the non-resonant process: $\eta\rightarrow\pi^{+}\pi^{-}e^{+}e^{-}$,
which is produced abundantly, as the corresponding branching ratio is 2.68$\times10^{-4}$. 

The full chain of generation-simulation-reconstruction-analysis was
repeated for each event set. The TOF and PID systems strongly reduce
the background already at the trigger level by a factor of order $\mathcal{\sim}10{}^{4})$.
Very generic requirements on the quality of reconstructed particles
were applied to the signal and background samples. In particular, no
cuts were applied to the reconstructed $e^{+}e^{-}$ vertex. Neutral
pions, decaying into $\gamma e^{+}e^{-}$ and $\gamma\gamma$, were 
reconstructed by considering all combinations of photons, electrons
and positrons with an invariant mass within 5 MeV from $\pi^{0}$
mass. This requirement is able to reject most of the combinatoric
background, where no $\eta$ mesons are present in the final state. Similarly, photons
converting into an $e^{+}e^{-}$ pair where reconstructed by requiring
that the invariant mass of the $e^{+}e^{-}$ system was lower than
5 MeV. Finally, the events were required to have a topology consistent
with a $\eta\rightarrow\pi^{+}\pi^{-}e^{+}e^{-}$ final state, and
an invariant mass compatible with the $\eta$ mass.

The total reconstruction efficiency for this process was found to
range  between $\sim$1.4\% and $\sim$3\%, strongly dependent on the
di-lepton invariant mass. The reconstruction efficiency was of order
$\mathcal{O}(10^{-9})-\mathcal{O}(10^{-9})$ for the Urqmd background. 

\begin{figure}[!ht]
\includegraphics[width=15cm]{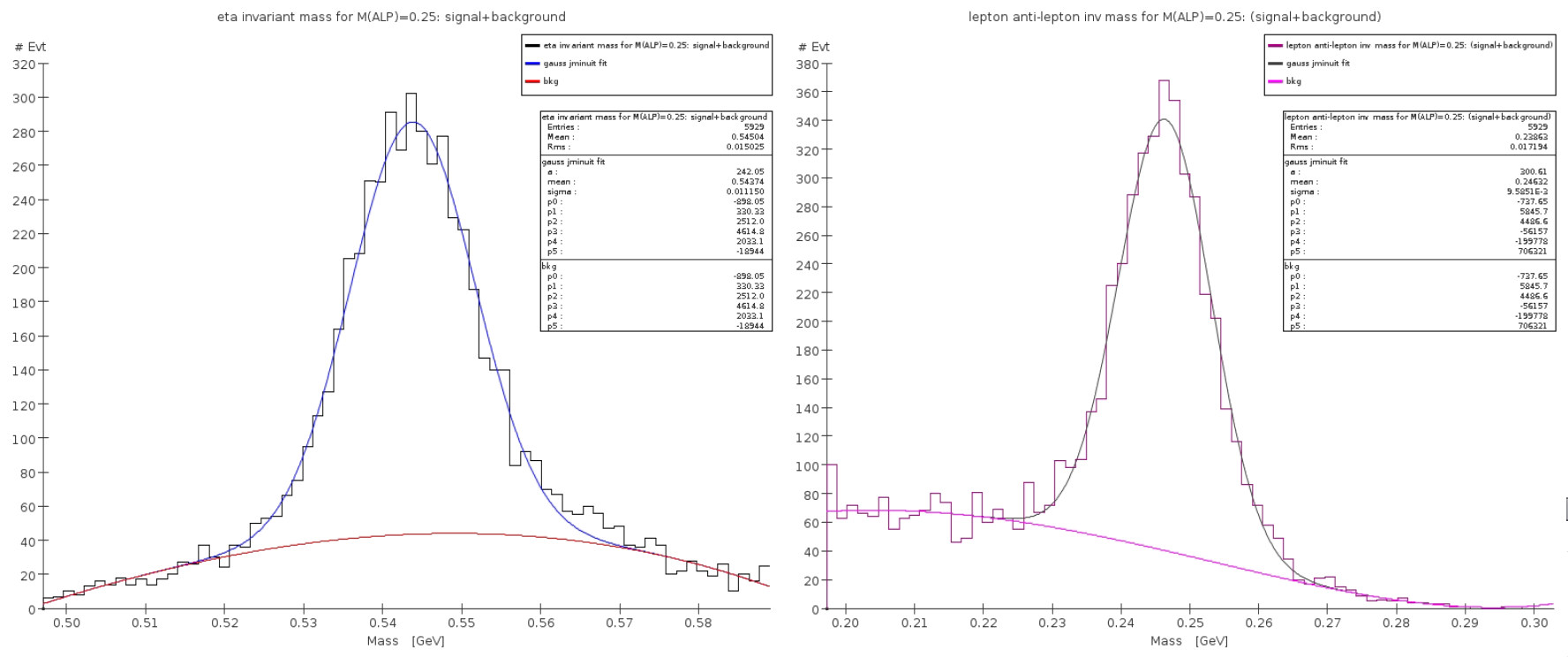}
\caption{Invariant mass of $\pi^{+}\pi^{-}e^{+}e^{-}$(left) and of the $e^{+}e^{-}$
system (right) for a pseudoscalar boson \emph{a} with mass M(\emph{a})=250 MeV. The plot includes
the Urqmd generated background. See text for an explanation of the
fitting procedure. }

\label{fig:eta2pippimalp_ee}
\end{figure}
For illustrative purposes, Fig.~\ref{fig:eta2pippimalp_ee} shows
the distribution of the invariant mass of the reconstructed $\pi^{+}\pi^{-}e^{+}e^{-}$ system, for
the event set with M(\emph{a})=250 MeV merged with the Urqmd generated background.
We assumed that BR($\eta\rightarrow\pi^{+}\pi^{-}a$)= $1.4\times10^{-4}$, while
 the $\eta$ sample consists of $5\times10^{8}$ events (corresponding to $4.5\times10^{-6}$
of the full integrated luminosity). The number of reconstructed signal
and background events was obtained form a fit to the $\eta$ meson
invariant mass using the sum of a gaussian and a 5th-order polynomial.
The integral of the fit function was used to extract the branching
ratio sensitivity for this process, according to Eqs. \ref{eq:Master1}-\ref{eq:MasterB}.
\begin{figure}[!ht]
\includegraphics[width=8cm]{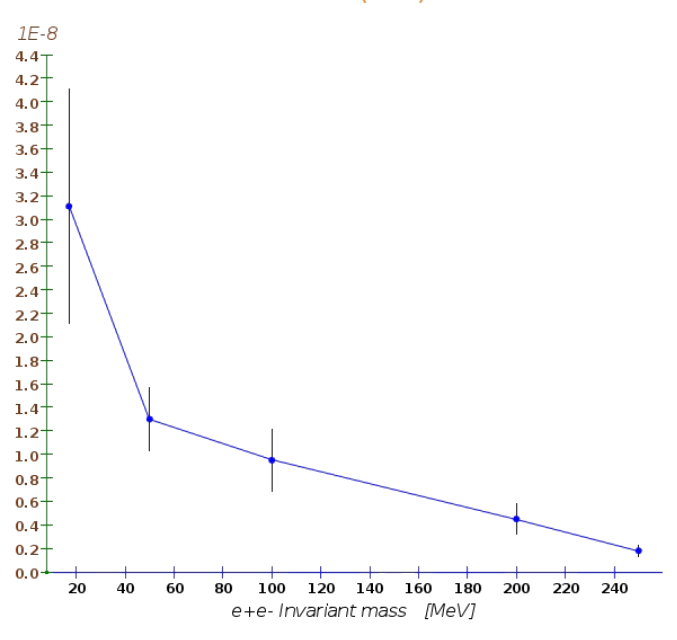} \caption{Branching ratio sensitivity for the process $\eta\rightarrow\pi^{+}\pi^{-}a\;with\;a\rightarrow e^{+}e^{-}$
as a function of the pseudoscalar boson mass M(\emph{a}).}

\label{fig:eta2pippimalp_ee_br}
\end{figure}
The resulting sensitivity is shown in Fig.~\ref{fig:eta2pippimalp_ee_br},
as a function of M(\emph{a}). As expected, the process
has a lower sensitivity for values of  M(\emph{a}) below 50
MeV, due to the large contribution of the $\gamma\rightarrow e^{+}e^{-}$
background. On the other side, the sensitivity is of order
$\mathcal{O}(10^{-9})$ for higher mass values. 


\subsubsection{\texorpdfstring{$\eta\rightarrow\pi^{0}\pi^{0}a$: {\it bump-hunt}}{} analysis}

This final state requires the reconstruction four particles, two of which (the neutral pions) decay further into at least two more particles. 
We expect, therefore, a considerably lower detection efficiency when compared to most of the other processes considered here, even just from purely geometrical considerations.
For the analysis of this final state, the pseudoscalar $a$ was generated in a mass
range between 17 MeV and 250 MeV. The full chain of generation-simulation-reconstruction-analysis
was repeated for each set of generated events. Very generic requirements
on the quality of reconstructed particles were  applied to the signal
and background samples.

We observed that the single largest background contribution to this process originates 
from the decay
$\eta\rightarrow\pi^{0}\pi^{0}\pi^{0}$, accounting for more than
50\% of the total background surviving the analysis cuts. That process
contaminates the signal from either a mis-reconstucted $\pi^{0}\rightarrow\gamma e^{+}e^{-}$
decay or from a $\pi^{0}\rightarrow\gamma\gamma$ decay followed by
a $\gamma\rightarrow e^{+}e^{-}$ pair production. However, that source
of background has a relatively unambiguous signature and it can be
reduced considerably with appropriate kinematic cuts. Consequently,
this is one of the channels with the largest branching ratio sensitivity,
among those considered in this work, especially for the higher values of  M(\emph{a}), where the surviving background is relatively tame. 

No cuts were applied to the reconstructed $e^{+}e^{-}$ vertex. Neutral
pions, decaying into $\gamma e^{+}e^{-}$ and $\gamma\gamma$, where
reconstructed by considering all combinations of photons, electrons
and positrons with an invariant mass within 5 MeV from $\pi^{0}$
mass. This requirement reduces considerably the combinatoric
background, where no $\eta$ mesons are present in the final state. Similarly, photons
converting into an $e^{+}e^{-}$ pair were reconstructed by requiring
that the invariant mass of the $e^{+}e^{-}$ system was lower than
5 MeV. Finally, the events were required to have a topology consistent
with a $\eta\rightarrow\pi^{0}\pi^{0}e^{+}e^{-}$ final state, and
an invariant mass compatible with the $\eta$ mass. 

The total reconstruction efficiency for this process was found to
range between $\sim$0.4\% and $\sim$2.8\%, strongly dependent on the
di-lepton invariant mass. The reconstruction efficiency was of order
$\mathcal{O}(10^{-13})-\mathcal{O}(10^{-10})$ for the Urqmd background.  The number of reconstructed signal and background events was obtained
from a fit to the $\eta$ meson invariant mass using the sum of a
Gaussian and a 5th-order polynomial. 

\begin{figure}[!ht]
\includegraphics[width=15cm]{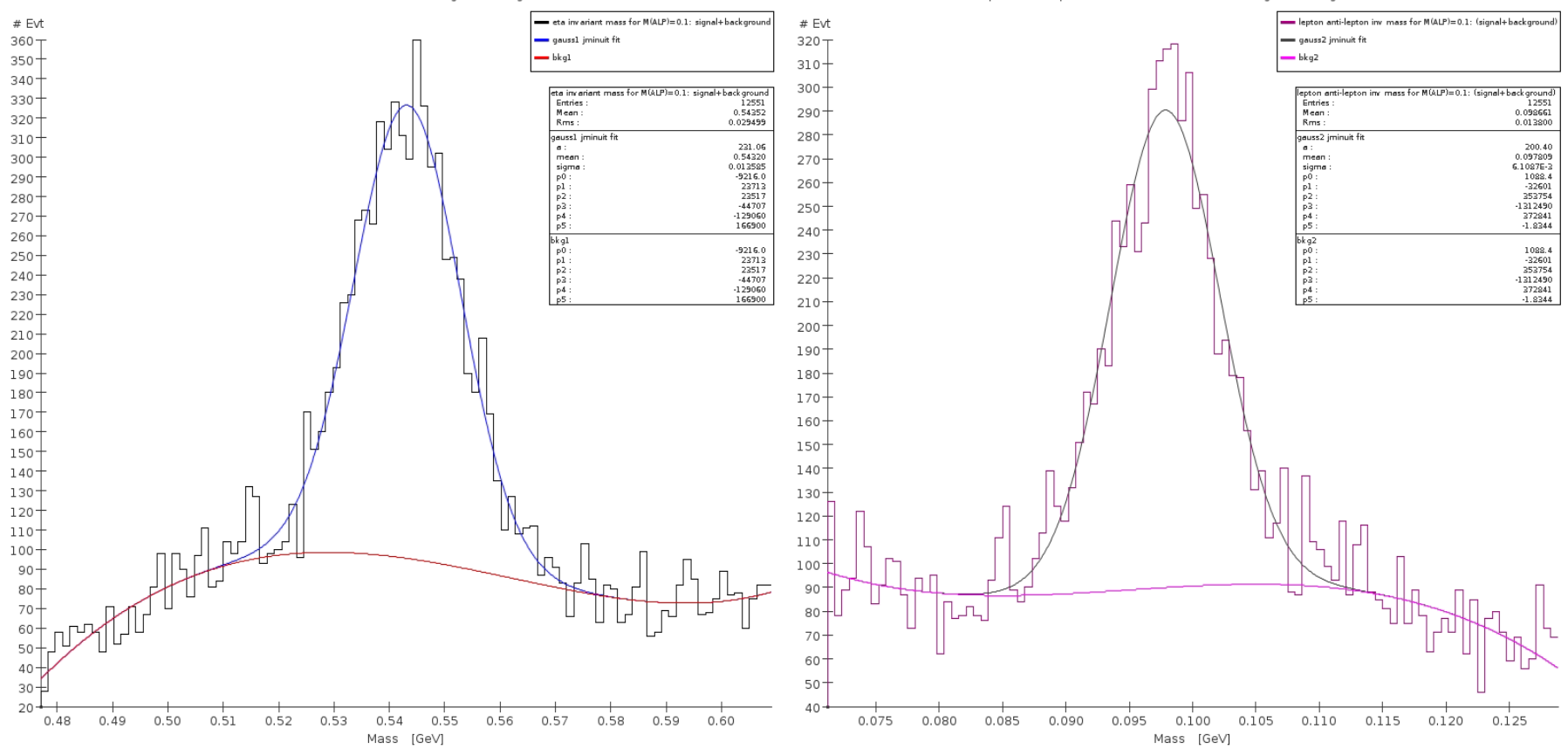}
\caption{Invariant mass of $\pi^{0}\pi^{0}e^{+}e^{-}$(left) and of the $e^{+}e^{-}$
system (right) for an \emph{a} with mass  M(\emph{a})=100 MeV. The plot includes
the Urqmd generated background. See text for an explanation of the
fitting procedure. }

\label{fig:eta2pi0pi0alp_ee}
\end{figure}
For illustrative purposes, Fig.~\ref{fig:eta2pi0pi0alp_ee}
shows the fit to the invariant mass of the reconstructed $\pi^{0}\pi^{0}e^{+}e^{-}$ for
an \emph{a} mass  M(\emph{a})=100 MeV, superimposed to the Urqmd generated background.
The plots assume a branching ratio BR($\eta\rightarrow\pi^{0}\pi^{0}a$) = $1.4\times10^{-4}$
and an $\eta$ sample of $5\times10^{8}$ (corresponding to $4.5\times10^{-6}$
of the full integrated luminosity).  

\begin{figure}[!ht]
\includegraphics[width=10cm]{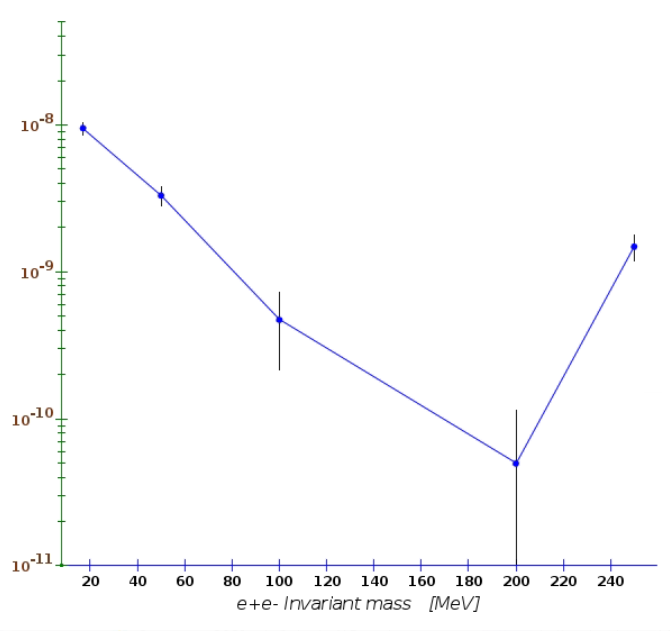} \caption{Branching ratio sensitivity for the process $\eta\rightarrow\pi^{0}\pi^{0}a\;with\;a\rightarrow e^{+}e^{-}$
as a function of the pseudoscalar boson \emph{a} mass.}

\label{fig:eta2pi0pi0alp_ee_br}
\end{figure}
The resulting sensitivity is shown in Fig.~\ref{fig:eta2pi0pi0alp_ee_br},
as a function of  \emph{M(a)}. As expected,
the curve has a lower sensitivity for values of \emph{M(a)} below 50 MeV,
due to the large contamination of the $\gamma\rightarrow e^{+}e^{-}$
background. The sensitivity extends in the $10^{-11}-10^{-9}$
range for higher \emph{M(a)} values. For values of \emph{M(a)} near to the
kinematic limits, the reconstruction efficiency, and then the sensitivity,
degrades again, as the recoiling $\pi^{0}\pi^{0}$ pair has a low
momentum and the photons from their decays have a lower probability to be
detected or reconstructed. 

\subsubsection{\texorpdfstring{$\eta\rightarrow\pi^{0}\pi^{0}a$: {\it detached-vertex}}{} analysis}

As stressed in the sections above, we expect that a pseudoscalar boson traveling
a measurable distance inside the detector will be disentangled more
easily from the underlying background,  resulting in an improved
sensitivity to that process. For the analysis of this channel, the pseudoscalar $a$ was
generated with a mass ranging between 17 MeV and 250 MeV. For each
value of the mass, the $c\tau$ of the pseudoscalar was varied from 20 mm
to 150 mm. A total of 20 event sets were generated and fully reconstructed.
The analysis for the kinematic variables follows the same guidelines
as for the $bump-hunt$ analysis. Further cuts on the $\chi^{2}$ of a
fit to the secondary vertex and on the distance between the primary and secondary vertexes
were applied. Since the $\beta\gamma$ boost of the particle depends
on the mass of the pseudoscalar particle $a$, those cuts were individually optimized for each 
event set. 

\begin{figure}[!ht]
\includegraphics[width=8cm]{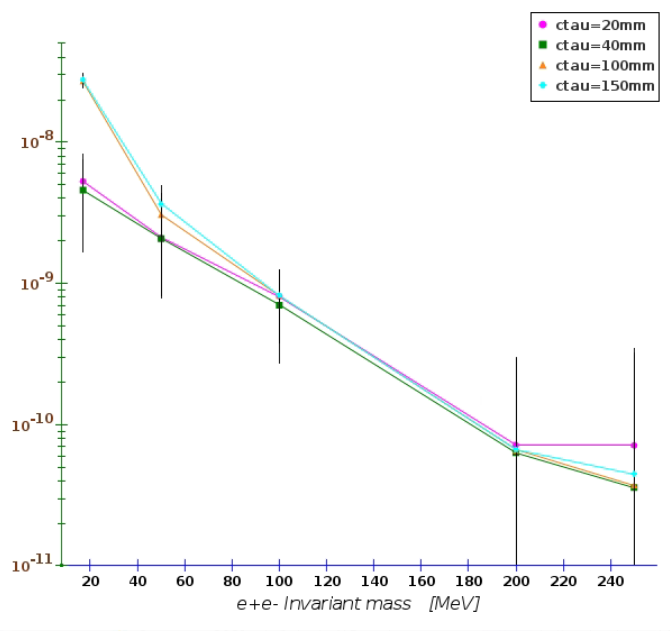} \caption{Branching ratio sensitivity for the process $\eta\rightarrow\pi^{0}\pi^{0}a\;with\;a\rightarrow e^{+}e^{-}$
as a function of the mass \emph{M(a)} and $c\tau$ of the a long-lived pseudoscalar
$a$.}

\label{fig:eta2pi0pi0ALP_ee_br-vtx}
\end{figure}

The total reconstruction efficiency for this process, with the additional
vertex cuts, was found to range between $\sim$0.2\% and $\sim$1.4\%
for the signal, and of order $\mathcal{O}(10^{-13})-\mathcal{O}(10^{-11})$
for the Urqmd background. The resulting branching ratio sensitivity
is shown in Fig.~\ref{fig:eta2pi0pi0ALP_ee_br-vtx}, as a function
of the invariant mass of the pseudoscalar. 
The effect of the cuts rejecting 
the $\pi^{0}\rightarrow\gamma\gamma$ background is clearly visible
for low values of \emph{M(a)}, where the sensitivity is degraded below 50 MeV. 
Furthermore, the $\beta\gamma$
boost for the case \emph{M(a)}=17 MeV is such that the  secondary vertex
is often located in the outermost region of the detector, where the 
charged track reconstruction efficiency is lower.
The region of higher mass values is relatively
free from combinatorics and Standard Model induced background, and
the $\frac{S}{\sqrt{B}}$ ratio is enhanced. The resulting branching ratio sensitivity, for that mass region,
extends below $10^{-10}$.

\subsection{Radiative decay: \texorpdfstring{$a\rightarrow\gamma\gamma$}{} final states}

As already noted above, in some theoretical models the coupling of the pseudoscalar to leptons
is suppressed. In that case, the preferred decay mode of \emph{a}
is into two photons. Therefore, in this work, we have  also included the study to
the radiative decay mode of the \emph{a} boson. The process considered is:
\begin{itemize}
\item $p+Li\rightarrow\eta+X\;with\;\eta\rightarrow\pi^{+}\pi^{-}a\;and\;a\rightarrow\gamma\gamma$
\end{itemize}
From the experimental point of view, this final state is affected
by a background complementary to the  leptonic decay mode. 
In this case, in fact, the feed-trough from the process $\gamma\rightarrow e^{+}e^{-}$
is almost non-existent, and the low region of\emph{M(a)} is expected to exhibit 
a higher sensitivity 
than for charged decay modes. 
On the other side, the $\eta\rightarrow\pi^{+}\pi^{-}\pi^{0}$
process contributes largely, and in an irreducible way, to the background
in the mass region around the $\pi^{0}$ mass. In the remaining mass
region, most of the background is due to combinatorics associated to QCD processes,
and to $\eta\rightarrow\pi^{+}\pi^{-}\pi^{0}$ events, when
one of the pions is mis-reconstructed. The second largest background
is due to the non-resonant process $\eta\rightarrow\pi^{+}\pi^{-}\gamma$, 
which survives the event selection when a second photon is found in the event faking
the $\eta\rightarrow\pi^{+}\pi^{-}\pi^{0}$ kinematics.

The study of the $a\rightarrow\gamma\gamma$ decay is restricted,
in the present work, to the $bump-hunt$ analysis only, since the reconstruction
of a detached $\gamma\gamma$ vertex is beyond the current capabilities
of the reconstruction software. Later studies will consider also a
$detached-vertex$ analysis, since the high-granularity of the ADRIANO2 detector 
could allow to reconstruct the direction and the impinging point of a photon. 
The
pseudoscalar boson \emph{a} was generated in a mass range between
17 MeV and 250 MeV. The boson was decayed promptly in \emph{GenieHad},
by setting the vertex of the two photons to the $\eta$ meson production point.
The full chain of generation-simulation-reconstruction-analysis
was repeated for each set of generated events. Very generic requirements
on the quality of reconstructed particles were applied to signal
and background samples.

Neutral pions, decaying into $\gamma e^{+}e^{-}$ and $\gamma\gamma$,
where reconstructed by considering all combinations of photons, electrons
and positrons with an invariant mass within 5 MeV from $\pi^{0}$
mass. This requirement is able to reject most of the combinatoric
background, where no $\eta$ mesons are present in the final state. Similarly, photons
converting into a $e^{+}e^{-}$ pair where reconstructed by requiring
that the invariant mass of the $e^{+}e^{-}$ system was lower than
5 MeV. The events were, finally, required to have a topology consistent
with a $\eta\rightarrow\pi^{+}\pi^{-}\gamma\gamma$ final state, and
an invariant mass compatible with the $\eta$ mass. 

The final reconstruction efficiency for this process was found to
range between $\sim$6\% and $\sim$14\%, strongly dependent on the $\gamma\,\gamma$
invariant mass. The reconstruction efficiency was of order $\mathcal{O}(10^{-10})-\mathcal{O}(10^{-8})$
for the Urqmd background. 

\begin{figure}[!ht]
\includegraphics[width=9cm]{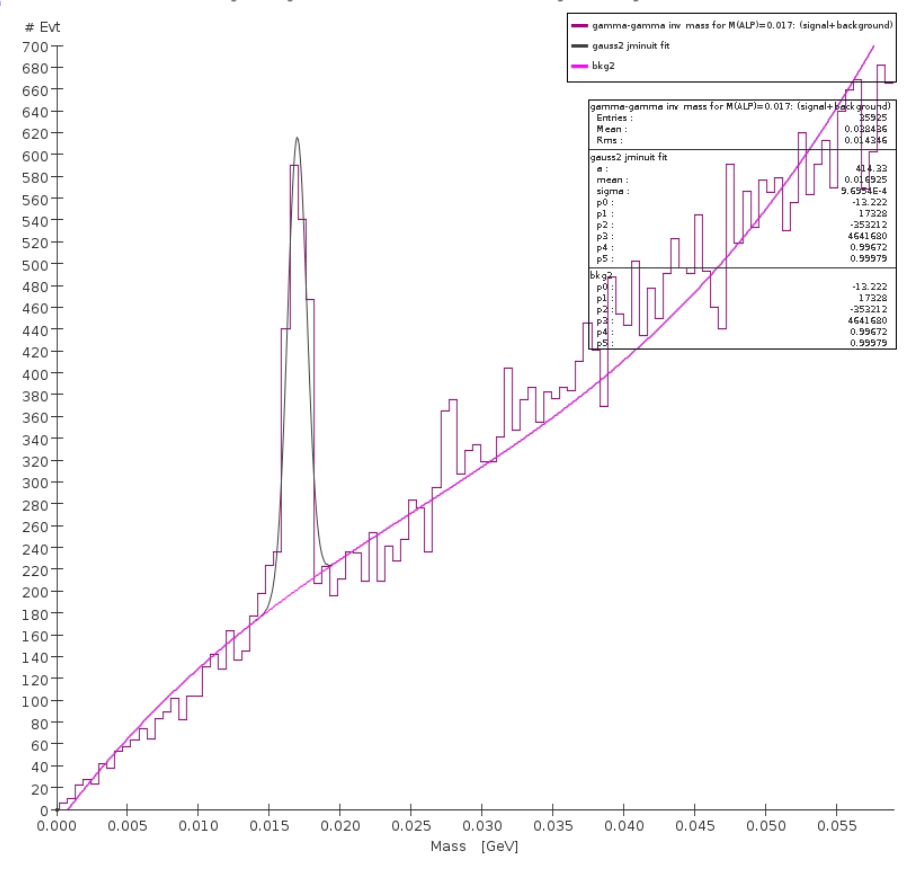}
\caption{Invariant mass of the $\gamma\,\gamma$ system (right) for a pseudoscalar boson \emph{a} 
with mass \emph{M(a)}=17 MeV. The plot includes the Urqmd generated background. See text for an explanation of the fitting procedure.}

\label{fig:eta2pippimalp_gg}
\end{figure}
For illustrative purposes, Fig.~\ref{fig:eta2pippimalp_gg} shows
the fit to the invariant mass of the reconstructed $\gamma\gamma$ system
for a mass of the pseudoscalar \emph{M(a)}=17 MeV, merged  with the Urqmd generated background.
The plot assumes a branching ratio BR($\eta\rightarrow\pi^{+}\pi^{-}a$)=$1.4\times10^{-4}$
and an $\eta$ sample consisting of $5\times10^{8}$ events (corresponding to $4.5\times10^{-6}$
of the full integrated luminosity). The number of reconstructed signal
and background events was obtained form a fit to the $\eta$ meson
invariant mass using the sum of a Gaussian and of a 5th-order polynomial.
The integral of the fitted function is used to extract the branching
ratio for this process. 

\begin{figure}[!ht]
\includegraphics[width=8cm]{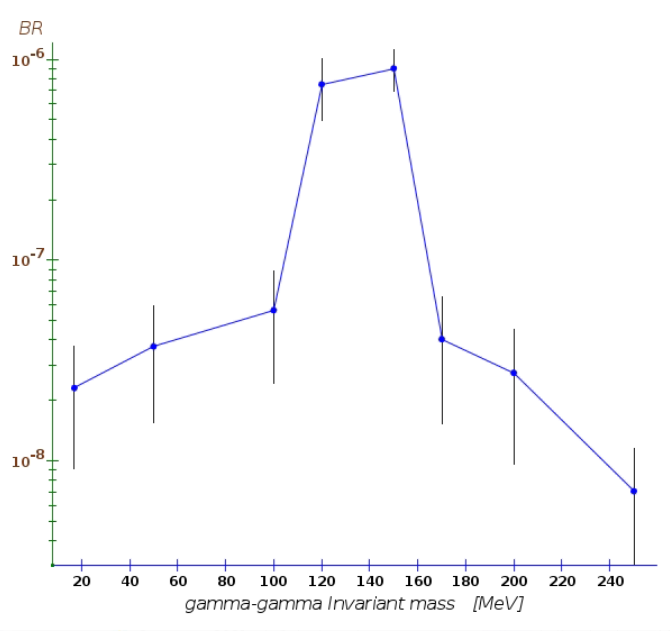} \caption{Branching ratio sensitivity for the process $\eta\rightarrow\pi^{+}\pi^{-}a\;with\;a\rightarrow\gamma\gamma$
as a function of the mass of the pseudoscalar boson \emph{a}}.

\label{fig:eta2pippimalp_gg_br}
\end{figure}
The sensitivity obtained from this analysis is shown in Fig.~\ref{fig:eta2pippimalp_gg_br},
as a function of the pseudoscalar boson \emph{a}  mass. As expected, the region around
the $\pi^{0}$ mass has a much lower sensitivity, due to the large and irreducible
background deriving from the non-resonant $\eta\rightarrow\pi^{+}\pi^{-}\pi^{0}$ process. 

\subsection{Sensitivity to selected theoretical models}
QCD axions have been suggested as 
a natural solution to the strong CP problem~\cite{Weinberg1978}. Current QCD axion models are restricted to the sub-eV range masses for the axion (although special variants of the QCD axion are also viable in the MeV mass range; see~\cite{Alves:2017avw}). A generalization of the minimal model to axion-like particles (ALPs)~\cite{Jaec10} allows for models with more massive particles. ALPs can be explored in REDTOP via $\eta/\eta^{\prime}\rightarrow\pi^{0}\pi^{0}a$ and $\eta/\eta^{\prime}\rightarrow\pi^{+}\pi^{-}a$ processes. Depending on the model and the mass of the decaying meson, the ALP can be observed in $a\rightarrow\gamma\gamma$, $a\rightarrow\pi\pi\pi$, $a\rightarrow l^{+}l^{-}.$

Three models, related to the pseudoscalar portal, are currently under consideration by REDTOP: the \emph{piophobic axion model}, the \emph{axion-like particles with quark dominance}, and \emph{axion-like particles with gluon dominance}. 
It should be noted that all of these models require UV completion at an appropriate scale.

\subsubsection{Sensitivity to Axion-Like Particles}

For the discussion of this class of models,
we follow the formalism presented in Sec.~\ref{subsec:Axion-like-particles}.
The relevant parameters in
this model are the coupling constants 
$c_{GG}$, $c_{q}$  and the ALP decay constant $f_{a}$ defined in Eq. \ref{LalpY}.
The branching ratio sensitivities have been derived above for the processes $\eta\rightarrow\pi^{0}\pi^{0}a$
and $\eta\rightarrow\pi^{+}\pi^{-}a$,  for several decay modes
and different values of the decay length of the ALP \emph{a}. 
For the determination of REDTOP sensitivity to $c_{GG}$ and $c_{q}$, 
we made the conservative assumption that the \emph{a} decays only into Standard Model particles. In that case, the values of the coupling constants are also related to the width of \emph{a}, and, consequently, to the capability of the detector to reconstruct detached vertices.

\begin{figure}[!ht]
\includegraphics[width=7cm]{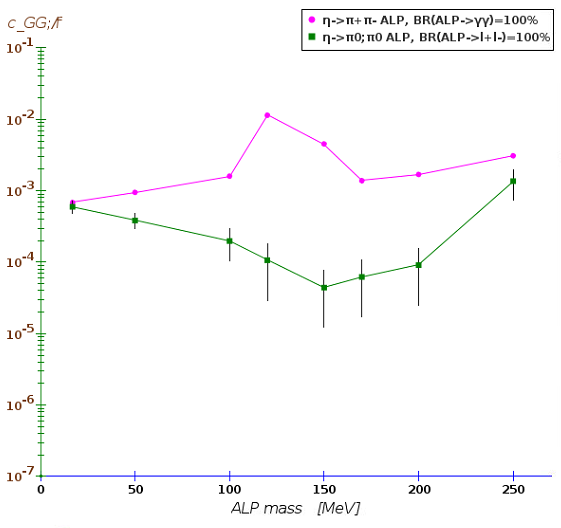} 
\includegraphics[width=7cm]{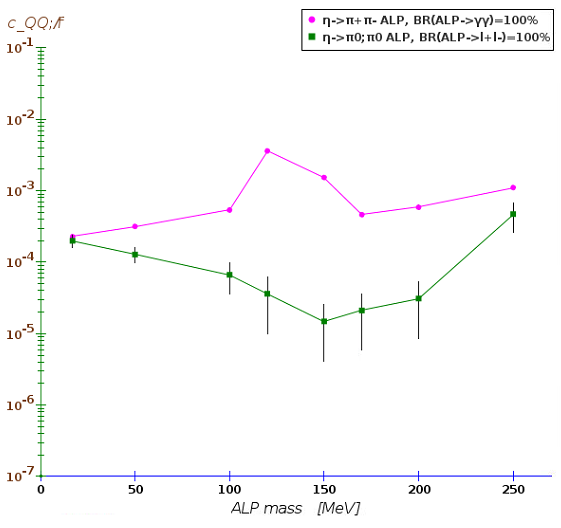} 
\caption{Sensitivity to $c_{GG}/f_{a}$ (left) and $c_{QQ}/f_{a}(=c_{q}/f_{a})$ (right) for the processes $\eta\rightarrow\pi^{+}\pi^{-}a$ and $\eta\rightarrow\pi^{o}\pi^{o}a$
as a function of the mass of a the ALP \emph{a}. The magenta curves refer to the decay $a\rightarrow e^{+}e^{-}$ while the green curves are for the case: $a\rightarrow \gamma\gamma$. See text for details of the analysis.}
\label{fig:eta2pipiALP_cgg_cqq}
\end{figure}

Using the branching ratios values shown in  Fig.~\ref{fig:eta2pippimalp_ee_br}
though \ref{fig:eta2pippimalp_gg_br}, the corresponding sensitivity
curves for $c_{GG}/f_{a}$ and  $c_{q}/f_{a}(=c_{QQ}/f_{a})$  are shown in Fig, \ref{fig:eta2pipiALP_cgg_cqq},
for the final states considered in this work.

\subsubsection{\emph{Piophobic QCD axion with mass of 17 MeV/\texorpdfstring{$c^{2}$}{} }}

Sensitivity studies for this process are performed separately and
with different selection cuts, since the underlying theoretical model (cf.\ Sec.~\ref{subsec:Searches-for-QCDaxion})  predicts precisely the branching ratio and the momentum distribution of the axion.
Furthermore, contrary to the models considered above, which are effective
interactions, and typically require UV completion at or below
the coupling constant scales, the QCD piophobic axion model does not require
external UV completion. 
The final states considered for this study are:
\begin{itemize}
\item $p+Li\rightarrow\eta+X\;with\;\eta\rightarrow\pi^{0}\pi^{0}a\;and\;a\rightarrow e^{+}e^{-}$
\item $p+Li\rightarrow\eta+X\;with\;\eta\rightarrow\pi^{+}\pi^{-}a\;and\;a\rightarrow e^{+}e^{-}$
\end{itemize}
Three separate sets
with $\sim$350,000  events were generated, each corresponding to a benchmark parameter
sets listed in Table I of Ref.~\cite{Spier2021}. A forth sample, with the parameters corresponding to
the Standard Model prediction, was also generated. Each set  was generated with a matrix
element obtained from Eq. (53) of Ref.\cite{Spier2021} and $M(a)$=17
MeV. The corresponding $\eta$ sample ranges between $1.9\times10^{8}$ and
$4.2\times10^{8}$ (or, $3.8\times10^{-6}$- $7.7\times10^{-6}$ of the full integrated luminosity), as the branching ratio for this process
depends on the values of the parameter set. 

The full chain
of generation-simulation-reconstruction-analysis was repeated for
each set of events. Very generic requirements on the quality
of reconstructed particles was applied to the signal and background
samples. 
Neutral pions, decaying into $\gamma e^{+}e^{-}$ and $\gamma\gamma$,
where reconstructed by considering all combinations of photons, electrons
and positrons with an invariant mass within 5 MeV from $\pi^{0}$
mass. This requirement is able to reject most of the combinatoric
background, where no $\eta$ mesons are present in the final state. Similarly, photons
converting into a $e^{+}e^{-}$ pair where reconstructed by requiring
that the invariant mass of the $e^{+}e^{-}$ system was lower than
5 MeV. The events were, finally, required to have a topology consistent
with a $\eta\rightarrow\pi^{0} e^{+} e^{-}$ final state, and
an invariant mass compatible with the $\eta$ mass. 

The reconstruction efficiencies for this process and for
the Urqmd generated background are summarized in Table~\ref{table:qcdaxioneff}
along with the sensitivity obtained by using Eq.~\eqref{eq:MasterS-final}
for  3.3$\times 10^{18}$ POT.

\begin{figure}[!ht]
\includegraphics[width=7cm]{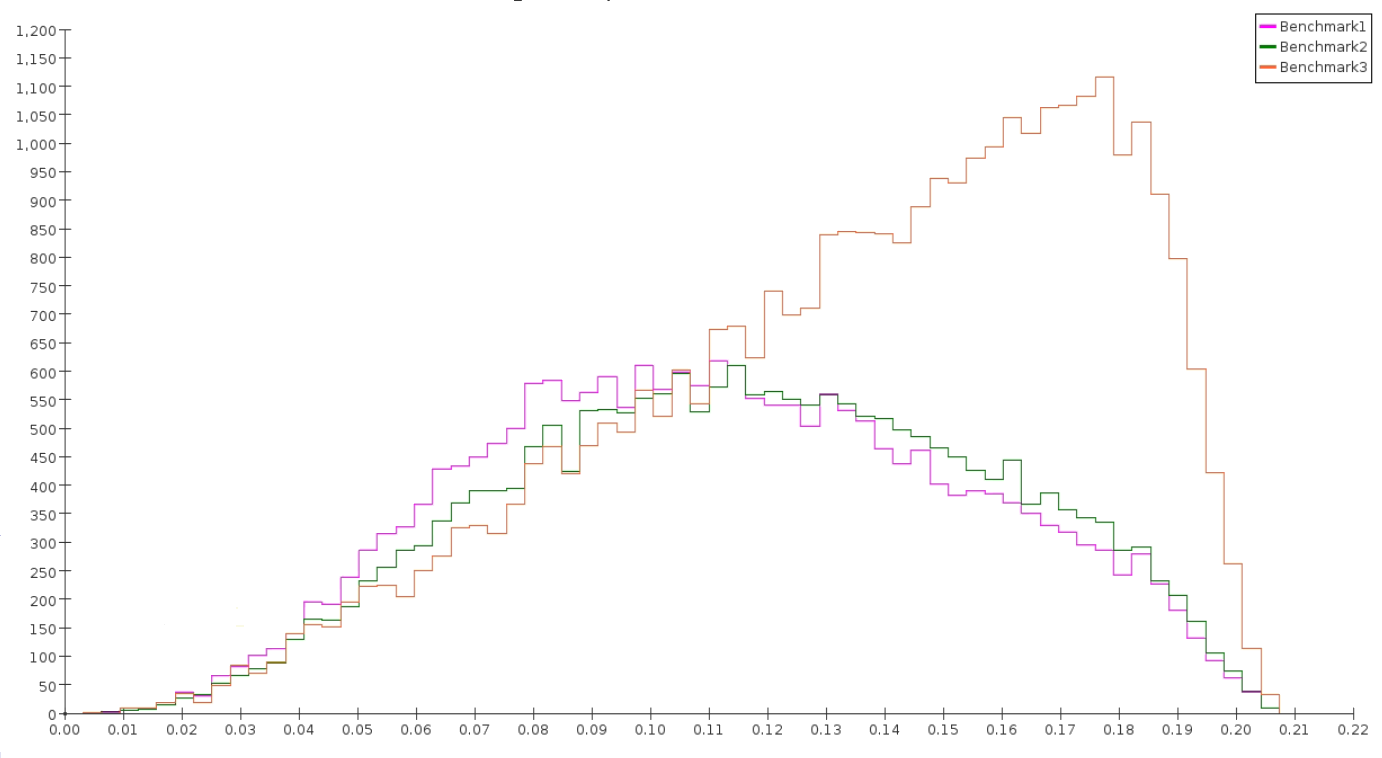} \includegraphics[width=7cm]{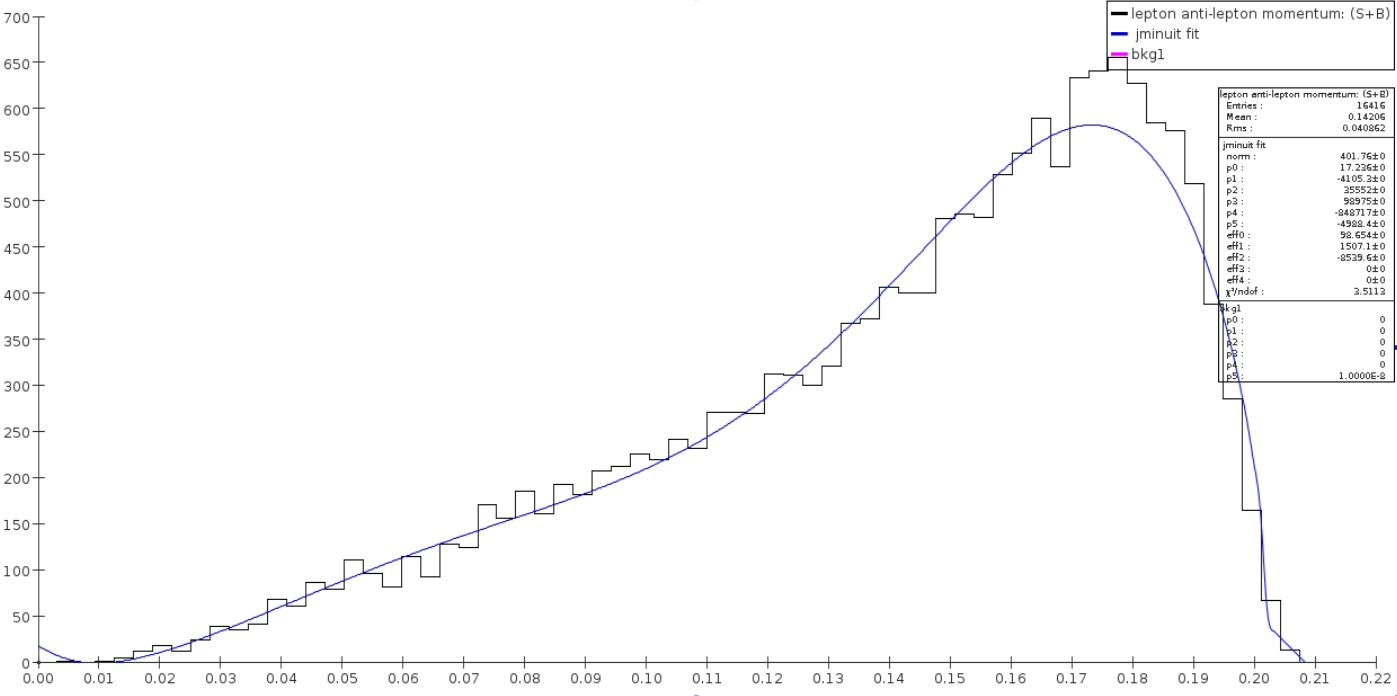}
\caption{The axion momentum in the $\eta$ CM system for the three parameter
sets considered in the study (left). Fit to the distribution corresponding to the
benchmark set \#3 (right) from Ref.~\cite{Spier2021}. See text for
an explanation of the fitting procedure. }
\label{fig:qcd-axion-plots-1}
\end{figure}

\begin{table}[!ht]
\centering\textcolor{black}{\scriptsize{}}%
\begin{tabular}{|c|c|c|c|c|c|c|c|c|}
\hline 
\textit{\textcolor{black}{\scriptsize{}Process}} & \textcolor{black}{\scriptsize{}Benchmark} & \textit{\textcolor{black}{\scriptsize{}Trigger}} & \textit{\textcolor{black}{\scriptsize{}Trigger}} & \textit{\textcolor{black}{\scriptsize{}Trigger}} & \textit{\textcolor{black}{\scriptsize{}Reconstruction}} & \textcolor{black}{\scriptsize{}Analysis} & \textbf{\textcolor{black}{\scriptsize{}Total}} & \textcolor{black}{\scriptsize{}BR}\tabularnewline
 & \textcolor{black}{\scriptsize{}set} & \textcolor{black}{\scriptsize{}L0} & \textcolor{black}{\scriptsize{}L1} & \textcolor{black}{\scriptsize{}L2} &  &  &  & \textcolor{black}{\scriptsize{}sensitivity}\tabularnewline
\hline 
\hline 
\textcolor{black}{\scriptsize{}$\eta$$\rightarrow\pi^{+}\pi^{-}a\;;\;a\rightarrow e^{+}e^{-}$} & \textcolor{black}{\scriptsize{}B1} & \textcolor{black}{\scriptsize{}$55.28\%$} & \textcolor{black}{\scriptsize{}$21.81\%$} & \textcolor{black}{\scriptsize{}$76.41\%$} & \textcolor{black}{\scriptsize{}75.12\%} & \textcolor{black}{\scriptsize{}42.94\% } & \textcolor{black}{\scriptsize{}$2.97$\%} & \textcolor{black}{\scriptsize{}$2.07\times10^{-8}$}\tabularnewline
\hline 
\textcolor{black}{\scriptsize{}$\eta$$\rightarrow\pi^{+}\pi^{-}a\;;\;a\rightarrow e^{+}e^{-}$} & \textcolor{black}{\scriptsize{}B2} & \textcolor{black}{\scriptsize{}56.15\% } & \textcolor{black}{\scriptsize{}22.32\% } & \textcolor{black}{\scriptsize{}76.76\% } & \textcolor{black}{\scriptsize{}75.12\% } & \textcolor{black}{\scriptsize{}42.83\% } & \textcolor{black}{\scriptsize{}3.10\% } & \textcolor{black}{\scriptsize{}$1.98\times10^{-8}$}\tabularnewline
\hline 
\textcolor{black}{\scriptsize{}$\eta$$\rightarrow\pi^{+}\pi^{-}a\;;\;a\rightarrow e^{+}e^{-}$} & \textcolor{black}{\scriptsize{}B3} & \textcolor{black}{\scriptsize{}59.67\% } & \textcolor{black}{\scriptsize{}23.06\% } & \textcolor{black}{\scriptsize{}79.81\% } & \textcolor{black}{\scriptsize{}76.14\% } & \textcolor{black}{\scriptsize{}44.03\% } & \textcolor{black}{\scriptsize{}3.68\% } & \textcolor{black}{\scriptsize{}$1.67\times10^{-8}$}\tabularnewline
\hline 
\textcolor{black}{\scriptsize{}Urqmd } &  & \textcolor{black}{\scriptsize{}$21.7\%$} & \textcolor{black}{\scriptsize{}$1.7\%$} & \textcolor{black}{\scriptsize{}$22.2\%$} & \textcolor{black}{\scriptsize{}$0.26$\%} & \textcolor{black}{\scriptsize{}1.04\% } & \textcolor{black}{\scriptsize{}$2.31\times10^{-6}$\%} & \tabularnewline
\hline 
\end{tabular}\caption{Reconstruction efficiencies for $\eta$$\rightarrow\pi^{+}\pi^{-}a\;;\;a\rightarrow e^{+}e^{-}$
for the piophobic axion model~\cite{Spier2021} and for the
Urqmd generated background}
\label{table:qcdaxioneff}
\end{table}
The $\eta$ content in each group is extracted with a fit using the
sum of a Gaussian and a 5th-order polynomial, and by integration of
the Gaussian signal after background subtraction. The signal and background
content for the events corresponding to the benchmark set \#1, is
shown in Fig.~\ref{fig:qcd-axion-plots-1}. The statistical error from the fit
is used to derive the uncertainty
of the branching ratio. The errors for the three benchmark sets are
summarized in the third column of Table~\ref{table:qcdaxionfit}.

\begin{figure}[!ht]
\includegraphics[width=12cm]{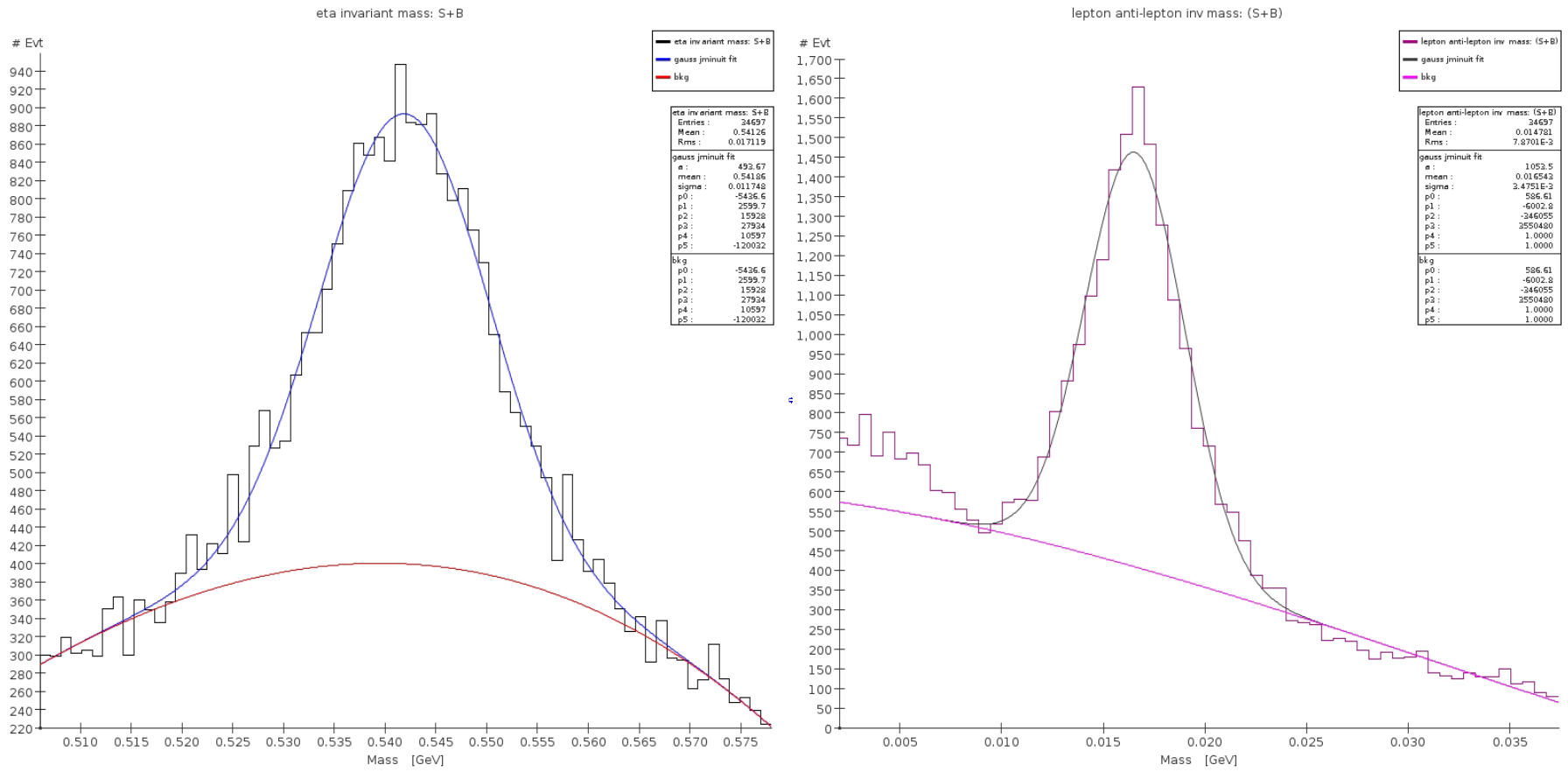} 

\caption{Invariant mass of $\pi^{+}\pi^{-}e^{+}e^{-}$(left) and of the $e^{+}e^{-}$
system (right) for a piophobic axion. The plots includes the Urqmd
generated background. See text for an explanation of the fitting procedure. }

\label{fig:qcd-axion-plots}
\end{figure}

Final cuts of, respectively 16 MeV and 4 MeV were applied around the
invariant mass of the $\pi^{+}\pi^{-}e^{+}e^{-}$ systems and of the
reconstructed axion. Those cuts aim at increasing the purity of the signal
before performing the final fit to extract the model parameters. 
In order to estimate
the sensitivity of the experiment to this model, the distribution of the axion momentum
is fitted using the the theoretical
matrix element from Ref.~\cite{Spier2021}, combined with a second
degree polynomial describing the background. 
To take into account the non-uniform acceptance of the detector, a multiplicative polynomial
efficiency was used in the fit. 
The analysis was repeated for each of the three sets of events generated. 
The fit to the axion momentum distribution corresponding
to the benchmark set \#3  is shown on the right side of Fig.
\ref{fig:qcd-axion-plots-1}. The resulting $\chi^{2}$ probability
for the three sets is summarized in Table~\ref{table:qcdaxionfit}.

\begin{table}[ht]
\centering%
\begin{tabular}{|c|c|c||c|}
\hline 
\textit{POT} & $\eta$$\rightarrow\pi^{+}\pi^{-}a\;;\;a\rightarrow e^{+}e^{-}$ & BR  & $\chi^{2}$/ndof \tabularnewline
 & Benchmark set & statistical error & \tabularnewline
\hline 
\hline 
1.1$\times10{}^{13}$ & B1 & 1.5\% & 3.2\tabularnewline
\hline 
1.3$\times10{}^{13}$ & B2 & 1.4\% & 2.9\tabularnewline
\hline 
5.6$\times10{}^{12}$ & B3 & 2.1\% & 3.5\tabularnewline
\hline 
\end{tabular}\caption{Goodness of fit of the P$_{axion}$ distribution using the matrix
element from Ref.~\cite{Spier2021} }
\label{table:qcdaxionfit}
\end{table}

It is worth noting that an interesting possibility of a QCD axion
with small $f_a$ opens additional avenues for $a$ searches. In particular,
the property of piophobia, together with a 17 MeV mass imply 
a rather large decay rate of a neutral pion to three axions, $\pi^0\to 3a$.
This decay has a branching ratio of $10^{-3}$~\cite{Hostert:2020xku},
and will create three electron-positron pairs as each $a$ decays.
Therefore, one can expect $2\pi 3 (e^+e^-)$ decay mode of $\eta$
with the branching of $\sim 10^{-5}-10^{-7}$. Moreover, if $a$'s
are discovered in a singles production channel, the exploration of
multiple $a$ production can support or refute the hypothesis of the
QCD axion, as nonlinear self-interaction terms of $a$ are necessary
for a QCD axion and are fixed by the requirements of the Peccei-Quinn
invariance. 


\section{Sensitivity to the Heavy Neutral Lepton portal\label{subsec:Sensitivity-to-the-heavylepton}}

\subsection{\texorpdfstring{$\eta\rightarrow\pi^{0}H\;;\;H\rightarrow\nu N_{2}\;;\;N_{2}\rightarrow N_{1}h'\;;\;h'\rightarrow e^{+}e^{-}$}{}:
Detached vertex analysis}

The studies presented in this Section are based on the model discussed
in Sec.~\ref{subsec:Scalar-portal-Models}. The benchmark parameters of that model are such that the process is kinematically allowed only for $\eta^{\prime}$
mesons. To test the branching ratio sensitivity at REDTOP, the 
values listed in Table~\ref{tab-REDTOP} are used instead:

\textcolor{black}{\begin{table}[ht!] \begin{center}  \begin{tabular}
{|c|c|c|c|c|}   
\hline   $m_{N_1}$& $m_{N_2}$ & $m_{N_3}$ &$y_{e(\mu)}^{h'}\!\!\times\!\! 10^{4}$&$y_{e(\mu)}^{H}\!\!\times\!\! 10^{4}$ \\   \hline    85\,MeV  & $130$\,MeV & $10$\,GeV &$0.23(1.6)$&$2.29(15.9)$ \\   \hline     \hline   $m_{h'}$& $m_{H}$ &$\sin\delta$  &$y_{\nu_{\!i 2}}^{h'(H)}\!\!\times\!\! 10^{3}$&$\lambda_{N_{\!12}}^{h'(H)}\!\!\times\!\! 10^{3}$ \\[0.05cm]   \hline    17\,MeV  & 250\,MeV &$0.1$ &$1.25(12.4)$&$74.6(-7.5)$ \\   \hline 
\end{tabular} \caption{Benchmark parameters for REDTOP.} \label{tab-REDTOP}  
\vspace{-0.5cm}
\end{center} 
\end{table}} 

The branching ratios of the decay modes of $H$ are recomputed according to the new parameters. The new values are as follow:  
\begin{eqnarray} &&\!\!\!\!\!\!\!\!\!\!\!\!\!\!\!\!\!\!\!\!\!\!{\rm BR}(H \rightarrow \nu_i N_2)= 96.93\%,~{\rm BR}(H \rightarrow N_1 N_2)= 2.90\%,~{\rm BR}(H \rightarrow \mu^+ \mu^-)= 0.15\%,\label{eq:REDTOP-1} \\ &&\!\!\!\!\!\!\!\!\!\!\!\!\!\!\!\!\!\!\!\!\!\!{\rm BR}(H \rightarrow e^+ e^-)= 0.021\%,~~\! {\rm BR}(H \rightarrow \gamma \gamma)= 5.36 \times 10^{-6}\%. \label{eq:REDTOP-2}
\end{eqnarray} 

For the parameters in Table~\ref{tab-REDTOP}, the total decay width of $H$, $h'$, and $N_2$ are $2.52\times 10^{-6}$~GeV, $3.56\times 10^{-13}$~GeV, and $2.07\times 10^{-5}$~GeV, respectively.
Hence all the particles will decay inside the REDTOP. However, the
width of $h'$ corresponds to $c\tau=620~\mu m$ so that the $e^{+}e^{-}$
pair has a detached secondary vertex, eventually resolvable with an
appropriate choice of a vertex detector.

The event generation for
this process is based on the model described in Sec.~\ref{subsec:CP-Violation-from-Dalitz}
for a reduced $\eta$ sample corresponding to $\sim$2.7$\times$$10^{13}$~POT (namely,
to $\sim$2.7$\times$$10^{-5}$ of the integrated luminosity foreseen for
the experiment). The full chain of generation-simulation-reconstruction-analysis
was repeated for each set of the parameters and for the parameters
corresponding to the Standard Model prediction. The largest background
contaminating the signal is due to the decays $\gamma\rightarrow e^{+}e^{-}$
and $\pi^{0}\rightarrow\gamma e^{+}e^{-}$, paired to an extra
photon, usually generated from the decay of a $\pi^{0}$ or an $\eta$
meson, which mimic a $\pi^{0}$.
 Very generic requirements on the
quality of reconstructed particles was applied to the signal and background
samples. 
Neutral pions, decaying
into $\gamma e^{+}e^{-}$ and $\gamma\gamma$, where reconstructed
by considering all combinations of photons, electrons and positrons
with an invariant mass within 5 MeV from $\pi^{0}$ mass. This requirement
is able to reject most of the combinatoric background with no $\eta$
mesons in the final state. Similarly, photons converting into a $e^{+}e^{-}$
pair where reconstructed by requiring that the invariant mass of the
$e^{+}e^{-}$ system was lower than 5 MeV.  
From the experimental point of view, this final state
suffer from the presence of three missing particles. Therefore, no
constraint could be imposed on the invariant mass of the detected
particles. On the other side, the presence of a fully reconstructed
$e^{+}e^{-}$ pair with a detached vertex could help to reject the
Standard model background. Therefore,  on the $\chi^{2}$ from the fit of two charged tracks to a common vertex and on the distance between the
primary and secondary vertexes were applied. 
The goal of those cuts was to remove events
with particles originating from the $\eta$ production point. 
The reconstruction efficiencies for this process and for the
Urqmd generated background are summarized in Table~\ref{table:recoeff_HNL}.

\begin{table}[!ht]
\centering\textcolor{black}{\scriptsize{}}%
\begin{tabular}{|c|c|c|c|c|c|c|c|}
\hline 
\textit{\textcolor{black}{\scriptsize{}Process}} & \textit{\textcolor{black}{\scriptsize{}Trigger}} & \textit{\textcolor{black}{\scriptsize{}Trigger}} & \textit{\textcolor{black}{\scriptsize{}Trigger}} & \textit{\textcolor{black}{\scriptsize{}Reco}} & \textit{\textcolor{black}{\scriptsize{}Analysis}} & \textbf{\textcolor{black}{\scriptsize{}Total}} & \textbf{\textcolor{black}{\scriptsize{}BR}}\tabularnewline
 & \textcolor{black}{\scriptsize{}L0} & \textcolor{black}{\scriptsize{}L1} & \textcolor{black}{\scriptsize{}L2} &  &  &  & \textbf{\textcolor{black}{\scriptsize{}Sensitivity}}\tabularnewline
\hline 
\hline 
\textcolor{black}{\makecell{\scriptsize{}$\eta\rightarrow\pi^{0}H\;;\;H\rightarrow\nu N_{2}\;;$ \\ \scriptsize{}$\;N_{2}\rightarrow N_{1}h'\;;\;h'\rightarrow e^{+}e^{-}$}}
& \textcolor{black}{\scriptsize{}$38.5\%$} & \textcolor{black}{\scriptsize{}$22.6\%$} & \textcolor{black}{\scriptsize{}$80.5\%$} & \textcolor{black}{\scriptsize{}$91.1$\%} & \textcolor{black}{\scriptsize{}$19.6\%$} & \textcolor{black}{\scriptsize{}$1.3$\%} & \textbf{\scriptsize{}$2.7\times10^{-7}\pm7\times10^{-9}$}\tabularnewline
\hline 
\textcolor{black}{\scriptsize{}Urqmd } & \textcolor{black}{\scriptsize{}$21.7\%$} & \textcolor{black}{\scriptsize{}$1.7\%$} & \textcolor{black}{\scriptsize{}$22.2\%$} & \textcolor{black}{\scriptsize{}$47.7$\%} & \textcolor{black}{\scriptsize{}$0.17\%$} & \textcolor{black}{\scriptsize{}$6.6\times10^{-5}$\%} & \tabularnewline
\hline 
\end{tabular}\caption{Reconstruction efficiencies and branching ratio sensitivity for $\eta\rightarrow\pi^{0}H\;;\;H\rightarrow\nu N_{2}\;;\;N_{2}\rightarrow N_{1}h'\;;\;h'\rightarrow e^{+}e^{-}$
and for the Urqmd generated background with the parameters used in
Table~\ref{table:recoeff_HNL}.}
\label{table:recoeff_HNL}
\end{table}

\subsection{Sensitivity to selected theoretical models}

For the studies presented in this work, we have considered only one theoretical model, which is discussed below. 

\subsubsection{\label{subsec:Two-Higgs-doublet-model-scalar-1}Two-Higgs doublet
model }

The parameters $\lambda_u$ and $\lambda_d$ for this model, along
with the branching ratios predicted for decay channels $\eta^{(\prime)} \to \pi^0 S$
are discussed in Sec.~\ref{subsec:Scalar-portal-Models}. For the Two-Higgs
doublet model considered in this work~\cite{Abdallah:2020vgg}, the
branching ratio, in the assumption that $\lambda_u = \lambda_d$,
is predicted to be of order $\mathcal{O}(10^{-13})$, which is below
REDTOP sensitivity in the present run. The situation is different
when $\lambda_{u}\neq\lambda_{d}$. 
In that case, assuming the values for the
branching ratios of the $H$ defined in Eqs.~(\ref{eq:REDTOP-1})$,$(\ref{eq:REDTOP-2}), and the
branching ratios of the $N_{2}$ and $h'$   defined in Eqs.~(\ref{eq:REDTOP-3})$,$(\ref{eq:REDTOP-4}) we derive the sensitivity
to $(\lambda_u-\lambda_d)^2$ case from the branching ratio 
sensitivity derived in Sec.~\ref{subsec:Sensitivity-to-the-heavylepton}
and from Eq.~(\ref{eq:BR__eta_decay_to_scalar-Approx}). The final value
is summarized in Table~\ref{table:delta-lambda-sensitivity-1} 

\begin{figure}[!ht]
\includegraphics[scale=0.4]{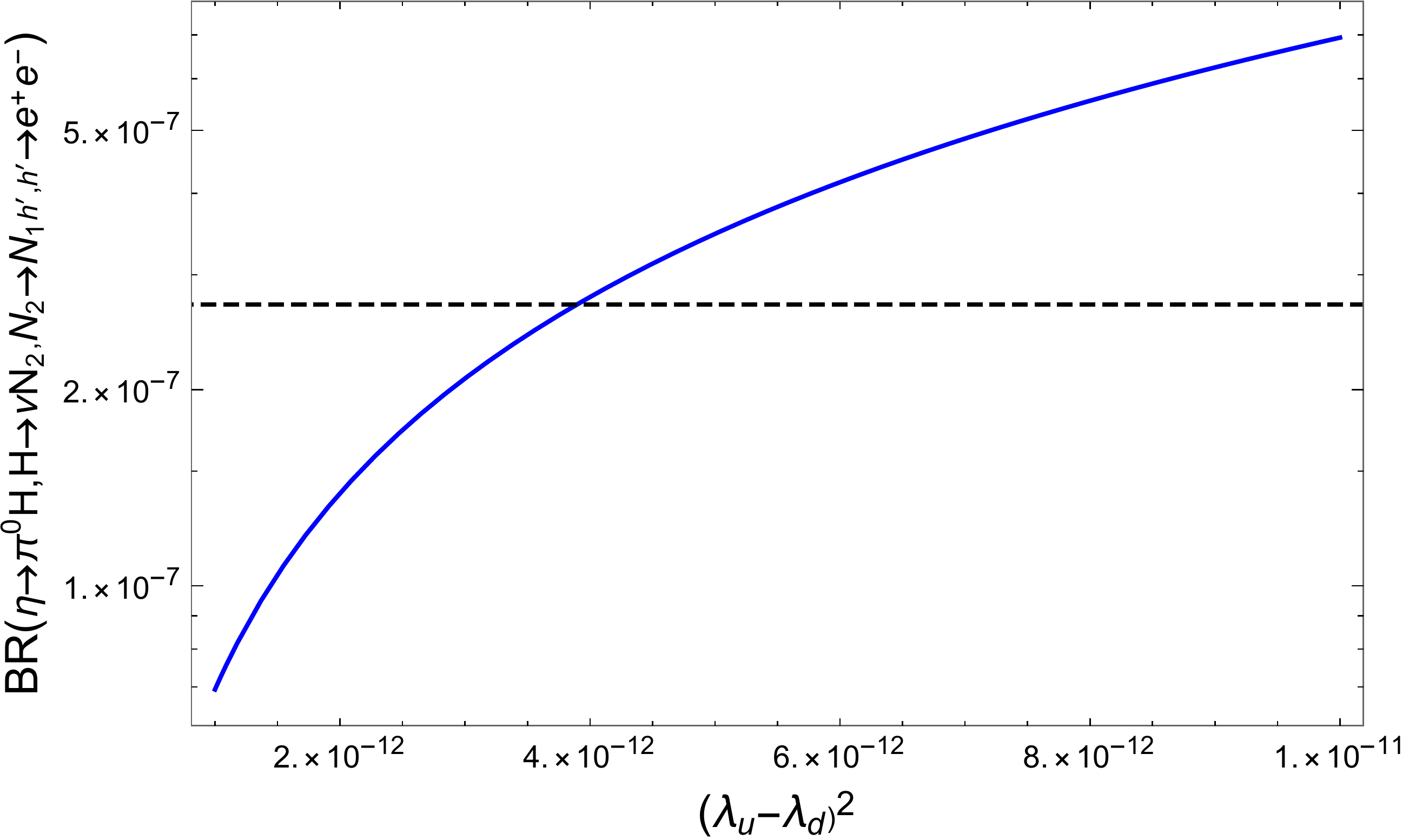} \caption{Branching ratio for the process \textcolor{black}{\footnotesize{}$\eta\rightarrow\pi^{0}H\;;\;H\rightarrow\nu N_{2}\;;\;N_{2}\rightarrow N_{1}h'\;;\;h'\rightarrow e^{+}e^{-}$}
predicted by the Two Higgs Doublet model~\cite{Abdallah:2020vgg}
as a function of $(\lambda_u-\lambda_d)^2$. The dashed line
corresponds to the experimental limit for REDTOP with an integrated
luminosity of 3.3$\times10^{18}$~POT.}

\label{fig:sensitivity-HNL}
\end{figure}

\begin{table}[!ht]
\centering\textcolor{black}{\scriptsize{}}%
\begin{tabular}{|c|c|c|c||c|}
\hline 
\textit{\textcolor{black}{\scriptsize{}Process}} & \textcolor{black}{\scriptsize{}$m_{h'}$} & \textcolor{black}{\scriptsize{}$m_{N_2}$} & \textit{\textcolor{black}{\scriptsize{}Analysis}} & \textcolor{black}{\scriptsize{}$(\lambda_{u}-\lambda_{d})^{2}$}\tabularnewline
 & \textcolor{black}{\scriptsize{}{[}MeV{]}} & \textcolor{black}{\scriptsize{}{[}MeV{]}} &  & \textcolor{black}{\scriptsize{}sensitivity}\tabularnewline
\hline 
\hline 
\textcolor{black}{\scriptsize{}$\eta\rightarrow\pi^{0}H\;;\;H\rightarrow\nu N_{2}\;;\;N_{2}\rightarrow N_{1}h'\;;\;h'\rightarrow e^{+}e^{-}$} & \textcolor{black}{\scriptsize{}17 } & \textcolor{black}{\scriptsize{}250 } & \textcolor{black}{\scriptsize{}detached vertex} & \textcolor{black}{\scriptsize{}$3.89\times10^{-12}$}\tabularnewline
\hline 
\end{tabular}\caption{Sensitivity to $(\lambda_{u}-\lambda_{d})^{2}$
for the Two-Higgs doublet
model~\cite{Abdallah:2020vgg}.}
\label{table:delta-lambda-sensitivity-1}
\end{table}
 and shown in Fig.~\ref{fig:sensitivity-HNL} superimposed
to the prediction of the theoretical model. 

\subsection{Concluding remark}

The reconstruction of the process $\eta\rightarrow\pi^{0}H\;;\;H\rightarrow\nu N_{2}\;;\;N_{2}\rightarrow N_{1}h'\;;\;h'\rightarrow e^{+}e^{-}$
is particularly challenging at REDTOP, because the decay chains contains
two missing particles: namely, the two neutral leptons $N_{1}$ and $N_{2}$.
Furthermore, the decay length of the $h'$ scalar is less than 1 mm,
which is resolved poorly by the fiber-tracker detector implemented
in the current studies. All that is reflected in a considerably
lower sensitivity to this channel, when compared to most of the other
processes analyzed in this work, where all the particles in the final
state are visible. A better vertex detector, such as the ITS3 option
discussed in Sec.~\ref{par:Vertex-detector-option-II}) which has the ability to
resolve with a higher efficiency the detached vertex of the $h'$
scalar, could improve considerably REDTOP sensitivity to this channel. 

\section{Tests of Conservation Laws\label{subsec:Tests-of-Conservation-Laws}}

\subsection{\emph{CP-}violation studies from Dalitz plot mirror asymmetry in \texorpdfstring{$\eta\rightarrow\pi^{+}\pi^{-}\pi^{0}$}{}}
The event generation for this process is based on the model described
in Sec.~\ref{subsec:CP-Violation-from-Dalitz}.
Several event samples were generated, corresponding to different values of the parameters
consisting of $\sim1.1\times10^{8}$ $\eta$ mesons from 3.3$\times10^{12}$~POT (corresponding to 1.0$\times10^{-6}$ of the total integrated
luminosity foreseen for the experiment).

As discussed in detail in Sec.~\ref{subsec:CP-Violation-from-Dalitz}),
the matrix element generating the asymmetry, in the Gardner-Shi parametrization
scheme (cf.\ Ref.\cite{gardner2019patterns}), depends on two
complex parameters $\bar{\alpha}$ and $\bar{\beta}$. In order to
estimate the sensitivity of REDTOP to this process, we have scanned
the real and the imaginary parts of $\bar{\alpha}$ and $\bar{\beta}$
in steps of 1.5 $\sigma$ around the values obtained by authors from
a fit to Kloe-II data (eq. 14 in Ref.\cite{gardner2019patterns}).
An additional event set was generated, with the parameters corresponding
to the Standard Model prediction.

The full chain of generation-simulation-reconstruction-analysis was
repeated for each event set. Very generic requirements on the
quality of reconstructed particles was applied to the signal and background
samples. Neutral pions, decaying into $\gamma e^{+}e^{-}$ and $\gamma\gamma$,
where reconstructed by considering all combinations of photons, electrons
and positrons with an invariant mass within 5 MeV from $\pi^{0}$
mass. Finally, the events were required to have a topology consistent
with a $\eta\rightarrow\pi^{+}\pi^{-}\pi^{0}$ final state, and an
invariant mass compatible with the $\eta$ mass. 
A final kinematic
cut of, respectively 6 MeV and 5 MeV around the invariant mass of
the three pions and of the reconstructed $\pi^{0}$ meson was applied
to increase the purity of the signal used for the analysis of the
Dalitz plot. The largest background for this channel was found to originate
from combinatorics $p+Li\rightarrow\pi^{+}\pi^{-}\pi^{0}+X$, when
the invariant mass of the three pions  survived the kinematic cuts 
and it was consistent with the decay of the $\eta$ meson.
The reconstruction efficiencies for this process and for the Urqmd
generated background are summarized in Table~\ref{table:recoeff_eta3pi}

\begin{table}[!ht]
\centering\textcolor{black}{\small{}}%
\begin{tabular}{|c|c|c|c|c|c|c|}
\hline 
\textit{\textcolor{black}{\small{}Process}} & \textit{\textcolor{black}{\small{}Trigger}} & \textit{\textcolor{black}{\small{}Trigger}} & \textit{\textcolor{black}{\small{}Trigger}} & \textit{\textcolor{black}{\small{}Reconstruction}} & Analysis & \textbf{\textcolor{black}{\small{}Total}}\tabularnewline
 & \textcolor{black}{\small{}L0} & \textcolor{black}{\small{}L1} & \textcolor{black}{\small{}L2} &  &  & \tabularnewline
\hline 
\hline 
\textcolor{black}{\small{}$\eta$$\rightarrow\pi^{+}\pi^{-}\pi^{0}$} & \textcolor{black}{\small{}$48.6\%$} & \textcolor{black}{\small{}$10.9\%$} & \textcolor{black}{\small{}$82.7\%$} & \textcolor{black}{\small{}$96.3$\%} & \textcolor{black}{\small{}$6.46$\%} & \textcolor{black}{\small{}$2.7\times10^{-2}$\%}\tabularnewline
\hline 
\textcolor{black}{\small{}Urqmd } & \textcolor{black}{\small{}$21.7\%$} & \textcolor{black}{\small{}$1.7\%$} & \textcolor{black}{\small{}$22.2\%$} & \textcolor{black}{\small{}$4.3$\%} & \textcolor{black}{\small{}$0.93$\%} & \textcolor{black}{\small{}$3.2\times10^{-5}$\%}\tabularnewline
\hline 
\end{tabular}\caption{Reconstruction efficiencies for $\eta\rightarrow\pi^{+}\pi^{-}\pi^{0}$
and for the Urqmd generated background}
\label{table:recoeff_eta3pi}
\end{table}
\begin{figure}[!ht]
\includegraphics[width=9cm,height=8cm]{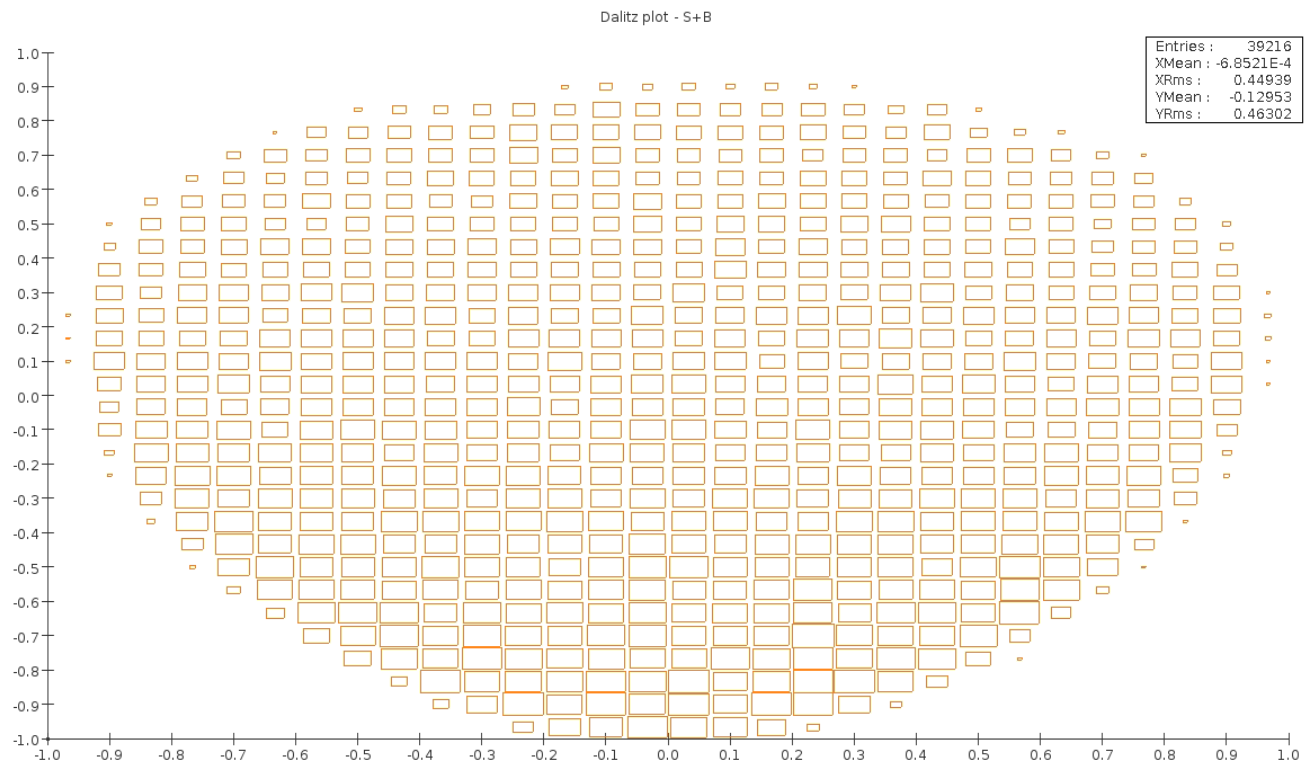} \caption{Dalitz plot of $\eta\rightarrow\pi^{+}\pi^{-}\pi^{0}$  combined with the Urqmd background.}

\label{fig:eta3pi_mass}
\end{figure}
A total of 18 Dalitz plots were generated, corresponding to the \emph{CP}-violating
parameters described above. The Dalitz plot of $\eta\rightarrow\pi^{+}\pi^{-}\pi^{0}$
for the set of parameters corresponding to Kloe-II is shown in Fig.
\ref{fig:eta3pi_mass}.


The parameters obtained from the fits to the Standard Model configuration, where the \emph{CP}-violating parameters are all zero, are summarized
in Table~\ref{table:recoeff_eta3pi-fit}. The same binning of the Dalitz plot as in KLOE experiment was used with 372 bins in the $(X,Y)$ plane. 
\begin{table}[!ht]
\centering\textcolor{black}{\small{}}%
\begin{tabular}{|l|c|c|c|c||c|}
\hline 
\textit{\textcolor{black}{\small{}\#Rec. Events}} & {\textcolor{black}{\small{}Re($\alpha$)}} & {\textcolor{black}{\small{}Im($\alpha$)}} & {\textcolor{black}{\small{}Re($\beta$)}} & {\textcolor{black}{\small{}Im($\beta$)}} & \textbf{\textcolor{black}{\small{} p-value}}\tabularnewline
\hline 
\textcolor{black}{\small{}$10^8$ (no-bkg)} & \textcolor{black}{\small{}$3.3\times10^{-1}$} & \textcolor{black}{\small{}$3.7\times10^{-1}$} & \textcolor{black}{\small{}$4.4\times10^{-4}$} & \textcolor{black}{\small{}$5.6\times10^{-4}$} & \textcolor{black}{\small{}$17$\%}\tabularnewline
\textcolor{black}{\small{}Full stat. (no-bkg)} & \textcolor{black}{\small{}$1.9\times10^{-2}$} & \textcolor{black}{\small{}$2.1\times10^{-2}$} & \textcolor{black}{\small{}$2.5\times10^{-5}$} & \textcolor{black}{\small{}$3.2\times10^{-5}$} & \textcolor{black}{\small{}$17$\%}\tabularnewline
\textcolor{black}{\small{}Full stat. (100\%-bkg)} & \textcolor{black}{\small{}$2.3\times10^{-2}$} & \textcolor{black}{\small{}$3.0\times10^{-2}$} & \textcolor{black}{\small{}$3.5\times10^{-5}$} & \textcolor{black}{\small{}$4.5\times10^{-5}$} & \textcolor{black}{\small{}$16$\%}\tabularnewline
\hline 
\end{tabular}\caption{Sensitivities -- statistical uncertainties for the two complex Dalitz plot CPV parameters  $\alpha$ and $\beta$ according to Ref.
\cite{gardner2019patterns}. The generated distribution is the Standard Model configuration i.e., $\alpha,\beta=0$. The fit reproduces the generated input, with the p-values given in the last column, within the statistical uncertainties given for the real and imaginary parts of the parameters as given in the columns 2--5.}
\label{table:recoeff_eta3pi-fit}
\end{table}
All parameters are consistent with the generated values, within the
fit error. The latter have the same order of magnitude as those obtained
from a fit to KLOE data, using a comparable event size (cf.\ Ref.\cite{gardner2019patterns}).
The projected sensitivities for the event sample obtained with the full
integrated luminosity of $3.3\times10^{18}$ POT (corresponding to $3\times10^{10}$ reconstructed $\eta\to\pi^+\pi^-\pi^0$ events) are given for case of no-background and 100\%-background contamination.
They are much smaller than from the KLOE result and potentially sufficient to detect non-zero values
of the \emph{CP}-violating parameters.

\begin{singlespace}
\subsection{\emph{CP-}violation studies from the asymmetry of the decay planes of \texorpdfstring{$\eta\rightarrow\pi^{+}\pi^{-}e^{+}e^{-}$\uline}{}{\label{subsec:CP-violation-from-asymmetry-in-pipiee}}}
\end{singlespace}

The formal aspects of this process, including the relationship between
the \emph{CP-}violating parameters and the observable, has been discussed
in details in Sec \ref{subsec:CP-violation-in-eta2pipiee}

\begin{figure}[!ht]
\includegraphics[width=10cm]{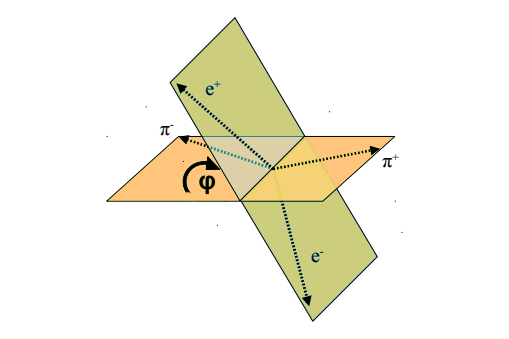} 

\caption{Definition of the dihedral angle $\phi$ for the $\eta\rightarrow\pi^{+}\pi^{-}e^{+}e^{-}$
decay in rest frame of the the $\eta$ meson.}

\label{fig:eta2pipiee_phi}
\end{figure}

This measurement has been previously attempted by the WASA-at-COSY
experiment~\cite{WASA07} using an event sample containing  3 \texttimes{}
10$^{7}$ $\eta$ mesons. The sensitivity to this process is determined
by examining at the asymmetry: 

\begin{equation}
A_{\phi}=\frac{N(\sin\phi\cos\phi>0)-N(\sin\phi\cos\phi<0)}{N(\sin\phi\cos\phi>0)+N(\sin\phi\cos\phi<0)}
\end{equation}

where $\phi$ is the angle between the decay planes of the lepton-antilepton
pair and the two charged pions (see Fig.~\ref{fig:eta2pipiee_phi}).
Five event samples, each corresponding to 6$\times10^{12}$ POT
(or $2\times10^{8}$ $\eta$-mesons), were generated using $GenieHad$
with $<sign(sin(\phi)cos(\phi)>$ in the range {[}0, $4\times10^{-2}${]}.
The full chain of generation-simulation-reconstruction-analysis was
repeated for each set of the parameters and for the parameters corresponding
to the Standard Model prediction. The largest background contaminating
the signal originates  from the decays $\gamma\rightarrow e^{+}e^{-}$
and $\pi^{0}\rightarrow\gamma e^{+}e^{-}$, simultaneously with the presence of two extra charged
pions  generated in  the beam-target collision. Very
generic requirements on the quality of reconstructed particles were
applied to signal and background samples. 

Neutral pions, decaying into $\gamma e^{+}e^{-}$ and $\gamma\gamma$,
where reconstructed by considering all combinations of photons, electrons
and positrons with an invariant mass within 5 MeV from $\pi^{0}$
mass. Successfully reconstructed $\pi^{0}$'s were removed from the
event. Similarly, the photons converting in the detector material
were reconstructed and removed from the event. Finally, the events
were required to have a topology consistent with a $pi^{+}\pi^{-}e^{+}e^{-}$
final state, and an invariant mass compatible with the $\eta$ mass.
The above selection rules were able to remove a large fraction of background
events.

The reconstruction efficiencies for the signal, averaged over the
five event sets, and for the Urqmd generated background are summarized
in Table~\ref{table:pippimee_eff} along with the sensitivity to the
branching ratio obtained  using Eq.~\eqref{eq:MasterS-final} for the full integrated beam current of
$3.3\times10^{18}$ POT.

\begin{figure}[!ht]
\includegraphics[width=7cm,height=7cm]{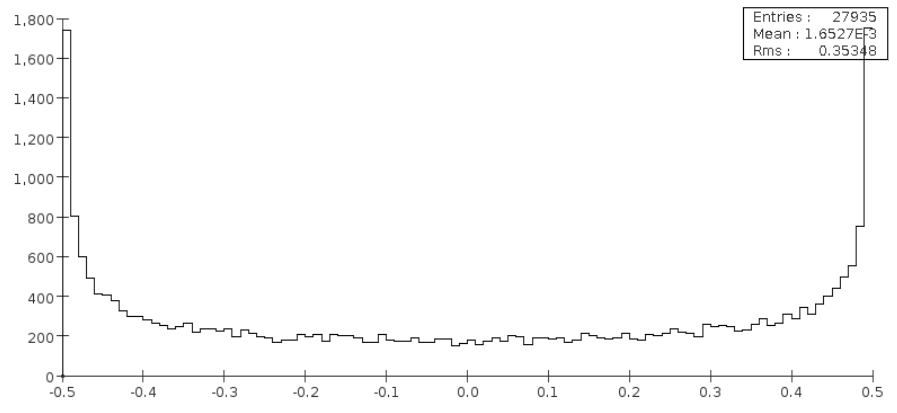}
\includegraphics[width=7cm]{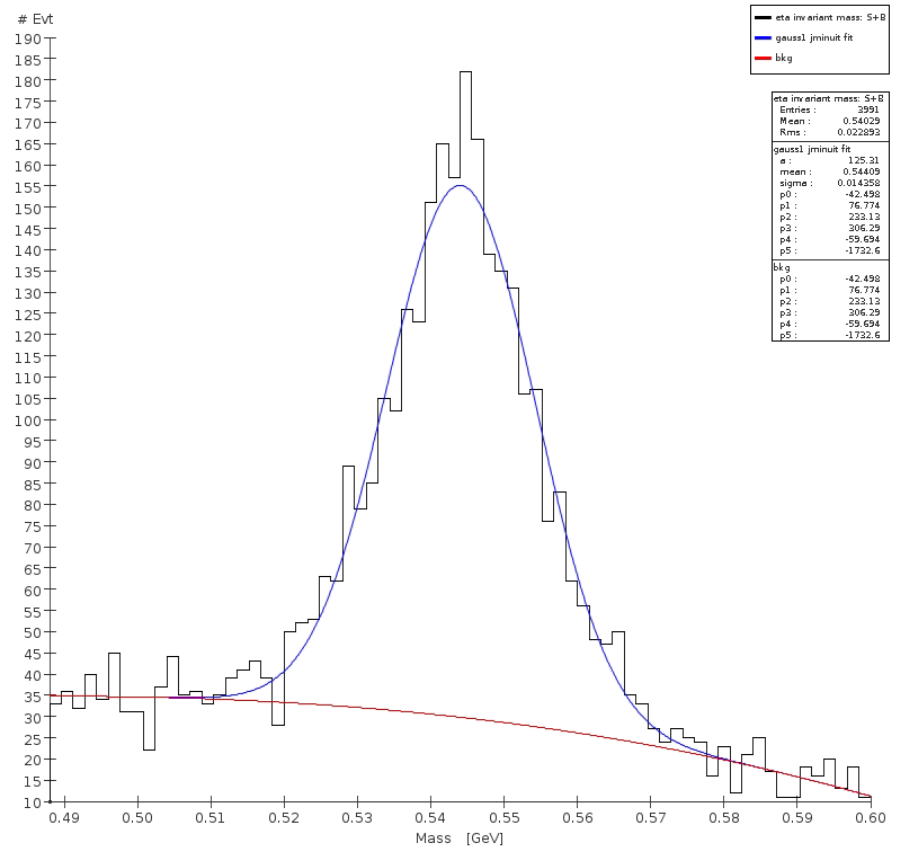} 

\caption{$sin\phi cos\phi$ distribution (left) and invariant mass of the {$\pi^{+}\pi^{-}e^{+}e^{-}$}
system (right) for the set of events with $<sign(sin(\phi)cos(\phi)>=0$.
The histograms include the Urqmd background. See text for an explanation
of the fitting procedure.}

\label{fig:eta2pipimee_mass}
\end{figure}

\begin{table}[!ht]
\centering\textcolor{black}{\small{}}%
\begin{tabular}{|c|c|c|c|c|c|c|c|c|}
\hline 
\textit{\textcolor{black}{\small{}Process}} & \textcolor{black}{\small{}Benchmark} & \textit{\textcolor{black}{\small{}Trigger}} & \textit{\textcolor{black}{\small{}Trigger}} & \textit{\textcolor{black}{\small{}Trigger}} & \textit{\textcolor{black}{\small{}Reco}} & \textcolor{black}{\small{}Analysis} & \textbf{\textcolor{black}{\small{}Total}} & \textcolor{black}{\small{}BR}\tabularnewline
 & \textcolor{black}{\small{}set} & \textcolor{black}{\small{}L0} & \textcolor{black}{\small{}L1} & \textcolor{black}{\small{}L2} &  &  &  & \textcolor{black}{\small{}sensitivity}\tabularnewline
\hline 
\hline 
\textcolor{black}{\small{}$\eta$$\rightarrow\pi^{+}\pi^{-}e^{+}e^{-}$} & \textcolor{black}{\small{}average.} & \textcolor{black}{\small{}$64.0
$} & \textcolor{black}{\small{}$34.2\%$} & \textcolor{black}{\small{}$83.1\%$} & \textcolor{black}{\small{}78.9\%} & \textcolor{black}{\small{}55.8\% } & \textcolor{black}{\small{}$8.0$\%} & \textcolor{black}{\small{}$4.9\times10^{-9}$}\tabularnewline
\hline 
\textcolor{black}{\small{}Urqmd } &  & \textcolor{black}{\small{}$21.7\%$} & \textcolor{black}{\small{}$1.7\%$} & \textcolor{black}{\small{}$22.2\%$} & \textcolor{black}{\small{}$0.26$\%} & \textcolor{black}{\small{}4\% } & \textcolor{black}{\small{}$8.6\times10^{-6}$\%} & \tabularnewline
\hline 
\end{tabular}\caption{Reconstruction efficiencies for the $\eta\rightarrow\pi^{+}\pi^{-}e^{+}e^{-}$
final state and for the Urqmd generated background. The values are
averaged over the five benchmark sets.}
\label{table:pippimee_eff}
\end{table}
The $sin\phi cos\phi$ distribution for the events belonging to the
$<sign(sin(\phi)cos(\phi)>=0$ set and surviving the analysis cuts
is shown in Fig.~\ref{fig:eta2pipimee_mass} along with the invariant
mass of the reconstructed $\pi^{+}\pi^{-}e^{+}e^{-}$ systems.
The combinatorics background is well fitted under the $\eta$ peak. 

\begin{figure}[!ht]
\includegraphics[width=8cm,height=7cm]{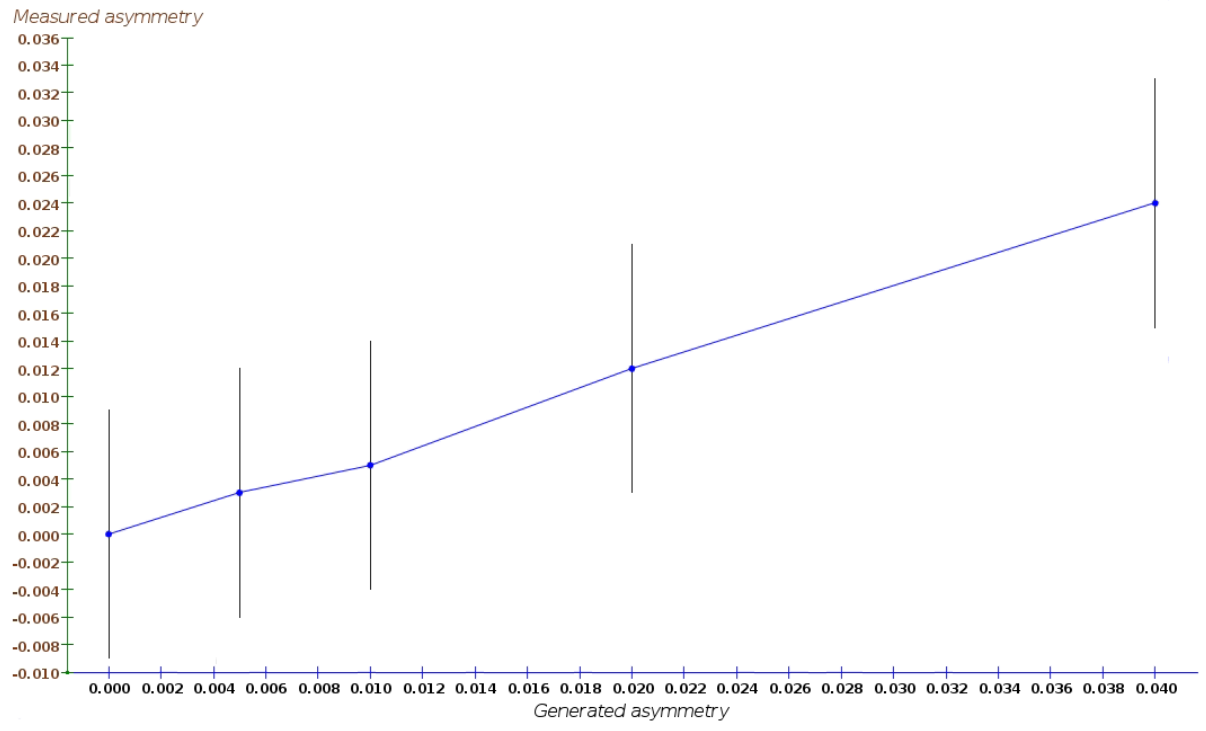} 

\caption{Sensitivity to $sin\phi cos\phi$ asymmetry for $\eta\rightarrow\pi^{+}\pi^{-}e^{+}e^{-}$.
The errors in the plot are statistical only.}

\label{fig:eta2pipimee_asymm}
\end{figure}

For the next step of the analysis, the data from each 
set were divided into separate $sin\phi cos\phi$>0
and $sin\phi cos\phi<0$ groups. The $\eta$ content in each group
was extracted with a fit using the sum of a Gaussian and a 5th-order
polynomial. The measured
asymmetry is plotted in Fig.~\ref{fig:eta2pipimee_asymm} vs the
true value for each of the five event sets. The diluting effect of
the background is clearly observed since the measured asymmetry is
attenuated by a factor of $\sim65\%$. We conclude that the measured
asymmetry for the samples considered is larger than the statistical
error (which is the dominant component in the measurements by
WASA and KLOE experiments) for values of the true asymmetry larger
than approximately $1\times10^{-2}$. This estimate is consistent
with the results obtained by the WASA  with an event sample a factor 100 smaller, as their measurement has a ten-times larger statistical error. When the total integrated luminosity
of $3.3\times10^{18}$ POT foreseen for REDTOP is taken into account,
we  expect that the statistical error will be reduced by more than
two orders of magnitude, corresponding to a contribution to the $A_{\phi}$ sensitivity
smaller than $\sim10^{-4}$. Therefore, we conclude that the uncertainty
on the measured asymmetry will be, most likely, dominated by the systematic
error.

\begin{singlespace}

\subsection{\emph{CP-}violation studies in the decays\textcolor{black}{{} }\texorpdfstring{$\eta\rightarrow\mu^{+}\mu^{-}$}{}}
\end{singlespace}

A detailed discussion on probing \emph{CP-}violation in leptonic $\eta$  decays can be found in Sec.~\ref{sec:CPVetaTo2L} and Sec.~\ref{subsec:CP-violation-in-eta-mumu}
along with the formulae relating the Wilson coefficients (carrying
the \emph{CP-}violating parameters) to the observable quantities.

The sensitivity to this process is studied by measuring the polarization
of the muon reconstructed from the following process:
\begin{itemize}
\item $p+Li\rightarrow\eta+X\;with\;\eta\rightarrow\mu^{+}\mu^{-}$
\end{itemize}
The two relevant observables are the two asymmetries defined in Eqs.(\ref{eq:AL},\ref{eq:AT}), that carry the information from the muon polarization. 
%
%
%
%
%
The $\eta\rightarrow\mu^{+}\mu^{-}$ process is
considered as the golden mode to explore $CP-violation$ since the operator
responsible for the asymmetry has the largest coupling to the observables. 
In order to measure the angles $\theta$ and $\phi$,
a full reconstruction of the $\mu^{\pm}\rightarrow e^{\pm}\nu\overline{\nu}$
decay needs to be performed. This capability will be implemented soon
in REDTOP Offline software. For the moment, only a rough estimate
of the expected polarization efficiency will be made.

The event generation for this process is based on the model described
in Sec.~\ref{sec:CPVetaToPi02L}. A sample consisting of $\sim8.6\times10^{9}$
$\eta$ mesons from 2.6$\times10^{14}$ POT (corresponding to 7.8$\times10^{-5}$
of the total integrated luminosity foreseen for the experiment). The
full chain of generation-simulation-reconstruction-analysis described
in Sec.~\ref{sec:Simulation-strategy} was repeated for each event
set. Very generic requirements on the quality of reconstructed particles
were applied to the signal and background samples. The main background
for this channel is due to the non-resonant decay $\eta\rightarrow\gamma\mu^{+}\mu^{-}$
which occurs with a relatively large branching ratio of $3.1\times10^{-4}$.
The second largest background  originates from charged
pions, mis-identified as muons. With the present detector configuration, the
$\pi$/$\mu$ mis-identification probability was estimate to be $\sim$3.5\%
(or, $\sim$0.12\% for mis-identifying both leptons). Since the probability
of generating two charged pions in the target is almost 11\% (see,
for example, Fig.~\ref{fig:urqmd_multiplicity}), we expect that about
$\sim3.3\times10^{11}$ events could possibly fake a $\eta\rightarrow\mu^{+}\mu^{-}$
process. In order to reduce such background, we impose  the stringent requirement
that no photons should be detected in the calorimeter.

\begin{figure}[!ht]
\includegraphics[width=7cm]{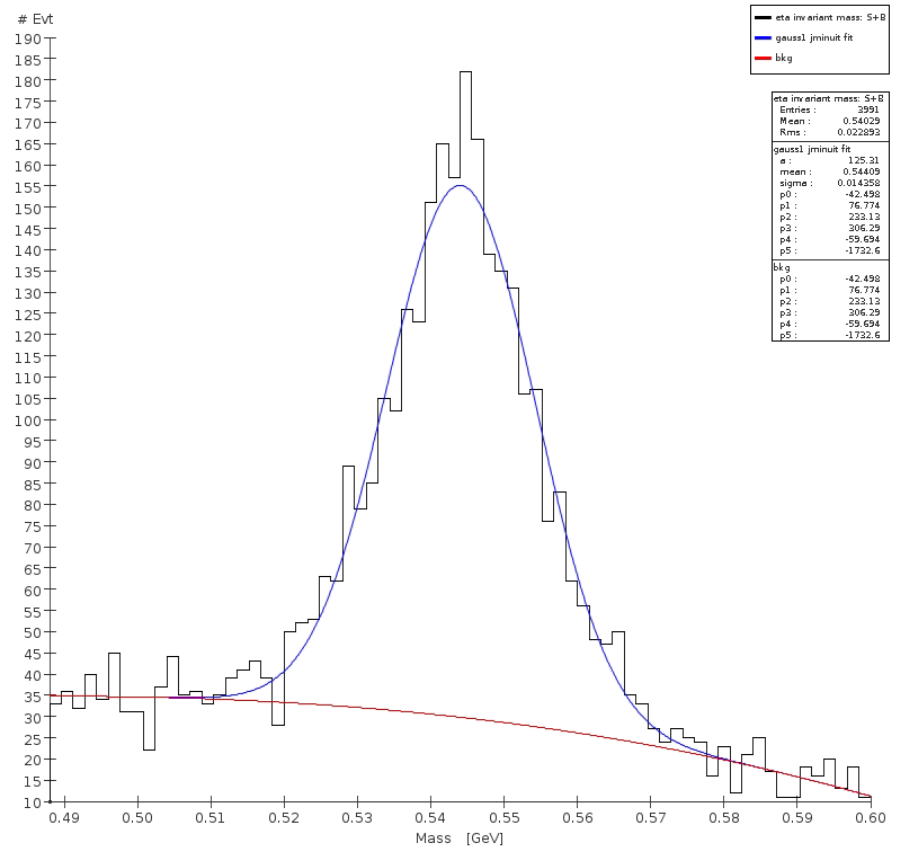} 

\caption{Invariant mass of the $\mu^{+}\mu^{-}$ system. The plot
includes the Urqmd background. See text for an explanation of the
fitting procedure.}

\label{fig:eta2leptonantilepton_mass}
\end{figure}

The reconstruction
efficiencies for this process for all reconstruction steps and for
the Urqmd background are summarized in Table~\ref{table:recoeff_leptonantilepton}.
The requirements discussed above drastically reduce the background, although at the cost of a reconstruction efficiency for the signal which is about ten times smaller than most of the other processes studied.
The invariant mass distribution of the reconstructed $\mu^{+}\mu^{-}$
systems for the signal and the Urqmd background is shown in Fig.~\ref{fig:eta2leptonantilepton_mass}.
The $\eta$ content was extracted with a fit using the sum of a Gaussian
and a 5th-order polynomial.  

The resulting branching ratio sensitivity is summarized in Table~\ref{table:recoeff_leptonantilepton}. 

\begin{table}[!ht]
\centering\textcolor{black}{\scriptsize{}}%
\begin{tabular}{|c|c|c|c|c|c|c|c|}
\hline 
\textit{\textcolor{black}{\scriptsize{}Process}} & \textit{\textcolor{black}{\scriptsize{}Trigger}} & \textit{\textcolor{black}{\scriptsize{}Trigger}} & \textit{\textcolor{black}{\scriptsize{}Trigger}} & \textit{\textcolor{black}{\scriptsize{}Reconstruction}} &  & \textbf{\textcolor{black}{\scriptsize{}Total}} & \textcolor{black}{\scriptsize{}Branching ratio}\tabularnewline
 & \textcolor{black}{\scriptsize{}L0} & \textcolor{black}{\scriptsize{}L1} & \textcolor{black}{\scriptsize{}L2} & \textit{\textcolor{black}{\scriptsize{}+ analysis}} &  &  & \textcolor{black}{\scriptsize{}sensitivity}\tabularnewline
\hline 
\hline 
\textcolor{black}{\scriptsize{}$\eta\rightarrow\mu^{+}\mu^{-}$} & \textcolor{black}{\scriptsize{}66.3\%} & \textcolor{black}{\scriptsize{}16.3\%} & \textcolor{black}{\scriptsize{}51.9\%} & \textcolor{black}{\scriptsize{}$69.6$\%} &  & \textcolor{black}{\scriptsize{}$3.9$\%} & \textcolor{black}{\scriptsize{}$2.7\times10^{-8}\pm3.0\times10^{-10}$}\tabularnewline
\hline 
\textcolor{black}{\scriptsize{}Urqmd } & \textcolor{black}{\scriptsize{}$21.7\%$} & \textcolor{black}{\scriptsize{}$1.7\%$} & \textcolor{black}{\scriptsize{}$22.2\%$} & \textcolor{black}{\scriptsize{}$8.6\times10^{-3}$\%} &  & \textcolor{black}{\scriptsize{}$7.0\times10^{-6}$\%} & \textcolor{black}{\scriptsize{}-}\tabularnewline
\hline 
\end{tabular}\caption{Reconstruction efficiencies for $\eta\rightarrow\mu^{+}\mu^{-}$and
for the Urqmd background. Errors are statistical only and are obtained
assuming 2.6$\times10^{14}$ POT.}
\label{table:recoeff_leptonantilepton}
\end{table}
The sensitivity to the Wilson coefficients is derived through the
asymmetry in Eq.~(\ref{eq:AL}) %
and will be limited by the statistical error on the signal diluted by the
background Incidentally, we observe that similar considerations apply also to the 
asymmetry defined in Eq.~(\ref{eq:AT}), that is, nevertheless, less sensitive to new physics. 
The values are obtained from Table~\ref{table:recoeff_leptonantilepton}
for 2.6$\times10^{14}$ POT, after scaling for the total integrated
luminosity foreseen for REDTOP (cf.\ Table~\ref{tab:Expected-yield-at}.
{} Combining them, we expect for the statistical error: 
\begin{equation}
\Delta(A_{L})=\frac{\sqrt{N}_{\eta\to\mu^+\mu^-} +\sqrt{N}_{\textrm{bkg}}}{N_{\textrm{\ensuremath{\eta\to\mu^{+}\mu^{-}}}}}.\label{eq:delta-a-l}
\end{equation}

Taking the expected $\eta$ and background events from Table~\ref{tab:Expected-yield-at} and the efficiencies above, we obtain $N_{\eta\to\mu^+\mu^-} = N_{\eta}\times \textrm{BR}(\eta\to\mu^{+}\mu^{+})\times\epsilon_{\textrm{reco}}\times\epsilon_{\textrm{pol}}$, together with $N_{\textrm{bkg}} = N_{ni} \times\epsilon_{\textrm{reco}}  \times\epsilon_{\textrm{pol}}$. Assuming a conservative $\epsilon_{\textrm{pol}} =  50\%$ estimate both for signal and background, we obtain $\Delta(A_{L})=2.7\times10^{-3}$.
 Comparing to Eq. (\ref{eq:AL}),
we find 
\begin{equation}
\Delta(c_{\ell equ}^{1122})=0.1\times10^{-1},\quad\Delta(c_{\ell edq}^{1122})=0.1,\quad\Delta(c_{\ell edq}^{2222})=6.6\times10^{-2}\label{eq:error-on-c2222},
\end{equation}
that for the $c_{\ell edq}^{2222}$ coefficient corresponds to the
same order of precision obtained from nEDM bounds.

\begin{singlespace}

\subsection{\emph{CP-}violation studies in the decays  \texorpdfstring{$\eta\rightarrow\gamma\mu^{+}\mu^{-}$}{}\label{subsec:CP-studies-in_gammamumu}}
\end{singlespace}

Polarization studies in Dalitz decays allow probing different aspects of discrete symmetries. Refer to Sec.~\ref{sec:muon-polarimetry} for a detailed discussion on that topic. Both, longitudinal or transverse polarization studies provide interesting tests of the SM.
In the following, we focus on probing \emph{CP-}violation in Dalitz decays of the $\eta$  meson using the SMEFT, which details can be found in Ref.~\cite{Sanchez-Puertas:2018tnp} 

The sensitivity to this process is studied by measuring the polarization of the muon reconstructed from the following process:
\begin{itemize}
\item $p+Li\rightarrow\eta+X\;with\;\eta\rightarrow\gamma\mu^{+}\mu^{-}$
\end{itemize}

The  relevant observables are the two asymmetries defined in Eqs.(\ref{eq:ALetapi0},\ref{eq:ATetapi0}), which carry the information from the muon polarization.

\begin{figure}[!ht]
\includegraphics[width=8cm,height=7cm]{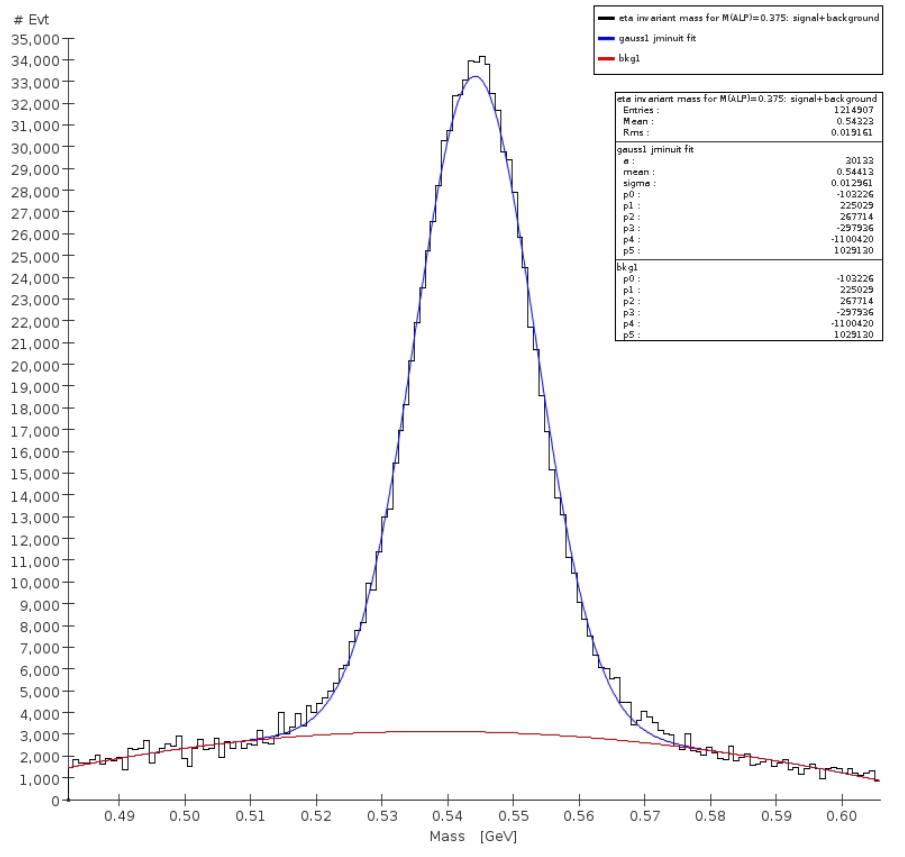}
\caption{Invariant mass of $\eta\rightarrow\gamma\mu^{+}\mu^{-}$ for for 4.5 $\times10^{14}$ POT. The plot includes
the Urqmd generated background. See text for an explanation of the
fitting procedure. }

\label{fig:eta2gammamumu}
\end{figure}

The event generation for this process is based on the model described
in Sec.~\ref{sec:CPVetaToPi02L}. A sample consisting of $1.5\times10^{10}$
$\eta$ mesons from 4.5$\times10^{14}$ POT (corresponding to 1.3$\times10^{-4}$
of the total integrated luminosity foreseen for the experiment). The
full chain of generation-simulation-reconstruction-analysis for this
process follows closely that described in Sec.~\ref{subsec:Sensitivity-to-the-vector} 
since the final states are, essentially, the same.  The invariant mass of the reconstructed
$\gamma\mu^{+}\mu^{-}$ system
was fitted using the sum of a Gaussian and a 5th-order polynomial.
The integral of the fitted function was used to extract the branching
ratio sensitivity for this process.
Fig.~\ref{fig:eta2gammamumu} shows
the distribution of the invariant mass of the reconstructed $\eta\rightarrow\gamma\mu^{+}\mu^{-}$ system, for 4.5 $\times10^{14}$ POT along with the results of the fit.

The reconstruction efficiency for the various steps of the analysis, along with the  final branching ratio sensitivity extracted with the analysis, are summarized in Table~\ref{table:recoeff_gammaleptonantilepton}. 
\begin{table}[!ht]
\centering\textcolor{black}{\scriptsize{}}%
\begin{tabular}{|c|c|c|c|c|c|c|c|}
\hline 
\textit{\textcolor{black}{\scriptsize{}Process}} & \textit{\textcolor{black}{\scriptsize{}Trigger}} & \textit{\textcolor{black}{\scriptsize{}Trigger}} & \textit{\textcolor{black}{\scriptsize{}Trigger}} & \textit{\textcolor{black}{\scriptsize{}Reconstruction}} &  & \textbf{\textcolor{black}{\scriptsize{}Total}} & \textcolor{black}{\scriptsize{}Branching ratio}\tabularnewline
 & \textcolor{black}{\scriptsize{}L0} & \textcolor{black}{\scriptsize{}L1} & \textcolor{black}{\scriptsize{}L2} & \textit{\textcolor{black}{\scriptsize{}+ analysis}} &  &  & \textcolor{black}{\scriptsize{}sensitivity}\tabularnewline
\hline 
\hline 
\textcolor{black}{\scriptsize{}$\eta\rightarrow\gamma\mu^{+}\mu^{-}$} & \textcolor{black}{\scriptsize{}80.6\%} & \textcolor{black}{\scriptsize{}64.6\%} & \textcolor{black}{\scriptsize{}94.3\%} & \textcolor{black}{\scriptsize{}$92.9$\%} &  & \textcolor{black}{\scriptsize{}$45.6$\%} & \textcolor{black}{\scriptsize{}$1.93\times10^{-9}\pm0.9\times10^{-11}$}\tabularnewline
\hline 
\textcolor{black}{\scriptsize{}Urqmd } & \textcolor{black}{\scriptsize{}$21.7\%$} & \textcolor{black}{\scriptsize{}$1.7\%$} & \textcolor{black}{\scriptsize{}$22.2\%$} & \textcolor{black}{\scriptsize{}$4.7\times10^{-3}$\%} &  & \textcolor{black}{\scriptsize{}$4.7\times10^{-6}$\%} & \textcolor{black}{\scriptsize{}-}\tabularnewline
\hline 
\end{tabular}\caption{Reconstruction efficiencies for $\eta\rightarrow\gamma\mu^{+}\mu^{-}$and
for the Urqmd background. Errors are statistical only and are obtained
assuming 4.5 $\times10^{14}$ POT.}
\label{table:recoeff_gammaleptonantilepton}
\end{table}
The sensitivity to the Wilson coefficients is derived for the longitudinal asymmetry, see Ref.~\cite{Sanchez-Puertas:2018tnp}, that
will be limited by statistical error on the signal diluted by the
background. The values are obtained from Table~\ref{table:recoeff_gammaleptonantilepton}
for 4.5$\times10^{14}$ POT, and scaled for the total luminosity foreseen at REDTOP.

Combining them, we expect for the statistical error $\Delta(A_{L})$  of the asymmetry: 
\begin{equation}
\Delta(A_{L})=\frac{\sqrt{N}_{\textrm{\ensuremath{\eta\to\gamma\mu^{+}\mu^{-}}}}+\sqrt{N}_{\textrm{bkg}}}{N_{\textrm{\ensuremath{\eta\to\mu^{+}\mu^{-}}}}}.\label{eq:delta-a-l-0}
\end{equation}

Taking the expected $\eta$ and background events from Table~\ref{tab:Expected-yield-at} and the efficiencies above, we obtain $N_{\eta\to\gamma\mu^+\mu^-} = N_{\eta}\times \textrm{BR}(\eta\to\gamma\mu^{+}\mu^{-})\times\epsilon_{\textrm{reco}}\times\epsilon_{\textrm{pol}}$, together with $N_{\textrm{bkg}} = N_{ni} \times\epsilon_{\textrm{reco}}  \times\epsilon_{\textrm{pol}}$. Assuming a conservative $\epsilon_{\textrm{pol}} =  50\%$ estimate both for signal and background, we obtain $\Delta(A_{L})=1.4\times10^{-5}$. Comparing to the longitudinal asymmetries defined in Ref.~\cite{Sanchez-Puertas:2018tnp}, one could achieve, for the Wilson coefficients responsible for the asymmetries, the following sensitivity:
\begin{equation}
\Delta(c_{\ell equ}^{1122})=2.6,\quad\Delta(c_{\ell edq}^{1122})=2.6,\quad\Delta(c_{\ell edq}^{2222})=1.7\label{eq:error-on-c1122}.
\end{equation}

\begin{singlespace}

\subsection{\emph{CP-} violation studies from the asymmetry of the decay planes of \texorpdfstring{$\eta\rightarrow\mu^{+}\mu^{-}e^{+}e^{-}$}{}}
\end{singlespace}

Testing $CP$-violation in pseudoscalar mesons via a search for asymmetries
in double Dalitz decay has the advantage that it does not require
to measure the polarization of the leptons. A recent theoretical model
\citep{Sanchez-Puertas:2018tnp} assumes that the asymmetry in this
process arises from heavy physics. As a consequence, the Standard
Model effective field theory (SMEFT) can be applied to relate the
Wilson coefficients (carrying the \emph{CP-}violating parameters)
and the observable quantities. A detailed discussion of that model
can be found in Sec.~\ref{subsec:Double-Dalitz-decays}.

The sensitivity to $CP$-violation is explored by examining the following
two asymmetries:

\begin{equation}
A_{sin\Phi cos\Phi}=\frac{N(\sin\phi\cos\phi>0)-N(\sin\phi\cos\phi<0)}{N(\sin\phi\cos\phi>0)+N(\sin\phi\cos\phi<0)}
\end{equation}

\begin{equation}
A_{sin\Phi}=\frac{N(\sin\phi>0)-N(\sin\phi<0)}{N(\sin\phi>0)+N(\sin\phi<0)}
\end{equation}

where $\phi$ is the angle between the decay planes of the $e^{+}e^{-}$
 and  the $\mu^{+}\mu^{-}$ pairs (see also Fig.~\ref{fig:eta2pipiee_phi}).
The reason for probing different asymmetries is due to the fact that they are proportional
to different operators. More specifically, amongst all the possible
invariants describing the 4-body kinematics ($A4-A8$, see, for example,
Appendix A in~\cite{Kampf:2018wau}:), it is only $A8$, proportional
to $sin(\phi)$, that changes sign under $P$ (while it is even under
$C$). Of course, $A8$ might come accompanied by additional invariants,
such as $A7$, that would produce an overall asymmetry also in $sin(\phi)cos(\phi)$.

Several events sets, each corresponding to 7.4$\times10^{13}$
POT (or $2.4\times10^{9}$ $\eta$-mesons), were generated with $GenieHad$
with $<sign(sin(\phi)cos(\phi)>$ ranging from 0 to $4\times10^{-2}$
and $<sign(sin(\phi)>$ ranging from 0 to $4\times10^{-2}$. The
full chain of generation-simulation-reconstruction-analysis was repeated
for each event set. The largest background for this final
state originates from the decays $\eta\rightarrow\pi^{+}\pi^{-}\pi^{0}$
and $\eta\rightarrow\pi^{+}\pi^{-}\gamma$ followed by $\gamma\rightarrow e^{+}e^{-}$
and $\pi^{0}\rightarrow\gamma e^{+}e^{-}$, where both pions where
mis-identified as muons. Another important source of background
is due to the combinatorics from two mis-identified charged pions,
generated by nuclear interaction of the beam on the target, and
from a spurious photon, usually generated from the decay of a $\pi^{0}$ or $\eta$
meson, which then converts in the materials of the detector. The excellent
particle identification of REDTOP keep such background to a very low
levels, as shown in Table~\ref{table:mupmumee_eff}. Very generic
requirements on the quality of reconstructed particles were applied
to the signal and background samples. Neutral pions, decaying into $\gamma e^{+}e^{-}$
and $\gamma\gamma$, where reconstructed by considering all combinations
of photons, electrons and positrons with an invariant mass within
5 MeV from $\pi^{0}$ mass. This requirement is able to reject most
of the combinatoric background, where no $\eta$ mesons are present in the final
state. Similarly, photons converting into a $e^{+}e^{-}$ pair where
reconstructed by requiring that the invariant mass of the $e^{+}e^{-}$
pair was lower than 5 MeV. 

The average reconstruction efficiency for all signal sets and for the Urqmd
background is summarized in Table~\ref{table:mupmumee_eff} along
with the sensitivity to the branching ratio obtained by using Eq.
\ref{eq:MasterS-final} for $3.3\times10^{18}$ POT.

\begin{figure}[!ht]
\includegraphics[width=7cm,height=7cm]{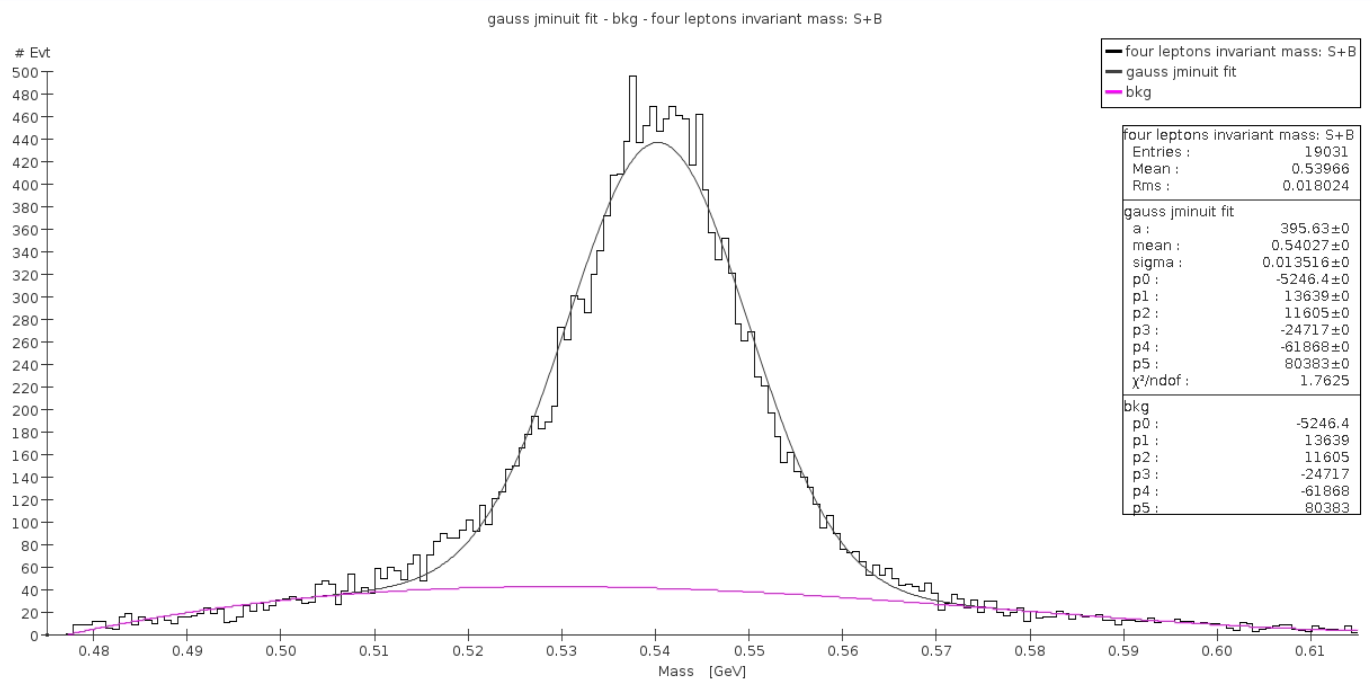} 

\caption{
Invariant mass of
the $\mu^{+}\mu^{-}e^{+}e^{-}$ system. The plot includes
the Urqmd generated background. See text for an explanation of the
fitting procedure.}

\label{fig:eta2mumumee_mass}
\end{figure}

\begin{table}[!ht]
\centering\textcolor{black}{\small{}}%
\begin{tabular}{|c|c|c|c|c|c|c|c|}
\hline 
\textit{\textcolor{black}{\small{}Process}} & \textcolor{black}{\small{}Benchmark} & \textit{\textcolor{black}{\small{}Trigger}} & \textit{\textcolor{black}{\small{}Trigger}} & \textit{\textcolor{black}{\small{}Trigger}} & \textit{\textcolor{black}{\small{}Reco}} & \textcolor{black}{\small{}Analysis} & \textbf{\textcolor{black}{\small{}Total}}\tabularnewline
 & \textcolor{black}{\small{}set} & \textcolor{black}{\small{}L0} & \textcolor{black}{\small{}L1} & \textcolor{black}{\small{}L2} &  &  & \tabularnewline
\hline 
\hline 
\textcolor{black}{\small{}$\eta$$\rightarrow\mu^{+}\mu^{-}e^{+}e^{-}$} & \textcolor{black}{\small{}average.} & \textcolor{black}{\small{}$80.4$\%} & \textcolor{black}{\small{}$57.0$\%} & \textcolor{black}{\small{}$20.4$\%} & \textcolor{black}{\small{}$16.6$\%} & \textcolor{black}{\small{}$58.2$\%} & \textcolor{black}{\small{}$0.9$\%}\tabularnewline
\hline 
\textcolor{black}{\small{}Urqmd } &  & \textcolor{black}{\small{}$21.7\%$} & \textcolor{black}{\small{}$1.7\%$} & \textcolor{black}{\small{}$22.2\%$} & \textcolor{black}{\small{}$1.0\times10^{-4}$\%} & \textcolor{black}{\small{}$36.4$\%} & \textcolor{black}{\small{}$3.0\times10^{-8}$\%}\tabularnewline
\hline 
\end{tabular}\caption{Reconstruction efficiencies for $\eta\rightarrow\mu^{+}\mu^{-}e^{+}e^{-}$
final state and for the Urqmd generated background. The values are
averaged over the benchmark sets.}
\label{table:mupmumee_eff}
\end{table}
The invariant mass of the $\mu^{+}\mu^{-}e^{+}e^{-}$ systems for the events surviving
the cuts is shown in Fig.~\ref{fig:eta2mumumee_mass}. The combinatorics
background is nicely fitted under the $\eta$ peak. 

\begin{figure}[!ht]
\includegraphics[width=7cm]{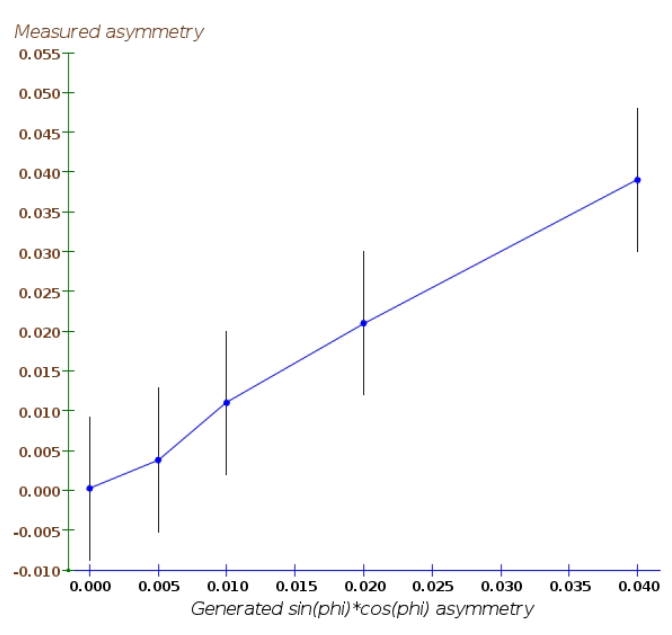} \includegraphics[width=7cm]{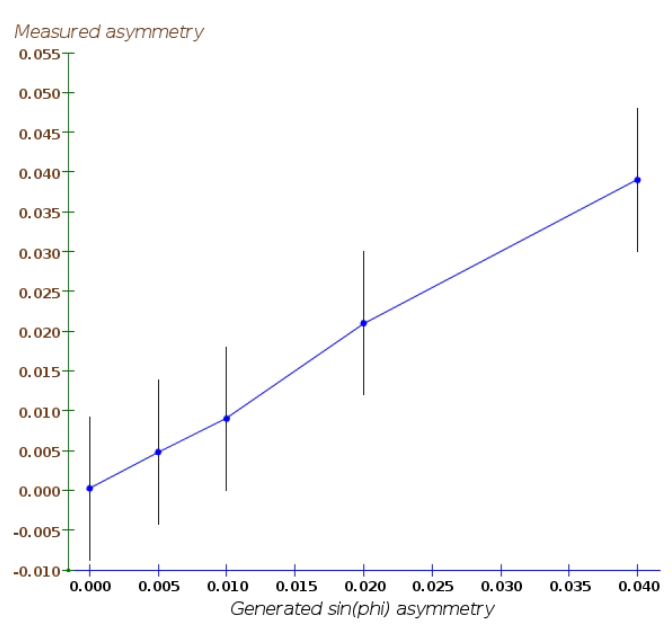} 

\caption{Sensitivity to $sin\varphi cos\varphi$ (left) and $sin\varphi$ (right)
asymmetries for $\eta\rightarrow\mu^{+}\mu^{-}e^{+}e^{-}$. }

\label{fig:eta2mumumee_asymm}
\end{figure}

The data for each event set were divided into separate $sin\varphi cos\varphi$>0 and
$sin\varphi cos\varphi<0$ groups and $sin\varphi$>0 and $sin\varphi<0$.
The $\eta$ content in each group was extracted from a fit to the
sum of a Gaussian function and a 5th-order polynomial and by integrating
the Gaussian signal after background subtraction. The 
asymmetries extracted with such procedure are plotted in Fig.~\ref{fig:eta2mumumee_asymm}
vs the Monte Carlo generated value. The diluting effect of the background
on the asymmetry is much reduced compared to the $\eta\rightarrow\pi^{+}\pi^{-}e^{+}e^{-}$
case. In fact, almost no degradation can be observed. We conclude
that the measured asymmetry for the samples considered is larger than
the statistical error (which is the dominant component in previous
measurements by WASA and KLOE experiments) for values of the true
asymmetry larger than approximately $1\times10^{-2}$. When extrapolating
the results to the full integrated luminosity of $3.3\times10^{18}$
POT foreseen for REDTOP, we estimate that the statistical error will
be reduced by about than two orders of magnitude, corresponding to
a contribution to the sensitivity smaller than $\simeq10^{-4}$. Therefore,
we conclude that the uncertainty on the measured asymmetry will, most
likely, be dominated by the systematic error.

Using the results in Sec.~\ref{subsec:Double-Dalitz-decays}, we obtain the following sensitivities for the full integrated luminosity
\begin{equation}\label{eq:DDasym}
\Delta c_{\ell edq}^{2222} = 8, \quad
\Delta c_{\ell edq}^{2222} = 5, \quad     \Delta \epsilon_1 = 5\times 10^{-4}, \quad     \Delta \epsilon_2 = 0.3. 
\end{equation}

\begin{singlespace}

\subsection{\emph{CP-}violation studies in the decay \texorpdfstring{$\eta\rightarrow\pi^{0}\mu^{+}\mu^{-}$}{}}
\end{singlespace}

A detailed discussion on probing \emph{CP-}violation in semi-leptonic $\eta$ decays can be found in Sec.~\ref{sec:CPVetaToPi02L}
along with the relationship between the Wilson coefficients (carrying
the \emph{CP-}violating parameters) and the observable quantities.

The sensitivity to this process is studied by measuring the polarization
of the muon reconstructed from the following process:
\begin{itemize}
\item $p+Li\rightarrow\eta+X\;with\;\eta\rightarrow\pi^{0}\mu^{+}\mu^{-}$
\end{itemize}
The two relevant observables are the two asymmetries defined in Eqs.(\ref{eq:ALetapi0},\ref{eq:ATetapi0}), that carry the information from the muon polarization.
%
%
The event generation for this process is based on the model described
in Sec.~\ref{sec:CPVetaToPi02L}. A sample consisting of $\sim2.3\times10^{8}$
$\eta$ mesons from 7$\times$$10^{12}$ POT (corresponding to 2$\times$$10^{-6}$
of the integrated luminosity foreseen for the experiment). The
full chain of generation-simulation-reconstruction-analysis for this
process follow closely that described in Sec.~\ref{subsec::pi0hmumu-Bump-hunt-analysis}
since the final states considered for the two analyses are, essentially, the same. The invariant mass of the reconstructed
$\pi^{0}\mu^{+}\mu^{-}$ system was fitted using the sum of a Gaussian and a 5th-order polynomial. The
integral of the fit function was used to extract the branching ratio sensitivity for this process.

The reconstruction efficiency for the various steps of the analysis, along with the resulting
branching ratio sensitivity are summarized in Table~\ref{table:recoeff_leptonantilepton-1}. 

\begin{table}[!ht]
\centering\textcolor{black}{\scriptsize{}}%
\begin{tabular}{|c|c|c|c|c|c|c|c|}
\hline 
\textit{\textcolor{black}{\scriptsize{}Process}} & \textit{\textcolor{black}{\scriptsize{}Trigger}} & \textit{\textcolor{black}{\scriptsize{}Trigger}} & \textit{\textcolor{black}{\scriptsize{}Trigger}} & \textit{\textcolor{black}{\scriptsize{}Reconstruction}} &  & \textbf{\textcolor{black}{\scriptsize{}Total}} & \textcolor{black}{\scriptsize{}Branching ratio}\tabularnewline
 & \textcolor{black}{\scriptsize{}L0} & \textcolor{black}{\scriptsize{}L1} & \textcolor{black}{\scriptsize{}L2} & \textit{\textcolor{black}{\scriptsize{}+ analysis}} &  &  & \textcolor{black}{\scriptsize{}sensitivity}\tabularnewline
\hline 
\hline 
\textcolor{black}{\scriptsize{}$\eta\rightarrow\pi^{0}\mu^{+}\mu^{-}$} & \textcolor{black}{\scriptsize{}64.1\%} & \textcolor{black}{\scriptsize{}36.7\%} & \textcolor{black}{\scriptsize{}91.4\%} & \textcolor{black}{\scriptsize{}$73.2$\%} &  & \textcolor{black}{\scriptsize{}$15.7$\%} & \textcolor{black}{\scriptsize{}$9.4\times10^{-9}\pm1.3\times10^{-10}$}\tabularnewline
\hline 
\textcolor{black}{\scriptsize{}Urqmd } & \textcolor{black}{\scriptsize{}$21.7\%$} & \textcolor{black}{\scriptsize{}$1.7\%$} & \textcolor{black}{\scriptsize{}$22.2\%$} & \textcolor{black}{\scriptsize{}$1.6\times10^{-2}$\%} &  & \textcolor{black}{\scriptsize{}$1.3\times10^{-5}$\%} & \textcolor{black}{\scriptsize{}-}\tabularnewline
\hline 
\end{tabular}\caption{Reconstruction efficiencies for $\eta\rightarrow\pi^{0}\mu^{+}\mu^{-}$and
for the Urqmd background. Errors are statistical only and are obtained
assuming 7$\times10^{12}$ POT.}
\label{table:recoeff_leptonantilepton-1}
\end{table}
The sensitivity to the Wilson coefficients is derived through the most sensitive
asymmetry, {\color{blue} Eq.~(\ref{eq:ALetapi0}),} that %
will be limited by statistical error on the signal diluted by the
background. The values are obtained from Table~\ref{table:recoeff_leptonantilepton-1}
for 7.2$\times10^{12}$ POT, after scaling for the total integrated
luminosity foreseen for REDTOP (cf.\ Table~\ref{tab:Expected-yield-at}.

Combining them, we expect for the statistical error: 
\begin{equation}
\Delta(A_{L})=\frac{\sqrt{N}_{\textrm{\ensuremath{\eta\to\pi^{0}\mu^{+}\mu^{-}}}}+\sqrt{N}_{\textrm{bkg}}}{N_{\textrm{\ensuremath{\eta\to\mu^{+}\mu^{-}}}}}.\label{eq:delta-a-l-1}
\end{equation}

Taking: $N_{\eta\to\pi^{0}\mu^{+}\mu^{-}}=N_{\eta}\times \textrm{BR}(\eta\to\pi^{0}\mu^{+}\mu^{+})\times\epsilon_{\textrm{reco}}\times\epsilon_{\textrm{pol}}=1.0\times10^{4}$,
with $N_{\eta}=1.1\times10^{14}$ , $\textrm{BR}=1.2\times10^{-9}$~\cite{Escribano:2020rfs}
$\epsilon_{\textrm{reco}}=15.7$\% , $N_{bkg}=1.3\times10^{-7}\times.5\times10^{16}\times\epsilon_{\textrm{pol}}$=1.6$\times$10$^{9}$,
$\epsilon_{\textrm{pol}}=50\%$ for the $\mu^{\pm}\rightarrow e^{\pm}\nu\overline{\nu}$
decay,  
we obtain: $\Delta(A_{L})=4.0$. Comparing to Eq.~(\ref{eq:ALetapi0}),
we find, for the Wilson coefficients responsible for the asymmetries, the following sensitivity:
\begin{equation}
\Delta(c_{\ell equ}^{1122})=21, \quad\Delta(c_{\ell edq}^{1122})=21,\quad\Delta(c_{\ell edq}^{2222})=200.
\end{equation}

\begin{singlespace}

\subsection{Lepton flavor violation in the decay \texorpdfstring{$\eta\rightarrow e^{+}\mu^{-}$}{}+
c.c.}
\end{singlespace}

This study requires the reconstruction of an electron/positron and a muon in the final state. 
It is especially challenging since a pion has a non-null probability to fake a muon.
Considering the large number of pions generated in the proton-target scattering, 
we expect that the sensitivity to measure 
the branching ratio for this process is considerably 
lower than all others considered so far.

A sample consisting of $\sim8.6\times10^{9}$ $\eta$ mesons from
2.6$\times$$10^{14}$ POT (corresponding to 7.8$\times$$10^{-5}$
of the full integrated luminosity foreseen for the experiment). The
full chain of generation-simulation-reconstruction-analysis described
in Sec.~\ref{sec:Simulation-strategy} was repeated for each event
set. Very generic requirements on the quality of reconstructed particles
were applied to the signal and background samples. The main background
for this channel is due to the non-resonant decay $\eta\rightarrow\gamma\mu^{+}\mu^{-}$
which occurs with a relatively large branching ratio of $3.1\times10^{-4}$
occurring in the presence of a gamma conversion: $\gamma\rightarrow e^{+}e^{-}$.
The second largest background contribution originates from charged
pions mis-identified as muons. With the present detector layout, the
$\pi$/$\mu$ mis-identification probability was estimates to be $\sim$3.5\%.
In order to reduce such background, we imposed the requirement that
no photons should be detected in the calorimeter. 

Neutral pions, decaying
into $\gamma e^{+}e^{-}$ and $\gamma\gamma$, where reconstructed
by considering all combinations of photons, electrons and positrons
with an invariant mass within 5 MeV from $\pi^{0}$ mass. Similarly,
photons converting into a $e^{+}e^{-}$ pair where reconstructed by
requiring that the invariant mass of the $e^{+}e^{-}$ system was
lower than 5 MeV. Reconstructed $\pi^{0}$ 's and $\gamma$'s were
removed from the event. Finally, the events were required to have
a topology consistent with a $\eta\rightarrow e^{\pm}\mu^{\mp}$ final
state, and an invariant mass compatible with the $\eta$ mass. 

\begin{figure}[!ht]
\includegraphics[width=8cm,height=7cm]{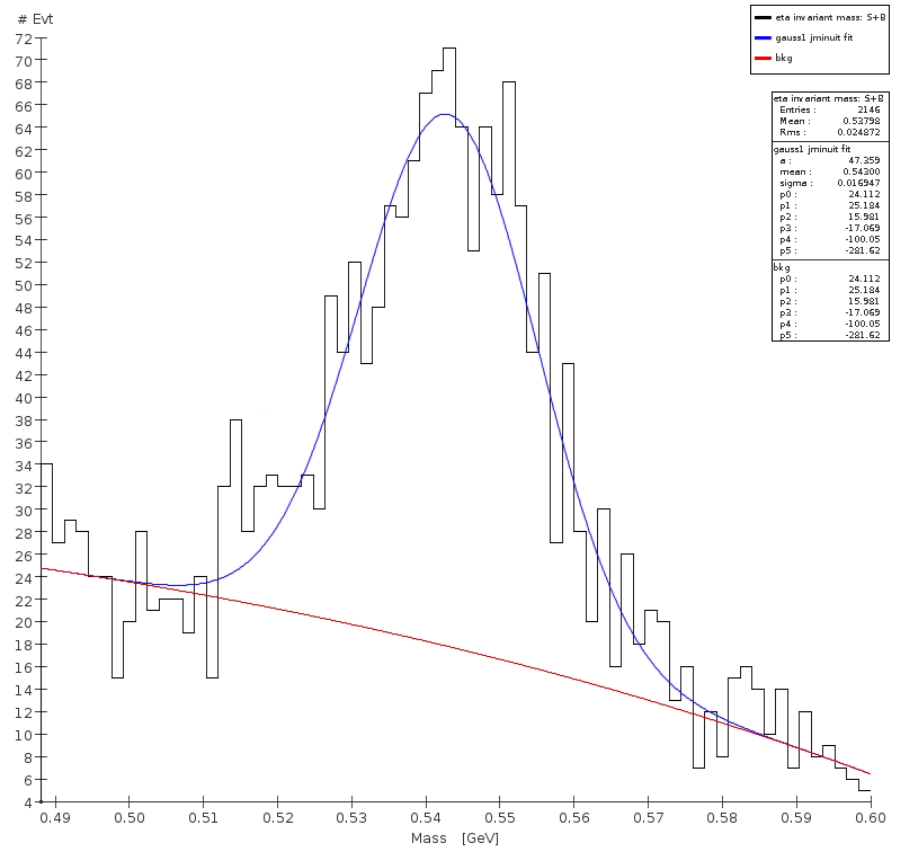} 

\caption{Invariant mass of $e^{+}\mu^{-}$ + c.c . The plot includes the Urqm
generated background. See text for an explanation of the fitting procedure.}

\label{fig:eta2CPV}
\end{figure}

The reconstruction efficiencies for this process and for the Urqmd
background is summarized in Table~\ref{table:recoeff_CPV}. For illustrative
purposes, Fig. \ref{fig:eta2CPV} shows the fit to the invariant
mass of the reconstructed $e^{\pm}\mu^{\mp}$ system along with the
Urqmd generated background, assuming BR($\eta\rightarrow e^{\pm}\mu^{\mp}$)=
$1.4\times10^{-4}$ and an $\eta$ sample of $5\times10^{8}$ (corresponding
to $4.5\times10^{-6}$ of the full integrated luminosity). The number
of reconstructed signal and background events was obtained form a
fit to the $\eta$ meson invariant mass using the sum of a Gaussian
and a 5th-order polynomial. The integral of the fitted function is
used to extract the branching ratio for this process.

The resulting branching ratio sensitivity is summarized in Table~\ref{table:recoeff_CPV}.

\begin{table}[!ht]
\centering\textcolor{black}{\small{}}%
\begin{tabular}{|c|c|c|c|c|c|c|c|}
\hline 
\textit{\textcolor{black}{\small{}Process}} & \textit{\textcolor{black}{\small{}Trigger}} & \textit{\textcolor{black}{\small{}Trigger}} & \textit{\textcolor{black}{\small{}Trigger}} & \textit{\textcolor{black}{\small{}Reco}} &  & \textbf{\textcolor{black}{\small{}Total}} & \textcolor{black}{\small{}Branching ratio}\tabularnewline
 & \textcolor{black}{\small{}L0} & \textcolor{black}{\small{}L1} & \textcolor{black}{\small{}L2} &  &  &  & \textcolor{black}{\small{}sensitivity}\tabularnewline
\hline 
\hline 
\textcolor{black}{\small{}$\eta\rightarrow e^{+}\mu^{-}+c.c.$} & \textcolor{black}{\small{}$79.3
$} & \textcolor{black}{\small{}$21.3\%$} & \textcolor{black}{\small{}$89.7\%$} & \textcolor{black}{\small{}$14.0\%$} &  & \textcolor{black}{\small{}$2.1$\%} & \textcolor{black}{\small{}$1.4\times10^{-7}\pm2\times10^{-9}$}\tabularnewline
\hline 
\textcolor{black}{\small{}Urqmd } & \textcolor{black}{\small{}$21.7\%$} & \textcolor{black}{\small{}$1.7\%$} & \textcolor{black}{\small{}$22.2\%$} & \textcolor{black}{\small{}$0.01$\%} &  & \textcolor{black}{\small{}$8.2\times10^{-6}$\%} & \tabularnewline
\hline 
\end{tabular}\caption{Reconstruction efficiencies for $\eta\rightarrow e^{+}\mu^{-}+c.c.$
and for the Urqmd generated background}
\label{table:recoeff_CPV}
\end{table}

\subsubsection{Concluding remark}

As stressed above, the reconstruction of the process $\eta\rightarrow e^{+}\mu^{-}+c.c.$
suffers from a large combinatorics background due to misidentified
pions generated in the target by the primary proton and gamma converting
in the fiber tracker. The excellent 4-momentum resolution of the detector
only partially mitigate this problem. From the study performed, we
conclude that the tagging of the $\eta$ meson could considerably
improve the branching ratio sensitivity, as most of the combinatorics
background could be identified and rejected. Another improvement would
arise from reducing the material budget in the vertex detector, where
the gamma conversion is reconstructed with lower efficiency. In that
respect, the adoption of the ITS3 option (cf.\ Sec.~\ref{par:Vertex-detector-option-II})
would be, in fact, an attractive solution.

\subsection{Test of Lepton Flavor Universality }
Leptonic and semileptonic decays of the $\eta$ mesons have relatively clear signatures in REDTOP, and they can be disentangled with good efficiency, from the large hadronic background. Furthermore, the decay rate into electrons and muons are only slightly affected by the different phase space. Therefore, those decays represent an excellent opportunity to probe Lepton Flavor Universality (LFU). In this work, we have considered two groups of processes: $\eta \to  \ell_1 \bar \ell_1 \ell_2 \bar \ell_2$, and $\eta \to  \gamma \ell \bar \ell$.

\subsubsection{\texorpdfstring{$\eta\rightarrow4\:leptons$}{} decays }
These constitute yet another set of interesting decays. Within the SM, these decays include $\eta^{(\prime)} \to \ell_1 \bar \ell_1 \ell_2 \bar \ell_2$, where $\ell_i$ can be either of $e$ or $\mu$. Interestingly, the amplitude for these decays is dominated by the very same dynamics---the $\eta^{(\prime)} \to \gamma \gamma$ vertex---that also describes the decays to two leptons, discussed in Sec.~\ref{sec:LFU}. In the case of 4 leptons, both the photons are off-shell and produce $\ell \bar \ell$ pairs (``Dalitz pairs''). It is interesting that, although all of the decay channels $e \bar e$, $e \bar e \gamma$, $\mu\bar \mu$, $\mu \bar \mu \gamma$, and $\ell_1 \bar \ell_1 \ell_2 \bar \ell_2$ all probe the very same underlying form factor, the different decays entail somewhat different ``effective dilepton masses'' and thereby probe this form factor in somewhat different kinematic regions for the off-shell photons. In practice, the contributions from large ``effective masses'' are negligible in the $e \bar e$ case, hence e.g. the $\eta^{(\prime)} \to e \bar e e \bar e$ and $\eta^{(\prime)} \to e \bar e \mu \bar \mu$ cases can usefully be inferred from the $\eta^{(\prime)} \to e \bar e$ and $\eta^{(\prime)} \to \mu \bar \mu$ ones~\cite{Jarlskog:1967fpu}. Besides, the contribution from non-zero $k_{1,2}^2 \neq 0$ ($k_i$ denoting the photon momenta in the $\eta^{(\prime)} \to \gamma \gamma$ vertex) are actually in the ballpark of 30\% in the muon case~\cite{Jarlskog:1967fpu}. In the presence of LFV New Physics, one may add to the above decays the further channels $\eta^{(\prime)} \to \ell_1 \bar \ell_2 \ell_2 \bar \ell_2$, where again $\ell_i$ can be either of $e$ or $\mu$. Considering for definiteness the LQ case already discussed in Sec.~\ref{sec:LFU} the underlying diagrams for these channels are similar: $t$-channel LQ exchange, further radiating an off-shell photon. Then the LQ vertices yield two oppositely charged leptons of different flavour, and the photon produces the two remaining like-flavoured leptons. Hence we see that, as is the case for the SM amplitude, also the NP amplitude is---for the 2-lepton and 4-lepton cases alike---dominated by the very same building block. In the presence of abundant data, a joint analysis of all channels is thus warranted. As already mentioned earlier, the NP amplitude tends to be sensitive to {\em light} NP, not exceeding $\sim 20$ GeV.  

A detailed discussion on possible ways to probe Lepton Universality $\eta^{(\prime)} \to \ell_1 \bar \ell_1 \ell_2 \bar \ell_2$, where $\ell_i$ can be either of $e$ or $\mu$  can be found
in Sec.~\ref{sec:LFU}.

Only the decays of the $\eta\rightarrow e^{+}e^{-}e^{+}e^{-}$ has
been experimentally observed, with a measured branching ratio of $2.4\times10{}^{-5}$ ~\citep{Zyla:2020zbs},
corresponding to an expected yield at REDTOP of $2.6\times10^{9}$
produced events. 
An estimate for the other two processes can be made following Ref.~\citep{Kampf:2018wau}, for which we foresee a yield of order $\mathcal{O}(10^{7})$   and $\mathcal{O}(10^{4})$, respectively, for the $\eta\rightarrow e^{+}e^{-}\mu^{+}\mu^{-}$ and
$\eta\rightarrow\mu^{+}\mu^{-}\mu^{+}\mu^{-}$ decays. The studies
performed on these processes have been carried using different $\eta$-meson
samples. The corresponding POT are summarized in the second column
of Table~\ref{table:staterrot_eta4lepton}. The full chain of generation-simulation-reconstruction-analysis
was repeated for each event set. Very generic requirements on the
quality of reconstructed particles were applied to the signal and
background samples. The final states with muons are relatively
background free, as the combinatorics can hardly mimic the  
PID and kinematics of the signal.

The main background for the $\eta\rightarrow e^{+}e^{-}e^{+}e^{-}$
channel is due to the non-resonant decay $\eta\rightarrow\gamma e^{+}e^{-}$
which occurs with a relatively large branching ratio of $3.1\times10^{-4}$,
accompanied by the gamma conversion: $\gamma\rightarrow e^{+}e^{-}$.
The second largest background contribution originates from events
with multiple neutral pions decaying as: $\pi^{0}\rightarrow\gamma e^{+}e^{-}$. 

In order to reduce the combinatoric background, neutral pions, decaying
into $\gamma e^{+}e^{-}$ and $\gamma\gamma$, where reconstructed
by considering all combinations of photons, electrons and positrons
with an invariant mass within 5 MeV from $\pi^{0}$ mass. Similarly,
photons converting into a $e^{+}e^{-}$ pair where reconstructed by
requiring that the invariant mass of the $e^{+}e^{-}$ system was
lower than 5 MeV. Reconstructed $\pi^{0}$ 's and $\gamma$'s were
removed from the event. Finally, the events were required to have
a topology consistent with a 
$\eta^{(\prime)} \to \ell_1 \bar \ell_1 \ell_2 \bar \ell_2$ in the final
state, and an invariant mass compatible with the $\eta$ mass. 

The reconstruction efficiencies for the three processes and for the
Urqmd generated background are summarized in Table~\ref{table:recoeff_eta4leptons}

\begin{table}[!ht]
\centering\textcolor{black}{\small{}}%
\begin{tabular}{|c|c|c|c|c|c|c|}
\hline 
\textit{\textcolor{black}{\small{}Process}} & \textit{\textcolor{black}{\small{}Trigger}} & \textit{\textcolor{black}{\small{}Trigger}} & \textit{\textcolor{black}{\small{}Trigger}} & \textit{\textcolor{black}{\small{}Reconstruction}} & \textit{\textcolor{black}{\small{}Analysis}} & \textbf{\textcolor{black}{\small{}Total}}\tabularnewline
 & \textcolor{black}{\small{}L0} & \textcolor{black}{\small{}L1} & \textcolor{black}{\small{}L2} &  &  & \tabularnewline
\hline 
\hline 
\textcolor{black}{\small{}$\eta$$\rightarrow e^{+}e^{-}e^{+}e^{-}$} & \textcolor{black}{\small{}$96.1\%$} & \textcolor{black}{\small{}$80.7\%$} & \textcolor{black}{\small{}$15.5\%$} & \textcolor{black}{\small{}$63.3$\%} & \textcolor{black}{\small{}$61.2$\%} & \textcolor{black}{\small{}$4.5$\%}\tabularnewline
\hline 
\textcolor{black}{\small{}$\eta\rightarrow e^{+}e^{-}\mu^{+}\mu^{-}$} & \textcolor{black}{\small{}$80.4$\%} & \textcolor{black}{\small{}$57.0$\%} & \textcolor{black}{\small{}$20.4$\%} & \textcolor{black}{\small{}$16.6$\%} & \textcolor{black}{\small{}$52.8$\%} & \textcolor{black}{\small{}$0.8$\%}\tabularnewline
\hline 
\textcolor{black}{\small{} $\eta\rightarrow\mu^{+}\mu^{-}\mu^{+}\mu^{-}$} & \textcolor{black}{\small{}$45.1$\%} & \textcolor{black}{\small{}$31.9$\%} & \textcolor{black}{\small{}$25.5$\%} & \textcolor{black}{\small{}$61.3$\%} & \textcolor{black}{\small{}$40.5$\%} & \textcolor{black}{\small{}$0.9$\%}\tabularnewline
\hline 
\textcolor{black}{\small{}Urqmd } & \textcolor{black}{\small{}$21.7\%$} & \textcolor{black}{\small{}$1.7\%$} & \textcolor{black}{\small{}$22.2\%$} & \textcolor{black}{\small{}$0.9-8.2\times10^{-4}$\%} & \textcolor{black}{\small{}17.6\%-$30.7$\%} & \textcolor{black}{\small{}$0.7-6.7\times10^{-7}$\%}\tabularnewline
\hline 
\end{tabular}\caption{Reconstruction efficiencies for $\eta\rightarrow4\:leptons$ and for
the Urqmd generated backgrounds}
\label{table:recoeff_eta4leptons}
\end{table}

\begin{figure}[!ht]
\includegraphics[width=7cm,height=7cm]{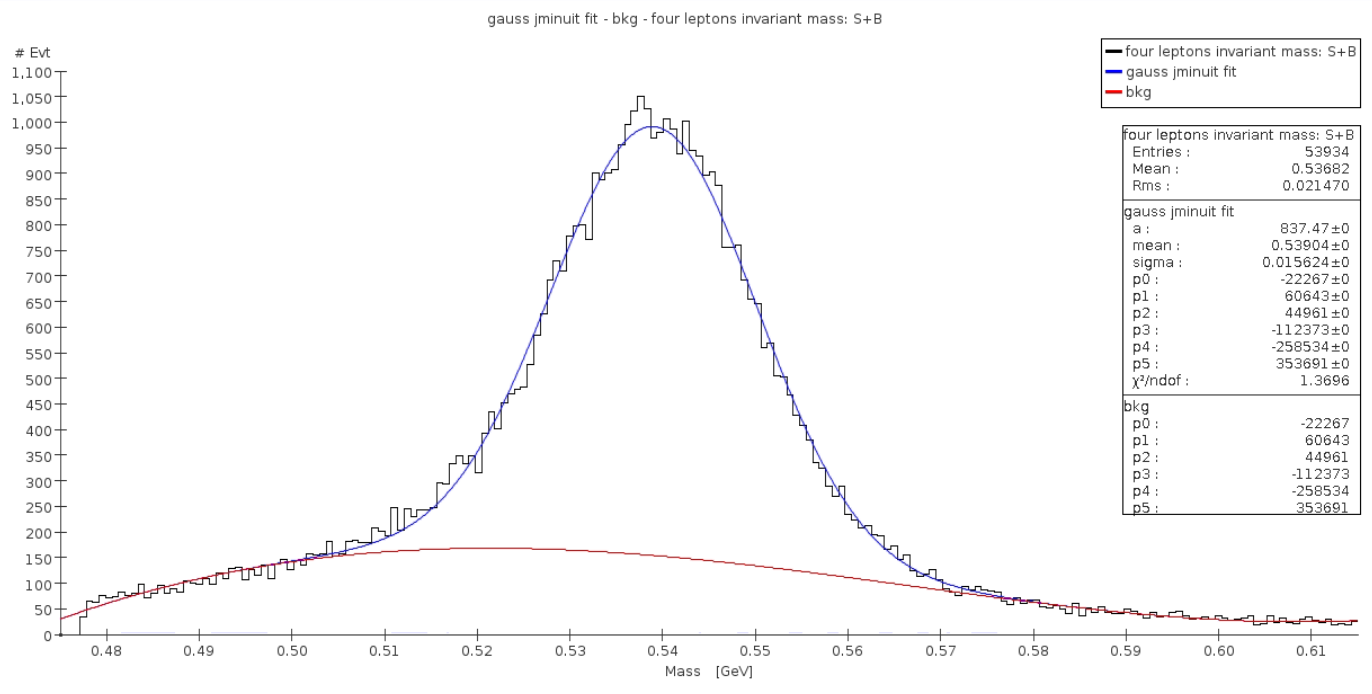} \includegraphics[width=7cm,height=7cm]{pictures/eta2mumuee} 

\includegraphics[width=7cm,height=7cm]{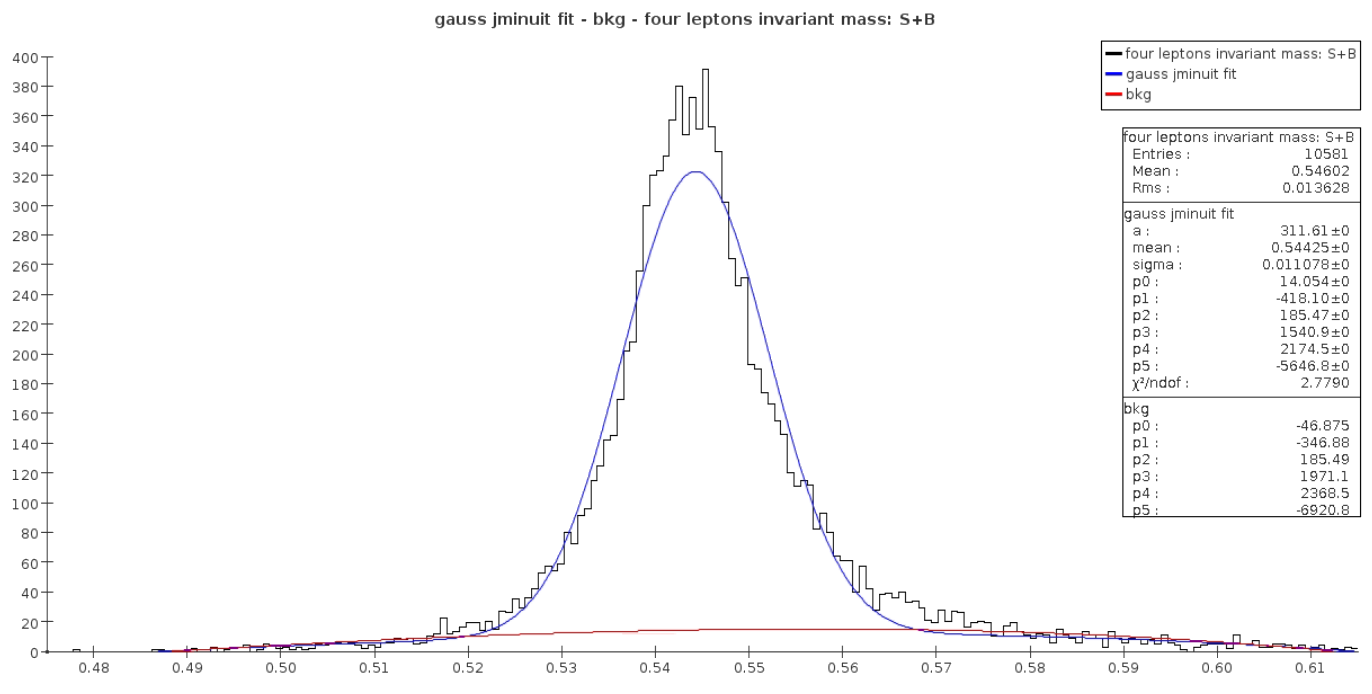} \caption{Invariant mass for f $\eta\rightarrow4\:leptons$ and for the Urqmd
generated background for the POT indicated in Table~\ref{table:staterrot_eta4lepton}.
$\eta\rightarrow e^{+}e^{-}e^{+}e^{-}$ (top left), $\eta\rightarrow e^{+}e^{-}\mu^{+}\mu^{-}$(top
right) and $\eta\rightarrow\mu^{+}\mu^{-}\mu^{+}\mu^{-}$(bottom).}

\label{fig:eta4leptonsi_mass}
\end{figure}

The number of reconstructed signal and background events was obtained
form a fit to the invariant mass of the four leptons, using the sum of a
Gaussian and a 5th-order polynomial. The integral of the fitted function
is used to extract the branching ratio for the considered process.
Fig.~\ref{fig:eta4leptonsi_mass} shows the invariant mass distribution
of the four leptons, along with the fitting curve. 
The statistical errors obtained from the fit to each final state are
summarized in Table. \ref{table:staterrot_eta4lepton}.

\begin{table}[!ht]
\centering\textcolor{black}{\small{}}%
\begin{tabular}{|c|c|c||c|}
\hline 
\textit{\textcolor{black}{\small{}Process}} & POT & Signal events & \textbf{\textcolor{black}{\small{}Statistical error}}\tabularnewline
 &  &  & \tabularnewline
\hline 
\hline 
\textcolor{black}{\small{}$\eta$$\rightarrow e^{+}e^{-}e^{+}e^{-}$} &  $4.4\times10^{14}$ & 53,934 & \textbf{\textcolor{black}{\small{}0.5\%}}\tabularnewline
\hline 
\textcolor{black}{\small{}$\eta\rightarrow e^{+}e^{-}\mu^{+}\mu^{-}$} &  $1.6\times10^{15}$  & 18,841 & \textbf{\textcolor{black}{\small{}0.8\%}}\tabularnewline
\hline 
\textcolor{black}{\small{} $\eta\rightarrow\mu^{+}\mu^{-}\mu^{+}\mu^{-}$} &  $2.2\times10^{18}$  & 10,548 & \textbf{\textcolor{black}{\small{}1.0\%}}\tabularnewline
\hline 
\end{tabular}{\small\par}

\caption{Statistical error from the fit of $\eta\rightarrow4\:leptons$ and
Urqmd generated background using a gaussian and a 5th-order polynomial.
The POT corresponding to each data sample is indicated in the second
column.}
\label{table:staterrot_eta4lepton}
\end{table}
When the statistics of the full event sample is taken into account,
the projected statistical error for the three processes considered
here is expected to be of the order of $10^{-5}$.

\subsubsection{Test of Lepton Flavor Universality with \texorpdfstring{$\eta\rightarrow\gamma\:2\:leptons$}{}
decays }

From a theoretical point of view, similar considerations hold as in the case of 2 leptons discussed in Sec.~\ref{sec:LFU}. The presence of the additional photon implies additional SM contributions, but the main conclusions reached in the purely leptonic case still hold. One important feature of the radiative 2-leptons case with respect to the non-radiative counterpart is the fact that the chiral suppression inherent in the latter decay~\cite{Jarlskog:1967fpu} is lifted because of the additional photon. Ratios of radiative di-leptonic decay rates, where the numerator and the denominator only differ by the lepton flavour are then excellent tests of Lepton Universality~\cite{Guadagnoli:2016erb}, and, by the previous argument, they are expected to be very close to unity within the Standard Model.

The studies performed on these processes are based on
a  relatively small $\eta$-meson sample consisting of 4.65 $\times10^{6}$
events. That sample corresponds to $\sim4.2\times10^{-8}$ 
of the
integrated luminosity foreseen for the experiment. 
The full chain of generation-simulation-reconstruction-analysis
was repeated for each set of the parameters and for the parameters
corresponding to the Standard Model prediction. final state. Very
generic requirements on the quality of reconstructed particles was
applied to the signals and background samples. 

In order to reduce the combinatoric background, neutral pions, decaying
into $\gamma e^{+}e^{-}$ and $\gamma\gamma$, where reconstructed
by considering all combinations of photons, electrons and positrons
with an invariant mass within 5 MeV from $\pi^{0}$ mass. Similarly,
photons converting into a $e^{+}e^{-}$ pair where reconstructed by
requiring that the invariant mass of the $e^{+}e^{-}$ system was
lower than 5 MeV. Reconstructed $\pi^{0}$ 's and $\gamma$'s were
removed from the event. Finally, the events were required to have
a topology consistent with a $\eta\rightarrow\gamma  e^{+}  e^{-}$ or $\eta\rightarrow\gamma  \mu^{+}  mu^{-}$ final
states, and an invariant mass compatible with the $\eta$ mass. 

The reconstruction
efficiencies for the two processes and for the Urqmd generated background
are summarized in Table~\ref{table:recoeff_etag2leptons}

\begin{table}[!ht]
\centering{\small{}}%
\begin{tabular}{|c|c|c|c|c|c|c|}
\hline 
\textit{\small{}Process} & \textit{\small{}Trigger} & \textit{\small{}Trigger} & \textit{\small{}Trigger} & \textit{\small{}Reconstruction} &  & \textbf{\small{}Total}\tabularnewline
 & {\small{}L0} & {\small{}L1} & {\small{}L2} & {\small{}\& analysis} &  & \tabularnewline
\hline 
\hline 
{\small{}$\eta$$\rightarrow\gamma\,e^{+}e^{-}$} & {\small{}$80.6\%$} & {\small{}$64.6\%$} & {\small{}$94.3\%$} & {\small{}$92.8$\%} &  & {\small{}$45.6$\%}\tabularnewline
\hline 
{\small{}$\eta\rightarrow\gamma\,\mu^{+}\mu^{-}$} & {\small{}$63.2$\%} & {\small{}$42.1$\%} & {\small{}$90.6$\%} & {\small{}$79.1$\%} &  & {\small{}$20.2$\%}\tabularnewline
\hline 
{\small{}Urqmd } & \textcolor{black}{\small{}$21.7\%$} & \textcolor{black}{\small{}$1.7\%$} & \textcolor{black}{\small{}$22.2\%$} & {\small{}$1.1\times10^{-1}$\%$\:($$1.64\times10^{-3}$\%)} &  & {\small{}$9.0\times10^{-6}$\%$\:($$1.3\times10^{-7}$\%)}\tabularnewline
\hline 
\end{tabular}\caption{Reconstruction efficiencies for $\eta\rightarrow\gamma\,\:lepton-antilepton$
and for the Urqmd generated background (values in parentheses are
for the $\eta\rightarrow\gamma\,\mu\,\mu$ final state). }
\label{table:recoeff_etag2leptons}
\end{table}

\begin{figure}[!ht]
\includegraphics[scale=0.3]{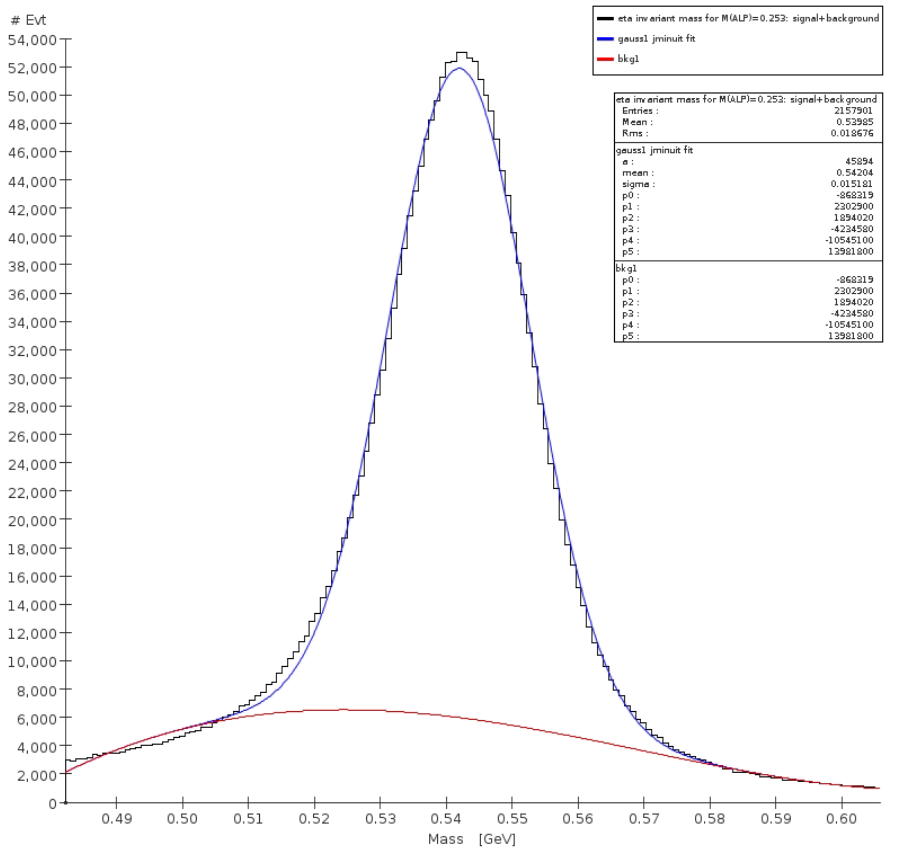} \includegraphics[scale=0.3]{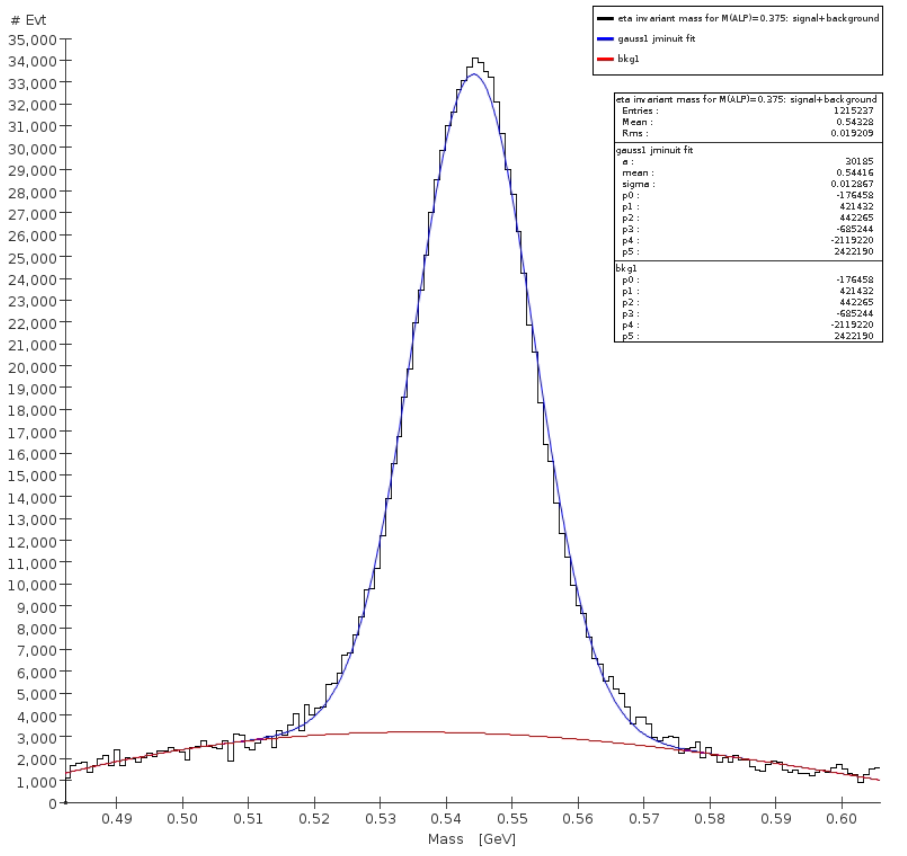}
\caption{Invariant mass for $\eta\rightarrow\gamma\:lepton-antilepton$ and
for the Urqmd generated background.}

\label{fig:etag2leptons_mass}
\end{figure}

The number of reconstructed signal and background events was obtained
form a fit to the invariant mass of the four leptons, using the sum of a
Gaussian and a 5th-order polynomial. The integral of the fitted function
is used to extract the branching ratio for the considered process.
Fig.~\ref{fig:etag2leptons_mass} shows the invariant mass distribution
of $\gamma\,\:lepton-antilepton$, with the fitting function superimposed. 

The statistical errors from the fit to each final state are summarized
in Table. \ref{table:staterrot_etag2lepton}

\begin{table}[!ht]
\centering{\small{}}%
\begin{tabular}{|c|c|c|c|c||c|}
\hline 
\textit{\small{}Process} & {\small{}POT} & {\small{}Signal events} & {\small{}Background events} & {\small{}$\frac{S}{\sqrt{B}}$} & \textbf{\small{}Statistical error}\tabularnewline
 &  &  &  &  & \tabularnewline
\hline 
\hline 
{\small{}$\eta$$\rightarrow\gamma\,e^{+}e^{-}$} & {\small{}$1.38\times10^{11}$} & {\small{}$2.13\times10^{6}$} & {\small{}$2.52\times10^{4}$} & {\small{}$1.3\times10^{4}$} & \textbf{\small{}0.09\%}\tabularnewline
\hline 
{\small{}$\eta\rightarrow\gamma\,\mu^{+}\mu^{-}$} & {\small{}$1.38\times10^{11}$} & {\small{}$8.84\times10^{5}$} & {\small{}$6.5\times10^{3}$} & {\small{}$3.5\times10^{3}$} & \textbf{\small{}0.14\%}\tabularnewline
\hline 
\end{tabular}{\small\par}

\caption{Statistical error from the fit of $\eta\rightarrow\gamma\,\:lepton-antilepton$
and Urqmd generated background using a gaussian and a 5th-order polynomial, for    1.38$\times10^{18}$ POT }
\label{table:staterrot_etag2lepton}
\end{table}
When the statistics of the full event sample for  3.3$\times10^{18}$ POT is taken into account,
the projected statistical error for the $\eta\rightarrow\gamma\,\:lepton-antilepton$
final state is expected to be of the order of $10^{-6}$, certainly
negligible compared to the systematic error.

\subsubsection{Remarks on LFU measurements with \texorpdfstring{$\eta$}{} mesons}

The theoretical discussion about the REDTOP sensitivity to LFU measurements deserves some further qualifications. As well-known, LFU measurements have recently come to the fore because of a coherent array of discrepancies in measurements of semi-leptonic $b \to s$ and $b \to c$ decays at LHCb and $B$ factories. The natural question for the present document is whether the putative beyond-SM LFU that these discrepancies imply could be tested in decays of light mesons, e.g. kaons~\cite{AlvesJunior:2018ldo}, and the $\eta^{(\prime)}$. On the latter there is a conspicuous gap in the literature, and on the other hand this possibility would be an important physics case for REDTOP. A positive answer to the question rests on addressing two challenges, that we would like to highlight here, and that definitely warrant further investigation.

The first issue is the fact that LHCb and $B$ factories access decays that directly probe---by definition---the operators affected by the putative $B$ anomalies, i.e., 4-fermion structures of the kind $(\bar{s} \Gamma b) (\bar{\ell} \Gamma^\prime \ell)$, where $\Gamma, \Gamma^\prime$ denote appropriate strings of Dirac matrices, and $\ell$ is any of the charged leptons. Conversely, in the case of $\eta^{(\prime)}$ we access operators with quarks (and leptons) of the light generations only. Relating the different quark sectors requires assumptions on the flavor structure of the couplings, and the suppression is ``by default'' CKM-like (see e.g.~\cite{Borsato:2018tcz}).

The second issue is that $\eta^{(\prime)}$ leptonic decays are dominated by the (long-distance) contribution with two intermediate photons, discussed around eq. (\ref{eq:P_to_ell_ell}). As a consequence, in order to have sensitivity to new physics, the latter must give a signal unambiguously larger than the uncertainty associated with such long-distance contribution. To this end, either the new-physics coupling must be large enough (but it cannot exceed $\sim 4 \pi$), and/or the new-physics mass scale $M$ has to be sufficiently small. As a reference, $Z$-boson exchange yields per-mil modifications of the rate~\cite{Gan:2020aco}, as already discussed in this document. This makes these decays way more suited as probes of {\em light}, MeV--GeV-scale new physics. This happens to be a very active field of research currently (see e.g.~\cite{Lanfranchi:2020crw} for a recent review), and has even been considered as a solution of $B$ anomalies~\cite{Datta:2017pfz,Sala:2017ihs,Datta:2017ezo,Alok:2017sui,Altmannshofer:2017bsz,Datta:2018xty,Darme:2020hpo,Darme:2021qzw}.

The task of pinning down such a possibility is daunting but not hopeless, and definitely worth pursuing given the huge opportunities it would open. Keeping in mind the above issues, the task at hand would be to study different flavour assumptions within an appropriate, general model of the new effects, and perform a joint analysis of the different flavour sectors. Note that the ``appropriate model''  cannot be the effective-theory framework used in many of the phenomenological interpretations of current $B$ anomalies, because light new physics has to be kept as dynamical degrees of freedom. However, the formalism to address this possibility in a general way exists (see e.g.~\cite{Arina:2021nqi}). One can thus study systematically the question under which circumstances the shifts to semi-leptonic operators from light new physics that can explain the $B$ anomalies would ``spill over'' to $\eta^{(\prime)}$ decays. Addressing this question goes clearly outside the scope of this document, but we reiterate its interest and wide applicability.

\section{Sensitivity to Non-perturbative QCD\label{sec:Sensitivity-to-Non-perturbative}}
\subsection{Form factor studies}


The studies on the $\eta$ and $\eta^{\prime}$ transition form factors can be of great importance in the evaluation of the dominant pseudoscalar-exchange contribution to the
HLbL scattering contribution to the prediction of the anomalous magnetic moment of the muon, as discussed in Sec.\ref{sec:non-perturbative-QCD}. A good precision in the determination of those form factors could unquestionably help in shedding more light on
this topic.

A detailed discussion on the $\eta$ form factors with semileptonic decays can be found in Sec.~\ref{sec:non-perturbative-QCD}.
The sensitivity is explored by looking at the kinematics of the di-muon
system reconstructed from the following process:
\begin{itemize}
\item $p+Li\rightarrow\eta+X\;with\;\eta\rightarrow\gamma\mu^{+}\mu^{-}$
\end{itemize}
The amplitude for such decay can be written as (cf.\ Sec.~\ref{sec:CPVetaTo2L}): 

\begin{equation}
\mathcal{M}=\epsilon^{\mu\nu\rho\sigma}\varepsilon_{\mu}^{*}k_{\nu}q_{\sigma}[\bar{u}(p_{-})\gamma_{\rho}v(p_{+})]F_{\eta\gamma^{*}\gamma^*}(q^{2},0)q^{-2}\label{eq:matrixelement_ff}
\end{equation}
where $q=p_+ + p_-$ is the sum of the the lepton momenta, $k$ is
the photon momenta, and $\varepsilon$ the photon polarization, and
$ F_{\eta\gamma^*\gamma^*}(q^2)$ the transition form factor characterizing
the $\eta$ structure. A phenomenologically useful (and simple) parametrization
for the form factor is the following (see Sec.~\ref{sec:non-perturbative-QCD} for details):

\begin{equation}
\tilde{F}_{\eta\gamma^{*}\gamma^*}(s,0)=\frac{\Lambda^{2}}{\Lambda^{2}-s}.\label{eq:eta_formfactor}
\end{equation}
where $\Lambda \simeq m_{\rho}\simeq 770$MeV. A recent measurement
by A2 Coll.~\cite{A2:2013wad} was based on of 2.2$\times$10$^{4}$
$\eta\rightarrow\gamma e^{+}e^{-}$ decays from a total of 3$\times$10$^{7}$
$\eta$ mesons produced in the $\gamma\,p\rightarrow\eta\,p$ reaction,
resulting in a statistical error on $\varLambda$ of $\sim8\%$. As
discussed in Sec.~\ref{sec:CPVetaTo2L}, the $\eta\rightarrow\gamma\mu^{+}\mu^{-}$ channel has
larger sensitivity than $\eta\rightarrow\gamma e^{+}e^{-}$.

\begin{figure}[!ht]
\includegraphics[width=9cm,height=8cm]{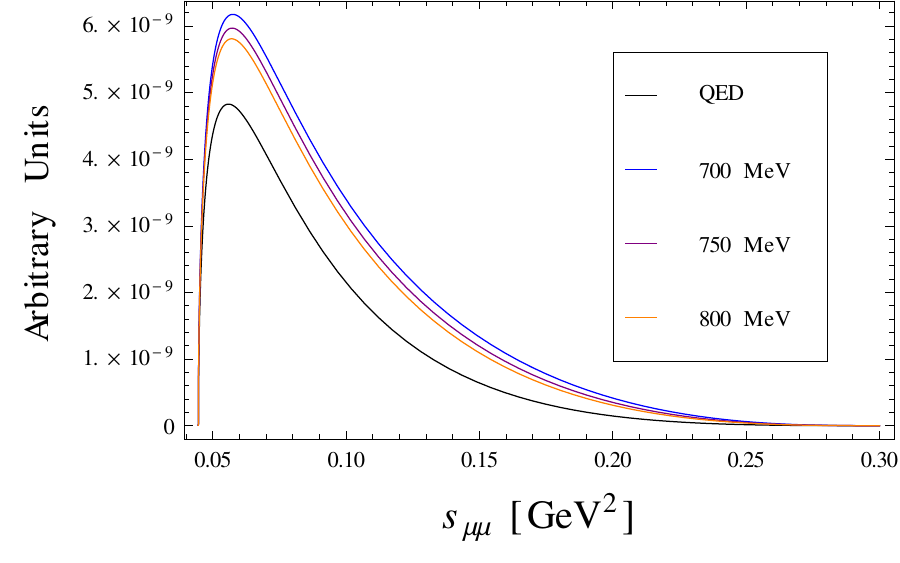} \caption{Generated spectrum $\frac{d\varGamma}{ds}$ for $\eta\rightarrow\gamma\mu^{+}\mu^{-}$
(s=$\mid P_{\mu^{+}}+P_{\mu^{-}}\mid^{2}$).}

\label{fig:forma-factors-generated}
\end{figure}

Five
event sets, each consisting of $\sim3.5\times10^{5}$ $\eta\rightarrow\gamma\mu^{+}\mu^{-}$
decays where generated with $GenieHad$. That statistics corresponds to 1.1$\times$10$^{9}$
$\eta$ mesons produced, or, $\sim1\times10^{-5}$
of the integrated luminosity foreseen for the experiment. 
The sets where
generated with the matrix element described in Sec.~\ref{sec:CPVetaTo2L}
and $\varLambda$ ranging from 660 MeV to 800 MeV. One extra sample
was generated with a point-like form factor, corresponding to a pure
QED process. The distribution of the $\nicefrac{d\Gamma}{ds}$, where
$s$ is the invariant mass squared of the $\mu^{+}\mu^{-}$ system
is shown in Fig.~\ref{fig:forma-factors-generated}. 

Different values of the parameter $\varLambda$ affect both the shape
of distribution as well as the total decay rate of the process. 

The analysis oh this final state proceeds according the same guidelines 
as those performed for the ~\emph{CP-}violation  in semileptonic $\eta$ decays(cf.\ Sec.~\ref{subsec:CP-studies-in_gammamumu}).
The full chain of generation-simulation-reconstruction-analysis was
repeated for each set of the parameters and for the parameters corresponding
to the Standard Model prediction. final state. Very generic requirements
on the quality of reconstructed particles was applied to the signals
and background samples. 

The largest background contribution
to this process is found to originate from mis-identified pions,
mistakenly reconstructed as muons. In fact, the $\pi$/$\mu$ mis-identification
probability for the detector considered in this work has a conservative
value of $\simeq 3.5\%$ (or, $\simeq 0.12\%$ for mis-identifying
both leptons). Since the probability of generating two charged pions
in the primary interaction is almost 11\% (see, also, Fig.~\ref{fig:urqmd_multiplicity}),
and the probability of having at least one $\pi^{0}$ is 58\%, we
expect that about $\sim1.9\times10^{11}$ events could potentially
fake a $\eta\rightarrow\gamma\mu^{+}\mu^{-}$ process. The second
largest background is due to process $\eta\rightarrow\gamma\pi^{+}\pi^{-}$,
when both pions are accidentally misidentified as muons and their kinematics is
compatible with an invariant mass of a genuine $\eta$ decay. 

The reconstruction efficiencies for this channel and for the Urqmd
 background are summarized in Table~\ref{table:recoeff_eta2gammamumu-ff}

\begin{table}[!ht]
\centering\textcolor{black}{\small{}}%
\begin{tabular}{|c|c|c|c|c|c|c|}
\hline 
\textit{\textcolor{black}{\small{}Process}} & \textit{\textcolor{black}{\small{}Trigger}} & \textit{\textcolor{black}{\small{}Trigger}} & \textit{\textcolor{black}{\small{}Trigger}} & \textit{\textcolor{black}{\small{}Reconstruction}} & \textit{\textcolor{black}{\small{}Analysis}} & \textbf{\textcolor{black}{\small{}Total}}\tabularnewline
 & \textcolor{black}{\small{}L0} & \textcolor{black}{\small{}L1} & \textcolor{black}{\small{}L2} &  &  & \tabularnewline
\hline 
\hline 
$\eta\rightarrow\gamma\mu^{+}\mu^{-}$  & \textcolor{black}{\small{}80.6\%} & \textcolor{black}{\small{}64.4\%} & \textcolor{black}{\small{}94.3\%} & \textcolor{black}{\small{}94.2\%} & \textcolor{black}{\small{}98.6\%} & 45.6\%\tabularnewline
\hline 
\textcolor{black}{\small{}Urqmd } & \textcolor{black}{\small{}$21.7\%$} & \textcolor{black}{\small{}$1.7\%$} & \textcolor{black}{\small{}$22.2\%$} & \textcolor{black}{\small{}0.014\%} & \textcolor{black}{\small{}47.3\%} & 4.7$\times10^{-6}$\tabularnewline
\hline 
\end{tabular}\caption{Reconstruction efficiencies for $\eta\rightarrow\gamma\mu^{+}\mu^{-}$
and for the Urqmd generated backgrounds}
\label{table:recoeff_eta2gammamumu-ff}
\end{table}

\begin{figure}[!ht]
\includegraphics[width=7cm,height=6cm]{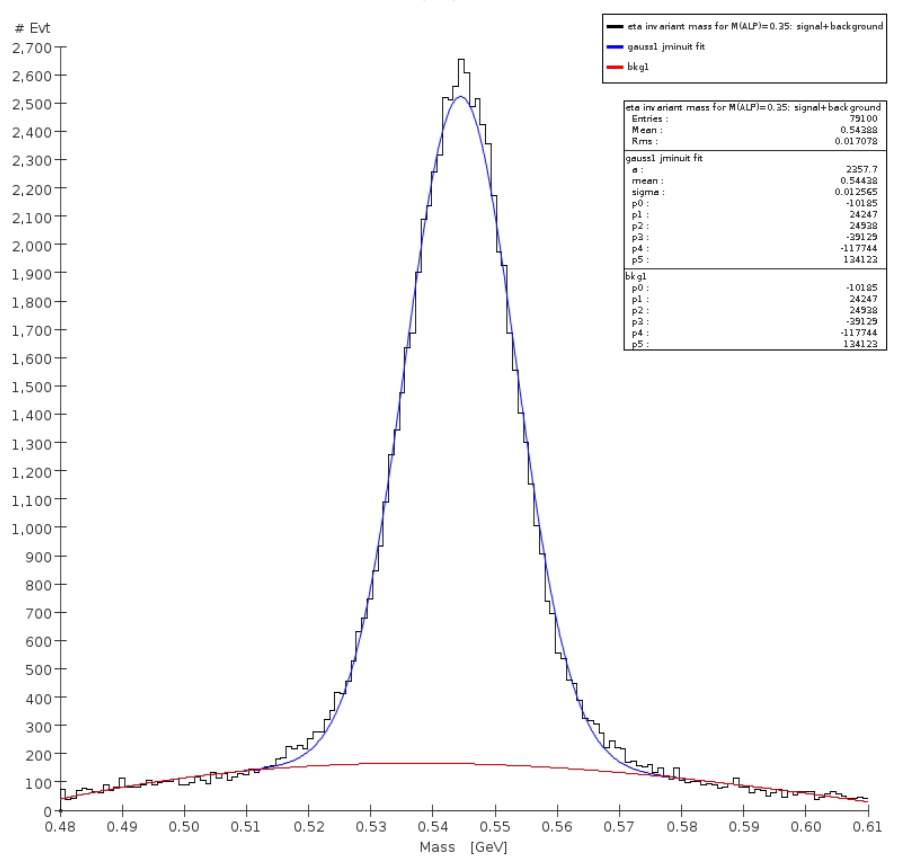} \includegraphics[width=7cm,height=6cm]{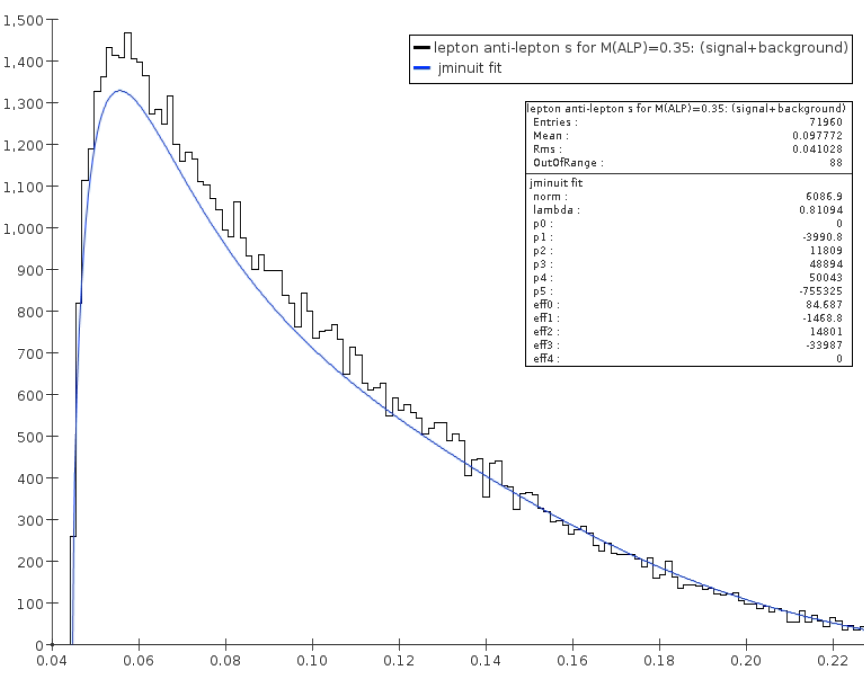} 

\caption{Invariant mass for $\eta\rightarrow\gamma\mu^{+}\mu^{-}$ and for
the Urqmd generated background (left) and fit to (s=$\mid P_{\mu^{+}}+P_{\mu^{-}}\mid^{2}$)
(right) for the set of events corresponding to $\varLambda=800\,MeV$.
See text for an explanation of the fitting procedure.}

\label{fig:formfactors_eta_mass}
\end{figure}

Fig.~\ref{fig:formfactors_eta_mass} shows the invariant mass distribution
of the reconstructed particles, fitted with a Gaussian and a 5th-order
polynomial. 

The sensitivity for the $\varLambda$ parameter is estimated with
two different approaches: a) by measuring the branching ratio
; b) by directly fitting the $\frac{d\varGamma}{ds}$ distribution.

\paragraph{Branching ratio method.}

In this approach, the uncertainty on the $\varLambda$ parameter is obtained
from the fact that: $BR(\eta\rightarrow\gamma\mu^{+}\mu^{-})\propto\mid\tilde{F}_{\eta\gamma^{*}\gamma}(s)\mid^{2}$ and  $\varLambda$  controls the value of $BR(\eta\rightarrow\gamma\mu^{+}\mu^{-})$ via the of  $\mid\tilde{F}_{\eta\gamma^{*}\gamma}(s)\mid^{2}$ from it.
Therefore, the sensitivity on $\mid\tilde{F}_{\eta\gamma^{*}\gamma}(s)\mid^{2}$
is derived by the uncertainty on the measurement of the branching ratio.

\begin{figure}[!ht]
\includegraphics[width=7cm]{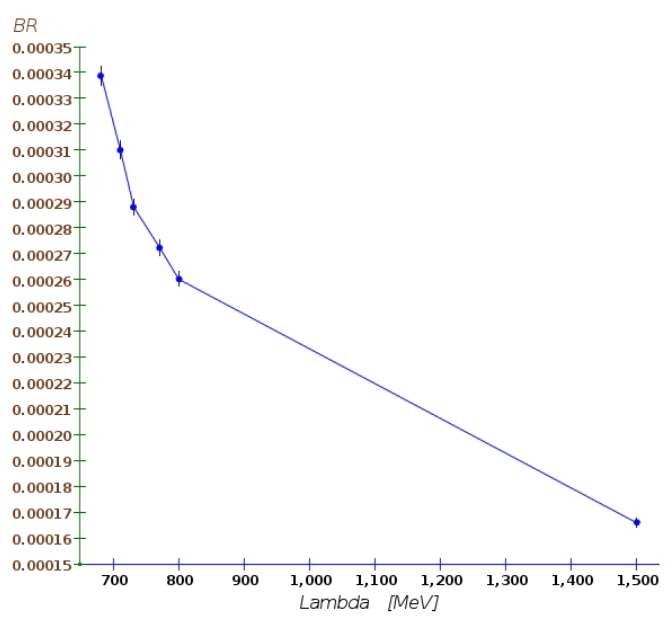} \includegraphics[width=7cm]{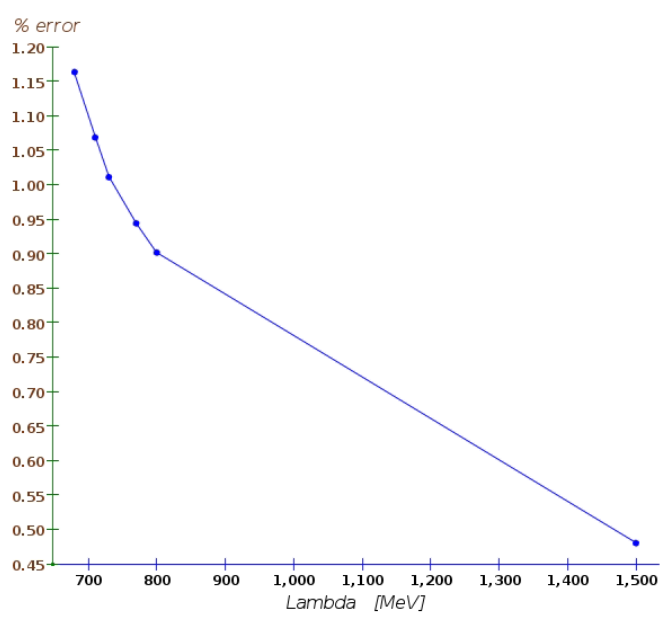} 

\caption{Branching ratio for the $\eta\rightarrow\gamma\mu^{+}\mu^{-}$ process
(left) and statistical error on the $\varLambda$ parameter obtained
with the branching ratio method (see text for details).  The curves are for 3.3$\times$10$^{15}$ POT}

\label{fig:formfactors_br_lambda}
\end{figure}

The branching ratios and the corresponding statistical errors are
obtained from the value of the reconstruction efficiency from the last column
of Table~\ref{table:recoeff_eta2gammamumu-ff} and from the fit to
the $\gamma\mu^{+}\mu^{-}$invariant mass, as described above. The branching
ratios are renormalized to the PDG value and to the central value
of the $\varLambda$ parameter measured by the by A2 Coll.~\cite{A2:2013wad}. 
The results obtained are shown in Fig.~\ref{fig:formfactors_br_lambda} for
the branching ratios (left) and for the statistical error on $\varLambda$
parameter (right) for 3.3$\times$10$^{15}$ POT, corresponding $\sim$10$^{-5}$
of the integrated luminosity foreseen for REDTOP.

\paragraph{Line shape fitting method.}

In this Approach, the uncertainty on the determination of the parameter $\varLambda$ is obtained
from a fit of the invariant mass squared of the di-muon system: $s$=$\mid P_{\mu^{+}}+P_{\mu^{-}}\mid^{2}$.
The fitting function is obtained by combining the amplitude in Eq.
\ref{eq:matrixelement_ff} with a three-body phase space distribution.
The background contribution is described with a polynomial function,
whose parameters are varied in the fit. The reconstruction efficiency
is expected to be non-uniform over the rang of kinematically allowed
values of \emph{s}, especially near near threshold. This effect is taken
into account by including a multiplicative polynomial function in
the fit. The results of the fit for the event set corresponding to
$\varLambda=800\,MeV$ is shown in the right plot of Fig.~\ref{fig:formfactors_br_lambda}.
We found that the determination of the parameter $\varLambda$ by
the line shape fitting method is  inferior to the branching
ratio method, due to the weak dependence of the fitting function on
$\varLambda$. 

In conclusion, the statistical uncertainty estimated with the
line shape fitting method is about one order of magnitude worse than the branching
ratio method. 

\subsubsection{Concluding remark}

We have examined REDTOP sensitivity to the determination of
the $\varLambda$ parameter used to describe the form factor of the
$\eta$ mesons. Two analyses where performed on a data sample corresponding
$\sim$10$^{-5}$ of the total expected statistics. The statistical
uncertainty on the measure of $\varLambda$  for a sample corresponding to 3.3$\times$10$^{15}$ POT was found to be $\sim1\%$
with the first method and one order of magnitude worse with the second
method. This corresponds to an improvement of  a factor $\sim8\times$ 
compared to the measurement
by the by A2 Coll.~\cite{A2:2013wad}.
When the statistics of the full event sample is taken into account,
the projected statistical error on the measurement of the $\varLambda$
parameter is expected to be of the order of $10^{-4}$, presumably
much smaller compared to the systematic error.

\section{Discussion of the results\label{subsec:Discussion-of-the-results}}

The studies presented in this manuscript suggest that,
with a integrated luminosity of 3.3$\times 10^{18}$ POT,  REDTOP has 
a very large potential 
for discovering New Physics by exploiting several rare processes.
In particular, the branching ratio sensitivity for all
decays of the $\eta$ meson where leptons are present in the final state, is in the range $\thickapprox10^{-9}-10^{-8}$. 
This follows from how the experiment as been optimized, in particular for what the trigger system is concerned,  having in mind a large  rejection  of the  hadronic
background in favor of the lighter and faster leptons.
The performance of the detector, along with an unprecedented   sample of $\eta$ mesons,
allow REDTOP to probe all four portals connecting the Standard Model with
the Dark Sector. 

Since, we do not know a priory if a new particle.
With a so called $"bump-hunt"$ analysis,  the background is  rejected with
purely kinematics   considerations. The level
of rejection obtained with this simplified approach is already sufficient to 
reach a sensitivity for exploring uncharted regions of the 
parameter space in many theoretical models. 
That is also an indication the detector requirements for REDTOP have 
been chosen wisely (see, for example, the discussion 
in  Sec.~\ref{subsec:Detector-requirements}. 

The second technique, the $\lyxmathsym{\textquotedblleft}detached-vertex\lyxmathsym{\textquotedblright}$ analysis, requires the presence of two reconstructed charged tracks with a common vertex, displaced from the primary vertex  of the event.
No Standard Model process is able of generating 
a similar event topology at the energies of interest for REDTOP.
Consequently, the background
rejection, is  much more effective than for the $bump-hunt$ analysis, when long-lived particles are searched.
For this technique to be successful, the target is split into thin foils, separated by enough distance to allow the long-lived particle, and its decay products, to leave undisturbed. A vertex detector with good position resolution is also an important ingredient.
The sensitivity improvement obtained with
the $\lyxmathsym{\textquotedblleft}detached\:vertex\lyxmathsym{\textquotedblright}$
analysis is  very good, but it could be improved further with a more performing vertex detector.
For this reason,  the Fiber Tracker option 
will probably be dropped in favor of the Wafer-scale silicon sensor technology (cf.\ Sec.
\ref{par:Vertex-detector-option-II}).

\medskip{}

Besides the exploration of new particles and fields, the studies performed  also indicate 
that REDTOP has a an excellent sensitivity to probe the conservation of discrete symmetries of the nature. 
This is achieved by studying the asymmetry in the distribution 
of certain  observables related to the $\eta$ decays.
We have demonstrated that, for nearly all  processes considered in this work, the statistical uncertainty on those asymmetries will be  much smaller than the typical systematic error. 
Therefore, it will be of paramount importance to have an excellent knowledge of the detector response, which is a key ingredient in reducing the systematic uncertainties. 
The very large event sample collected by REDTOP will help in that respect.

\section{Conclusions}

The $\eta$ and $\eta^{\prime}$ mesons are almost unique
in the particle universe. Carrying the same quantum number as the
Higgs boson (except for parity), and no Standard Model charges, they
represent an excellent laboratory for studying rare processes and
new physics in the MeV--GeV energy range. The present world sample is too small to test violations of conservation
laws or support searches for new particles that  couple weakly to the
Standard Model. 
A new $\eta/\eta$\textquoteright -factory is being
proposed: REDTOP. The goal is to produce $\mathcal{O}(10^{14})$ 
 $\eta$ mesons and $\mathcal{O}(10^{12})$ $\eta^{\prime}$ in a
few years of running. 
The sensitivity studies presented in this
work indicate that, with such a large data sample and with a detector
with REDTOP characteristics, the potential for discovering New Physics 
is very high for all four portals presently considered as links between dark matter and the Standard Model. 
Among 
existing and foreseen experiments, only SHiP at CERN has a similarly 
broad spectrum of research, although at a projected cost four times
larger.\medskip{}

As discussed in detail in Sec.~\ref{sec:Sensitivity-Studies-to-physics-BSM},
several outstanding anomalies could be investigated at REDTOP. The
detector has excellent sensitivity for bosons decaying into\textcolor{black}{\small{}
$e^{+}e^{-}$} pairs with invariant masses below 17 MeV. Several theoretical
models, across the Vector, Pseudoscalar, and Heavy Neutral Lepton portals,
predict the existence such a particle. 
Full simulations has demonstrated the excellent  sensitivity REDTOP to such models. 
One of those models~\cite{Spier2021},
assumes that such a boson is, in fact, the long-searched Peccei-Quinn
axion. 
We have demonstrated that REDTOP could  discover such a 
particle, with sensitivity to  
a  broad range
of values of model parameters.

The anomalies observed by  LSND and Miniboone  could
be explained by theoretical models with a new scalar particle and/or
heavy neutral leptons. Two such models have been studied in this
work: in both cases, the statistics and the performance
of REDTOP appear to be sufficient for the discovery
of such a particle. 
One of the models also explains the observed muon $g-2$ anomaly. 
In that context, we, also, presented
sensitivity studies on the form-factors of the $\eta$ meson. The precision
of that measurement is crucial for understanding $g-2$. Previous
measurements on $\eta$ form-factors were based on samples 
$10^{7}$ smaller than that foreseen for REDTOP, 
using detectors
with lower performance.

\medskip{}

Many sensitivity studies presented in this work
are related to the exploration of conservation laws in  nature.
$CP-$violation is one of the most important topics in particle physics, since it allows 
a model-independent search for new processes beyond the standard model. $CP$ symmetry
can be explored at REDTOP in a multitude of different ways, thanks
to large statistics and a well
known background. 
In all cases, $CP$-violation is observed via asymmetries
of observables related to the decays of the $\eta$ meson. An almost unique technique proposed in this
work is based on the measurement of the polarization of the muon,
copiously produced in several decays of the $\eta$. The muon is  tagged
by fully reconstructing its  decay chain. This could be achieved
either in the high-granularity, polarization-conserving, ADRIANO2 calorimeter
or in a dedicated polarimeter. Furthermore, the proposed technique opens
the road to a broader family of polarization measurements which will
be discussed in a future work.

Recent LHCb results have highlighted the potential of 
Lepton Universality measurements to find new physics. 
REDTOP is well suited to explore the possible implications of those measurements in light-quark currents, if the underlying new physics is light. To this end, REDTOP exploits the abundant semileptonic
decays of the $\eta$ meson and can, in principle, probe Lepton Universality with unprecedented precision. A dedicated discussion has been presented in this work, making the case for a sensitivity study based on a global analysis where light new physics is modeled in a general way.

This work also examined the related subject of the violation of lepton flavor in $\eta$ decays. 
REDTOP cannot, on this specific ground, compete with dedicated experiments based on muon beams. However the different mechanism of production of the muons might make this study interesting. 
\medskip{}

Although not discussed in this work, the composition of the $\eta$ and $\eta^{\prime}$
mesons has never been fully understood.
Prior experiments indicate
that only about 80\% of the $\eta$ and $\eta^{\prime}$ constituents is represented by quark
and gluons. The energy and momentum of the non-standard component
indicate that this has, in fact, a  non-zero mass. Unfortunately, as of this writing, there are
no theoretical models capable of  explaining
such measurements. 
The discrepancy could be associated, in fact,
to new particles or forces. REDTOP will try to address those aspects of  $\eta$  physics in future  studies.

\medskip{}

An important aspect of REDTOP is related to the detector technologies
chosen for the experiment. 
With an event rate of 700 kHz, a factor
almost 18 times larger than LHCb, REDTOP is the, collider-style experiment,
with the fastest Level-0 trigger ever considered. In spite of the
relatively low momentum of the charged tracks to be detected,
a realm once restricted to low-mass gaseous detectors, such
an event rate requires that the tracking system be based on solid state technologies.
Still, the material budgets needs to be appreciably smaller and more
radiation resistant than existing detectors. 
On the other side,
the, recently approved, EIC project has very similar requirements 
for its tracking systems, and the detector technologies explored are similar to those proposed for REDTOP.
Synergies and common activities
between the two programs are being discussed as of this writing.

New Physics is being searched for at REDTOP predominantly in
association with leptons. In a hadron-produced environment, particle
identification is, therefore of crucial importance. A relatively novel
 dual-readout technique, has been identified to push lepton-hadron
separation to a new level. The ADRIANO2 technology, currently
under development, adds time-of-flight capabilities to dual-readout,
for a full 5-D reconstruction of the showers. The ongoing R\&D on all such techniques, while crucial for the success
of REDTOP, will benefit several future experiments in high energy
and nuclear science.

\medskip{}

There is no doubt that REDTOP is a challenging experiment, requiring
at the same time, rapid response, high precision and radiation hardness.
On the other hand, the requirements on the accelerator are very tame, requiring only 30-60 W of proton beam. 
Several laboratories  have expressed  interest in
hosting the experiment.

Due to its relatively small size, the projected cost of the detector is only  \$M50-80. 
The  low cost required to adapt an existing accelerator 
and the availability of unused superconducting solenoid help to control the cost. 
The physics
case for a super-$\eta$/$\eta^{\prime}$ factory is very strong and well
supported by the theoretical community. 
The cost is well  justified by REDTOP's broad scope for exploration of new physics, making this experiment
very attractive. 

\medskip{}
When future low-energy, MegaWatt class accelerator facilities will
become available (PIPII, ESS, CSNS), REDTOP could be eventually upgraded
to a tagged $\eta$-factory experiment, with a similar $\eta$-meson
yield but with the obvious advantage of $\eta$-tagging. Therefore
REDTOP should be considered more like an exploration facility than
a single experiment, offering the opportunity of a precise observatory
on New Physics and a multi-year research program.

\begin{acknowledgments}
This paper and the research behind it would not have been possible
without the exceptional computing contribution provided by the Collaboration Support at the OSG
Fabric of Services~\cite{osg07,osg09}, the computing cluster at the
Northern Illinois Center for Accelerator and Detector Development
(NICADD) and the Northern Illinois Center for Research Computing and
Data (CRCD). We are grateful to the National Science Foundation for
supporting the OSG consortium under the Partnership to Advance Throughput Computing award \#2030508. 
 This document was prepared by the REDTOP Collaboration using the resources of the Fermi National Accelerator Laboratory (Fermilab), a U.S. Department of Energy, Office of Science, HEP User Facility. Fermilab is managed by Fermi Research Alliance, LLC (FRA), acting under Contract No. DE-AC02-07CH11359.
\end{acknowledgments}

\newpage

\section{\label{sec:The-Acceleration-Scheme}Appendix I: The Acceleration
Scheme}

Although the beam requirements for REDTOP are modest, none of the
existing HEP laboratories worldwide as a ready-to-go accelerator satisfying
all the conditions. The Collaboration has engaged in a broad exploration
and several possible hosting laboratories have been identified.

\subsection{Fermilab configuration.}

The low energy proton beam available at Fermilab has a pulsed structure
with 84 bunches and a fixed energy of 8 GeV. The buckets are distributed
to the accelerator complex and either accelerated or dumped in order
to create a secondary beam. On the other hand, REDTOP requires a Continuous
Wave (CW) proton beam with a user selectable energy, with a range
considerably lower than that available (cf.\ Sec.~\ref{sec:The-Experimental-Technique}).
The accelerator scheme\citep{Decel_2016} proposed for REDTOP foresee
the extraction of a single pulse from the booster (consisting of $\sim$4$\times$10$^{12}$
protons) which is subsequently injected in the Delivery Ring (former
debuncher in anti-proton production at Tevatron). The energy is, then
removed from the beam by operating the RF cavities in reverse, until
it reaches the required value. The time required for reaching 1.8
Gev is $\sim$5 seconds (about 3.3 for the 3.5 GeV case). Once the
desired energy is reached, the beam is kept circulating inside the
ring where the buckets relax adiabatically. Slow extraction to REDTOP
(located in the nearby AP50 hall) occur over $\sim$40 seconds. The
proposed accelerator scheme requires minimal changes to the existing
complex and involves no extra beam elements. In fact, the extra RF
cavity necessary to stabilize the energy removal process is already
available as a spare for the Mu2e experiment. Furthermore, the extraction
point to AP50 corresponds to a betatron phase advance of 270$^{o}$
compared to the location of the existing Mu2e electrostatic septum.
Consequently, the latter can be used also to shift the protons for
the final extraction via a Lambertson magnet. The total time to decelerate-debunch-extract
the beam is 51 sec, corresponding to a duty cycle $\sim$80\%.

With the above scheme, large beam losses in the Delivery Ring will
occur if beam is decelerated from injection at 8 GeV (corresponding
to $\gamma$=9.53) to <2 GeV ($\gamma$ = 3.13) through the natural
transition energy of the ring: $\gamma{}_{t}$ = 7.64. Transition
can, however, be avoided by using select quadrupole triplets to boost
$\gamma{}_{t}$ above beam $\gamma$ by 0.5 units throughout deceleration
until $\gamma{}_{t}$ = 7.64 and beam $\gamma$ = 7.14 (corresponding
to E$_{kin}$=5.76 GeV ). Below 5.76 GeV the Delivery Ring lattice
is reverted to the nominal design configuration. Optical perturbations
are localized within each triplet while the straight sections are
unaffected thereby keeping the nominal M3 injection beamline tune
valid\citep{Decel_2016}.

Achieving the full set of goals of the experiment will require production
and processing of at least $10^{13}$ $\eta$ meson decays. The goals
can be met by having a flux of $10^{11}$ protons per second on target,
and a duty factor of over 75\%. The Fermilab Booster routinely delivers
$4\times10^{12}$ protons per cycle at up to 15-Hz repetition rate.
Thus, there should be an abundance of protons to meet the REDTOP goals.
However, the Booster extraction kinetic energy is 8~GeV, well above
the $\sim$2~GeV and 3~GeV needed for REDTOP to produce, respectively,
$\eta$ and $\eta^{\prime}$ mesons. A deceleration stage will be implemented.

\subsection{CERN configuration.}

A preliminary study has been performed at CERN to provide a beam with
REDTOP's requirements. Several possible schemes have been considered,
the most promising corresponding to the extraction of a 1.8 GeV beam
from the ProtoSynchrotron (PS) which could then be delivered to the
East Hall, where the experiment could be housed. For the moment a
24 GeV/c proton beam is routinely slow-extracted into the CHARM and
IRRAD facilities along the T8 beam line with a maximum intensity of
6.5$\times$10$^{11}$ protons per extraction over 0.4 seconds. REDTOP
would require a much longer flat top of e.g. 10 seconds at $\sim$1.8
GeV kinetic energy (as it can be delivered directly from the PS Booster)
in a cycle of 9 basic periods(10.8 s). No show-stoppers have been
identified up to date, although more studies are required. However,
the PS the duty cycle cannot be much higher than 50\%, which would
already have a significant impact on the existing CERN physics program.
Consequently, the expected luminosity available to REDTOP at CERN
is expected to be $\sim\nicefrac{1}{10}$ of that available at Fermilab.

\subsection{BNL configuration.}

AT BNL the REDTOP detector could sit comfortably at the end of the
existing C4 extraction line. Logistics would be optimal, since the
experimental hall is well accessible and serviced. The beam would
be slowly extracted from the AGS using well known techniques. The
instrumentation electronics of the extraction line is no longer functional
and it needs to be refurbished at a very modest cost estimated of order ($\mathcal{O}$(100K\$)).
Altogether, BNL is an excellent candidate for hosting REDTOP.

\subsection{HIAF configuration.}
The REDTOP detector can be installed at the Multi-function terminal of the HIAF
accelerator. The beam extracted from the Booster Ring ($BRing$) fullfil all requirements for REDTOP
experiment (cf.\ Sec.~\ref{subsec:Beam-and-target}). The transfer beam line construction is already included in the HIAF
project. The maximum magnetic rigidity is 34 Tm, corresponding to a maximum  proton beam energy of 9.3 GeV. The facility is expected to be operational around 2025 for a run above the $\eta^{\prime}$ production threshold with a beam density equal or exceeding 1$\times$10$^{13}$ protons-per-spill.

\subsection{GSI configuration.}

As a possible alternative, the use of the SIS18 beam at GSI/FAIR could be considered, since the necessary energies and intensities could be provided by the existing accelerators. An assessment of the feasibility of the experiment in this laboratory will be started after the submission of this White Paper. 

\section{Appendix II: Tagged \texorpdfstring{$\eta$}{}-factory}

An upgraded version of the REDTOP, t-REDTOP~\cite{Arrington:2022pon}, could be run in a later
stage of the experiment at the PIP-II facility, currently under construction
at Fermilab. At the 800 MeV, high intensity (100KW-1MW), CW proton
beam, the production mechanism of the $\eta$-meson is substantially
different. The nuclear process providing the $\eta$-meson would be:

\begin{equation}
p+De\rightarrow\eta+^{3}He^{+}\label{eq:tagged_eta}
\end{equation}

Therefore, t-REDTOP requires a gaseous Deuterium target and an extra
detector to tag the $^{3}He^{+}$ ion. The production cross section
for the process (\ref{eq:tagged_eta}) is approximately five orders
of magnitude smaller than at 1.8 GeV and the gaseous target reduces
further the luminosity of the beam. However, the large intensity of
PIP-II more than compensate for that and the number of $\eta-mesons$
produced at t-REDTOP is expected to be in the range {[}$10^{13}/year-10^{14}/year${]}.
The biggest advantages of a tagged $\eta$-factory are the following:
\begin{itemize}
\item By tagging the production of the $\eta$-meson via the detection of
the $^{3}He^{+}$ ion, the combinatorics background from $non-\eta$
events, is greatly diminished. Consequently, the sensitivity of the
experiment to New Physics is increased by a factor proportional to
the square root of that reduction;
\item By measuring the momentum of the $^{3}He^{+}$ ion in process (\ref{eq:tagged_eta}),
the kinematics of the reaction is fully closed. That portends to a
better measurements of the kinematics of the particles detected, since
a 4-C kinematic fit could be applied;
\item Since the kinematics is closed, any long lived, dark particle escaping
detection could be identified using the \emph{missing 4-momentum}
technique. The latter is considerably more powerful than the, 1-,
missing p$_{t}$ or missing energy proposed by some recent experiments
searching for dark matter.
\end{itemize}
The disadvantages of a tagged-$\eta$ experiment are due mostly to
the larger complexity of the experiment, requiring an extra detector
at very small angles and to the necessity to control the halo of the
beam at a much higher level. Also, the mass of New Physics explored
will be lower, since no $\eta^{\prime}$-meson could be produced at the PIP-II
design energy.

\section{Appendix III: Radiation Damage and Detector Aging\label{sec:Appendix-III:-Radiation-Damage}}

The intense proton beam required for reaching REDTOP's physics goal
is also expected to generate a radiation halo that will age and could
potentially damage the detector. Descriptions of the detector components
are contained in Sec.~\ref{subsec:The-REDTOP-detector}.

\begin{figure}[!ht]
\begin{centering}
\includegraphics[scale=0.3]{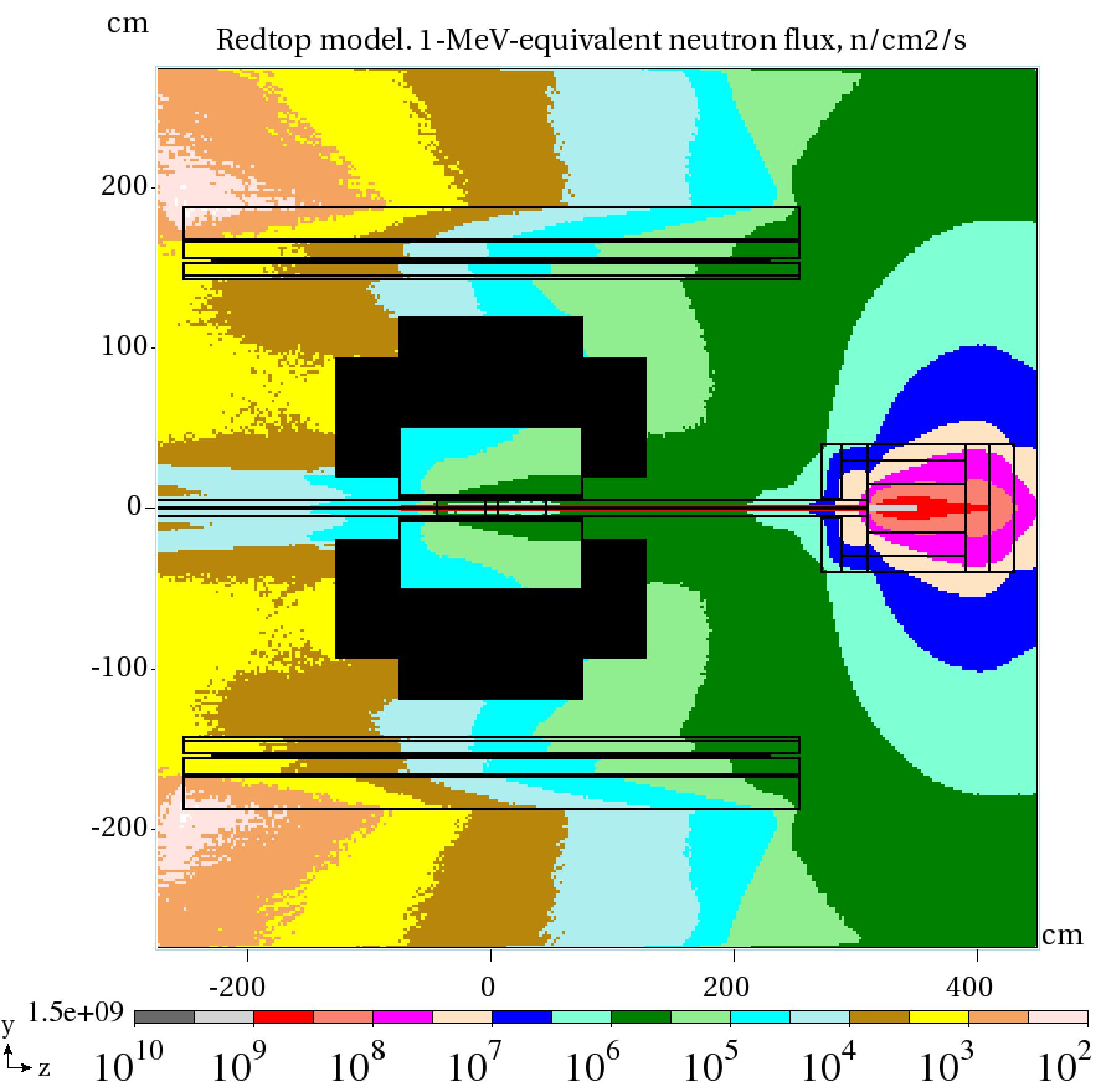} 
\par\end{centering}
\caption{``\textit{1-MeV equivalent neutron fluence}'' dose expected in the
REDTOP detector for a 30-W proton beam.}
\label{fig:MARS15_full} 
\end{figure}

An analysis of the radiation flux expected throughout the detector
has been performed with the MARS15 code~\cite{MARS15} and a geometrical
model that reflects our baseline detector layout, along with an aluminum-Borated
Polyethylene-barite beam dump. The result of a study for a 30-W proton
beam with 1.8-GeV energy impinging onto the beryllium target systems
is shown in Fig.~\ref{fig:MARS15_full}. The plot represents the
so-called ``\textit{1-MeV equivalent neutron fluence}'' (NEQF) dose,
which is commonly used to estimate the damage inflicted by radiation
to photo-sensors.

The areas of concern are the photo-sensors around the vessel of the
OTPC and the area around the beam pipe that is planned to be instrumented
with a scintillating fiber tracker (cf.\ Sec.~\ref{subsec:The-REDTOP-detector}).
The maximum NEQF flux
illuminating the photo-sensors is located on the forward endplate
of the of OTPC and is expected to be about $6\times10^{5}$/cm$^{2}$/s
(or $6\times10^{12}$/cm$^{2}$ integrated over 1 year).
The rate for the barrel is, in
average, a factor of 5-6 lower.
The currently recognized integrated limit for commercial SiPMs, the
current baseline technology adopted for the OTPC, corresponds to about
10$^{13}$/cm$^{2}$ NEQF. The value is uncomfortably close to the
estimated flux in the end plates. A consequence is an increase in
the dark count rate (DCR) of the sensors (typically, 50~kHz/mm$^{2}$
or 250 OTPC hits/evt for a 1-ns integration time). While the effect
of such noise is expected to be minimal on the reconstruction of \u{C}erenkov
rings, a bigger issue would come from the saturation of the online
systems which could compromise the data transfer of the compressed
data from the detector to the L0 trigger systems. We are aware of
the issue and plan to investigate it further. Possible roads we will
consider to mitigate an increased DCR are: 
\begin{itemize}
\item A larger bandwidth between the FEE and the L0 trigger systems; 
\item Cooling the SiPM's; 
\item Narrowing the integration time; 
\item Investigate new generation sensors (e.g. addition of trenches between
pixels to reduce cross talk); 
\item Using Multichannel plates for the forward endplate and SiPMs for the
remainder of the detector; 
\item Consider LAPPD instead of Sipm's, although the formers are not yet
commercially available. 
\end{itemize}
In addition to these items, the LHC Community is very active in this
area and we expect to benefit greatly from the R\&D performed by them.

\begin{figure}[!ht]
\begin{centering}
\includegraphics[width=0.7\textwidth]{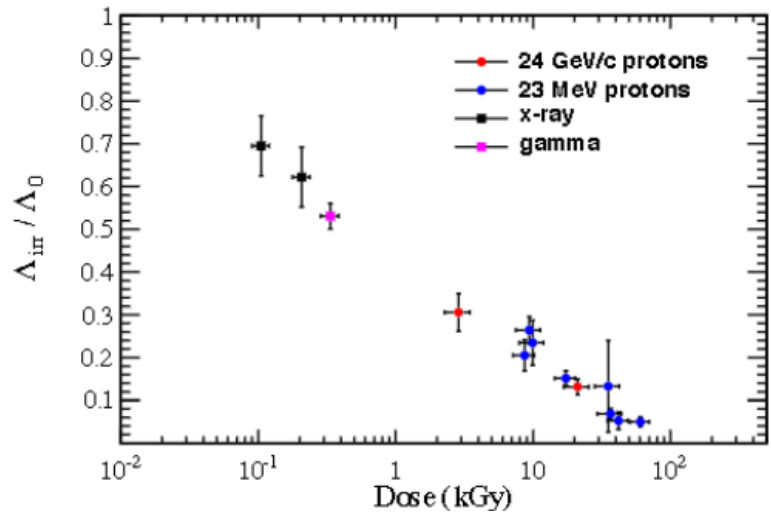} 
\par\end{centering}
\caption{Change in attenuation length of a KURARAY SCSF-78 fiber (\diameter 250~$\mu$m)
vs.\ integrated dose.}
\label{fig:fibertrackerdamage} 
\end{figure}

The area near the beam pipe, where the fiber tracker will be located,
is currently under study. 
Recent studies by LHCb~\cite{Joram15} where fibers have been tested
up to 60~kGy, is summarized in Fig.~\ref{fig:fibertrackerdamage},
where the change in attenuation length is plotted vs.\ the radiation
dose. In the case of REDTOP, we expect that the fiber tracker would
degrade considerably during a 1-year run, with the attenuation length
of a typical KURARAY SCSF-78 fiber (\diameter 250~$\mu$m), diminishing
from about 4~m to 10 ~cm. We are aware of the issues and plan to
investigate them further.

\section{Appendix IV: Radiation Shielding and Safety}

The experimental hall proposed for REDTOP, AP50, is part of Fermilab's
``Muon Campus'' where two experiments are or will be running (Muon
g-2 and Mu2e). All areas related to the Muon Campus are under close
scrutiny by the Fermilab Safety Dept., and a full radiation shielding
assessment has been recently completed and approved by the ESHQ Office.
According to the latter, a maximum beam intensity of $3.60\times10^{13}$
protons/hr at 8~GeV can be circulated and delivered in the accelerator
complex where AP50 is located, with the current radiation shielding,
while 2.20x10$^{16}$ protons/hr can be circulated with the modification
foreseen for Mu2e. The latter rate corresponds to a maximum beam power
circulated and dumped of 8~kW, to be compared with the 30~W proposed
to run REDTOP as an $\eta$-factory, and the 50~W proposed for a
later run as an $\eta^{\prime}$-factory. In both cases, the proposed beam
intensity is considerably smaller than the maximum allowed and it
requires no refurbishing of the existing radiation shielding.

\bibliographystyle{apsrev}
\bibliography{BibTeX/iopart-num/WhitePaper2021}

\end{document}